\documentclass[10pt,a4paper]{article}
\newif\ifquick
\quickfalse
\newif\ifauthors
\authorstrue
\usepackage[margin=20mm]{geometry}
\usepackage{iftex}
\ifPDFTeX
  \usepackage[T1]{fontenc}
  \usepackage[utf8]{inputenc}
  \usepackage{textcomp}
  \IfFileExists{tgtermes.sty}{\usepackage{tgtermes}}{}
  \IfFileExists{tgheros.sty}{\usepackage{tgheros}}{\usepackage[scaled=.92]{helvet}}
\else
  \usepackage{fontspec}
\fi
\usepackage{amsmath,amssymb}
\usepackage{graphicx}
\usepackage{tikz}
\usetikzlibrary{arrows.meta,positioning,calc}
\newcommand{\figurelabel}{\rmfamily\fontsize{9}{11}\selectfont}

\definecolor{figureblue}{HTML}{33769F}
\definecolor{figurebluefill}{HTML}{F3F8FB}
\definecolor{figuregreen}{HTML}{537741}
\definecolor{figuregreenfill}{HTML}{F3F7EF}
\tikzset{surveyfigure/.style={font=\figurelabel,line width=.5pt},
 surveyarrow/.style={->,line width=.65pt,draw=figureblue}}
\usepackage{booktabs,longtable,tabularx,array,ragged2e}
\usepackage{caption,float,enumitem}
\usepackage{needspace}
\usepackage[section]{placeins} 
\usepackage[breakable]{tcolorbox} 
\usepackage[numbers,sort&compress]{natbib}

\usepackage{xurl}
\usepackage{authblk}

\renewcommand\Affilfont{\small}

\makeatletter
\renewcommand\AB@affilsepx{\hspace{1.5em}\protect\Affilfont}
\makeatother
\ifquick\providecommand{\href}[2]{#2}\else
\usepackage[unicode,colorlinks=true,linkcolor=blue!45!black,citecolor=blue!45!black,urlcolor=blue!45!black]{hyperref}
\usepackage[protrusion=true,expansion=false]{microtype} 
\usepackage{bookmark}
\fi
\ifXeTeX\fi
\tcbset{surveybox/.style={breakable,colback=violet!3,colframe=violet!35!black,boxrule=.4pt,arc=0pt,left=7pt,right=7pt,top=6pt,bottom=6pt}}

\usepackage{titlesec}
\usepackage{fancyhdr}
\definecolor{surveyink}{HTML}{243746}
\definecolor{surveymuted}{HTML}{687078}
\titleformat{\section}{\normalfont\fontsize{13}{16}\selectfont\bfseries}{\thesection}{0.65em}{}
\titleformat{\subsection}{\normalfont\fontsize{11}{13}\selectfont\bfseries}{\thesubsection}{0.65em}{}
\titleformat{\subsubsection}{\normalfont\fontsize{10}{12}\selectfont\bfseries}{\thesubsubsection}{0.65em}{}
\titlespacing*{\section}{0pt}{3.5ex plus 1ex minus .2ex}{2.3ex plus .2ex}
\titlespacing*{\subsection}{0pt}{3.25ex plus 1ex minus .2ex}{1.5ex plus .2ex}
\titlespacing*{\subsubsection}{0pt}{3.25ex plus 1ex minus .2ex}{1.5ex plus .2ex}
\fancypagestyle{plain}{\fancyhf{}\fancyfoot[C]{\fontsize{9}{11}\selectfont\thepage}}
\ifquick\else
\hypersetup{linkcolor=surveyink,citecolor=surveyink,urlcolor=surveyink,
 pdftitle={Physical AI: A Survey of Models, Learning, Systems, and Open Challenges},
 pdfauthor={Bin Gao et al.}}
\fi
\makeatletter
\renewcommand{\@maketitle}{%
  \newpage\null\vskip 1em%
  \begin{center}%
    \let\footnote\thanks%
    {\fontsize{18}{21}\selectfont\bfseries\@title\par}%
    \vskip 1.4em%
    {\normalsize\lineskip .5em\begin{tabular}[t]{c}\@author\end{tabular}\par}%
  \end{center}%
  \par\vskip .5em%
}
\makeatother
\title{Physical AI: \\ A Survey of Models, Learning, Systems, and Open Challenges}
\ifauthors
\author[1]{Bin~Gao}
\author[2]{Anuj~Gautam}
\author[1]{Aritra~Bagchi}
\author[3]{Yunkai~Liang}
\author[3]{Wentao~Li}
\author[3]{Ronghui~Zhu}
\author[3]{Zhi~Zhou}
\author[4]{Yancan~Mao}
\author[5]{Xiao~Yan}
\author[5]{Xin~He}
\author[1]{Qingyun~Zou}
\author[1]{Wenshuo~Yue}
\author[1]{Dan~Chen}
\author[1]{Zhanglu~Yan}
\author[6]{Fei~Xu}
\author[7]{Huize~Li}
\author[8]{Weijie~Wang}
\author[9]{Tao~Ouyang}
\author[1]{Tulika~Mitra}
\affil[1]{\mbox{National University of Singapore, Singapore}}
\affil[2]{\mbox{Indian Institute of Technology Bombay, India}}
\affil[3]{\mbox{Sun Yat-sen University, China}}
\affil[4]{\mbox{Huazhong University of Science and Technology, China}}
\affil[5]{\mbox{Wuhan University, China}}
\affil[6]{\mbox{East China Normal University, China}}
\affil[7]{\mbox{University of Central Florida, USA}}
\affil[8]{\mbox{Fondazione Bruno Kessler, Italy}}
\affil[9]{\mbox{Central South University, China}}
\else
\author{}
\fi
\date{}
\newcommand{\email}[1]{\href{mailto:#1}{\mbox{#1}}}
\begin{document}\maketitle
\ifauthors\begingroup\renewcommand{\thefootnote}{}\footnotetext{\raggedright Email addresses:
\email{bingao.cs@gmail.com} (Bin Gao),
\email{gtmanuj04@gmail.com} (Anuj Gautam),
\email{bagchi.aritra@nus.edu.sg} (Aritra Bagchi),
\email{liangyk7@mail2.sysu.edu.cn} (Yunkai Liang),
\email{liwt89@mail2.sysu.edu.cn} (Wentao Li),
\email{zhurh8@mail2.sysu.edu.cn} (Ronghui Zhu),
\email{zhouzhi9@mail.sysu.edu.cn} (Zhi Zhou),
\email{maoyancan@hust.edu.cn} (Yancan Mao),
\email{yanxiaosunny@whu.edu.cn} (Xiao Yan),
\email{hexin.research@gmail.com} (Xin He),
\email{qingyunzou@u.nus.edu} (Qingyun Zou),
\email{wenshuoyue@nus.edu.sg} (Wenshuo Yue),
\email{danchen@nus.edu.sg} (Dan Chen),
\email{e0385109@u.nus.edu} (Zhanglu Yan),
\email{fxu@cs.ecnu.edu.cn} (Fei Xu),
\email{huize.li@ucf.edu} (Huize Li),
\email{wwang@fbk.eu} (Weijie Wang),
\email{ouyangtao@csu.edu.cn} (Tao Ouyang),
\email{dcstm@nus.edu.sg} (Tulika Mitra).}\endgroup\fi
\begin{abstract}
Physical AI aims to enable intelligent systems to perceive, reason, act, and learn through interaction with the physical world. Unlike purely digital intelligence, physical agents must operate under partial observability, uncertain dynamics, embodiment constraints, and real-time requirements, while their actions continuously reshape the environment and subsequent observations. Recent advances in foundation models and robot learning have expanded the capabilities of physical agents, yet achieving reliable autonomy requires more than improvements in individual models. This survey presents a comprehensive, system-level review of Physical AI, organized around the closed-loop process of physical interaction. We examine the foundations and recent developments in task representations and memory, predictive and action-generating models, data and learning methods, embodiment and control, computing systems, evaluation, and safety. Beyond reviewing individual advances, we investigate how these components interact and collectively shape the capabilities and limitations of Physical AI systems. Building on this integrated perspective, we identify fundamental challenges in generalization, resource-efficient learning and execution, reliable state tracking, safety, recovery, and continual adaptation. We argue that advancing Physical AI requires moving beyond isolated model performance toward integrated systems capable of \textbf{robust, efficient, safe, and autonomous physical interaction} under changing real-world conditions.
\end{abstract}

\section{Introduction}\label{sec:1}

\begin{figure}[!htbp]
\centering
\includegraphics[width=\linewidth]{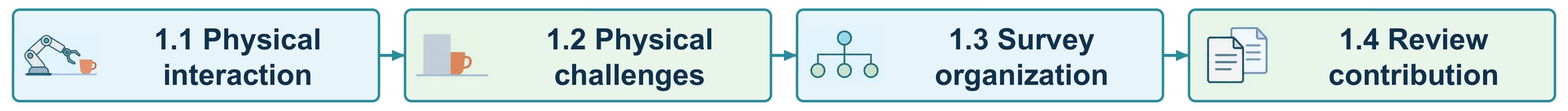}
\caption{Introduction map. Physical interaction motivates the challenges, organization, and contribution of the survey. The four panels follow the introductory subsections; arrows show the progression of the argument.}\label{fig:guide-01}
\end{figure}

The introduction moves from physical interaction (\S~\ref{sec:1.1}) and its challenges (\S~\ref{sec:1.2}) to the survey's scope and organization (\S~\ref{sec:1.3}), then its relationship to existing reviews and contributions (\S~\ref{sec:1.4}). Figure~\ref{fig:guide-01} maps this progression.

\FloatBarrier

\subsection{Physical AI and Physical Interaction}\label{sec:1.1}

Physical AI aims to enable intelligent systems to perceive, reason, act, and learn through interaction with the physical world. Physical agents must operate under partial observability, uncertain dynamics, embodiment constraints, and real-time requirements. Their actions change both the environment and the observations available for subsequent decisions, making physical interaction a closed-loop process. This survey focuses on robotic systems: a robot combines a physical body with the sensing, models, control, and computing needed for physical interaction. The body is its physical structure and mechanisms; the environment comprises the objects, surfaces, and other actors with which it interacts. To clarify these roles, the discussion follows an execution cycle, then distinguishes how pretrained models, feedback, training, and memory support it.

A task specifies a goal and constraints on reaching it. Sensors measure the body and environment, providing observations; representations encode this information for models that select actions. Controllers compute commands from selected actions and feedback, and actuators apply them to the body. New observations provide evidence of the resulting physical effects and inform the next decision. Task completion depends on those effects, including changes in objects and contact, rather than on the production of a plausible description or action sequence.

Pretrained models extend the knowledge available within this process. Language models can select among skills a robot can execute~\citep{arxiv_2204_01691}; vision--language--action (VLA) policies connect visual and linguistic knowledge to robot-action outputs~\citep{arxiv_2307_15818}. Visual pretraining supplies reusable representations~\citep{arxiv_2103_00020,arxiv_2203_12601}, while demonstrations across robots, simulated interaction, and feedback from real execution supply experience for learning physical behavior~\citep{arxiv_2310_08864,arxiv_2108_10470,arxiv_1011_0686,arxiv_2401_16013}. These advances in foundation models and robot learning expand the capabilities of physical agents, but reliable autonomy also depends on how they work with sensing, memory, control, and computing. Recognizing an object, selecting a feasible action, and learning from its outcome remain distinct requirements that must be coordinated through feedback.

\begin{figure}[!htbp]
\centering
\includegraphics[width=\linewidth]{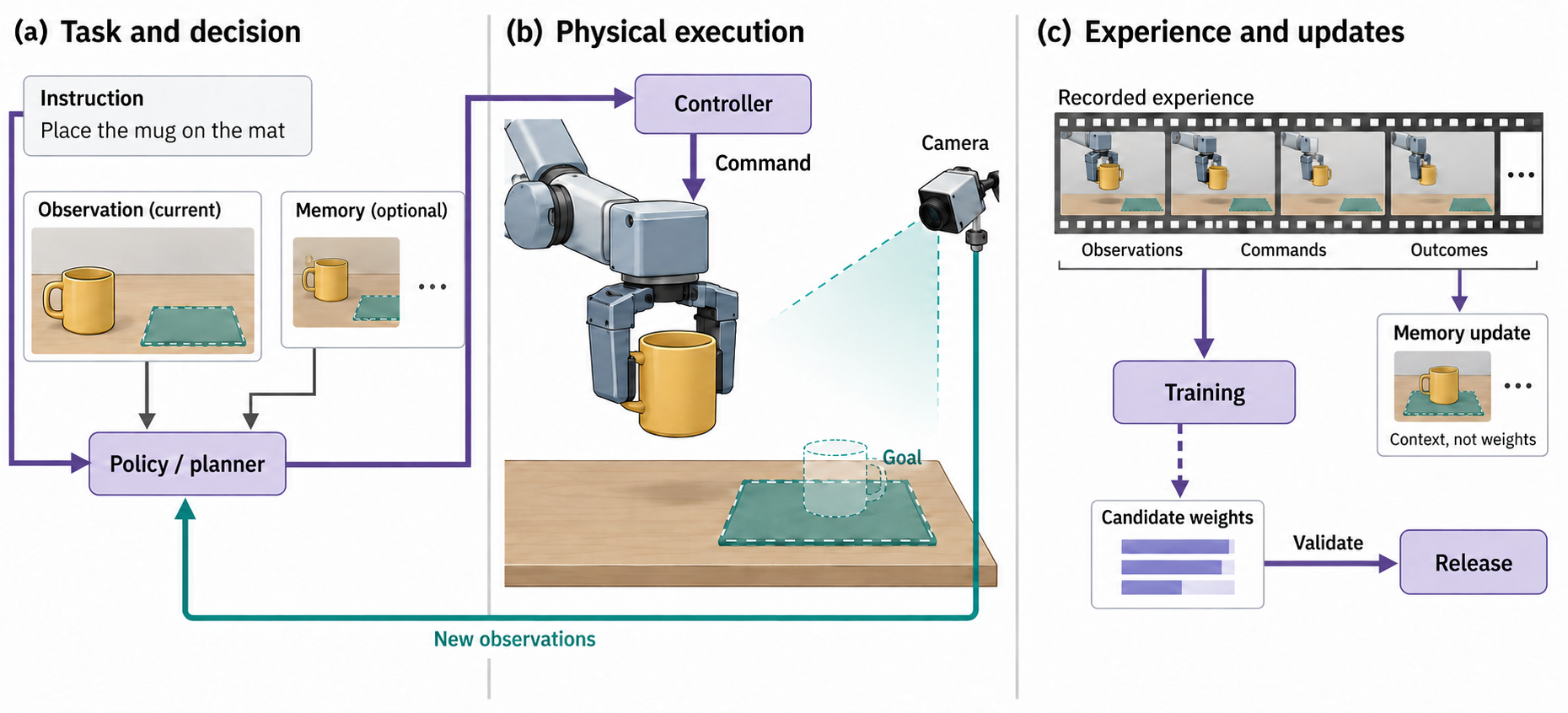}
\caption{Physical interaction and experience updates in a mug-placement example. (a) Instructions, observations, and optional memory inform action selection. (b) A controller executes requested actions; new observations inform subsequent decisions. The outlined mug is a goal, not an observed outcome. (c) Recorded experience can support training and memory maintenance through separate paths. Candidate weights require validation before release, and release does not itself activate a policy on a robot. Memory updates need not change model parameters.}\label{fig:overview-interaction}
\end{figure}

Figure~\ref{fig:overview-interaction} separates execution from model training. During execution, new observations can update estimates of physical state and task progress, and thereby change the next action, without changing model parameters. Model training uses records and supervision to update parameters for later decisions. Memory can also retain records or skills for later reuse without updating model parameters. The functions can share one model or be distributed across learned and engineered components.

\subsection{Challenges of Physical Tasks}\label{sec:1.2}

The interaction cycle creates difficulties in what the robot can know, what it can execute in time, how it acquires experience, and what remains useful when conditions change. The four challenges below separate these sources of difficulty, while the cup-task example shows why a single failure may involve several of them.

\noindent\textbf{Incomplete and Heterogeneous Information.} Cameras, joint encoders, and touch sensors measure different aspects of the body and environment at different rates, with different noise and delay. An image can provide evidence of an object's shape while leaving its mass, friction, or contact state uncertain. Sensor placement and the body's geometry also determine what can be observed. The system must therefore combine measurements with prior knowledge, retain useful history, and distinguish estimated quantities from confirmed task outcomes. Acting can itself provide new evidence: a small push can reveal how an object responds to force, while a reach may occlude it from the camera.

\noindent\textbf{Physical Feasibility and Real-time Response.} A selected action must be realizable by the body's joints, grippers, and support contacts. The body and environment continue to change during sensing and computation, so a geometrically appropriate target can become outdated before execution. The useful response time depends on the physical process: slow task planning and rapid contact correction can coexist, provided that their outputs remain consistent. Executing actions while computing the next decision is one way to accommodate inference latency~\citep{arxiv_2506_07339}, but it also requires deciding when new observations should replace an earlier action sequence.

\noindent\textbf{Costly Experience and Persistent Consequences.} Real attempts consume time, equipment, energy, and often human effort for supervision or resets. A failure may damage an object or leave the robot in a state from which the original plan cannot continue. The difficulty is both to learn from limited attempts and to make further attempts feasible. This requires usable records of observations, action commands, and observed outcomes, together with recovery, safe termination, or assistance when execution fails.

\noindent\textbf{Changes in Tasks, Environments, and Bodies.} A skill learned under one set of conditions may fail when the task, environment, or body changes. Recognizing a new cup does not establish how to grip it, and a motion feasible for one arm may be unreachable for another. Changes can also occur together, making it difficult to identify what must be adapted. Reusing experience therefore requires determining which representations and behaviors transfer, how much additional interaction is needed, and whether adaptation retains earlier capabilities.

Consider a robot instructed to ``find the cup, wash it, and place it at the specified location''; Figure~\ref{fig:overview-interaction} illustrates the placement stage. This hypothetical task provides a running example for distinctions that recur across the survey. Suppose the gripper closes without securing the cup. We denote this hypothetical missed-grasp event by E1; it records an outcome, not its cause. The position estimate may be stale, the grasp may be unsuitable, or execution may be late. Distinguishing these explanations requires evidence from the relevant components. Even successful motions do not establish that washing is complete: cleanliness needs its own observable criterion.

\subsection{Scope and Organization}\label{sec:1.3}

This survey presents a system-level review of Physical AI organized around the closed-loop process of physical interaction. It focuses on robotic systems that use learned representations or learned decision-making, together with the sensing, control, data, and computing components needed to complete physical tasks. Manipulation supplies the main detailed technical evidence. Examples from other robotic systems illustrate how bodies, dynamics, and outcome requirements alter the mechanisms under discussion. Classical estimation, planning, and control are included where they implement or support these systems. This scope is followed by the assessment dimensions used throughout the review and the chapter organization built around the functions needed for physical tasks.

The overall objective is sustained, reliable task completion under changing conditions. Four dimensions guide its assessment: \emph{robustness}, maintaining task performance under disturbances and changes; \emph{efficiency}, the time, data, computation, energy, and human effort required to acquire and use that capability; \emph{safety}, keeping risks within stated limits during learning and operation; and \emph{autonomy}, the extent to which task execution, recovery, and continued operation depend on human assistance. These dimensions can interact: fewer interventions may reduce cost, but withholding necessary assistance can undermine safety or completion. These assessment dimensions describe what must be measured, not assumed properties of current systems.

Incomplete observations, costly experience, and physical execution constraints affect several of these assessment dimensions at once. The review first examines the information and mechanisms needed for decisions, then how they are learned and physically executed, and finally how their combined behavior is assessed. This order separates questions for analysis; it does not prescribe a processing pipeline or require separate modules. For example, learning can update several components, while evaluation and safety apply throughout development and operation.
\begin{itemize}[leftmargin=2pt]
    \item \textbf{Information and Models.} To make decisions from incomplete and heterogeneous observations, the system must estimate and retain task-relevant information (\S~\ref{sec:2}), then use it to predict outcomes, plan, and generate actions (\S~\ref{sec:3}). These chapters examine what information each decision requires and how feedback can revise it.
    \item \textbf{Data and Learning.} To learn from costly and uneven experience, the system needs sources that supply observations, action correspondence (the mapping from source data to the robot's action interface), and feedback (\S~\ref{sec:4}). Learning methods use these signals to acquire behavior, adapt to new tasks, environments, and bodies, and retain earlier capabilities (\S~\ref{sec:5}). Simulation supplies one source of experience; its use for testing is treated separately in Section~\ref{sec:8}.
    \item \textbf{Physical Execution and Computing.} Embodiment describes how the body's geometry, dynamics, sensing, and actuation shape physical interaction (\S~\ref{sec:6}). Action interfaces and feedback control connect selected actions to that body. Timely response depends on where computation runs and how its results enter execution (\S~\ref{sec:7}). Computing systems also support experience collection and policy updates, linking execution costs to learning and adaptation.
    \item \textbf{Evaluation and Safety.} Evaluation tests completion under changing conditions, sustained performance, constraint violations, and resource and human costs (\S~\ref{sec:8}). Safety mechanisms address the consequences of physical action through constraints, monitoring, recovery, and human takeover (\S~\ref{sec:9}). Together, these chapters distinguish evidence about behavior from mechanisms that constrain it.
    \item \textbf{Open Challenges and Synthesis.} Open challenges connect the remaining limitations across the four assessment dimensions (\S~\ref{sec:10}), and the conclusion draws together the conditions for sustained task completion (\S~\ref{sec:11}).
\end{itemize}

Within this organization, mechanism comparisons ask what information a method uses, what it produces, and what assumptions its use requires. Reported results then establish which benefits have been tested and under what conditions. The open challenges use these dependencies to connect local improvements to system-level capability.
\subsection{Related Surveys and Contributions}\label{sec:1.4}

We compare the coverage of related reviews, then state this survey's contributions and study-selection approach.

Existing surveys approach learning-enabled robotics through complementary perspectives. Foundation-model reviews organize models for perception, decision-making, and control~\citep{arxiv_2312_07843,arxiv_2402_02385,psiris2026fmreview}, while broader embodied-AI reviews examine research targets and paradigms~\citep{arxiv_2407_06886,arxiv_2508_10399}. More focused reviews examine VLA and manipulation methods~\citep{ma2026vlasurvey,bai2026manipulation}, predictive models~\citep{wang2026wam,hou2026world}, and the data and learning methods used to acquire behavior~\citep{ye2026datapyramid,arxiv_2408_03539}. System dependencies are also established concerns: Grover et al.~\citep{grover2026gauntlet} connect sensing, memory, scheduling, power, and safety in edge deployment, and Brunke et al.~\citep{arxiv_2108_06266} connect learning-based control, safe reinforcement learning, and policy certification.

This survey brings these topics together to explain how their interaction shapes the capabilities and limitations of Physical AI systems. Its contributions are threefold: comparing mechanisms under their reported conditions, tracing dependencies between their interfaces throughout the physical interaction loop, and identifying the cross-component limitations that shape the research agenda. The comparisons distinguish measured results from expectations based on how a mechanism works. The interface analysis examines how action representations depend on control interfaces, how computation changes the timeliness of feedback, and how execution records support learning and recovery. The resulting synthesis connects challenges in generalization, resource-efficient learning and execution, reliable state tracking, safety, recovery, and continual adaptation. It identifies which dependencies still limit robustness, efficiency, safety, and autonomy, and what evidence is needed to establish progress.

We use a problem-driven narrative review, selecting studies for their relevance to physical action, identifiable mechanisms, and traceable evidence. Reported results are interpreted under each study's tasks, training access, and physical conditions.

\section{Task Information and Representations}\label{sec:2}

Incomplete and heterogeneous observations raise a basic question: what information does a robot need to complete a task, and how can it obtain and retain that information? The answer depends on the task goal, constraints, and the decision being made: selecting a target, correcting a grasp, and confirming completion require different evidence.

\begin{figure}[!htbp]
\centering
\includegraphics[width=\linewidth]{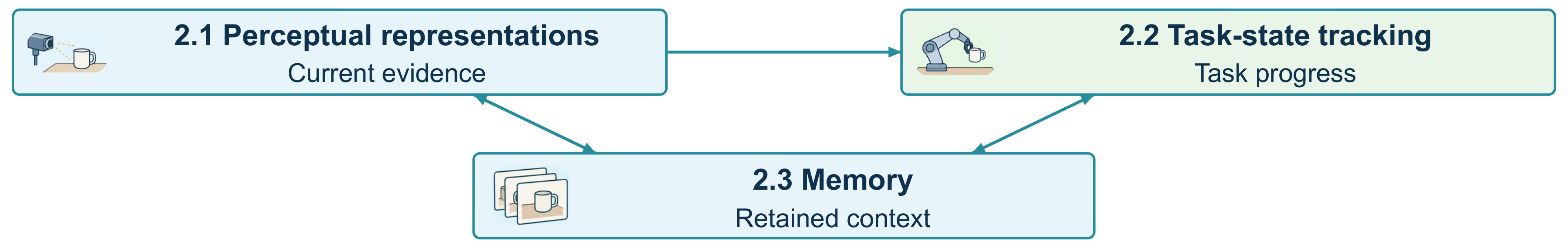}
\caption{Information for physical decisions. Perceptual representations provide current evidence, task-state tracking estimates progress, and memory retains context. Their connections express information exchange rather than a compulsory processing sequence.}\label{fig:guide-02}
\end{figure}

Observations provide partial measurements of the body and environment, from which a robot estimates physical quantities such as object poses and contact conditions. These estimates do not by themselves establish task progress, which also depends on task history and completion criteria. This section separates three related questions: how perceptual representations encode relevant information (\S~\ref{sec:2.1}), how task-state tracking uses available evidence to estimate progress (\S~\ref{sec:2.2}), and how memory retains and maintains information across time (\S~\ref{sec:2.3}). The order moves from representing observations to interpreting task progress and supporting later decisions; these functions can share representations.
Figure~\ref{fig:guide-02} organizes the chapter around information sufficient for the next decision: current perceptual evidence, estimated task progress, and retained context constrain one another rather than forming a compulsory serial pipeline.

\FloatBarrier

\subsection{Perceptual Representations}\label{sec:2.1}

To wash the correct cup, the robot must identify the target, locate its handle, track its motion, and detect slipping during grasping. The following account moves from what is identified to where and how interaction unfolds: semantic grounding links instructions to targets (\S~\ref{sec:2.1.1}), spatial representations retain geometry for action (\S~\ref{sec:2.1.2}), and temporal representations use history to resolve changing or hidden conditions (\S~\ref{sec:2.1.3}). The account then compares where and how multimodal fusion combines signals to affect a decision (\S~\ref{sec:2.1.4}), and closes with contact and force representations of local physical interaction (\S~\ref{sec:2.1.5}). These are complementary information needs; their sensing and processing costs must be weighed against their task value.

\begin{figure}[!htbp]
\centering
\includegraphics[width=\linewidth]{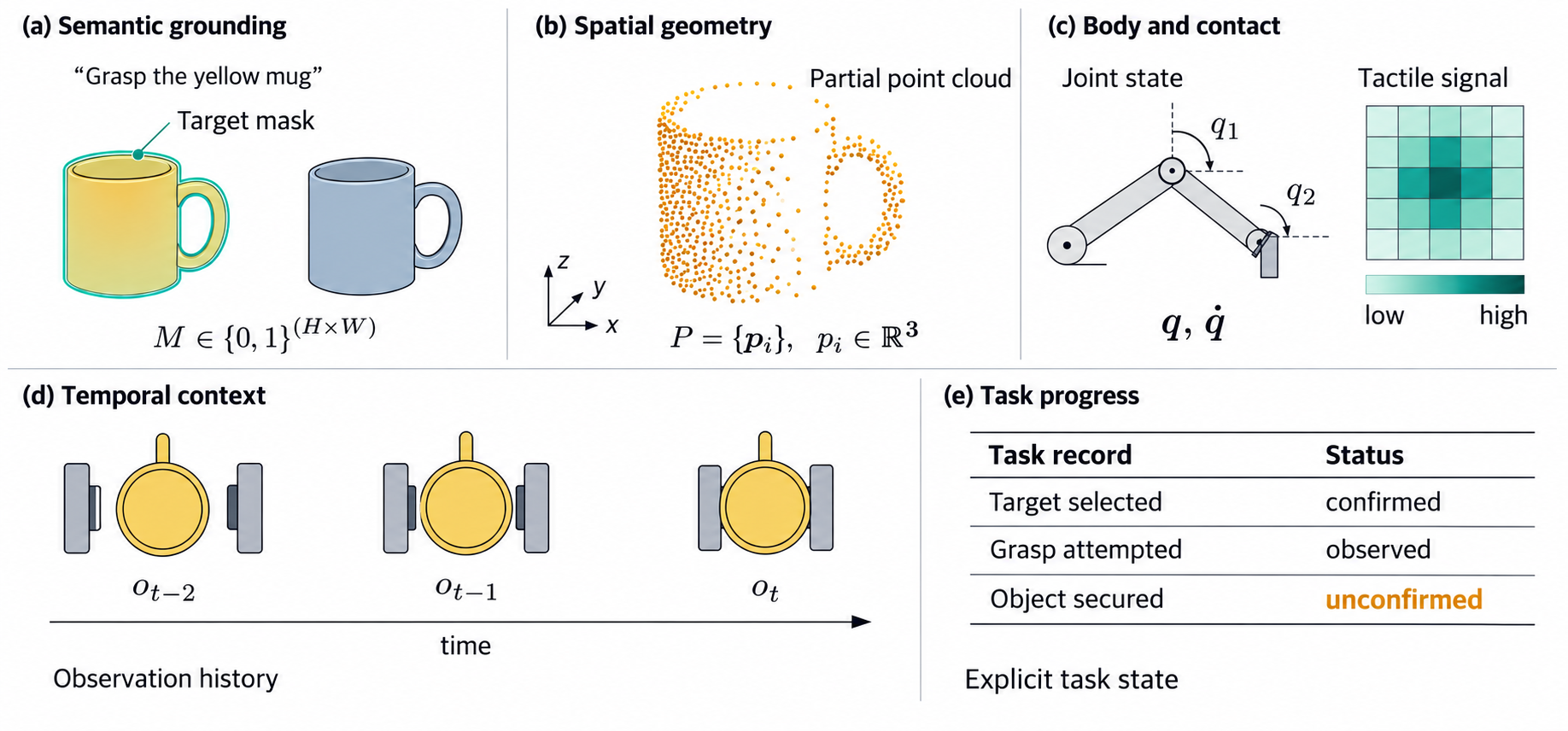}
\caption{Complementary information representations for a mug-grasping task. (a) A binary mask $M$ identifies the image region associated with the instruction; $H$ and $W$ are the image height and width. (b) A partial point cloud $P$ encodes observed 3D geometry; occluded surfaces remain unknown. (c) Joint angles $q$ and velocities $\dot q$ describe body configuration and motion, while tactile signals provide local contact evidence. (d) Observations $o_{t-2:t}$ record gripper closure over time. (e) An explicit task record distinguishes target selection and an attempted grasp from a confirmed secure grasp. Illustrations are schematic: tactile intensity does not directly specify force, and gripper closure alone does not confirm a secure grasp.}\label{fig:overview-information}
\end{figure}

Figure~\ref{fig:overview-information} illustrates these information needs through a mug-grasping example: a target mask, a partial point cloud, joint and tactile measurements, an observation history, and an explicit task record. These are complementary views of the information needed for a decision, rather than mutually exclusive representation classes or a required processing pipeline. The figure selects representative encodings; the following discussion also covers learned features, recurrent states, and other alternatives. Explicit forms such as masks, points, and task records can also be produced by learned models.

\subsubsection{Semantic Grounding}\label{sec:2.1.1}

Semantic grounding links instructions to objects, properties, and relations in a scene. Beyond recognizing categories, the robot must identify the target object among similar ones and preserve local or relational details, such as handles or occlusion, for action selection. 
Three key design choices are how visual features align with language, how much spatial detail they retain, and how they are integrated into the decision model. These functions can be combined within a single encoder.

\noindent\textbf{Vision--language Alignment.} Image--text learning provides broad semantic knowledge before robot-specific training. CLIP~\citep{arxiv_2103_00020} learns image--text correspondence from large-scale data, enabling recognition of categories and attributes beyond predefined robot action labels. SigLIP~\citep{zhai2023sigmoid} instead uses a pairwise sigmoid loss, avoiding CLIP's batch-wide normalization. Both improve visual--language alignment but do not directly determine how accurately a downstream policy localizes the target object.
The source of alignment also matters. R3M~\citep{arxiv_2203_12601} combines time-contrastive learning and video--language alignment on human videos to learn manipulation-relevant representations. Unlike image--caption pairs, videos capture temporal relationships and interactions, but transferring these features to robots still requires robot data or an appropriate action interface. The key perceptual question is whether pretraining captures appearance, language references, temporal relationships, or their combination.

\noindent\textbf{Spatially Resolved Features.} Identifying the target object does not necessarily reveal its parts, such as a cup's handle. Local features and segmentation provide finer spatial information. DINOv2~\citep{arxiv_2304_07193} learns self-supervised features for image- and pixel-level tasks, while SAM~\citep{kirillov2023segment} generates prompt-conditioned masks. Dense features require downstream interpretation, whereas masks explicitly identify image regions. Modular pipelines can combine open-vocabulary detection and segmentation: Grounding DINO~\citep{arxiv_2303_05499} integrates text and image features through a transformer to localize referring expressions as bounding boxes, which SAM can further refine into masks.
Spatial detail and language alignment can be combined. CLIPort~\citep{shridhar2021cliport} uses a CLIP-based semantic stream to identify what to manipulate and a Transporter-based spatial stream to determine where, fusing both to predict pick-and-place locations. Similarly, OpenVLA~\citep{arxiv_2406_09246} combines DINOv2 and SigLIP features in its visual encoder. However, neither provides a complete physical-state estimate: masks may isolate objects without revealing their constraints, while local features may capture handles without identifying feasible grasps.

\noindent\textbf{Language-model Integration.} The model that interprets the task must be able to use the visual features. PaLM-E~\citep{arxiv_2303_03378} projects visual and continuous-state information into a language model's input as multimodal sequences, so the model that interprets a request or produces a plan also processes the perceptual content. In action-generating VLAs, image and instruction information similarly enter a shared model, but an action interface determines how the resulting representation affects control. Placing perception inside a language model therefore specifies an integration point, not an alternative to language-aligned or spatially resolved features.
Selecting the wrong cup indicates a grounding failure, missing handle details reflects insufficient spatial granularity, and unusable perceptual information indicates an interface failure. Language-aligned features help identify the target, but grasp selection also requires local geometry and an action mapping.

\subsubsection{Spatial Representations}\label{sec:2.1.2}
Spatial representations help policies connect observed structures to possible actions. Grasping requires identifying an approach, aligning it with the robot’s workspace, and handling uncertainty at contact. Some methods explicitly represent geometry using depth, points, voxels, or spatial targets, while others infer spatial information from image features. The key difference lies in how geometry is represented and supervised, as 3D inputs can be inaccurate and image-based policies can still reason spatially.

\noindent\textbf{Explicit Geometry.} PerAct~\citep{arxiv_2209_05451} converts multiview RGB-D observations into a voxel grid and predicts language-conditioned keyframe actions, including translation, rotation, gripper state, and collision avoidance. A motion planner executes these keyframes rather than directly predicting low-level controls. By selecting translation targets within the voxel grid, PerAct aligns action coordinates with scene geometry, reducing the need to infer metric positions from images. However, its accuracy depends on voxel resolution and calibration, as discretization limits spatial precision and reconstruction errors can shift predicted targets.
3D Diffusion Policy~\citep{ze20243d} uses a compact point-cloud representation to condition continuous action sequences executed by a lower-level controller. Unlike PerAct, which selects keyframes from an explicit voxel grid for motion planning, it encodes 3D geometry into a latent representation. Both leverage measured 3D structure but differ in scene representation and action interface. Their performance also depends on depth quality, preprocessing, and training setup, so differences cannot be attributed to 3D input alone.
Explicit geometry is limited by sensor coverage, while higher voxel resolution or point density increases computational cost. RVT~\citep{arxiv_2306_14896} addresses this trade-off by reconstructing a point cloud from RGB-D observations and re-rendering it into virtual views containing color, depth, and world coordinates. A transformer processes these views and predicts heatmaps that are back-projected into 3D to select the gripper translation. This avoids a full voxel grid while preserving measured geometry. On RLBench, RVT reports higher success rates and lower training and inference costs than PerAct, although these results reflect differences in the overall architectures. RVT remains sensitive to depth errors, calibration, and occlusion. Ultimately, the required geometric resolution depends on task precision.

\noindent\textbf{Geometry Inferred from Images.} Image-based models rely more on learned spatial inference. SpatialVLM~\citep{chen2024spatialvlm} uses synthetic question--answer supervision to improve metric and relational reasoning, while RoboPoint~\citep{yuan2024robopoint} predicts instruction-conditioned image keypoints as spatial affordances. The former provides spatial knowledge, whereas the latter identifies actionable locations. MolmoAct~\citep{lee2025molmoact} instead predicts depth-aware perception tokens from images and instructions, followed by an editable image-space trajectory trace and actions conditioned on both. This incorporates depth reasoning without requiring measured depth as input.
Embodied-R1~\citep{yuan2025embodiedr1}, like RoboPoint, uses pointing to connect reasoning with action primitives. This allows target selection to be inspected separately from execution. However, converting image-space points into feasible motions still requires depth and calibration, learned mappings, or sensor-guided action primitives.
Explicit geometry reduces learned spatial inference but is limited by measurement errors, occlusion, and computational cost. Image-based methods rely more on learned spatial relationships and downstream action mapping.

\subsubsection{Temporal State Representations}\label{sec:2.1.3}

Similar images can represent different physical states, such as a moving versus stationary ball or a slipping versus firmly held cup. Contact history and occluded objects introduce further ambiguity. Temporal representations reduce this partial observability by preserving past information through either an explicit history window or an updated internal summary. Whereas stored history records earlier evidence, prediction estimates states or outcomes that have not yet been observed. Predictive supervision can enhance either approach independently of how history is stored.

\noindent\textbf{History Windows.} A history window provides the model with a bounded sequence of observations, preserving recent motion and information about occluded objects. Its effective horizon depends on sampling frequency and window size, while subsampling may miss brief contact events. History can also reveal conditions hidden from the current observation. For example, RMA~\citep{arxiv_2107_04034} uses the last 50 state--action pairs (about 0.5\,s) to estimate latent terrain, payload, and other conditions for quadruped locomotion. Rather than tracking object motion, this window captures body--environment interactions. The adaptation module updates the latent from a rolling window, while the policy reuses it between updates.
A history window loses events once they fall outside its span, while longer windows increase computational costs and make relevant observations harder to identify. They can also harm imitation learning through the copycat problem, where a policy repeats previous actions rather than predicting the next. Wen et al.~\citep{arxiv_2010_14876} address this causal confusion by adversarially removing previous-action information while preserving features useful for predicting the next action. Multiscale designs such as MEM mitigate window limitations by retaining recent details and selectively preserving older information.

\noindent\textbf{Recurrent State Updates.} A recurrent model updates a latent state with each new observation instead of repeatedly processing the full history. PlaNet~\citep{hafner2018latent} combines deterministic and stochastic transitions to maintain state estimates and represent uncertainty under partial observability. PlaNet uses this state for planning, while Dreamer uses a similar recurrent state for direct action selection. However, what information the state retains depends on the training objective, model capacity, and available training evidence.

Either history representation can be trained with predictive supervision. Predicting future observations or latent states can help representations capture temporal changes. GR-1~\citep{arxiv_2312_13139} jointly predicts future images and actions. However, the benefit depends on which changes are emphasized: background motion and impending grasp failure require different information.
Prediction can also compensate for execution delays. AHEAD~\citep{syed2026intercepting} estimates motion across observations to predict future patch tokens for a frozen VLA, providing its action decoder with a more up-to-date scene representation. Unlike predictive training losses, this prediction occurs during execution. Its effectiveness depends on delay, prediction accuracy, and uncertainty handling, and differs from action-conditioned prediction for planning.


\subsubsection{Multimodal Fusion}\label{sec:2.1.4}

\begin{figure}[!htbp]
\centering
\includegraphics[width=\linewidth]{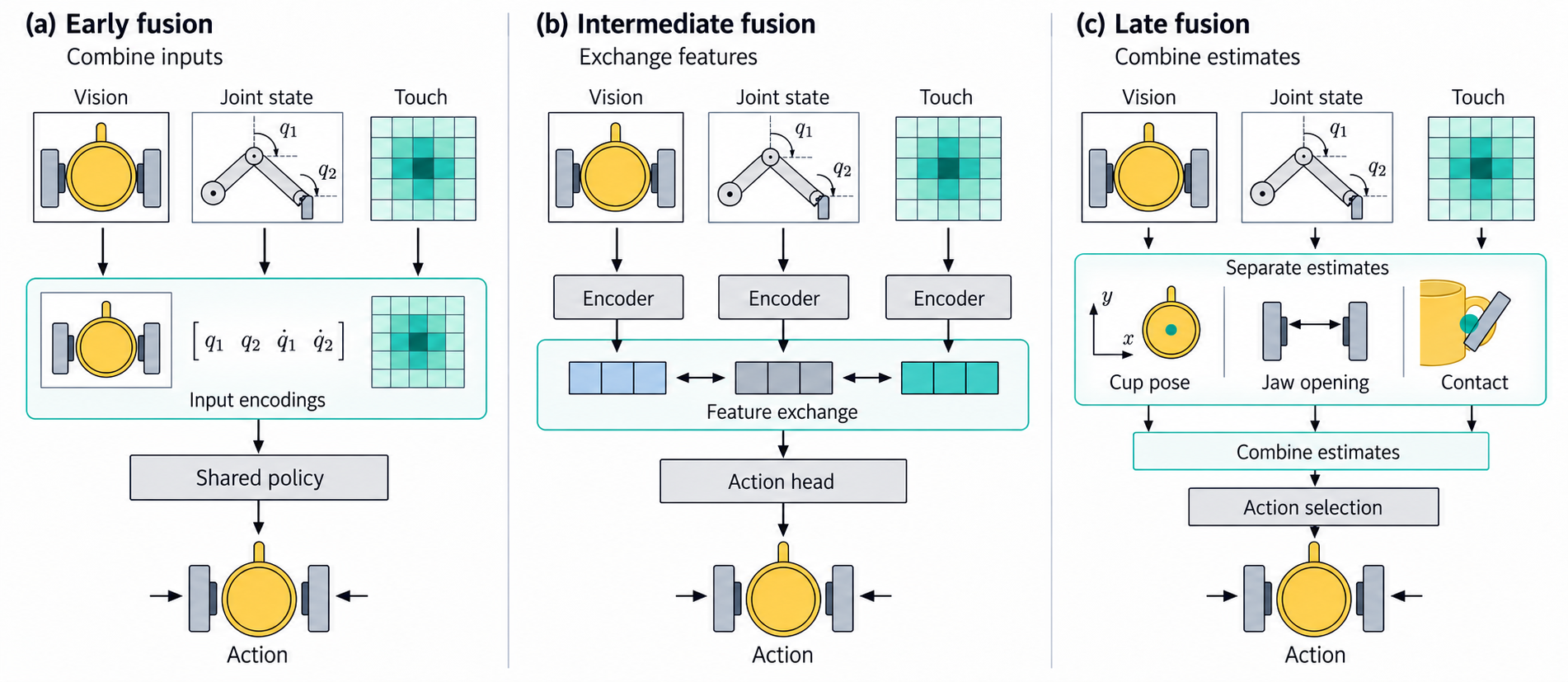}
\caption{Multimodal fusion locations in a mug-grasping example. The same visual, joint-state, and tactile inputs support three conceptual patterns. (a) Early fusion combines modality encodings for processing by a shared policy. (b) Intermediate fusion exchanges separately encoded features before action generation. (c) Late fusion combines modality-specific estimates of cup pose, jaw opening, and contact before action selection. These patterns may coexist within a system. Feature cells and tactile signals are schematic; the inward arrows illustrate closure commands, not confirmed grasp success.}\label{fig:information-fusion}
\end{figure}

During grasping, vision locates the cup, proprioception tracks hand configuration, and touch or force detects slipping. These signals arrive at different rates and contribute at different moments. Their influence depends on the fusion stage: input, internal features, or outputs. This subsection compares where signals meet; the following subsection examines what contact and force signals themselves encode. Figure~\ref{fig:information-fusion} compares input, feature, and estimate fusion for the same mug-grasping task. These patterns can coexist within a system.

\noindent\textbf{Early Fusion.} Input-level fusion allows modalities to interact through shared processing. PaLM-E adds visual and state encodings to language-model inputs~\citep{arxiv_2303_03378}; RT-2 and OpenVLA combine images and instructions to generate actions~\citep{arxiv_2307_15818,arxiv_2406_09246}; and Octo uses a transformer policy with a diffusion action head~\citep{arxiv_2405_12213}. Although all fuse modalities at the input, they differ in outputs, training, and embodiment interfaces. Shared processing lets task context shape sensory information use, but its cost depends on token count, attention structure, and feature recomputation. Attention masks control token interactions. For fast-changing signals like touch, responsiveness depends on when new measurements reach the model and affect actions. A slow backbone cannot provide corrections faster than its update rate.

\noindent\textbf{Intermediate Fusion.} Feature-level fusion allows separate encoders or branches to exchange information within a model. TransFuser uses attention to combine image and LiDAR features~\citep{arxiv_2104_09224}. In \(\pi_0\), an action expert combines a reusable vision--language prefix with proprioceptive state~\citep{arxiv_2410_24164}. Reactive Diffusion Policy similarly separates slow planning from fast tactile feedback~\citep{arxiv_2503_02881}. A latent diffusion policy generates action chunks from visual and tactile or force inputs at 1–2 Hz, while a fast decoder uses current tactile or force feedback to update actions at 24 Hz. This allows rapid corrections without rerunning the slower policy.
In Reactive Diffusion Policy's asymmetric tokenizer, tactile input reaches the decoder but not the action encoder. This lets the latent chunk represent a general feedback strategy while the decoder adjusts individual actions, such as lifting a peeler when contact force increases.
In Reactive Diffusion Policy's peeling, wiping, and bimanual lifting tasks, adding tactile input to a standard Diffusion Policy performed similarly to vision alone because action chunks remained open loop. The slow–fast design performed better across all three tasks. This shows that a modality's effectiveness depends on where it enters the model and how often its branch updates. New feedback can only correct an action if the model allows it to be revised.

\noindent\textbf{Late Fusion.} Output-level fusion combines results from separately processed modalities. Swift combines visual–inertial pose estimates with corrections from detected race gates~\citep{kaufmann2023swift}. Similarly, a manipulation system might confirm a grasp when gripper width, load, and vision agree. Outputs remain independently usable, but fusion is limited to the information they retain. Missing uncertainty, timing, or alternative hypotheses can reduce reliability, so the combining module must handle differences in coordinates, timing, validity, and missing or inconsistent inputs.

Fusion location determines where information is combined, while the update mechanism determines when new measurements can influence an action. The slow--fast policy example above illustrates how current contact measurements can revise actions while reusing slower context; this schedule is a separate design choice, not an automatic consequence of feature-level fusion. Across fusion stages, additional information helps only when it can influence decisions at the right time. Fusion cannot recover information lost during encoding or compensate for delayed feedback.

\subsubsection{Contact and Force Representations}\label{sec:2.1.5}

Contact information describes gripper--object interactions, including contact occurrence, load distribution, and potential slip. The discussion separates learning reusable tactile features from integrating contact measurements into a policy. These are complementary choices: an encoder determines what information is available, while the policy interface determines how and when it influences actions. Sensor frame, calibration, and sampling time are essential for interpretation.

\needspace{3\baselineskip}
\noindent\textbf{Learning Tactile Features.} Because labeled tactile data are scarce and sensor-specific, Sparsh~\citep{arxiv_2410_24090} uses self-supervised learning on unlabeled tactile images to train a reusable encoder; in its frozen-encoder evaluation, task-specific decoders are trained while the encoder is fixed. These small decoders predict three-axis contact forces from indentation data or detect slip from 80\,ms of tactile frames labeled using a friction-cone model. Output accuracy and physical meaning depend on task-specific labels and calibration, so performance validated on probe sliding may not directly generalize to object grasping.

\noindent\textbf{Contact Inputs for Policies.} ViTaL~\citep{zhao2025touch} combines egocentric vision and tactile sensing for local contact control after vision--language target localization. Its ablations demonstrate the benefit of tactile input in contact-rich tasks. While learned tactile embeddings can guide action selection, explicit physical contact estimates require corresponding supervision or calibration.
For wrist force input, ForceVLA~\citep{arxiv_2505_22159} integrates six-axis wrist force--torque sensing into a \(\pi_0\)-style policy by fusing a force token with vision--language features through a small mixture-of-experts module before the flow-matching action expert. Across five contact-rich tasks, it achieves 60.5\% average success, compared with 37.3\% without force and 40.2\% when force is appended to proprioception. Ablations show that post-VLM fusion outperforms earlier injection, highlighting that where contact information enters the policy matters as much as its presence.
For the wet cup, object identity and pose alone cannot detect slip. Tactile, force, and proprioceptive history provide the necessary feedback, but must reach the controller in time to prevent grasp failure. Contact therefore requires dedicated sensing and timely feedback updates.

Representations should preserve information needed to resolve task-relevant ambiguities. Part localization can identify candidate handles, while explicit geometry supports metric reasoning at the cost of depth, calibration, and computation. Image-conditioned control remains viable when reconstruction offers little benefit. Hidden contact requires timely tactile, force, visual, or proprioceptive feedback. Longer histories help only when past observations remain relevant, while multimodal fusion must account for differences in sensor timing and reliability. The choice therefore depends on the ambiguity to resolve, the available sensors, and when the resulting estimate must affect an action.
Open questions include robustness to calibration errors and occlusion, history relevance after unexpected changes, and sensor reliability under missing or delayed measurements. Active reobservation can refresh spatial information, while fast local controllers can process contact feedback without routing it through a shared backbone.


\subsection{Task State Representation and Tracking}\label{sec:2.2}

Task state is the actual progress toward the goal, including completed and remaining stages and the conditions required for the next action. It depends on both physical state and task history: a cup's pose describes its current physical configuration, while a confirmed washing event records earlier progress. The account first describes how progress and completion conditions can be represented, then how observations and memory update the task-state estimate and check conditions for subsequent actions. Figure~\ref{fig:overview-information}(e) illustrates an explicit task record: target selection is confirmed and a grasp attempt is observed, while whether the object is secured remains unconfirmed. The closure sequence in panel~(d) therefore does not by itself justify advancing to transport.
Task state can be represented by subtasks, confirmed predicates, event records, or learned summaries, each providing different information to the policy. Explicit records distinguish planned, attempted, and confirmed actions, while learned summaries may encode these distinctions implicitly. Crucially, action attempts do not establish completion: after the missed grasp E1, transport requires confirmation that the object is held.

Task-state updates combine current observations, memory, and completion criteria. MEM~\citep{torne2026mem} tracks task stages through text memory and short-horizon video, while Hi Robot~\citep{arxiv_2502_19417} revises high-level decisions using observations and feedback. Both preserve past progress for future decisions through different architectures.
Task-state tracking must account for uncertainty and changing conditions. A successful grasp may fail after slip, while a past cleaning event does not guarantee the object remains clean. Thus, tracking must distinguish historical events from current states. Memory maintains relevant records, planning uses updated estimates, and evaluation defines evidence of task completion.

\subsection{Memory}\label{sec:2.3}

Memory stores and retrieves observations, events, estimates, and reusable skills over time to support state estimation and behavior. It extends the observation history illustrated in Figure~\ref{fig:overview-information}(d) with mechanisms for storage, retrieval, maintenance, and reuse. The first three parts distinguish what is retained: past events for later decisions (\S~\ref{sec:2.3.1}), persistent information about objects and places (\S~\ref{sec:2.3.2}), and reusable behaviors (\S~\ref{sec:2.3.3}). Memory maintenance then addresses the shared problem of updating, retrieving, checking, or invalidating those records and skills (\S~\ref{sec:2.3.4}). These are complementary design dimensions: different memory types may share retrieval mechanisms but have distinct requirements for remaining useful.

\subsubsection{Episodic Memory}\label{sec:2.3.1}

Retrieving a cup after visiting another room may require recalling its location, a failed grasp, or completed steps. Episodic memory preserves relevant past observations and events for future decisions, requiring clear rules for storage, compression, and task-driven retrieval.

\noindent\textbf{Learned and Multiscale Memory.} A history window fails when relevant events fall outside its range. Learned memory mechanisms address this by retaining selected events in memory banks or compressed summaries. MemoryVLA~\citep{shi2025memoryvla} retrieves relevant perceptual and semantic entries and merges redundant ones during updates. MEM~\citep{torne2026mem} combines compressed video for short-term detail with text memory for task progress, enabling tasks lasting up to fifteen minutes and context-dependent manipulation. Their key difference lies in memory format: learned tokens in MemoryVLA versus readable text in MEM.

\noindent\textbf{Caption-based Retrieval.} ReMEmbR~\citep{arxiv_2409_13682} stores robot observations as video captions with spatial and temporal metadata. A vision--language model periodically generates captions, which are embedded and stored with the robot's position and timestamp. At query time, a language-model agent iteratively retrieves records by text, location, or time to answer questions or guide navigation. Retrieval keeps query time roughly stable across tested video lengths, but performance depends on caption accuracy. For example, a soda-machine query led the robot to a water fountain mislabeled as a ``silver machine'', illustrating how incomplete or misleading records can cause navigation errors.

\subsubsection{Spatial Memory}\label{sec:2.3.2}

A robot returning to the kitchen must recall the cup’s last known location, even if it is out of view or has moved. Spatial memory links observations to persistent objects and places, enabling retrieval across viewpoints. This requires suitable representations, cross-view associations, and updates when stored locations become outdated.
Dense maps and scene graphs provide complementary spatial representations. VLMaps~\citep{arxiv_2210_05714} projects vision--language features into a top-down grid, enabling language-based localization. ConceptGraphs~\citep{arxiv_2309_16650} instead builds an open-vocabulary 3D scene graph of objects and their semantic and spatial relations. SayPlan~\citep{rana2023sayplan} uses hierarchical scene graphs for building-scale planning, retrieving task-relevant regions while a classical planner handles navigation. However, constructing spatial representations does not ensure their validity over time: moved objects require re-association, changed relations must be updated, and unobserved regions may become outdated.

Maintaining spatial memory requires associating new observations with existing objects and updating outdated records. DynaMem~\citep{arxiv_2411_04999} addresses this using a voxel map that stores vision--language features, observation counts, source images, and timestamps. New RGB-D observations update the map, while depth evidence of free space removes outdated voxels within a 2\,m range. Object queries use voxel features or stored images, with exploration triggered when objects are not found. In dynamic environments, DynaMem achieves 70\% pick-and-drop success versus 30\% for static OK-Robot~\citep{liu2024okrobot}. However, moved objects remain at their old locations until new observations provide evidence for removal.

\subsubsection{Skill Memory}\label{sec:2.3.3}

Skill memory, or procedural memory, stores reusable behaviors such as policies, programs, or task procedures. Unlike state memory, it supports action reuse across tasks through retrieval and composition. Its effectiveness depends on selecting applicable skills and verifying that their preconditions match the current state.
LOTUS~\citep{arxiv_2311_02058} discovers skills from unsegmented demonstrations, expands a skill library, and trains a meta-controller to compose skills for new manipulation tasks. Voyager~\citep{wang2023voyager} similarly stores and reuses self-verified executable programs in Minecraft. However, retrieving a skill does not guarantee that its preconditions hold, nor does executing an action guarantee success.

A reusable skill requires defined inputs, parameters, preconditions, and completion criteria. Retrieval by task similarity alone may select inapplicable behaviors, so planners must verify their suitability against the current state. Skill libraries can grow through new modules or programs, but retaining them requires storage and maintenance. Effective reuse requires retrieving, composing, or acquiring skills as needed; library size alone does not guarantee applicability.

\subsubsection{Memory Maintenance}\label{sec:2.3.4}

An issued command does not guarantee successful execution. Recording ``cup held'' without confirming the grasp can mislead later planning. Memory maintenance must therefore check the evidence for stored claims, resolve conflicts, and invalidate outdated information.
Retrieval, recording, and maintenance have distinct failure modes: writing may treat inferred outcomes as observed, retrieval may miss relevant records, and maintenance may preserve stale information or make incorrect updates. When an object reappears elsewhere, data association must determine whether it is the same instance. Timestamps, provenance, confidence, and validity conditions help detect such conflicts. MemoryVLA merges redundant entries, while DynaMem removes voxels contradicted by new observations.
Memory updates should distinguish observations, inferred outcomes, and confirmation status. A ``cup held'' claim should be confirmed only by sufficient visual, gripper, or contact evidence. Failure explanations also depend on retained records. OpenEQA~\citep{majumdar2024openeqa} evaluates environmental understanding through recorded episodes and active exploration, but not memory-guided manipulation. Closed-loop evaluation, as in RoboMemArena~\citep{lei2026robomemarena}, is needed to assess how memory affects subsequent behavior.

\begin{figure}[!htbp]
\centering
\includegraphics[width=\linewidth]{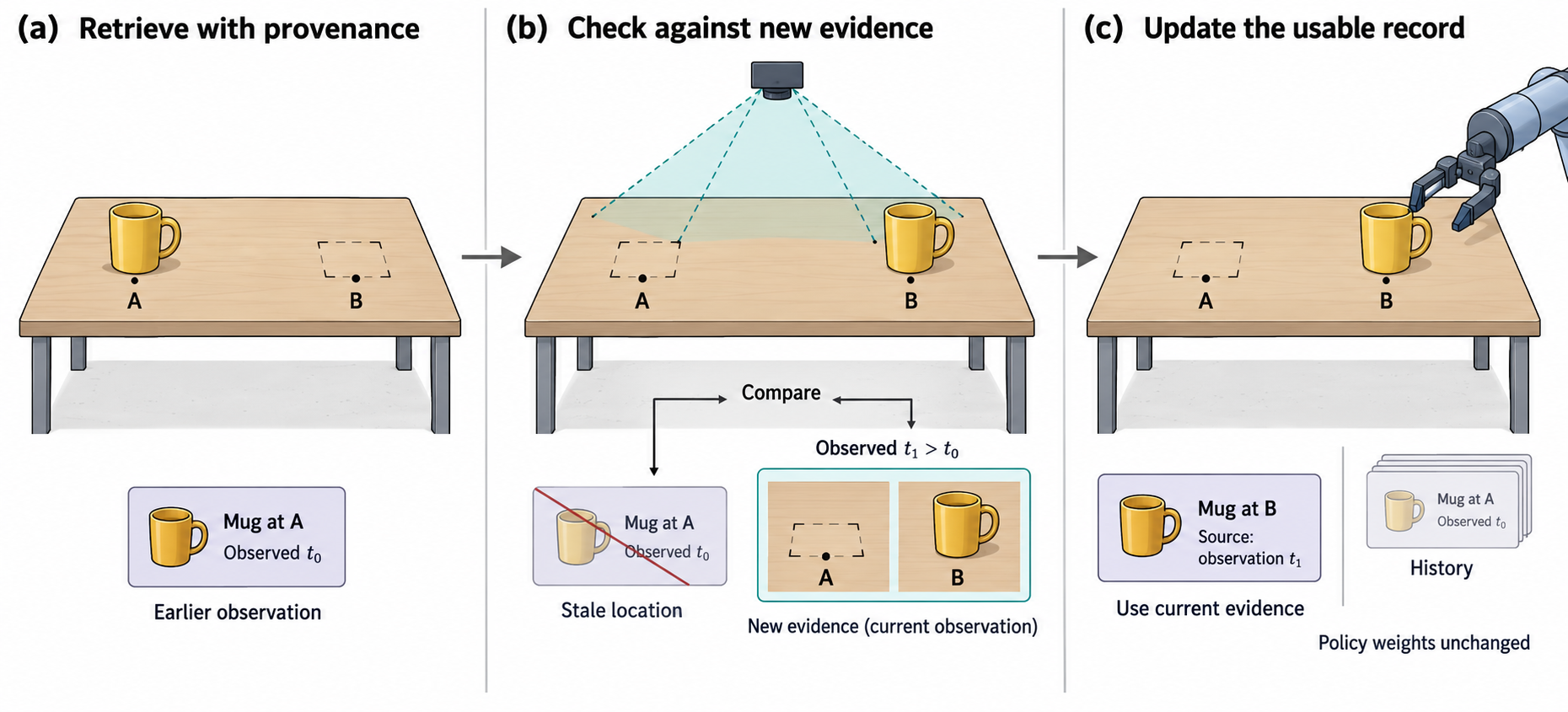}
\caption{Maintaining spatial memory when an object moves. (a) A retrieved location retains its observation time. (b) A later view reveals an empty location A and the mug at B, contradicting the old current-location claim. (c) The usable record is updated while the earlier observation can remain in history. This schematic assumes correct object association and visibility of both locations; an occluded location is not evidence of absence. The memory change does not require a policy-parameter update.}\label{fig:memory-maintenance}
\end{figure}

Figure~\ref{fig:memory-maintenance} illustrates the distinction between historical truth and a valid current-state claim: a mug observed at A can later be observed at B. Updating the usable location does not erase the earlier event or require retraining the policy. The update depends on identifying the same object and observing the relevant locations; lack of visibility alone cannot invalidate the old record.

Episodic, spatial, and skill memory are effective when stored information remains relevant and valid. Structured events compress long histories but may omit useful details. Task-local spatial memory reduces maintenance costs but requires updates when objects move. Skill libraries support reusable behaviors and modular updates, while shared policies reduce management overhead at the cost of greater reliance on shared parameters. In all cases, retrieval is useful only when stored facts or skill preconditions remain valid.
Task-state tracking uses past and new evidence to update progress estimates. Together with physical-state estimates and memory, these estimates provide the inputs for predicting outcomes and selecting actions in Section~\ref{sec:3}.

\section{Models for Prediction and Action}\label{sec:3}

Models use observations, state estimates, task conditions, and memory to predict outcomes and select actions. Predictive models predict state changes, policies select actions, and planners explore alternatives. These functions may be separate or share parameters. Their distinction depends on the output and its use, rather than the model architecture or sequence generation.

\begin{figure}[!htbp]
\centering
\includegraphics[width=\linewidth]{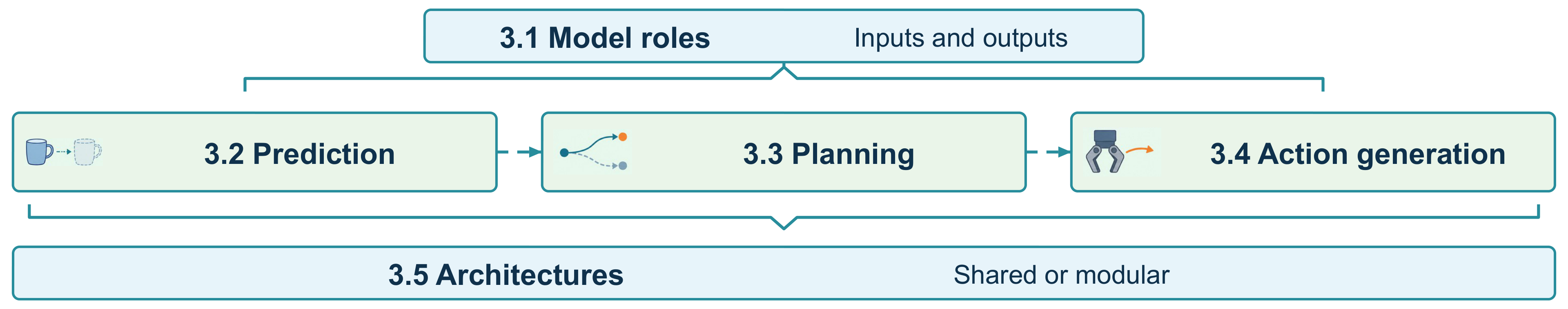}
\caption{Model roles and architectures. Prediction, planning, and action generation are distinguished by their inputs and outputs and can be organized in shared or modular architectures. Dashed arrows indicate possible dependencies; direct policies need not use explicit prediction or planning.}\label{fig:guide-03}
\end{figure}

The chapter first distinguishes model roles by their inputs and outputs (\S~\ref{sec:3.1}). It then examines three functions: predicting physical outcomes (\S~\ref{sec:3.2}), planning and selecting actions (\S~\ref{sec:3.3}), and generating actions at a policy interface (\S~\ref{sec:3.4}). The final subsection compares how architectures combine or separate these functions (\S~\ref{sec:3.5}). This progression establishes the functional distinctions before examining their integration; it does not require every system to use each function as a separate stage. Prediction may support planning, policy conditioning, or learning; observed physical effects provide evidence of actual outcomes.
Figure~\ref{fig:guide-03} maps the roles of prediction, planning, and action generation and the architectures that combine them.

\FloatBarrier

\subsection{Model Inputs, Outputs, and Roles}\label{sec:3.1}

An encoder converts observations, task conditions, and optional history into features. A world model predicts future states or observations of the body and environment, often conditioned on actions. A policy selects actions or action sequences, while a planner selects skills, proposes subgoals, or evaluates alternatives. These functional roles come first; the next paragraph relates vision--language and vision--language--action models to their inputs and outputs. Explicit prediction and planning are optional, as a policy can act without a separate world model or planner.

A vision--language model (VLM) processes visual and linguistic inputs, while a vision--language--action (VLA) policy generates robot actions, as in RT-2 and OpenVLA~\citep{arxiv_2307_15818,arxiv_2406_09246}. A VLA may share a model across modalities or add an action-generation component to a VLM backbone. The key distinction lies in the output: predicting future object states differs from generating robot actions, even when both use generative modeling.

\subsection{Predicting Physical Outcomes}\label{sec:3.2}

World models predict physical state changes and, when conditioned on actions, their consequences, such as whether a grasp remains stable during lifting. Prediction is distinct from action selection. Alternatively, a task-conditioned future can define a goal, with a controller or inverse model determining the actions to achieve it. We compare predictive representations in observation space, explicit physical variables, and learned latent features (\S~\ref{sec:3.2.1}); this representation choice is distinct from how predictions are conditioned or used.
World models serve multiple purposes: UniSim, Cosmos, and GAIA-1 generate future observations or simulations~\citep{yang2023interactive,nvidia2025cosmos,hu2023gaia1}; DreamGen generates training experience~\citep{jang2025dreamgen}; and WorldEval, Ctrl-World, and Veo-based simulation evaluate policies~\citep{li2025worldeval,guo2025ctrlworld,geminiroboticsteam2025evaluating}. These applications impose different requirements~\citep{hou2026world}. 

\subsubsection{Predictive Representations}\label{sec:3.2.1}

\begin{figure}[!htbp]
\centering
\includegraphics[width=\linewidth]{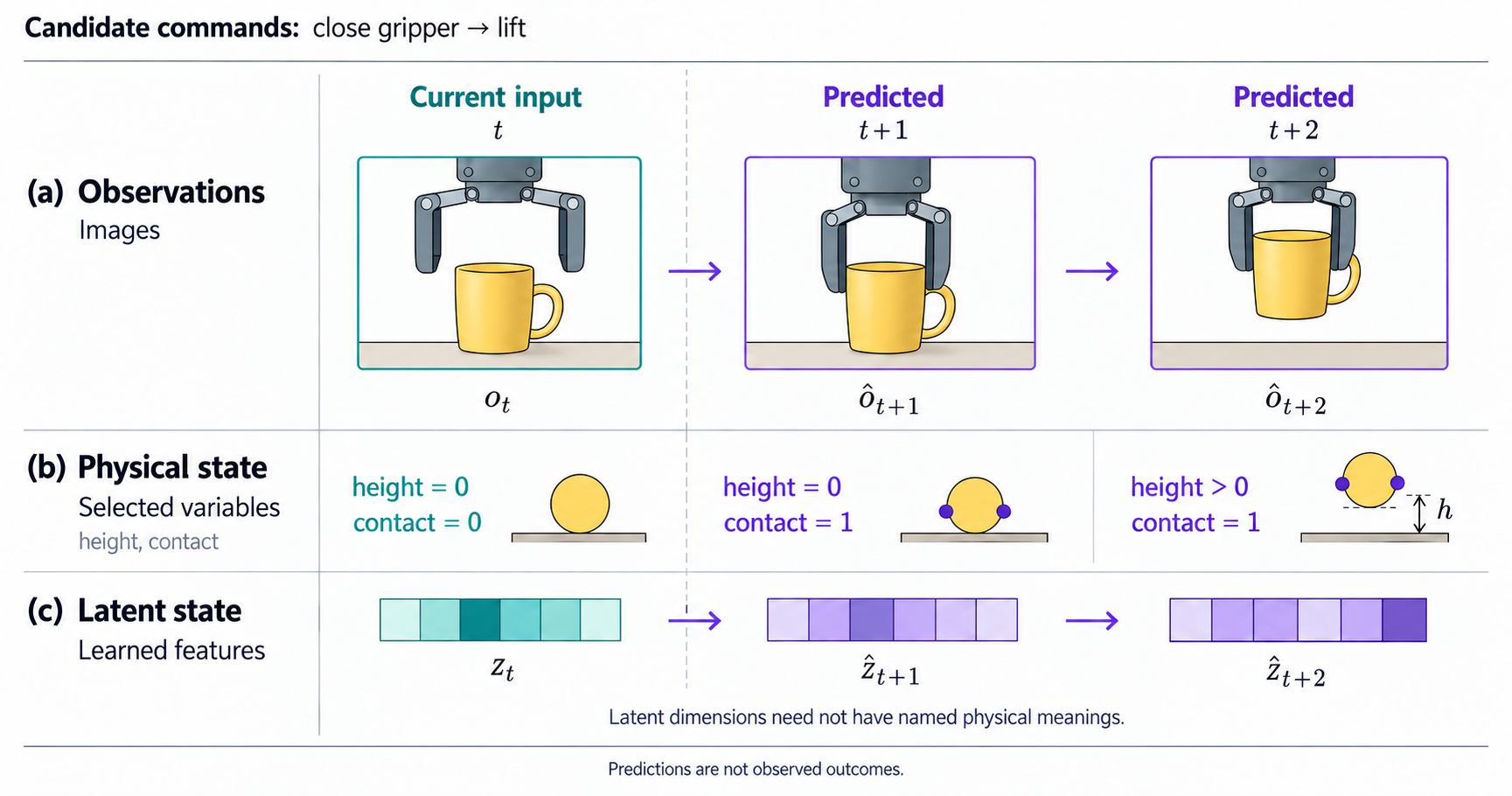}
\caption{Predictive representations of an illustrative action-conditioned grasp-and-lift sequence. Columns align current input with two predicted future times. (a) Observation-space prediction depicts future images. (b) Selected physical variables describe the mug--table gap and gripper contact. (c) Latent prediction evolves learned features without requiring named physical meanings for individual dimensions. Teal marks current observations, estimates, or encodings; purple and hats mark predictions. Contact alone does not establish grasp security, and new observations are needed to establish actual outcomes.}\label{fig:overview-prediction}
\end{figure}

The prediction space determines what information is available for decision-making. Observation-space models predict future measurements, structured models represent quantities such as object positions and relations, and latent models predict learned features. These approaches trade representation cost against task-relevant information, without determining whether the model supports planning, training, or evaluation. Figure~\ref{fig:overview-prediction} aligns the three prediction spaces at the current time and two hypothetical future times, conditioned on closing the gripper and lifting. Its structured readouts track mug--table separation and contact; neither predicted contact nor a plausible lifted image verifies a secure grasp. Candidate actions can condition consequence prediction, while a task goal can condition a desired future; the conditioning and downstream use are separate from the representation.

\noindent\textbf{Observation-space Prediction.} Generated images provide visual predictions for controllers, policies, or learners, as in UniPi's video plans, SuSIE's goal images, and DIAMOND's generated Atari experience~\citep{alonso2024diffusion}. However, visual plausibility does not guarantee physical feasibility, and similar predictions may require different control capabilities. Generating pixel-level detail also increases computational cost across predicted steps.

\noindent\textbf{Structured Physical Prediction.} A structured model predicts selected physical variables directly. RoboCraft~\citep{arxiv_2205_02909} represents deformable plasticine as particles reconstructed from RGB-D point clouds, with graph edges connecting nearby particles and gripper--object pairs. A graph network predicts state changes through repeated message passing, allowing forces to propagate through the material. Trained on about ten minutes of real interaction, the model supports model-predictive control to shape plasticine into targets unseen during training. However, particle states must be estimated from partially occluded point clouds using shape and physics priors. Since particles lack temporal correspondence, training compares point distributions rather than individual trajectories. Moreover, the model cannot predict interactions involving entities absent from its graph.

\noindent\textbf{Latent-state Prediction.} A latent world model encodes history as \(z_t=e_\phi(h_t)\) and predicts future states through \(z_{t+1}\sim p_\phi(z_{t+1}\mid z_t,a_t)\), where \(h_t\) is the interaction history or its summary, \(z_t\) is its learned encoding, \(a_t\) is an action, and \(\phi\) denotes model parameters. Prediction heads interpret these states as rewards, values, goal distances, or observations. PlaNet and Dreamer learn latent dynamics for control~\citep{hafner2018latent,arxiv_2301_04104}, while V-JEPA~2 and LeWorldModel support planning without rendering future observations~\citep{assran2025vjepa,maes2026leworldmodel}. Alternatively, DINO-WM~\citep{arxiv_2411_04983} freezes a pretrained DINOv2 encoder and trains a transformer to predict action-conditioned patch features. Its separately trained decoder visualizes predictions but is not used for planning.

Compression is useful when it preserves task-relevant information. Reward-based models may discard irrelevant appearance details, while goal-image planners require visual correspondence to track progress. Reconstructing unnecessary details increases model cost. Models trained on action-free video may need action-conditioned training to predict intervention outcomes, and one-step accuracy does not guarantee reliable multi-step predictions.
The prediction space should match its intended use: images for visual control or evaluation, structured variables for explicit physical interactions, and latent states when task progress can be assessed without pixels. Any information omitted from the representation also limits the predictions and downstream decisions.

\subsection{Planning and Action Selection}\label{sec:3.3}

Planning translates task goals into executable behavior through complementary operations. Skill-based planning selects or composes available behaviors (\S~\ref{sec:3.3.1}), while subgoal interfaces specify what a lower-level policy should achieve (\S~\ref{sec:3.3.2}). Planning with world models compares predicted consequences of candidate actions (\S~\ref{sec:3.3.3}), and feedback-driven replanning revises a choice when new evidence changes the task-state estimate (\S~\ref{sec:3.3.4}). This order moves from the plan's executable units and handoff to prospective evaluation and revision, without requiring every system to use every operation.

\subsubsection{Skill-Based Planning}\label{sec:3.3.1}

Skill-based planning composes executable robot behaviors by selecting predefined skills, generating programs that invoke perception and control functions, or searching symbolic actions with feasible continuous parameters. All approaches depend on available skills, valid execution conditions, and observable completion.

\noindent\textbf{Skill Selection.} Skill-based planning lets language models select from predefined robot behaviors. SayCan~\citep{arxiv_2204_01691} combines language-based skill relevance with learned value functions estimating feasibility, enabling long-horizon tasks on a mobile manipulator. SayPlan extends this approach to building-scale environments using 3D scene graphs~\citep{rana2023sayplan}. However, execution depends on current skill feasibility, and tasks requiring skills absent from the library remain unsupported.

\noindent\textbf{Code Generation.} Program-based planning supports loops, perception-dependent conditions, and numerical parameters beyond simple skill sequences. Code as Policies~\citep{arxiv_2209_07753} uses a language model to generate programs combining perception and control functions, enabling feedback loops, reactive control, and parameter selection. In Minecraft, Voyager stores generated programs as reusable skills~\citep{wang2023voyager}, while VoxPoser and ReKep generate code for geometric subgoals. Although code makes planning logic explicit and testable, reliability depends on runtime execution, available functions, and explicit checks of necessary preconditions.

\noindent\textbf{Symbolic Planning with Feasibility Checks.} Task and motion planning integrates feasibility into action search. PDDLStream~\citep{arxiv_1802_08705} combines symbolic actions and preconditions with samplers that generate and validate continuous parameters, including grasps, placements, and collision-free motions. It searches over symbolic plans, fills missing parameters through sampling, and revises plans when sampling fails. However, its guarantees depend on manually specified actions, conditions, and geometric models. Text2Motion~\citep{arxiv_2303_12153} similarly evaluates skill-sequence feasibility using learned skill-value functions and geometric reasoning.

\subsubsection{Subgoal Interfaces}\label{sec:3.3.2}

A lower-level policy needs an interpretable, actionable target. A planner can provide a verbal instruction, a geometric or visual goal, or a learned feature rather than just a skill name. These formats vary in interpretability and execution requirements, and can coexist within a hierarchy. The key questions are what is communicated, when it can be executed, and what feedback triggers a new subgoal.

\noindent\textbf{Language Subgoals.} Language can serve as an interface to lower-level policies. RT-H~\citep{arxiv_2403_01823} uses language-described motions (e.g., ``move arm forward'') as an intermediate representation between tasks and actions, enabling shared low-level behaviors and learning from human corrections. Hi Robot~\citep{arxiv_2502_19417} uses a vision--language model (VLM) to decompose complex tasks into executable steps based on instructions and user feedback, which are passed to a learned low-level policy across diverse robotic platforms and tasks.
\(\pi_{0.5}\)~\citep{intelligence2025pi05} integrates both levels in one model, predicting a language subtask (e.g., ``pick up the plate'') before generating conditioned action chunks, with less frequent high-level inference. Gemini Robotics~1.5~\citep{geminiroboticsteam2025geminib} interleaves actions with natural-language reasoning, supported by a separate model for planning and progress estimation. Language subgoals enable human interpretation and correction but require reliable grounding and execution feedback to detect mismatches.

Reasoning within a single policy does not necessarily establish a hierarchical interface. ECoT~\citep{zawalski2024robotic} adds task reasoning and visual grounding to OpenVLA without additional robot data. It reuses high-level reasoning for five steps while updating low-level reasoning and actions at each step, making inference cost dependent on reasoning depth and refresh rate.

\noindent\textbf{Geometric and Visual Subgoals.} Spatial targets reduce ambiguity. VoxPoser~\citep{huang2023voxposer} uses language-generated code and VLMs to construct 3D value maps that guide model-based trajectory planning without predefined motion primitives. ReKep~\citep{huang2024rekep} encodes tasks as 3D keypoint constraints optimized into end-effector poses. Similarly, MolmoAct and Embodied-R1 use image-space targets, while SuSIE generates goal images for goal-conditioned policies. These representations enable geometric verification before execution but cannot capture physical factors such as grip force or object slippage.

\noindent\textbf{Latent Subgoals.} Learned features also enable multi-rate hierarchies within a VLA. HiRT~\citep{zhang2024hirt} uses a low-frequency VLM to guide a high-frequency vision policy, doubling control frequency with comparable static-task success and improving dynamic-task success from 48\% to 75\%. Fast-in-Slow~\citep{chen2025fastinslow} reuses the slow VLM's final layers for fast control, sharing pretrained knowledge. Hi Robot and \(\pi_{0.5}\) similarly separate slow language planning from fast actions. However, asynchronous updates risk stale or inconsistent guidance. The slow policy periodically refreshes a subgoal or feature, while the fast policy uses that guidance together with the latest observation. Under periodic updates, guidance can age by one slow-policy update interval, with inference and communication adding further delay. UniFS~\citep{sun2026unifs} balances this staleness--efficiency trade-off by updating VLM layer groups at different frequencies, reducing average LIBERO inference latency from 36.5 to 17.8\,ms.


Skill names invoke predefined behaviors, programs specify control flow, language enables correction, geometric targets support verification, and latent features enable joint learning. Regardless of format, an effective interface requires executable goals and observable completion conditions. A valid unfinished step can continue, while advancing requires evidence of completion and support for the next step's preconditions. Level-specific success distinguishes planning from execution failures, while tracking subgoal age helps identify stale guidance at the interface.

\begin{figure}[!htbp]
\centering
\includegraphics[width=\linewidth]{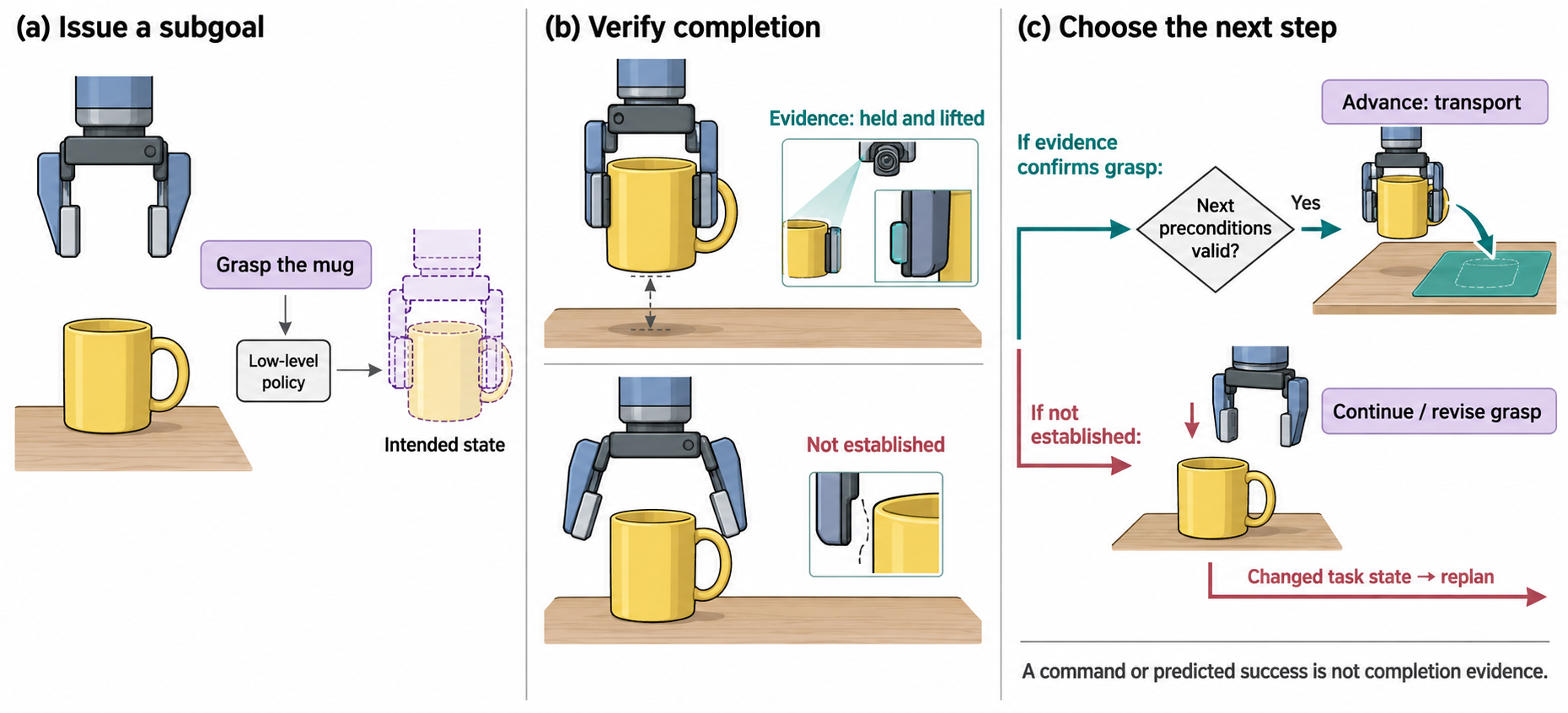}
\caption{An evidence-dependent subgoal handoff. (a) A grasp instruction specifies an intended state. (b) Observations distinguish a held-and-lifted mug from a grasp whose completion is not established; contact alone is insufficient. (c) Advancing to transport requires completion evidence and valid next-step preconditions. Otherwise the system may continue or revise the grasp, or replan when the task state changes. The scenes illustrate alternative branches, not a guaranteed sequence from a command to successful execution.}\label{fig:subgoal-handoff}
\end{figure}

Figure~\ref{fig:subgoal-handoff} makes this handoff concrete. A grasp instruction specifies what should happen; observed lifting and grasp evidence determine whether the planner can advance to transport. If completion is not established, the current skill may continue or be revised. A changed task state or invalid next-step precondition instead requires reassessing the plan. These branches separate a planning decision from evidence of its execution.

\subsubsection{Planning with World Models}\label{sec:3.3.3}

Prediction supports planning by either generating future goals for a controller to achieve or evaluating the outcomes of candidate actions. Both require grounding in current observations but impose different demands: goal prediction requires a controller capable of realizing the imagined state, while action evaluation requires a model that accurately captures action-dependent outcomes.

\noindent\textbf{Generated Plans and Subgoals.} UniPi~\citep{arxiv_2302_00111} generates task-conditioned videos for planning and extracts actions through inverse dynamics. This separates future-state prediction from action execution but introduces two failure modes: physically unreachable predictions and inaccurate action extraction. Evaluating the full pipeline is therefore essential to distinguish executable plans from merely plausible videos.
SuSIE~\citep{black2023zeroshot} generates visual subgoals using a pretrained image-editing model, which a goal-conditioned policy executes. Compared with video plans, subgoals offer greater control flexibility but may hide infeasible transitions. Its CALVIN and real-world manipulation results evaluate the combined system without isolating subgoal-generation quality from low-level execution capability.

\noindent\textbf{Candidate Evaluation.} Action-conditioned world models plan by predicting and ranking candidate action sequences. A representative reward-based objective is
\begin{equation}
a^\star_{t:t+H-1}=\arg\max_{a_{t:t+H-1}}\ \mathbb E_{p_\phi}\Big[\textstyle\sum_{j=0}^{H-1}\gamma^j\,\hat r_\phi(z_{t+j},a_{t+j})+\gamma^H\hat V(z_{t+H})\Big].\label{eq:plan}
\end{equation}
Here, \(\hat r_\phi\) predicts immediate reward, \(\hat V\) estimates value beyond horizon \(H\), and \(\gamma\) discounts future returns. The system replans after execution using new observations. Goal-conditioned planners may instead minimize predicted distance to a target. Both require reliable ranking of candidate outcomes.

Sampling-based planners approximate this optimization. PlaNet~\citep{hafner2018latent} uses the cross-entropy method (CEM), iteratively sampling action sequences and refitting a Gaussian to high-return candidates. TD-MPC2~\citep{arxiv_2310_16828} uses model predictive path integral (MPPI) control, weighting elite sequences by predicted return, incorporating a learned policy prior, and estimating value beyond the planning horizon. DINO-WM applies CEM to minimize predicted goal distance in frozen feature space. V-JEPA~2-AC~\citep{assran2025vjepa} similarly plans toward image goals using action-conditioned representations adapted from video pretraining.
Sampling-based optimization can exploit model errors by favoring actions with overestimated returns. Longer horizons and more candidates increase both error exposure and computational latency. LeWorldModel~\citep{maes2026leworldmodel} addresses computational cost through compact prediction. Effective planning therefore requires accurate ranking of relevant actions and sufficiently fast replanning to avoid stale observations.

\subsubsection{Feedback-Driven Replanning}\label{sec:3.3.4}

Hierarchical systems should correct errors at the appropriate level: minor offsets within a skill, changed targets through replanning, and ambiguous intent through human clarification. Each requires accurate task-state estimation, since issuing an action does not guarantee success.

\noindent\textbf{Local Correction.} Local correction avoids frequent replanning. Code as Policies~\citep{arxiv_2209_07753} supports feedback loops and reactive controllers, while VoxPoser~\citep{huang2023voxposer} generates closed-loop trajectories robust to perturbations. ReKep~\citep{huang2024rekep} continuously re-optimizes keypoint constraints for reactive control without task-specific data. However, local correction is limited to the current skill's goal: a moved cup can be tracked, but a fallen cup may require a new subgoal or skill.

\noindent\textbf{Replanning.} When a subtask becomes invalid, replanning requires reliable execution feedback. SayCan~\citep{arxiv_2204_01691} uses learned skill feasibility to guide selection but does not verify completion. Inner Monologue~\citep{huang2022inner} feeds success detection, scene descriptions, and human feedback into a language-model planner, improving long-horizon task completion. RoCo~\citep{mandi2023roco} uses environment feedback, including collision checks, to revise plans and waypoints. REFLECT~\citep{liu2023reflect} instead analyzes summarized multisensory experience after failure to improve subsequent attempts. Effective replanning requires feedback that changes future decisions, not merely periodic planner calls. Its reliability also depends on accurate outcome detection, since false success reports propagate incorrect task-state estimates.

\noindent\textbf{Help-seeking.} When uncertainty prevents reliable decisions, the system may seek human assistance. KnowNo~\citep{arxiv_2307_01928} uses conformal prediction to request help when multiple plans remain plausible. Hi Robot~\citep{arxiv_2502_19417} incorporates situated user feedback to revise hierarchical decisions, while RT-H accepts language-based motion corrections. KnowNo uses feedback to determine whether autonomous execution can proceed, whereas Hi Robot uses it to update subsequent choices. Effective help-seeking requires questions humans can answer and responses that the robot can translate into executable actions.

Across local correction, replanning, and help-seeking, response time determines which failures can still be corrected. Feedback operates at different timescales: language-model deliberation may take seconds, while physical control requires faster updates. Excessive communication and reasoning can therefore increase latency without improving success~\citep{wan2025ispass,wan2025reca}. Practical systems reduce this overhead through event-triggered reasoning, cached plans, compact memories, and reactive control~\citep{wan2025ispass}. ReKep exemplifies this by re-solving predefined constraints at control rate, while Fast-ThinkAct~\citep{huang2026fastthinkact} distills explicit reasoning into compact latent representations for faster inference. Reducing deliberation time is useful only if deviations are detected soon enough to trigger local correction, replanning, or help-seeking, so runtime monitoring must support the chosen correction timescale.

Across the planning interfaces considered here, effective planning requires knowing the controller's capabilities and when control returns. Skill libraries simplify selection and failure attribution but limit available behaviors. Flexible subgoals offer greater freedom, explicit constraints expose infeasibility, and latent interfaces reduce manual design at the cost of interpretability. These interfaces can coexist with clear completion conditions, allowing local feedback to correct routine errors and replanning to address significant deviations.
Checking skill preconditions can prevent failures from inappropriate selection more effectively than increasing planner complexity. Event-triggered or distilled replanning reduces deliberation cost, but its effectiveness depends on whether the robot resumes progress after deviations.

\subsection{Action Generation}\label{sec:3.4}

An action generator outputs actions at the policy interface, such as joint or end-effector targets, either as a single action or an action chunk. The execution system then converts these targets into control inputs at a specified layer. Since regression, token-based, diffusion, and flow-matching methods can all generate chunks, the action horizon must be specified independently of the generation method. We define the local policy as:

\begin{equation}
a_{t:t+H-1}\sim\pi_\theta(\cdot\mid h_t,g),\qquad 1\leq K\leq H.\label{eq:1}
\end{equation}

Here, \(h_t\) is the interaction history, \(g\) is the task condition, and \(\theta\) represents model parameters. \(H\) is the generated action horizon, while \(K\) is the number of steps executed before replanning. Actions may represent poses, joint targets, or other interface quantities, which are converted into control inputs \(u\) as described in Section~\ref{sec:6.3}. Deterministic regression is included as a special case. Equal horizons do not necessarily imply equal physical durations or feedback intervals, as these also depend on the interface frequency.

For a VLA policy, Eq.~\eqref{eq:1} typically involves a pretrained vision--language backbone that encodes visual, language, and optional proprioceptive inputs~\citep{beyer2024paligemma,zhai2023sigmoid}, followed by an action-generation head that produces an action chunk. This is a functional decomposition and does not imply separately trained or executed networks. We denote the task-conditioned representation by \(c\) and the action-head parameters by \(\psi\). Proprioception is optional and can be incorporated through either component. RT-2~\citep{arxiv_2307_15818} and OpenVLA~\citep{arxiv_2406_09246} use transformers for multimodal processing and action generation without requiring proprioception, whereas \(\pi_0\)~\citep{arxiv_2410_24164} and GR00T~N1~\citep{bjorck2025groot} employ separate action experts. Both input availability and architectural choices influence policy behavior~\citep{arxiv_2405_12213,geminiroboticsteam2025gemini}, as further examined by VLANeXt~\citep{wu2026vlanext} under a unified setup.

\begin{figure}[!htbp]
\centering
\includegraphics[width=\linewidth]{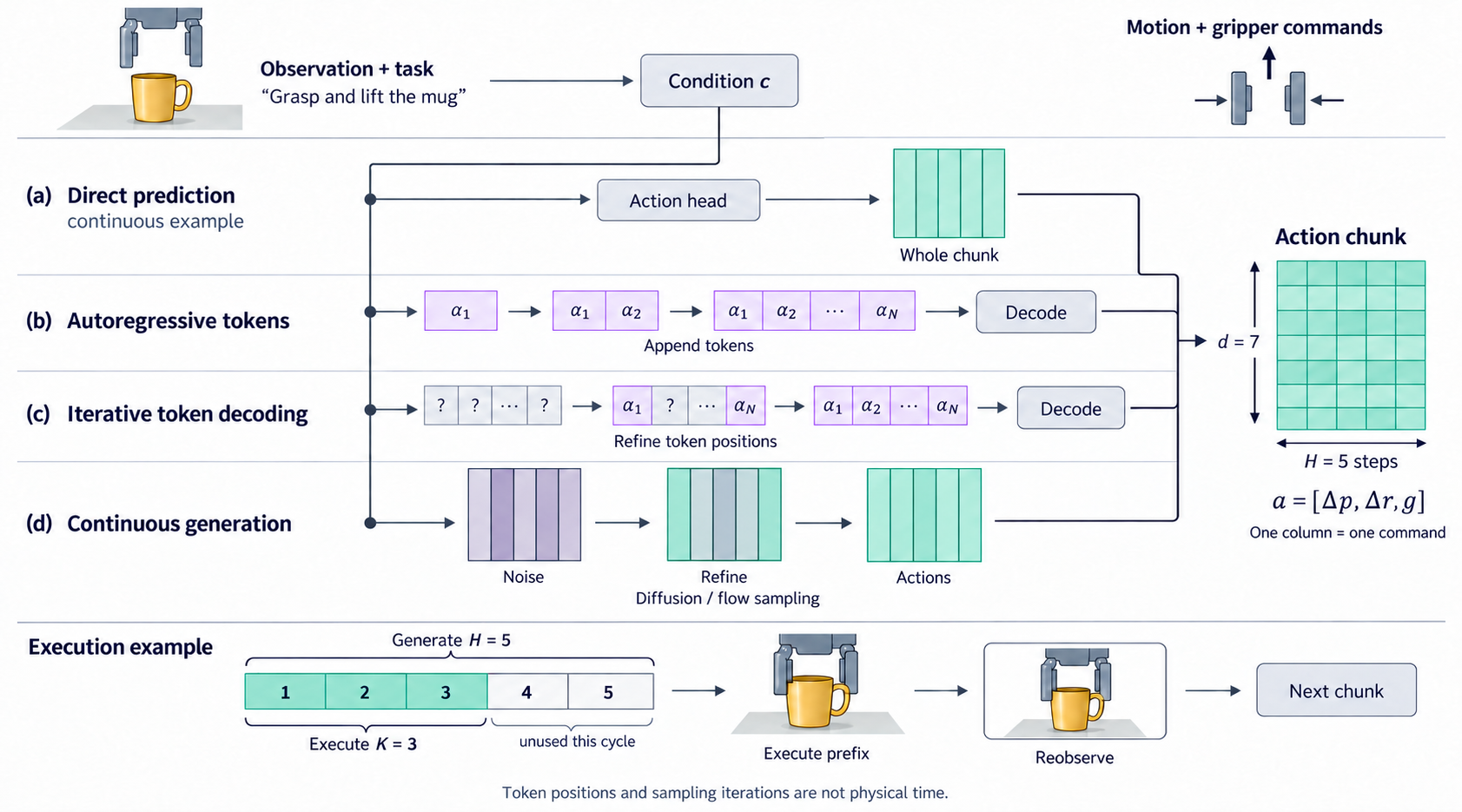}
\caption{Representative mechanisms for generating an action chunk from condition $c$: (a) direct continuous prediction, (b) autoregressive token decoding, (c) iterative token decoding, and (d) continuous diffusion or flow sampling. Discrete tokens are decoded into numerical commands. Small continuous sketches show command-vector slots; the shared output expands five commands into seven dimensions: three translation increments, three rotation increments, and one gripper command. The unlettered execution example uses $K=3$ of $H=5$ steps before reobservation; local control feedback may continue. Token positions, sampling iterations, and physical time are distinct.}\label{fig:overview-generation}
\end{figure}

Output representation and generation mechanism are separate choices (Figure~\ref{fig:overview-generation}). Action sequence modeling first distinguishes a generated chunk from its executed prefix (\S~\ref{sec:3.4.1}). An action chunk can be represented as continuous values, coordinate-wise discrete tokens, or temporally compressed tokens. Continuous commands may be regressed directly (\S~\ref{sec:3.4.2}) or generated iteratively (\S~\ref{sec:3.4.4}); discrete tokens may be predicted in parallel, autoregressively, or through iterative decoding (\S~\ref{sec:3.4.3}). Discrete outputs must be decoded into numerical commands. The encoding therefore determines neither the generation process nor the execution horizon. The figure shows representative mechanisms rather than every encoding--decoder pairing; its direct row is a continuous example, and its execution strip is a separate scheduling choice.

\subsubsection{Action Sequence Modeling}\label{sec:3.4.1}

Jointly generating an action chunk captures dependencies across steps, enabling coordinated behaviors such as obstacle avoidance and bimanual manipulation while reducing inference frequency. However, longer executed prefixes rely on increasingly outdated observations. In Eq.~\eqref{eq:1}, \(H\) defines the predicted horizon and \(K\) the executed prefix; selecting \(K\) and connecting successive chunks are execution-level decisions. At a fixed action rate, increasing \(K\) lengthens the interval before the next policy update, while increasing \(H\) alone does not set that interval. Local controller feedback may continue during prefix execution. Figure~\ref{fig:overview-generation}(a)--(d) shows generation alternatives, while the unlettered bottom strip illustrates prefix execution: the generated horizon $H$ and executed prefix $K$ need not be equal, and new observations inform the next request.

ACT~\citep{arxiv_2304_13705} uses a transformer and a training-time latent variable to model demonstration variability (\S~\ref{sec:5.2.2}), but adopts the prior-mean latent at inference rather than sampling behaviors. In contrast, Diffusion Policy can sample multiple valid behaviors~\citep{arxiv_2303_04137}. OpenVLA-OFT~\citep{arxiv_2502_19645} combines chunking, parallel decoding, and an L1 objective. Thus, temporal modeling and multimodal behavior selection are distinct: chunking alone does not resolve regression-induced mode averaging.
Attention access determines which information each output position can use. Next-token generation uses an available prefix, whereas a bidirectional refinement step can access the current noisy chunk; both can be conditioned on \(c\). These dependencies, reusable conditioning, and repeated head computations determine chunk-generation cost (Section~\ref{sec:7.2.3}).

\subsubsection{Direct Action Prediction}\label{sec:3.4.2}

Direct action prediction maps task-conditioned representations to numerical control targets, either as single actions or sequences. Its effectiveness depends on whether the regression objective and conditioning can preserve valid behaviors when demonstrations contain multiple solutions.
Let \(\mathbf a=a_{t:t+H-1}\) denote a demonstrated action chunk of dimension \(H\times d\), and \(\hat{\mathbf a}_\theta(h_t,g)\) its prediction. OpenVLA-OFT~\citep{arxiv_2502_19645} uses the L1 regression objective:

\begin{equation}
\mathcal L_{\mathrm{reg}}(\theta)=\mathbb E_{(h_t,g,\mathbf a)\sim\mathcal D}\big\|\hat{\mathbf a}_\theta(h_t,g)-\mathbf a\big\|_1 .\label{eq:reg}
\end{equation}

The L1 norm sums absolute errors over all \(H\cdot d\) entries; replacing it with squared L2 error yields the L2 objective. When demonstrations disagree, L1 favors the coordinate-wise median, whereas L2 favors the mean. This deterministic formulation is a special case of Eq.~\eqref{eq:1}, while probabilistic policies use the likelihood objective in Eq.~\eqref{eq:2a} (\S~\ref{sec:5.2.1}).
Regression performance depends on how well conditioning determines behavior. When demonstrations contain distinct strategies, L2 may average incompatible trajectories, while L1 may combine coordinates into an invalid chunk. Neither inherently preserves multiple modes, motivating latent-variable, discrete-sequence, and continuous distributional methods. Evaluating mode averaging requires distinct strategies under matched conditioning, with the loss, \(H\), \(K\), and action-interface rate specified.


\subsubsection{Token-Based Action Generation}\label{sec:3.4.3}

Token-based policies discretize continuous actions into tokens, generate them with a sequence model, and decode them into control commands. RT-2~\citep{arxiv_2307_15818} and OpenVLA~\citep{arxiv_2406_09246} leverage VLM vocabularies and training objectives for action prediction. Preserving distinct valid behaviors depends on conditioning, the learned distribution, and decoding. Tokenization must balance control precision and generation cost, while binning, token order, and temporal compression determine the dependencies the model learns.
These designs separate action representation from generation. Representation determines whether tokens encode individual coordinates or compressed action segments, while generation determines token dependencies through autoregressive, parallel, or iterative decoding. Hybrid architectures may also combine token-based training with continuous action generation.

\noindent\textbf{Action Tokenization.} RT-2~\citep{arxiv_2307_15818} and OpenVLA~\citep{arxiv_2406_09246} autoregressively generate coordinate tokens for a single action (\(H=1\)), encoding six end-effector displacement components and a gripper command; RT-2 additionally predicts episode termination.
Per-dimension quantization produces long, redundant token sequences, particularly for high-frequency dexterous actions. FAST~\citep{arxiv_2501_09747} addresses this using temporal discrete cosine transforms, coefficient quantization, and byte-pair encoding, while retaining autoregressive decoding. This compresses smooth trajectories into fewer tokens. With comparable backbones and training data, \(\pi_0\)-FAST matches flow-matching \(\pi_0\) using roughly five times fewer training GPU-hours, though autoregressive inference remains sequential. Encoding choices therefore affect token count and reconstruction error as well as the sequence length presented to the decoder.

\noindent\textbf{Token Decoding.} For the task-conditioned representation \(c\) and action tokens \(\alpha_{1:N}\), the autoregressive factorization is

\begin{equation}
p_\theta(\alpha_{1:N}\mid c)=\prod_{j=1}^{N}p_\theta(\alpha_j\mid\alpha_{<j},c).\label{eq:ar}
\end{equation}

Here, each token \(\alpha_j\) depends on preceding tokens \(\alpha_{<j}\) and \(c\), and the generated sequence is decoded into numerical actions. For coordinate-wise tokenization, \(N=H\cdot d\), while temporal compression reduces token count. Although this formulation supports longer chunks, chunking is optional; RT-1~\citep{arxiv_2212_06817} also discretizes actions without autoregressive token decoding. Tokenization supports heterogeneous outputs but introduces quantization error; autoregressive decoding adds sequential computation. As \(H\) increases, token count and decoding latency grow, with early prediction errors potentially propagating through the sequence. When trained with teacher forcing, the decoder conditions on demonstrated prefixes but must use its own generated prefixes at inference, introducing a further source of error beyond quantization.

Action tokens can also be generated in parallel. PD-VLA~\citep{song2025pdvla} retains an autoregressive model but updates all tokens in a chunk through parallel fixed-point iterations. Discrete Diffusion VLA~\citep{liang2025discrete} instead starts with fully masked action tokens and iteratively refines them within the VLM backbone using bidirectional attention, fixing high-confidence tokens while re-masking uncertain ones. On LIBERO-Goal, instruction paraphrasing reduced its success by only 0.8 points, compared with 8.0 for parallel discrete decoding. The authors attribute this robustness to sharing the pretrained VLM's token space and cross-entropy objective.

Hybrid training connects these choices to continuous action generation: token targets can train a backbone whose features condition a separate continuous head. Adding a newly initialized flow-matching or diffusion expert to a pretrained VLM may slow training and disrupt its knowledge. Knowledge insulation~\citep{driess2025knowledge} addresses this by training the backbone on FAST tokens and vision--language data while blocking expert gradients, allowing the flow-matching expert to generate continuous actions. Thus, discrete and continuous outputs are handled by separate components. Token-based designs can also be categorized by token type: language, code, latent action, or raw action~\citep{zhong2025survey}.

\subsubsection{Continuous Generative Policies}\label{sec:3.4.4}

Continuous generative policies sample action chunks from a learned conditional distribution, capturing multiple strategies without averaging them, though physical feasibility is not guaranteed. Diffusion heads reverse a gradual noising process, while flow-matching heads learn a velocity field along a noise-to-data path (Eq.~\eqref{eq:fm}). Standard implementations use iterative sampling but differ in training objectives and generation dynamics; distilled variants can reduce the number of sampling steps (\S~\ref{sec:7.2.3}).

\noindent\textbf{Diffusion Heads.} Diffusion Policy~\citep{arxiv_2303_04137} models visuomotor control as conditional denoising diffusion, learning a conditional score through denoising and iteratively refining noisy action sequences for receding-horizon execution. It reports an average improvement of 46.9\% across 12 tasks from four manipulation benchmarks, attributing its gains to multimodal action modeling, high-dimensional action handling, and stable training. Training uses noised actions and denoising supervision; inference generates actions through reverse sampling.
Diffusion-transformer heads demonstrate the scalability of action modules. RDT-1B~\citep{liu2024rdt1b} uses a 1.2B-parameter diffusion transformer pretrained on multi-robot data with a unified action space and fine-tuned on over 6,000 bimanual episodes, enabling skill learning from 1--5 demonstrations. CogACT~\citep{li2024cogact} attaches a separate diffusion action transformer to a 7B VLM, showing favorable scaling and outperforming OpenVLA by over 35 and 55 percentage points in simulation and real-world tasks, respectively. However, these gains cannot be attributed solely to head size.

\noindent\textbf{Flow-matching Heads.} Flow-matching heads, as in \(\pi_0\)~\citep{arxiv_2410_24164}, learn a velocity field that transports Gaussian noise into action chunks. Built on a pretrained VLM, \(\pi_0\) is trained on single-arm, dual-arm, and mobile manipulators for tasks including laundry folding, table cleaning, and box assembly. Given an action chunk \(\mathbf a\) (Eq.~\eqref{eq:reg}), noise \(\epsilon\sim\mathcal N(0,I)\), and the interpolated state \(\mathbf a^\sigma=\sigma\epsilon+(1-\sigma)\mathbf a\) at \(\sigma\in[0,1]\), a velocity-matching objective for this linear interpolation is

\begin{equation}
\mathcal L_{\mathrm{FM}}(\psi)=\mathbb E_{\sigma,\epsilon,(h_t,g,\mathbf a)\sim\mathcal D}\big\|v_\psi(\mathbf a^\sigma,\sigma,c,s_t^{\mathrm{prop}})-(\epsilon-\mathbf a)\big\|^2 .\label{eq:fm}
\end{equation}

Here, \(v_\psi\) predicts velocity conditioned on the task representation \(c\) and optional proprioceptive input \(s_t^{\mathrm{prop}}\) describing the body's configuration and motion, while \(\mathcal D\) denotes demonstrations and \(\epsilon-\mathbf a\) is the target velocity. Sampling integrates \(\mathbf a^{\sigma-\Delta\sigma}=\mathbf a^\sigma-\Delta\sigma\,v_\psi(\mathbf a^\sigma,\sigma,c,s_t^{\mathrm{prop}})\) from \(\sigma=1\) to \(0\) over \(N_\sigma\) Euler steps, with \(\Delta\sigma=1/N_\sigma\). When \(c\) remains fixed, only the action head is repeatedly evaluated. Thus, \(N_\sigma\) controls expert evaluations and can be reduced at deployment to trade sample quality for latency without retraining, although total latency also includes encoding and runtime overhead.
GR00T~N1~\citep{bjorck2025groot} applies the same objective to humanoid control, combining a vision--language module with a flow-matching diffusion-transformer action module. Trained jointly on robot trajectories, human videos, and synthetic data, it enables language-conditioned bimanual manipulation on the Fourier GR-1 humanoid.

Under the Gaussian linear path, velocity, noise, and clean-action predictions are equivalent parameterizations away from singular endpoints, e.g., \(\epsilon=\mathbf a^\sigma+(1-\sigma)v\). Thus, diffusion and flow matching overlap, differing mainly in path and noise schedule, loss weighting, and sampling rules rather than inherently stochastic versus smooth control.

The choice between these generative objectives and direct regression requires task-level evidence. Representing multiple behaviors is not always necessary. On LIBERO, OpenVLA-OFT reports similar average success rates for continuous L1 regression and diffusion heads~\citep{arxiv_2502_19645}. This suggests that the need for richer action distributions depends on the task and whether chunking and output design already address key challenges, without ruling out their benefits for contact-rich, dynamic, or multimodal tasks.

Action representations and generators should match the task's behavioral complexity. Direct regression avoids iterative generation but may collapse incompatible trajectory modes unless richer conditioning or strategy selection resolves ambiguity. Token-based methods leverage sequence modeling at costs shaped by precision, compression, and decoding. Continuous generative policies preserve multiple behaviors but incur sampling overhead. ACT lies between these approaches, modeling latent variation during training while using a fixed latent at inference.
The key question is how much distributional complexity remains necessary after conditioning and strategy selection. Separately, adaptive prefixes, event-triggered updates, and local feedback can improve responsiveness while retaining chunked execution.

\subsection{Model and System Architectures}\label{sec:3.5}

Once methods are assigned to each role, the architectural choice is how to connect them: a shared network can jointly process information and generate actions, while separate components expose interfaces for independent updates and execution rates. In the cup task, planning, VLA control, and memory can be unified or modular. The discussion begins with how policies and planners connect to action paths and update at different rates (\S~\ref{sec:3.5.1}), then examines how future prediction can be coupled to action generation during training or inference (\S~\ref{sec:3.5.2}). It closes by comparing architectures against differences in scale, training coverage, action interfaces, and evaluation conditions (\S~\ref{sec:3.5.3}).

\subsubsection{Policy and Planner Organization}\label{sec:3.5.1}

Planners can select skills or generate executable programs, as in SayCan and Code as Policies~\citep{arxiv_2204_01691,arxiv_2209_07753}. Alternatively, shared models can generate action tokens, as in RT-2 and OpenVLA~\citep{arxiv_2307_15818,arxiv_2406_09246}, or condition continuous action experts, as in \(\pi_0\)~\citep{arxiv_2410_24164}, where the backbone and expert are coupled through attention within one transformer.
Components may operate at different rates. GR00T~N1 passes features from a slower vision--language module to a faster action module~\citep{bjorck2025groot}, while \(\pi_{0.5}\) and Gemini Robotics~1.5 use explicit subtask text or reasoning~\citep{intelligence2025pi05,geminiroboticsteam2025geminib}. Text provides interpretable intermediate decisions, whereas latent features rely on downstream interpretation; both may become stale before execution.

\subsubsection{World--Action Models}\label{sec:3.5.2}

World--action models connect future observation or state prediction with action generation. Video provides broad predictive supervision, while robot interaction grounds actions. These targets can be combined through shared training, joint generation, or predictive features conditioning a policy. The architecture determines whether future prediction is required at inference. The prediction spaces in Figure~\ref{fig:overview-prediction} can support different roles: evaluating candidate actions, conditioning a policy, or supplying training targets and imagined experience. A policy may also select actions directly without an explicit future prediction at inference.

\noindent\textbf{Shared Training Targets.} GR-1~\citep{arxiv_2312_13139} pretrains a GPT-style transformer on future-frame prediction and fine-tunes it to jointly predict images and actions from language, visual history, and proprioception. Video prediction shapes the representation, but only actions are needed at inference. GR-2~\citep{cheang2024gr2} extends this approach with 38 million video clips, while Seer~\citep{tian2024predictive} conditions inverse-dynamics action prediction on predicted future visual states, requiring prediction at inference. WorldVLA~\citep{cen2025worldvla} jointly generates image and action tokens autoregressively, making action quality sensitive to output ordering, masking, and prediction errors. Thus, the benefits of predictive supervision depend on training objectives, attention structure, and inference dependencies, rather than parameter sharing alone. Ablating a predictive training objective requires retraining; bypassing an optional inference branch tests computation under fixed learned parameters.

\noindent\textbf{Joint Generation.} UWM~\citep{zhu2025unified} unifies action and video diffusion in one transformer with modality-specific timesteps. Setting a timestep to zero conditions on that modality, holding it at maximum noise approximately marginalizes it, and denoising generates it, enabling policy, forward-dynamics, inverse-dynamics, and video-generation tasks. Cosmos Policy~\citep{kim2026cosmos} extends pretrained video generation to actions, future observations, and values, using predictions to evaluate candidates at inference. While sharing representations across roles, these designs risk correlated errors: agreement among a policy, simulator, and evaluator sharing the same flawed prediction provides weaker evidence than independent observation.

\noindent\textbf{Predictive Features as Policy Input.} Policies can use predictive features without generating future images. VPP~\citep{hu2024video} learns actions from video-model representations, while the Unified Video Action model~\citep{li2025unified} and GigaWorld-Policy~\citep{ye2026gigaworldpolicy} allow action inference without video generation. LAPA and Genie~\citep{ye2024latent,bruce2024genie} learn latent actions from action-free videos but still require embodiment-specific action mapping. Fast-WAM~\citep{fastwam2026} finds that video co-training can help without inference-time generation. On real-robot towel folding, its inverse-dynamics variant achieves the highest success among its variants, while action-only inference improves completion time; pretrained $\pi_{0.5}$ outperforms both in the reported comparison. In contrast, AHEAD uses execution-time prediction to address delays in a frozen policy. These results suggest that prediction should be used where it informs decisions, without establishing a universal deployment rule.

Across these designs, evaluation should establish whether prediction improves learning, action selection, or data efficiency, specifying what is predicted, which component uses it, how real observations correct errors, and what computation remains at inference~\citep{wang2026wam}. Images retain visual detail but incur generation and interpretation costs; compact latent states or physical quantities reduce computation but may omit critical contact information. Compare these representations under a fixed decision procedure. Online search adds latency and model-error risks, while imagined learning shifts simulation to training, where short rollouts and real corrections can limit errors. Predictive supervision can support action-only inference, as Fast-WAM suggests, although its real-world evidence remains limited. Prediction is therefore most useful online when it changes consequential decisions; it cannot verify task completion. The following discussion examines model scale, training coverage, and architectural comparisons.

\subsubsection{Architecture Comparisons}\label{sec:3.5.3}

VLA performance depends on more than parameter count, requiring a balance of semantic capacity, action precision, horizon, control frequency, and robustness. Compact models such as TinyVLA, SmolVLA, and VLA-Adapter~\citep{wen2024tinyvla,shukor2025smolvla,wang2025vlaadapter} depend on suitable action interfaces and training distributions. Larger models such as RT-2 and \(\pi_{0.5}\)~\citep{arxiv_2307_15818,intelligence2025pi05} leverage broader web, robot, and subtask data, though their gains cannot be attributed solely to size. Cross-embodiment approaches include body-specific tokenizers (HPT), cross-body training (CrossFormer), soft prompts (X-VLA), and latent actions (UniVLA)~\citep{wang2024scaling,doshi2024scaling,zheng2025xvla,bu2025univla}.
Blind randomized evaluations of large behavior models based on Diffusion Policy~\citep{trilbmteam2025careful} show that multitask pretraining improves success, robustness, and data efficiency on new tasks, with gains increasing under the tested scale and diversity settings. However, these findings do not establish universal advantages for particular backbone sizes or action heads.

Representative developments span planning and grounding, robot policies and resources, and world or world--action models. Early modular systems linked language reasoning to executable skills (SayCan, Code as Policies, Inner Monologue)~\citep{arxiv_2204_01691,arxiv_2209_07753,huang2022inner}. Later work expanded shared robot data and reusable policies (Open X-Embodiment, Octo, OpenVLA)~\citep{arxiv_2310_08864,arxiv_2405_12213,arxiv_2406_09246}, followed by advances in continuous action experts, broader co-training, deployment learning, prediction, and memory (\(\pi_0\), \(\pi_{0.5}\), \(\pi^{*}_{0.6}\), Cosmos Policy, MEM)~\citep{arxiv_2410_24164,intelligence2025pi05,intelligence2025pi06,kim2026cosmos,torne2026mem}. This historical progression does not imply replacement, causal lineage, or performance ranking; related surveys cover these overlapping developments~\citep{psiris2026fmreview,ma2026vlasurvey,bai2026manipulation}.


The reviewed architectures differ in training data, action interfaces, and deployment conditions, preventing a definitive ranking. High LIBERO scores demonstrate progress under specific protocols but do not establish open-world instruction following, recovery, or cross-embodiment robustness. Comparing shared and modular designs therefore requires reporting the data used to train each role, the action interface, feedback timing, and the physical conditions under which completion is measured.
Two questions remain: what information and prediction horizon are needed for robust decisions across changing objects, scenes, and embodiments; and whether geometry, memory, prediction, and multimodal action modeling justify their costs over simpler alternatives under matched data and computation. Section~\ref{sec:4} examines the data needed to train and compare these learned roles.

\section{Data for Physical AI}\label{sec:4}

Robots learn skills from experience, but available data is limited and uneven. Real robot interactions capture observations, actions, and outcomes, yet are costly to collect and cover only specific situations. Images, human videos, simulation, and synthetic data can expand coverage, but each provides only partial information. Their learning value depends on which fields are recorded or inferred, how actions correspond to the target robot, and whether outcomes can be checked.

\begin{figure}[!htbp]
\centering
\includegraphics[width=\linewidth]{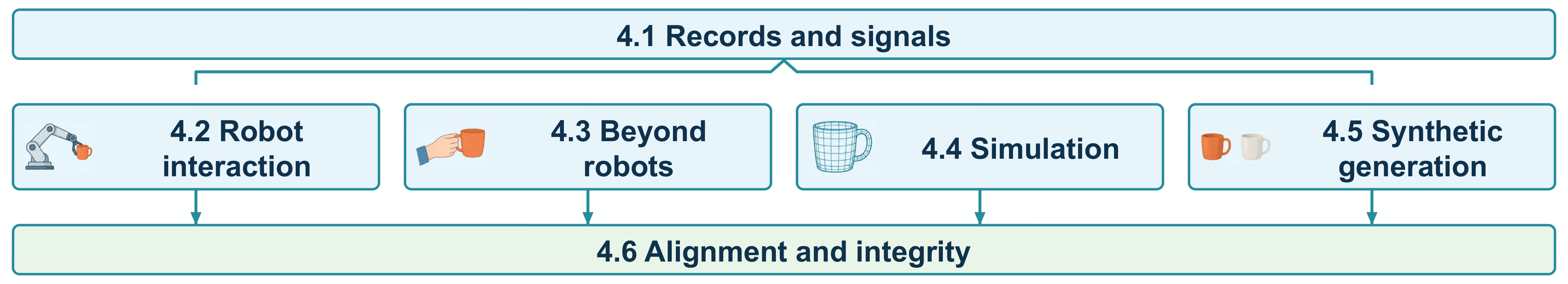}
\caption{Data sources and record quality. A common record schema organizes robot interaction, non-robot sources, simulation, and synthetic generation. Alignment and integrity apply across all four sources, whose categories may overlap and whose available fields may differ.}\label{fig:overview-data}
\end{figure}

These sources contribute different aspects of a skill: for example, a labeled image identifies a cup, a teleoperated demonstration shows how to scrub it, and a failed grasp followed by correction teaches recovery. The chapter first identifies record fields and learning signals (\S~\ref{sec:4.1}), providing a common basis for comparing robot interaction data (\S~\ref{sec:4.2}) with sources beyond robot interaction (\S~\ref{sec:4.3}). It then examines how simulation (\S~\ref{sec:4.4}) and synthetic generation (\S~\ref{sec:4.5}) expand available records, before addressing the alignment and integrity checks needed to use these records (\S~\ref{sec:4.6}). The organization moves from what a record contains to how it is obtained and what makes it usable; the collection and generation methods can overlap.
Figure~\ref{fig:overview-data} connects the chapter's three organizing questions: what a record contains, which sources can supply it, and which checks make it usable for learning.

\FloatBarrier

\subsection{Data Records and Learning Signals}\label{sec:4.1}

A recorded robot trajectory typically contains timestamped observations, action commands, and task conditions, and may include states, rewards, outcomes, or operator interventions. The discussion first identifies these fields and their learning signals, then compares their availability across data sources in Table~\ref{tab:recordfields}, before considering situation coverage, label validity, and collection cost. Observations support representation learning and prediction, actions supervise policies, outcomes evaluate consequences, and interventions capture corrections. Not all datasets contain every field, and recorded commands, measured motion, and inferred actions must remain distinguishable. Action correspondence defines how source actions or motions map to the target robot's action interface.

Table~\ref{tab:recordfields} summarizes typical fields across sources. Images, annotations, and human videos support visual and temporal learning~\citep{arxiv_2103_00020,kirillov2023segment,assran2025vjepa,arxiv_2110_07058}; pose and handheld-tool data add motion information~\citep{hoque2025egodex,zheng2026egoscale,arxiv_2402_10329,wang2024dexcap}; and robot datasets provide actions and body-specific outcomes~\citep{arxiv_2310_08864,arxiv_2308_12952,arxiv_2403_12945,fang2023rh20t,wu2024robomind}. Simulation and generation further expand coverage under different validity assumptions~\citep{arxiv_2310_17596,jiang2024dexmimicgen,arxiv_2406_02523,jang2025dreamgen}. Dataset scale, physical correspondence, and collection cost require explicit evaluation. This source hierarchy connects the data-pyramid perspective of GR00T~N1~\citep{bjorck2025groot} with the broader survey by Ye et al.~\citep{ye2026datapyramid}.

\begin{center}\begin{minipage}{\linewidth}
\captionof{table}{Data sources, recorded fields, and learning uses. Availability is dataset-specific; inferred, mapped, and absent quantities must not be treated as directly recorded robot actions.}\label{tab:recordfields}
\small\renewcommand{\arraystretch}{1.15}
\begin{tabularx}{\linewidth}{@{}>{\RaggedRight\arraybackslash}p{0.19\linewidth}*{3}{>{\RaggedRight\arraybackslash}X}@{}}
\toprule
Source & Observed / recorded & Action correspondence & Learning use \\
\midrule
Image--text & Pixels, text; no motion interval & Robot action absent & Visual and language representations \\
Human video & Frames and timestamps & Actions usually inferred & Temporal features and future prediction \\
Pose / handheld tool & Motion / tool pose; optional force & Requires robot mapping & Motion supervision after body mapping \\
Robot demonstrations & Observations, commanded actions & Recorded for that body & Imitation and action-conditioned prediction \\
Intervention logs & Policy/expert segments and state & Labels may be partial & Corrective imitation and outcome learning \\
Simulation / generation & Configured states or synthesized views & Depends on the generator & Policy and model training under generator assumptions \\
\bottomrule
\end{tabularx}
\end{minipage}\end{center}

The key question is what missing information each source provides. Diverse object appearances improve recognition, while corrective demonstrations cover failure states often absent from successful trajectories. Shared representations and embodiment-aware interfaces help integrate heterogeneous data, as shown by HPT and Octo~\citep{wang2024scaling,arxiv_2405_12213}. Alignment ensures consistency across recordings, while selection and mixture weights control training exposure. Dataset utility depends on both situation coverage and available action and outcome labels, whereas collection cost includes human effort and discarded data.

\subsection{Robot Interaction Data}\label{sec:4.2}

Robot data collection aims to capture missing behaviors and failure states while reducing human effort. The parts are organized by who controls the robot and what supervision the record supplies: demonstrations provide expert actions (\S~\ref{sec:4.2.1}), autonomous collection captures policy-visited states and their outcomes (\S~\ref{sec:4.2.2}), and interventions record deviations, corrections, and control handoffs (\S~\ref{sec:4.2.3}). In the cup task, these correspond to human-guided washing, autonomous attempts, and operator corrections when the cup slips. A single collection process can combine all three.

\subsubsection{Demonstration Collection}\label{sec:4.2.1}

A demonstration is shaped by what the control interface allows and what the recording preserves. Three factors matter: the human control interface, diversity across scenes, tasks, and operators, and recorded data channels.

\noindent\textbf{Demonstration Interfaces.} In leader--follower teleoperation, operator motions are mapped to robot actions. ALOHA enables low-cost bimanual demonstrations for training ACT~\citep{arxiv_2304_13705}, while ALOHA~2 improves hardware robustness and ergonomics~\citep{alohateam2024aloha}. GELLO uses a kinematically matched input device across robot platforms~\citep{wu2023gello}. Matching leader and follower joint structures makes motion mapping explicit and facilitates coordinated control.
Other interfaces expand operator perception and control. Open-TeleVision~\citep{arxiv_2407_01512} combines stereo VR feedback with head, wrist, and hand pose retargeting for humanoid teleoperation. Mobile ALOHA~\citep{arxiv_2401_02117} integrates waist-guided base motion with bimanual arm control. Both require explicit mappings from operator motion to robot commands, including coordinate frames, scaling, constraints, and feedback, as these determine how human motions translate into robot actions.
Human motion following can also produce robot demonstrations. HumanPlus~\citep{arxiv_2406_10454} first trains a motion-following policy on human data, then collects demonstrations through real humanoid execution. Unlike human pose data, these records capture actual robot behavior, but should preserve deviations between intended and executed motions introduced by the following controller.

\noindent\textbf{Collection Organization.} Once an interface is established, collection can scale across scenes, tasks, operators, and robot platforms. DROID emphasizes diversity across sites and operators~\citep{arxiv_2403_12945}, while BridgeData~V2 covers varied environments, tasks, and objects on a low-cost platform~\citep{arxiv_2308_12952}. Lin et al.~\citep{lin2024data} show that generalization scales roughly as a power law with environment and object diversity, whereas additional demonstrations within the same setting yield diminishing returns.
AgiBot World combines large-scale collection with human verification~\citep{agibotworldcontributors2025agibot}, while Open X-Embodiment aggregates data across heterogeneous robots~\citep{arxiv_2310_08864}. However, differences in calibration, timing, annotations, operator practices, and action interfaces remain. Platform, site, task, and operator diversity therefore require distinct consideration and alignment.

\noindent\textbf{Recorded Modalities.} Images and action commands capture only part of an interaction. Motion measurements verify execution, contact sensing reveals visually ambiguous events, and task and outcome labels support different learning objectives. RH20T~\citep{fang2023rh20t} combines visual, force, audio, and action data across robots and viewpoints. The value of these channels depends on calibration, synchronization, and their availability during training or deployment. Timestamps, termination causes, and intervention records help interpret mismatches between actions, measurements, and outcomes, while missing data should be explicitly marked.

Demonstration datasets are most useful when they expand behavioral, environmental, or sensing coverage rather than merely increasing trajectory counts. Repeated demonstrations may yield diminishing returns, while collection costs include setup, calibration, resets, and rejected data alongside recording time.

\subsubsection{Autonomous Collection}\label{sec:4.2.2}

Expert demonstrations provide target behavior but rarely cover the states a deployed policy encounters, including unexpected failures and recovery situations. Autonomous collection captures these states but requires task selection, outcome labeling, scene resets, and safety constraints without constant human supervision.

\noindent\textbf{Task Selection.} AutoRT~\citep{arxiv_2401_12963} automates task selection in unfamiliar environments. A vision--language model describes the scene, a language model proposes tasks, and a robot constitution filters proposals based on safety and embodiment constraints. Tasks are matched to the capabilities of the selected collection policy, including teleoperation, scripted control, or RT-2. However, filtering rejected only 17 of 31 unsafe proposals, requiring human oversight. The dataset includes both successes and failures, with success rates of 82\% for teleoperation, 21\% for scripted policies, and below 5\% for RT-2.

\noindent\textbf{Outcome Labeling and Resets.} QT-Opt~\citep{arxiv_1806_10293} demonstrates scalable autonomous collection through simple outcome labeling and resets. Seven robot arms collected over 580,000 grasp attempts in roughly 800 robot-hours. Success was automatically determined by comparing bin images before and after grasping, while dropping objects back into the bin served as a reset. Collection shifted from scripted to learned policies once success reached 50\%. Fine-tuning with 28,000 on-policy grasps improved success on unseen objects from 87\% to 96\%, although the image-based test misclassified some successful grasps. More complex tasks, such as washing a cup, lack such simple success checks and resets.
For tasks without simple engineered labels or resets, these functions can be learned. RoboCat~\citep{arxiv_2306_11706} enables autonomous collection across tasks using learned success detectors and reset policies. Fine-tuned policies collect task-specific data, while vision-based reward models trained on crowd-sourced annotations detect success. Reset policies reuse complementary tasks, such as gear removal after insertion, to restore initial states. Human operators initiate and terminate collection, and stored episodes train successive generalist models, enabling improvement across generations rather than during execution.

\noindent\textbf{Collection Coupled to Learning.} When the learner drives collection, as in SERL and HIL-SERL~\citep{arxiv_2401_16013,arxiv_2410_21845}, policy updates continuously change the collected experience. Episodes may end in success, recoverable failure, unsafe termination, or reset, each providing distinct information. Recording policy versions, termination causes, and reset procedures is therefore essential for interpreting outcomes.

Autonomous collection scales only when outcome labeling and resets are reliable. QT-Opt embeds both in the task, RoboCat learns them, and AutoRT relies on human supervision. Remaining reset and supervision effort must therefore be included in collection costs.

\subsubsection{Intervention Records}\label{sec:4.2.3}

Human takeovers can obscure policy failures unless intervention records distinguish human and autonomous control. In the cup task, an operator may prevent a drop, reposition the cup, or complete the wash, each providing different supervision. Records should preserve autonomous, intervention, and resumed segments, along with takeover triggers, proposed and executed actions, and outcomes. These fields enable corrective learning and analysis of control authority and physical response.
Sirius~\citep{arxiv_2211_08416} learns from deployment logs using weighted behavior cloning. This requires distinguishing policy and operator actions, preserving intervention-triggering states, and identifying correction purposes. Protective withdrawals and task-recovery demonstrations provide different learning signals, making intervention segmentation and intent essential.
HIL-SERL~\citep{arxiv_2410_21845} integrates demonstrations, interventions, and autonomous experience, highlighting the need to preserve action provenance and outcomes. Human takeovers may or may not complete a task, so reward labels, termination causes, resets, and control handoffs are essential for distinguishing these cases.
These records identify expert supervision, autonomous failures, and resumed autonomy.

\begin{figure}[!htbp]
\centering
\includegraphics[width=\linewidth]{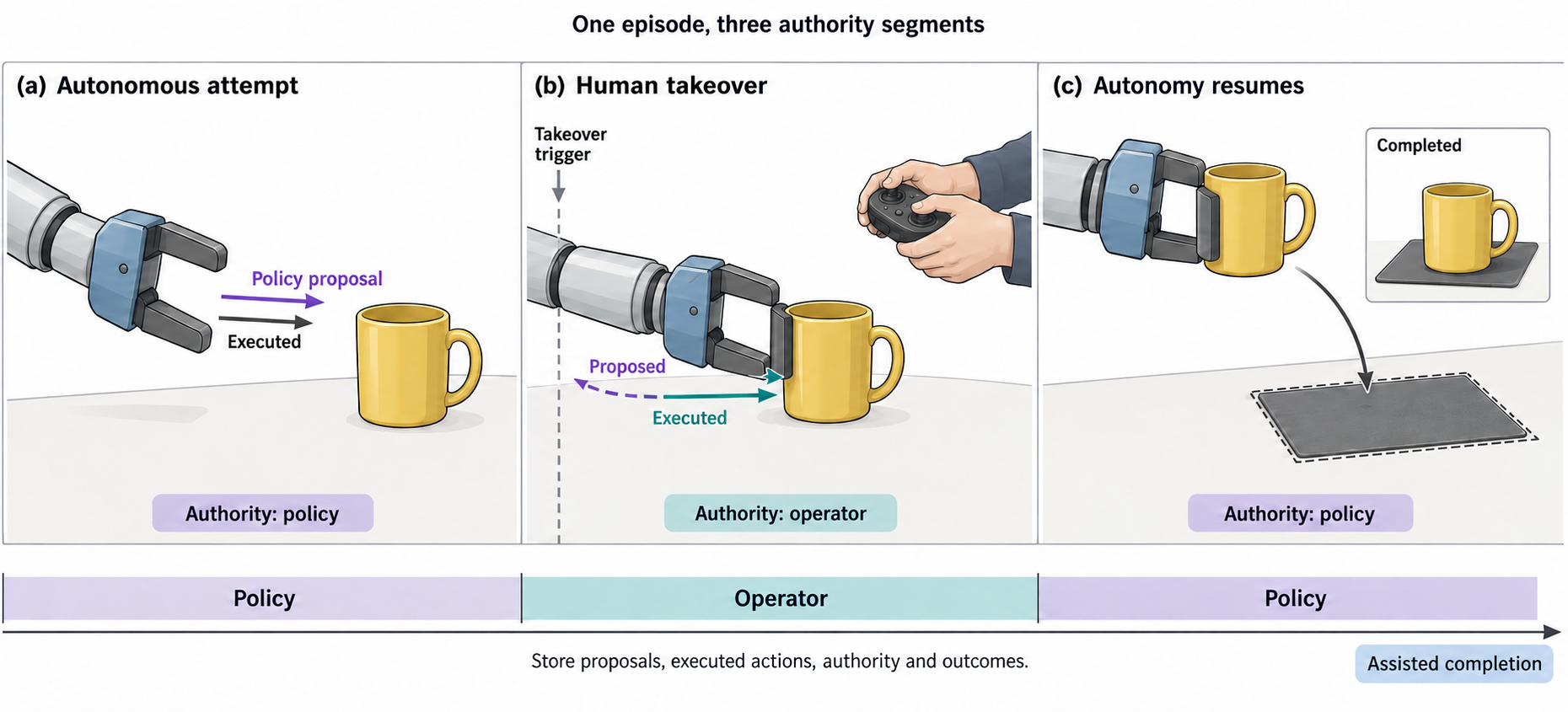}
\caption{Action provenance within one assisted episode. (a) An autonomous attempt misses the mug. (b) An operator takes control and corrects the robot motion, so the policy proposal and executed action differ. (c) Autonomous control resumes and the task is completed. The authority log preserves all three segments, together with takeover triggers, proposed and executed actions, and outcomes. Completion after an intervention is an assisted episode even when the final segment is autonomous; human teleoperation here is distinct from physically repositioning the object.}\label{fig:intervention-provenance}
\end{figure}

Figure~\ref{fig:intervention-provenance} follows an autonomous miss, a teleoperated correction, and resumed autonomy within one episode. During takeover, the proposed policy motion and the action actually executed must remain separately identifiable. A successful final segment does not retroactively make the episode autonomous; the intervention remains part of both its supervision and its evaluation record.

Demonstrations, autonomous collection, and interventions provide complementary experience at different human and robot costs. Demonstrations teach behaviors difficult to discover autonomously, with the control interface determining what can be shown and recorded. Autonomous collection exposes policy failures but requires reliable outcome checks and resets. Interventions target deviations with expert corrections, requiring clear control and recovery records. Fewer interventions do not necessarily indicate improvement if failures increase. The appropriate collection method therefore depends on the missing behavior, reliability of outcome checks and resets, and total collection cost, rather than episode count alone.

\subsection{Data Beyond Robot Interaction}\label{sec:4.3}

Human data broadens coverage but lacks directly executable robot commands. The discussion moves from image--text corpora that support object and instruction understanding (\S~\ref{sec:4.3.1}), to human videos that also reveal temporal task structure (\S~\ref{sec:4.3.2}), and then to motion and tool records that measure human movement more directly (\S~\ref{sec:4.3.3}). In the cup task, these sources can support recognition, task steps, or movement respectively; using them for robot actions requires additional correspondence, with increasing demands but no strict ranking of sources. Measured tool motion, mapped robot targets, and recorded robot outcomes must remain distinct (Table~\ref{tab:recordfields}); these are part of the source-to-record alignment problem in Figure~\ref{fig:overview-data}.

\subsubsection{Image--Text Corpora}\label{sec:4.3.1}

Image--text corpora provide object recognition and spatial understanding without robot execution, helping robots identify targets such as ``the cup'' and their locations. However, they lack the robot states, action commands, forces, and timing needed to learn executable behavior directly.
CLIP learns visual representations aligned with language from image--text pairs~\citep{arxiv_2103_00020}, supporting perception and learned rewards. DataComp~\citep{arxiv_2304_14108} shows the importance of data filtering: under fixed training code and compute, a filtered subset of 1.4 billion pairs improved zero-shot ImageNet accuracy by 3.7 points over a comparable CLIP model. However, benchmark gains do not guarantee preservation of fine-grained features needed for robotic control, such as cup handles or wet rims, which require robot-level evaluation.

Web captions often lack the metric spatial information needed for robotic control. SpatialVLM~\citep{chen2024spatialvlm} generates spatial question--answer pairs using object detection, depth estimation, and segmentation, enabling quantitative reasoning and reward annotation, though accuracy depends on depth estimates. RT-2~\citep{arxiv_2307_15818} and \(\pi_{0.5}\)~\citep{intelligence2025pi05} integrate web vision--language data with robot training. In \(\pi_{0.5}\), removing web data had little effect on familiar mock-home tasks but reduced generalization to unseen object categories, highlighting the importance of data mixture weights.
Evaluation should therefore distinguish recognition and instruction following on novel objects from physical execution. Web data provides semantic knowledge but not embodiment-specific control parameters, such as grasping force.

\subsubsection{Human Activity Videos}\label{sec:4.3.2}

Human activity videos capture motion, contact, object changes, and temporal order absent from static images, but lack executable robot commands. First- and third-person views provide complementary interaction information. Videos can support representation learning, subgoal extraction, and action inference, each requiring progressively stronger correspondence to the target robot.

\noindent\textbf{Representation Learning.} Ego4D~\citep{arxiv_2110_07058} provides diverse egocentric daily-life videos for understanding past, present, and future activities, with some recordings including audio, gaze, 3D meshes, and multi-camera views. R3M~\citep{arxiv_2203_12601} uses Ego4D for representation learning through time-contrastive and video--language objectives. Such videos capture task sequences and scene changes without robot action labels, reducing reliance on robot execution while still requiring capture, consent, annotation, and curation efforts.

\noindent\textbf{Plans and Subgoals.} MimicPlay~\citep{wang2023mimicplay} learns latent plans from human play videos to guide a low-level controller trained on limited robot demonstrations. Across 14 long-horizon real-world tasks, it improves success, generalization, and robustness over imitation baselines. Human videos provide goal sequences, while robot demonstrations teach their execution, as further reviewed by Ma et al.~\citep{ma2026robot}. This transfer depends on whether the robot can physically achieve the demonstrated goals.

\noindent\textbf{Action Recovery.} Visible hand motion does not directly translate into robot actions. Two approaches infer action-like signals from unlabeled videos. Latent-action methods, such as LAPA~\citep{ye2024latent}, encode visual changes and ground them in robot actions using limited demonstrations (Section~\ref{sec:5.1.2}). UniVLA~\citep{bu2025univla} further uses DINOv2 features and language guidance to suppress irrelevant visual variation. Point-tracking methods, such as ATM~\citep{arxiv_2401_00025}, predict future point trajectories from images and instructions, then use limited robot demonstrations to train policies. ATM outperformed video-pretraining baselines across over 130 tasks, including transfer from human videos with relatively small domain gaps. However, latent codes and 2D tracks remain pseudo-labels that require grounding in robot actions.

\subsubsection{Motion and Tool Records}\label{sec:4.3.3}

Unlike videos that capture motion as pixels, hand-tracking and handheld-tool records directly measure human motion. Hand tracking captures body and finger movements, while handheld grippers record task-space trajectories and tool states. Both require temporal alignment, coordinate mapping, and feasibility checks for robot execution. Unlike robot demonstrations collected through real-time motion following, these records originate from human or handheld-device actions.

\noindent\textbf{Tracked Hands and Gloves.} DexCap~\citep{wang2024dexcap} tracks wrist and finger motion with camera-based localization and occlusion-resistant electromagnetic sensing, and it also records 3D observations of the scene. Its DexIL algorithm replicates the human actions on a robot hand through inverse kinematics and point-cloud imitation. It is evaluated on six dexterous tasks, with optional human-in-the-loop correction during rollouts. The measurement is direct, but each motion must still be retargeted to a robot hand with different proportions and joints.

Hand-motion records can also be collected at larger scale with egocentric tracking. Head-mounted capture enables scalable hand-motion collection without gloves or external trackers. EgoDex~\citep{hoque2025egodex} pairs egocentric video with 3D hand tracking across tabletop tasks, evaluating trajectory prediction rather than robot execution. EgoScale~\citep{zheng2026egoscale} trains a VLA model on 20,854 hours of action-labeled human video, finding log-linear scaling of validation loss and a correlation with real-robot performance. After human--robot alignment and task adaptation, it improves success by approximately 54 percentage points over no pretraining on a 22-DoF hand. Even its one-robot-demonstration setting uses additional aligned human data. Thus, human-data scaling reduces robot demonstration needs but still depends on alignment quality and interaction coverage.

\noindent\textbf{Whole-body Motion Capture.} For humanoids, whole-body motion capture provides movement data. AMASS~\citep{arxiv_1904_03278} unifies marker-based datasets using a common parametric body model but lacks object interaction and force information. Transferring these motions to robots requires retargeting for body proportions, joint limits, and balance.

\noindent\textbf{Handheld Gripper Tools.} UMI~\citep{arxiv_2402_10329} collects manipulation demonstrations using handheld grippers with wrist cameras, linking human recordings to robot execution through relative trajectories and latency matching. Its policies transfer across robot arms and generalize to unseen environments and objects. Shared gripper geometry reduces retargeting needs, but calibration, motion feasibility, and timing alignment remain essential. Actions beyond the instrumented tool require additional sensing or annotation for supervision.

Non-robot sources differ in the knowledge they supply and the correspondence needed for transfer. Web data support semantics; human videos provide representations, subgoals, or inferred actions; motion and tool records offer more direct action correspondence when geometry matches. All require task-level validation, with robot-specific mapping, calibration, or target data as needed. Missing force and contact measurements may require robot interaction, and human demonstrations may omit failures caused by robot geometry and control. Source selection should therefore follow the missing knowledge and supervision needed for executable behavior. Simulation provides a complementary route through controlled interactions that deliberately vary configurations, contacts, and challenging states.

\subsection{Data Collection in Simulation}\label{sec:4.4}

Simulation supports perception, task planning, and policy learning, each requiring different levels of fidelity. The discussion separates the physics simulation that governs transitions and contact (\S~\ref{sec:4.4.1}) from environment construction that determines objects, layouts, tasks, and coverage (\S~\ref{sec:4.4.2}). Slip-sensitive manipulation depends on accurate contact dynamics, while household tasks benefit from diverse objects and layouts. Under fixed compute, more detailed physics limits the number and diversity of simulated interactions. These task-dependent fidelity requirements call for validation of physical transitions, task outcomes, action interfaces, and control timing as well as visual appearance.

\subsubsection{Physics Simulation}\label{sec:4.4.1}

Small physics errors can cause sim-to-real failures, such as a grasp that succeeds in simulation but slips on the real robot. Physics simulators model dynamics, contact, actuation, and sensing under explicit assumptions. Their required fidelity depends on the task: friction and compliance are critical for contact-rich manipulation but less important for free-space motion.
Rigid-body simulators model motion and contact through discrete integration, friction models, and constraint solvers. Their accuracy depends on friction, compliance, collision geometry, and actuator latency, while policies may exploit numerical artifacts absent in reality. More complex interactions require deformable and fluid simulation. SoftGym~\citep{arxiv_2011_07215} uses particle-based models for cloth, rope, and liquids, but simplifies grasping through particle attachment or direct object control. OmniGibson~\citep{li2024behavior1k} supports fluids, deformable objects, and cloth, enabling over half of BEHAVIOR-1K's 1,000 activities that would otherwise be unsupported. However, cleaning remains rule-based: dirt particles disappear upon tool contact, and soaking depends on liquid-particle contact. Thus, simulated cleaning success does not necessarily reflect realistic scrubbing forces or physical cleaning dynamics.

GPU-based frameworks expand data collection: Isaac Gym integrates physics and learning; Isaac Lab adds rendering, sensors, and collection tools; ManiSkill3 combines GPU physics and rendering for manipulation; MuJoCo Playground builds on MJX with batched environments~\citep{arxiv_2108_10470,nvidia2025isaac,tao2024maniskill3,zakka2025mujoco}. Their throughput does not remove physics errors. Resource bottlenecks are discussed in \S~\ref{sec:7.6.1}.
Simulators provide privileged states and controllable parameters for automatic labeling, curricula, and teacher--student training, as demonstrated in quadrupedal locomotion~\citep{lee2020quadrupedal,rudin2021walk} and Rubik's-cube manipulation~\citep{openai2019solving}. Domain randomization expands training diversity~\citep{tobin2017domain}, while Eureka uses language models and simulator code to generate reward functions~\citep{ma2023eureka}. These applications determine which states and parameters simulators must expose.

SimOpt~\citep{arxiv_1810_05687} calibrates simulator parameters using paired real and simulated trajectories, requiring data that captures relevant dynamics. Lin et al.~\citep{lin2025simtoreal} extend this to humanoid arms and hands by matching joint trajectories across real and simulated robots with varied dynamics parameters. Using under four minutes of real data, eight of ten policies trained with the best-matching model transferred a real grasp, compared with none using the worst model. However, calibration primarily captures robot dynamics, while hand--object contact remains approximated through primitive shapes.
Simulation evaluation should distinguish observation, state-transition, and action-execution gaps, as each requires different transfer methods. Simulator adequacy depends on whether task-critical errors remain within policy tolerance, verified through paired real-world tests rather than simulation throughput or visual realism alone.

\subsubsection{Environment Construction}\label{sec:4.4.2}

Accurate physics alone cannot ensure generalization without diverse environments. Assets, layouts, robots, sensors, initial conditions, and success criteria define training coverage. Environments can be generated procedurally or constructed from real-scene references; reference-based construction may reproduce a particular scene or preserve its relevant affordances using similar assets.

\noindent\textbf{Procedural Scenes.} ProcTHOR~\citep{arxiv_2206_06994} procedurally generates diverse indoor environments by varying layouts, lighting, object placements, colors, and textures. Scaling training environments from 10 to 10,000 houses improved downstream performance, highlighting the value of scene diversity for household tasks.

\noindent\textbf{Construction from Real-scene References.} RialTo~\citep{torne2024reconciling} reconstructs real scenes as digital twins, transfers real demonstrations into simulation through inverse distillation, and improves policies with simulated reinforcement learning before hardware deployment. This enables additional practice with varied object poses, distractors, and disturbances without further real-world collection. However, transfer depends on accurate scene geometry and articulation, and the digital twin remains specific to the reconstructed environment.
A reference can also guide construction without an exact replica. ACDC~\citep{dai2024automated} generates ``digital cousins'' from a single RGB image by matching detected objects to similar simulation assets while preserving geometric and semantic affordances. In cabinet manipulation, an exact digital twin achieved 100\% simulated but only 25\% real-world success; stronger randomization improved real success to 55\%, while digital cousins reached 90\%. Simulation suites such as LIBERO~\citep{arxiv_2306_03310} and RoboCasa~\citep{arxiv_2406_02523} also provide standardized benchmarks.

Simulator design balances physical fidelity, environment coverage, and calibration cost. Scene diversity, including digital cousins, can reduce overfitting to simulation-specific details more effectively than exact reconstruction, but adequate contact physics remains necessary. Detailed modeling matters when contact or actuation errors affect decisions; targeted calibration and robust control may address specific discrepancies more efficiently. Hardware-in-the-loop testing and real-world trials complement these choices by revealing timing, interface, and dynamics errors.

\subsection{Synthetic Data Generation}\label{sec:4.5}

Synthetic data reduces collection costs only when generated action and outcome labels remain valid. The preceding subsection examines interactive simulation environments; here, the organizing question is how training records are created or transformed, including methods that use simulation. The discussion covers transforming existing demonstrations (\S~\ref{sec:4.5.1}), generating scenes and appearances (\S~\ref{sec:4.5.2}), or producing new behavioral videos or trajectories (\S~\ref{sec:4.5.3}). For the cup task, these change positions, visual conditions, or behavior respectively. This order distinguishes reuse of an existing behavior, modification of its observation context, and generation of behavioral records; depending on the method, actions may be preserved, executed after transformation, or inferred from generated observations. Greater visual diversity supplies useful action supervision only when correspondence remains valid.

\subsubsection{Demonstration Augmentation}\label{sec:4.5.1}

A single demonstration covers only one object pose, such as washing a cup at a fixed position. Demonstration augmentation extends this behavior to new poses by preserving the gripper–object relationship rather than copying the original actions. The pipeline identifies invariant coordinate relations and contact constraints, transforms the reference trajectory, and verifies that the resulting behavior remains successful.

\noindent\textbf{Transformation with Execution Validation.} MimicGen~\citep{arxiv_2310_17596} divides demonstrations into object-centered subtasks, adapts trajectories to new object poses, and retains successful executions. From roughly 200 human demonstrations, it generates over 50,000 demonstrations across 18 tasks, varying scenes, objects, and robot arms. Policies trained on this data perform well on long-horizon and precision tasks, including multi-part assembly, and compare favorably with collecting additional human demonstrations.

Generation alternates between trajectory transformation and execution at each subtask boundary. For example, after grasping a cup at a new position, its pose is re-observed before moving it to the sink. Each source segment is transformed for the current scene, executed, and followed by a fresh state estimate before transforming the next segment. Successful full-task executions are retained as generated demonstrations. Because later transformations depend on earlier executions, object poses must be estimated at every boundary. Key limitations include invalid transformation assumptions (e.g., shape-dependent grasps), low success rates under difficult initial conditions, and simulator-to-real-world gaps. Success filtering also biases the generated dataset toward already feasible behaviors. Importantly, the method stores executed trajectories rather than transformed action sequences, making it an augmentation of experience rather than labels.

The same augmentation principle extends to coordinated and dexterous manipulation. DexMimicGen~\citep{jiang2024dexmimicgen} extends demonstration generation to humanoids with dexterous hands, producing 21,000 demonstrations from 60 sources across simulated bimanual tasks with varying coordination demands. It also evaluates generation and policy-learning choices and demonstrates a real-to-sim-to-real pipeline for humanoid can sorting. Like MimicGen, it evaluates data quality through downstream policy performance. However, bimanual manipulation imposes additional constraints: transforming one arm's motion restricts the other's, reducing the space of feasible configurations.

\noindent\textbf{Editing Without Execution.} DemoGen~\citep{arxiv_2502_16932} eliminates real-world execution during generation. From a single human demonstration, it transforms contact segments with the object, replans free-space motions, and updates point clouds accordingly. This enables computation-only augmentation but relies on planning and controller assumptions rather than execution-based success filtering. Single-view point clouds further limit appearance consistency, causing performance gains to saturate.

Overall, demonstration augmentation generalizes existing skills to new conditions but depends on accurate object poses, subtask segmentation, and transferable action relations. It cannot directly handle new grasp or contact strategies, and subtask recombination alone does not create fundamentally new skills.

\subsubsection{Scene and Appearance Generation}\label{sec:4.5.2}

Visual editing is inexpensive, but preserving action validity is challenging. Appearance changes, such as recoloring a counter, can preserve the original action, while geometric changes, such as altering a cup’s shape, may invalidate its grasp. Two approaches introduce visual diversity under this constraint: editing recorded episode images and generating assets for simulated scenes.

ROSIE~\citep{arxiv_2302_11550} uses text-guided diffusion to edit robot demonstration images while preserving recorded actions. It detects and masks target regions, excluding the robot arm, and replaces objects, containers, or backgrounds, updating instructions when needed. In one experiment, 779 can-placement episodes were edited to replace a drawer with a sink, enabling a policy to achieve 60\% success on real sink placement versus 0\% without augmentation. However, unchanged actions remain valid only if the edited geometry supports the original motion. ROSIE modifies appearance rather than motion or physics, and frame-wise editing can introduce temporal inconsistencies.

RoboCasa~\citep{arxiv_2406_02523} diversifies simulated kitchens using text-to-3D object assets and text-to-image textures while preserving physics-based interaction. However, generated geometry and materials can affect contact dynamics and sim-to-real transfer, requiring physical validation beyond visual inspection. Across these appearance-generation approaches, the added visual diversity still relies on recorded or simulated actions for supervision.

\subsubsection{Synthetic Behavioral Data}\label{sec:4.5.3}

A realistic robot video does not necessarily provide executable actions. Synthetic behavioral data must link observations to actions over time, specifying task conditions, motion sources, action labels, and validation procedures. Two main approaches differ in how actions are obtained: executing generated tasks in simulation or inferring actions from generated videos.

\noindent\textbf{Simulator-executed Behavior.} RoboGen~\citep{arxiv_2311_01455} uses language models to generate tasks, configure scenes, select assets, decompose tasks, and define rewards. It then solves subtasks through reinforcement learning, motion planning, or trajectory optimization in simulation. Although the resulting actions are executable in simulation, task specifications may be incorrect: 6 of 155 generated tasks contained faulty decompositions or rewards. Large-scale skill verification and the sim-to-real gap remain key limitations.

\noindent\textbf{Video-generated Behavior.} DreamGen~\citep{jang2025dreamgen} generates photorealistic robot videos across tasks and environments, then recovers pseudo-actions using latent action or inverse-dynamics models to train policies. From teleoperation data for a single pick-and-place task, a humanoid learns 22 new behaviors across seen and unseen environments. DreamGen evaluates data through downstream policy success, with benchmark scores correlating positively with RoboCasa performance. Unlike execution-based filtering in \S~\ref{sec:4.5.1}, this does not validate individual trajectories. The method still requires robot data for action inference and manually supplied initial frames, while generation cost and task complexity remain limitations.

The two approaches have distinct trade-offs: simulation provides executable actions but inherits specification errors and sim-to-real gaps, whereas video generation offers realistic visual diversity but relies on inferred actions without verified physics. Ultimately, data quality depends on validation and downstream real-world performance, not generation volume alone.

Across demonstration augmentation, observation generation, and behavioral generation, the key issue is action-label validity. Demonstration augmentation extends existing skills but cannot supply missing contact strategies without additional source behavior. Observation generation adds visual diversity, while geometry changes require action revalidation. Behavioral generation offers greater novelty but requires action inference, filtering, and physical validation; direct collection may remain more efficient for complex contacts. Generated records should retain their source, transformation, and action-label provenance so that curation can distinguish valid variations from unverified new behavior.

\subsection{Data Alignment and Record Integrity}\label{sec:4.6}

Data preparation first aligns observations, actions, and outcomes across space, time, and embodiment (\S~\ref{sec:4.6.1}), then checks record quality to distinguish valid failures from corrupted or unsupported labels (\S~\ref{sec:4.6.2}). This order matters because selection and sampling determine what the learner sees only after correspondence and integrity are established. Each record must also match its training objective: an image labeled ``cup'' supports recognition but lacks the actions and next states needed for imitation or dynamics learning.

\subsubsection{Data Alignment}\label{sec:4.6.1}

Action vectors of equal length may differ in physical meaning, reference frame, or control frequency, which normalization alone cannot resolve. Before combining datasets, alignment must establish consistent semantics across four dimensions: spatial alignment (coordinate frames), action alignment (command meanings), temporal alignment (measurement–execution timing), and embodiment metadata (robot configuration).

Combining 15\,Hz measured base-frame poses with 10\,Hz commanded camera-frame displacements requires spatial, action, temporal, and embodiment alignment. Coordinate transforms reconcile reference frames, while measured motion must remain distinct from commanded actions. Streams are synchronized before resampling, and embodiment metadata distinguish grippers from dexterous hands. Resampling, discretization, and cross-embodiment mapping may irreversibly lose contact details, precision, or finger motion, so these conversions must remain traceable in training records.

Formats, metadata, and versioning support data alignment and traceability. Open X-Embodiment~\citep{arxiv_2310_08864} standardizes datasets from 21 institutions but still requires compatible action and sensor fields, as illustrated by Re-Mix's normalization and discretization. Similarly, DROID's multiview data~\citep{arxiv_2403_12945} may not all be used by a given policy. Each policy-specific view must distinguish decision-time inputs from supervision targets and record target availability and action source separately. LeRobot~\citep{cadene2026lerobot} integrates timestamps and embodiment metadata into its robot-learning framework, enabling consistent training and deployment. However, metadata make misalignment detectable rather than impossible. Version tracking must preserve repairs, data subsets, transformations, and separate paper, code, and data versions and usage terms, since citing the same dataset does not guarantee identical training data.

\subsubsection{Data Quality}\label{sec:4.6.2}

A dropped cup and a frozen camera may both yield unsuccessful episodes, but only the former provides useful behavioral information. Quality processing must distinguish corrupted records (e.g., sensor failures, timestamp errors, calibration drift, or invalid actions) from informative task failures, interventions, and recoveries. Integrity checks should precede outcome labeling, preserving valid failures that reveal the limits of robot competence while excluding unusable data.

BridgeData~V2~\citep{arxiv_2308_12952} supports both offline reinforcement learning and imitation, while MimicGen~\citep{arxiv_2310_17596} retains successful executions for behavior cloning. Quality criteria therefore depend on training objectives rather than a universal success threshold: failed episodes can support reward learning, while their successful segments may still provide imitation targets.

Failures, interventions, and safety constraints reveal competence boundaries, especially when their causes are recorded. RoboMIND~\citep{wu2024robomind} provides 107k teleoperated trajectories across 479 tasks and four embodiments, plus 5k cause-annotated real-world failures for policy correction. REFLECT~\citep{liu2023reflect} automates failure explanation using hierarchical multisensory summaries and language models. Learned success detectors, as used in RoboCat, can scale outcome labeling but require evaluation of false accepts and rejects to ensure reliable data selection.

Cleaning must preserve three labels: record integrity, task outcome, and control authority. Human interventions should also distinguish pre-intervention, intervention, and post-intervention segments. This enables reuse for imitation, reward learning, and failure detection without attributing human corrections to the policy, while supporting sustained-execution evaluation.

Alignment and integrity checks establish usable observation–label relationships, while selection and sampling determine the learner's data distribution. Valid failures may remain informative without corrective actions, whereas corrupted records require repair or exclusion.

Different data sources provide different supervision: visual data support representation learning, robot interactions provide action–outcome relationships, and generated data expand coverage only when these relationships remain valid. The resulting training mixture must preserve action source, outcome evidence, and the transformations applied to each record; Section~\ref{sec:5} examines how learning methods use these signals.

\section{Learning and Adaptation}\label{sec:5}

\begin{figure}[!htbp]
\centering
\includegraphics[width=\linewidth]{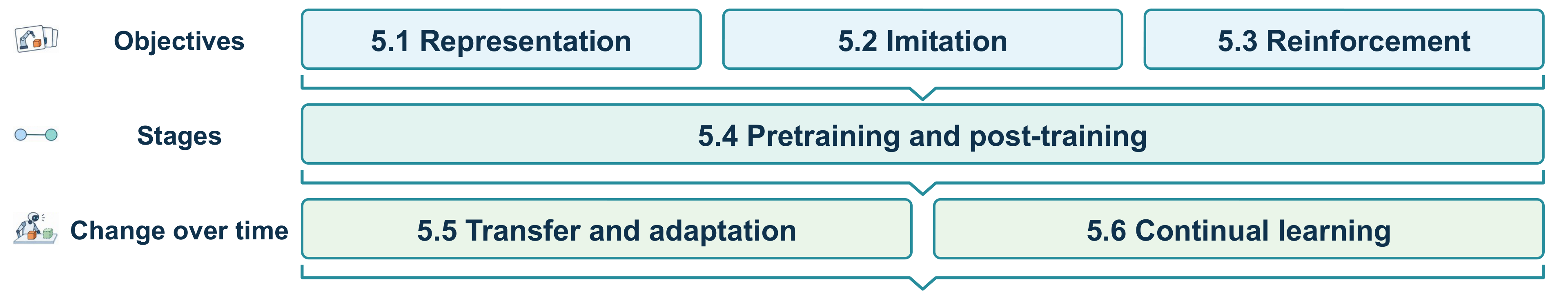}
\caption{Three axes of learning and adaptation. Objectives describe what is learned, stages locate pretraining and post-training, and change over time distinguishes transfer and adaptation from continual learning. The rows organize complementary choices, not successive steps.}\label{fig:guide-05}
\end{figure}

Training updates representations, predictions, policies, or value models from experience. This differs from updating execution-time state or memory without changing parameters. Figure~\ref{fig:guide-05} separates three axes of learning and adaptation.
Representation learning uses visual or predictive targets (\S~\ref{sec:5.1}), imitation fits demonstrated actions (\S~\ref{sec:5.2}), and reinforcement learning optimizes rewards (\S~\ref{sec:5.3}). The same records can support several objectives.
Pretraining and post-training describe when objectives are applied (\S~\ref{sec:5.4}). Transfer identifies changes in tasks, environments, or bodies (\S~\ref{sec:5.5}), while continual learning concerns accumulation and retention across successive changes (\S~\ref{sec:5.6}). A policy can therefore be pretrained by imitation, refined with rewards, and updated with replay to retain earlier skills.

\FloatBarrier

\subsection{Representation Learning}\label{sec:5.1}

\begin{figure}[!htbp]
\centering
\includegraphics[width=\linewidth]{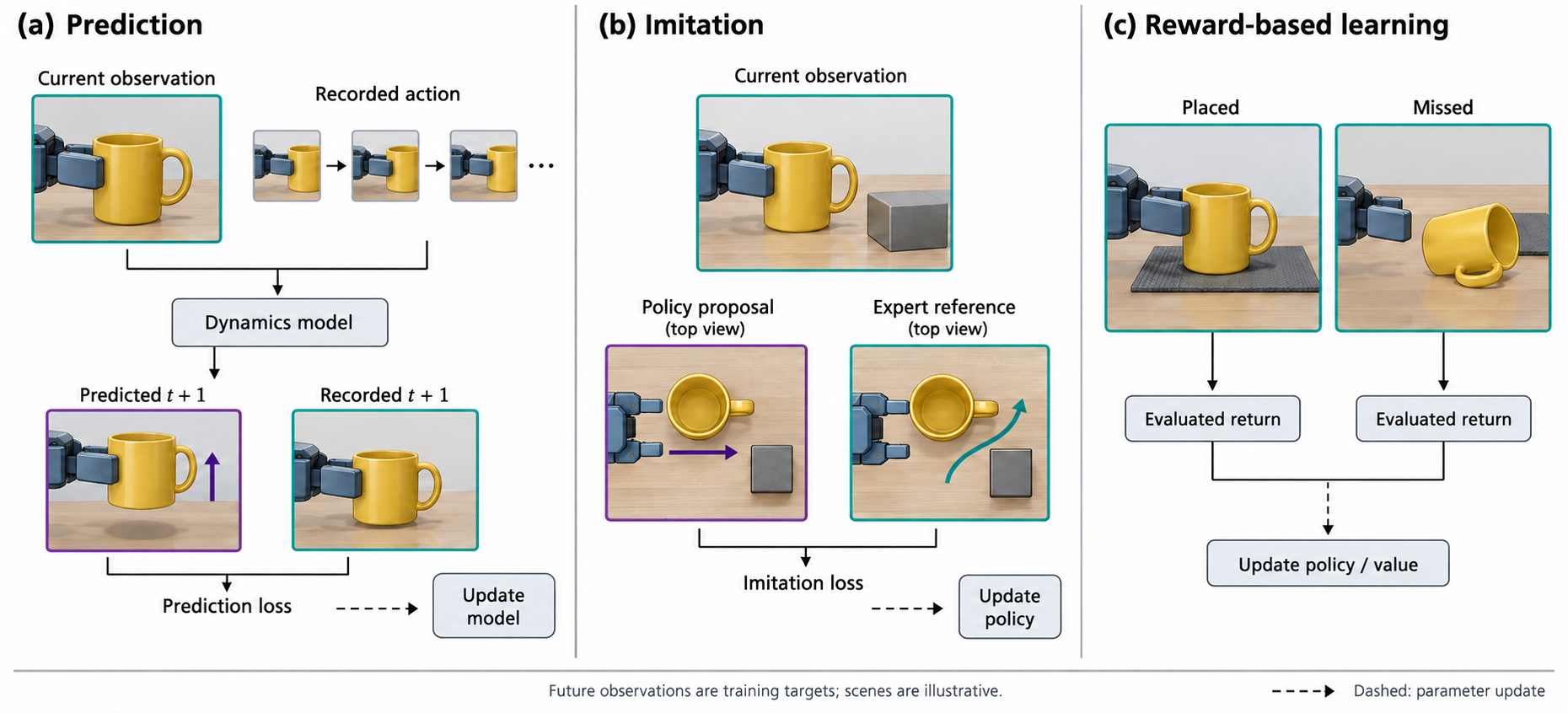}
\caption{Learning objectives illustrated through a mug task. (a) Predictive learning compares a future prediction with an observation recorded at the same future time. (b) Imitation compares a policy proposal with an expert reference under the same current information; arrows schematically depict motion. (c) Reward-based learning uses evaluated outcomes to update policies or value models. Dashed arrows denote parameter updates. Future observations are training targets, not deployment inputs. Objectives can be combined across training stages; the scenes are illustrative rather than experimental results.}\label{fig:overview-learning}
\end{figure}

Representation learning moves from general image and image--text objectives (\S~\ref{sec:5.1.1}) to temporal and action-conditioned objectives (\S~\ref{sec:5.1.2}). The comparison asks what each signal adds for control, separately from encoder adaptation and deployment-time prediction.
Figure~\ref{fig:overview-learning} compares predictive, imitation, and reward-based objectives through their targets and parameter updates. These objectives can be combined at different training stages.

\subsubsection{General Visual Objectives}\label{sec:5.1.1}

Web-image pretraining learns features that may not capture grasp-relevant details. Two common approaches, image-text alignment and image-only self-supervised learning, preserve different types of information.

\noindent\textbf{Vision--language Correspondence.} CLIP and SigLIP learn visual-semantic representations from image-text pairs~\citep{arxiv_2103_00020,zhai2023sigmoid}. However, action learning must still connect objects to robot actions, and localization accuracy depends on the spatial detail retained by the encoder.

\noindent\textbf{Self-supervised Visual Structure.} Self-supervised methods like DINOv2 learn visual features without captions or action labels~\citep{arxiv_2304_07193}, while masked reconstruction encourages spatial information by predicting missing patches. MVP~\citep{arxiv_2210_03109} uses this objective to pretrain a vision transformer on 4.5 million images, then freezes the encoder and trains a controller with few demonstrations. Its comparisons show better transfer than CLIP, supervised ImageNet pretraining, or training from scratch on tested tasks, highlighting the importance of pretraining objectives beyond encoder capacity. However, demonstrations are still needed to identify action-relevant features.

The control benefit of either objective depends on adaptation and evaluation. Two studies qualify these gains. Hansen et al.~\citep{arxiv_2212_05749} find that a simple encoder trained from scratch with random-shift augmentation can match frozen pretrained representations across diverse tasks. They attribute pretrained models' limitations to domain gaps, which fine-tuning can reduce, while freezing mainly saves training cost. Similarly, the VC-1 and CortexBench study~\citep{arxiv_2303_18240} finds no universally superior encoder across 17 tasks, though task-specific adaptation substantially improves performance. Thus, encoder comparisons depend on adaptation protocols and demonstration budgets. For control, representations must preserve action-relevant distinctions, not merely object categories.

\subsubsection{Interaction and Predictive Objectives}\label{sec:5.1.2}

Interaction data support learning temporal structure and learning the relation between actions and state changes. Within either group, predictions may use pixels or latent features; the prediction space is a separate choice from whether actions are observed. Force or touch signals provide additional contact supervision when recorded.

\noindent\textbf{Temporal Structure from Observations.} R3M~\citep{arxiv_2203_12601} learns visual representations from human videos using time-contrastive learning, video--language alignment, and sparsity regularization. Its frozen encoder supports robot learning from 20 demonstrations. In contrast, GR-1~\citep{arxiv_2312_13139} uses video generation pretraining and retains future-image prediction during robot fine-tuning. While R3M learns temporal relationships between representations and GR-1 predicts future visual content, both provide temporal information without directly learning robot motor commands.
Temporal prediction can also operate in feature space. V-JEPA~2 predicts masked video representations, with an action-conditioned variant for image-goal planning~\citep{assran2025vjepa}. Avoiding unpredictable pixel details can reduce irrelevant modeling, but latent features must still retain action-critical geometry.

\noindent\textbf{Relations Between Actions and State Changes.} With aligned state--action transitions, forward models predict \(p_\phi(z_{t+1}\mid z_t,a_t)\) (\S~\ref{sec:3.2.1}), while inverse models infer actions from surrounding states or observations. Forward prediction supports action evaluation and imagined experience, as in PlaNet and Dreamer~\citep{hafner2018latent,arxiv_2301_04104}. Inverse prediction can turn unlabeled video into action-labeled data. Both depend on action representations, such as joint angles or end-effector motions, and the time interval between states.
Inverse or latent-action prediction can supply labels for further learning. VPT trains inverse dynamics on labeled Minecraft gameplay, then labels a larger video corpus for imitation~\citep{arxiv_2206_11795}. LAPA learns discrete latent actions from frame transitions and maps them to robot commands using limited trajectories~\citep{ye2024latent}. Such labels expand training access but can lose embodiment-specific detail; LAPA underperforms true action supervision on fine-grained tasks such as grasping.

Across visual and interaction objectives, general pretraining supplies reusable visual knowledge; interaction data and encoder adaptation address missing control-relevant distinctions. Predictive targets help when they capture consequences beyond action labels. Comparisons should hold parameter updates and deployment-time prediction fixed to isolate this benefit, rather than treating richer reconstruction as inherently better for control.

\subsection{Imitation Learning}\label{sec:5.2}

Imitation learning trains policies from reference behavior. The account starts with behavior cloning, which fits actions in existing demonstrations (\S~\ref{sec:5.2.1}), then examines objectives for preserving alternative valid behaviors under similar conditions (\S~\ref{sec:5.2.2}). Interactive imitation adds expert supervision at states the learned policy visits (\S~\ref{sec:5.2.3}). This sequence separates errors in fitting available actions, averaging distinct strategies, and policy-induced distribution shift; the methods can be combined.

\subsubsection{Behavior Cloning}\label{sec:5.2.1}

Behavior cloning learns a policy from demonstrated observation--action pairs and task conditions. Demonstrations capture hard-to-specify behaviors, such as how firmly to grasp a wet cup. Supervision can use trajectories, action chunks, task segments, or labeled skills, with targets predicting either individual actions or chunks (Eq.~\eqref{eq:1}).
For probabilistic policies, behavior cloning minimizes the negative log-likelihood of demonstrated actions:

\begin{equation}
\mathcal L_{\mathrm{BC}}(\theta)=-\mathbb E_{(h,g,a)\sim\mathcal D}\big[\log\pi_\theta(a\mid h,g)\big].\label{eq:2a}
\end{equation}

Here, \(\mathcal D\) denotes demonstrations, \(h\) the history, \(g\) the task condition, and \(a\) a single action or chunk (Eq.~\eqref{eq:1}). Deterministic policies use regression losses, while diffusion and flow matching use specialized objectives (Eq.~\eqref{eq:fm}). All variants share a limitation: training covers demonstrated states, not unseen states arising from the policy's own errors.
Deployment is sequential, so each action affects subsequent states. Small errors, such as a misaligned grasp, can lead to unseen observations, causing policy-induced distribution shift. Thus, low imitation error does not guarantee closed-loop success, which depends on whether demonstrations cover the states encountered during execution.

The robomimic study~\citep{arxiv_2108_03298} compares six offline algorithms across five simulated and three real multi-stage manipulation tasks. Results highlight the importance of algorithm design, demonstration quality, and training duration, as low validation loss does not guarantee task success. Offline human demonstrations also solve tasks that the compared reinforcement-learning methods could not. Imitation failures should therefore be diagnosed in terms of data coverage, history or chunk length, training choices, and deployment distribution shift rather than supervised learning itself.
The data-collection study by Lin et al.~\citep{lin2024data} distinguishes diversity across environments and objects from repeated demonstrations under the same conditions. It suggests prioritizing new conditions when additional repetitions yield diminishing returns, depending on deployment coverage. Coverage alone is insufficient, however: demonstrations may contain errors and varying operator styles, so validation loss must be complemented by full-task rollouts that assess sustained progress.

\subsubsection{Learning Multimodal Behavior}\label{sec:5.2.2}

Demonstrations may contain distinct valid strategies, such as grasping a cup by its handle or rim. Averaging these actions can produce an invalid grasp. While action representations determine what a policy can express, multimodal objectives preserve alternative strategies, and chunking improves temporal coordination without preventing averaging~\citep{arxiv_2304_13705,arxiv_2303_04137}. The objectives below preserve alternatives through latent variables, denoising or transport, discrete mixtures, or an energy function. Decoding must also maintain strategy consistency.

\noindent\textbf{Latent-variable Models.} ACT learns a conditional variational model of action sequences~\citep{arxiv_2304_13705}, but fixes its latent variable to the prior mean during evaluation. Its success therefore does not demonstrate sampling of distinct strategies.

\noindent\textbf{Diffusion and Flow Objectives.} Diffusion Policy instead generates action sequences through iterative denoising~\citep{arxiv_2303_04137}. Both offer richer action distributions than direct prediction, but preserving distinct valid strategies depends on conditioning, training, and test-time decoding.
The \(\pi_0\) action expert uses flow matching (Eq.~\eqref{eq:fm}, \S~\ref{sec:3.4.4}) to learn a velocity field between Gaussian noise and demonstrated action chunks~\citep{arxiv_2410_24164}. At inference, it integrates this field in ten steps from sampled noise, allowing different samples to produce distinct strategies rather than averaged actions. However, the objective does not guarantee coverage or sampling of every demonstrated strategy.

\noindent\textbf{Discrete Mixtures.} Behavior Transformers~\citep{arxiv_2206_11251} address action averaging by clustering actions into \(k\) bins, predicting a bin from observation history, and regressing a continuous offset.

\noindent\textbf{Energy Models.} Implicit behavioral cloning~\citep{arxiv_2109_00137} instead learns an energy function that favors demonstrated actions, representing multimodal mappings through inference-time optimization. On the Insert-Blue task (1\,mm tolerance), it achieves 83.3\% versus 6.7\% for MSE regression.

Action sequence length and overlapping-prediction aggregation are separate design choices (\S\S~\ref{sec:3.4.1} and~\ref{sec:7.3.3}). Multimodal policies should be evaluated by strategy-specific success and consistency rather than average trajectory error. Representing multiple strategies does not prevent mid-execution switching, while interactive supervision can address failures caused by distribution shift.

\subsubsection{Interactive Imitation}\label{sec:5.2.3}

Interactive imitation adds expert supervision at learner-visited states. Algorithms can query labels, robots can request help, or humans can take control; each route trades coverage, safety, and expert effort.

\noindent\textbf{Expert Queries.} DAgger aggregates expert labels on states reached by the learner and retrains the policy~\citep{arxiv_1011_0686}. Its improved error bounds rely on no-regret and reduction assumptions and achievable loss at visited states, rather than guaranteeing finite-data performance.

\noindent\textbf{Robot-gated Requests.} Labeling every state requires substantial expert effort. ThriftyDAgger~\citep{arxiv_2109_08273} reduces this burden by requesting interventions at novel states, identified through policy-ensemble disagreement, or risky states, identified through estimated task-completion probability. Thresholds control the intervention rate, making effectiveness dependent on the accuracy of both estimates.

\noindent\textbf{Human-gated Takeover.} HG-DAgger lets experts intervene when the policy struggles and learns an uncertainty-based safety threshold. BC-Z scales human-gated collection across tasks, using corrections to teach recovery~\citep{kelly2018hgdagger,jang2022bcz}.

After collection, intervention records can be selected and weighted independently of who requested help. Sirius uses intervention history to approximate trust and weight behavior-cloning samples~\citep{arxiv_2211_08416}. Thus, collection authority, query timing, and sample weighting are separate choices; aggregation helps only when corrections are accurate and cover relevant failures.

Across the imitation-learning methods, richer distributions and temporal conditioning improve fitting to available behavior, while corrective demonstrations supply missing supervision at deployment states. Both need consistent execution. When expert actions are costly but outcomes are verifiable, reward-based learning offers another source of improvement.

\subsection{Reinforcement Learning}\label{sec:5.3}

Imitation learning fits demonstrated actions, while reinforcement learning (RL) optimizes outcomes through rewards. The discussion first asks how rewards are constructed and checked (\S~\ref{sec:5.3.1}), then separates learning from existing interaction records (\S~\ref{sec:5.3.2}) from learning with new policy interactions (\S~\ref{sec:5.3.3}). Policy learning in imagined environments uses predictive models to add training experience while remaining limited by model coverage (\S~\ref{sec:5.3.4}). These routes differ in available feedback and interaction cost, and can be combined with demonstrations and corrections. Tang et al.~\citep{arxiv_2408_03539} survey practical challenges in real-robot RL.
The RL objective maximizes expected return:

\begin{equation}
J(\theta)=\mathbb E_{\tau\sim\pi_\theta}\!\left[\sum_{t=0}^{T-1}\gamma^t r_t\right].\label{eq:2b}
\end{equation}

Here, \(\tau\) is a trajectory generated by the policy and environment, \(T\) the horizon, \(r_t\) the reward, and \(\gamma\in[0,1]\) the discount factor. Unlike behavior cloning (Eq.~\eqref{eq:2a}), RL learns from outcomes rather than action labels and can optimize rewards using either existing logs or new interactions.

\subsubsection{Reward Construction}\label{sec:5.3.1}

Rewards may use measured state, fitted judgments, or pretrained models. Dense feedback can aid exploration, but every source can omit relevant conditions or introduce exploitable errors.

\noindent\textbf{Functions of Measured State.} Learning to Walk in Minutes combines terms for tracking, motion, effort, collisions, and contact~\citep{rudin2021walk}. Such terms are cheap in simulation, but unmeasured conditions remain unprotected: object distance can be computed from poses, whereas cleanliness requires other evidence.

\noindent\textbf{Judgments Fitted to Examples.} HIL-SERL fits a success classifier and adds examples to correct errors. VICE instead updates its classifier adversarially against policy-visited negatives; PEBBLE learns graded rewards from pairwise preferences and relabels replay as judgments change~\citep{arxiv_2410_21845,arxiv_1805_11686,arxiv_2106_05091}. These differ in label collection and how they respond to changing policy behavior.
Extending example-based rewards to robot video, RoboReward combines natural failures, counterfactual negatives, and partial-progress examples from robot video~\citep{arxiv_2601_00675}. Its trained models outperform larger VLMs on tested short-horizon tasks, emphasizing failure coverage rather than model size alone.

\noindent\textbf{Pretrained Models as Reward Sources.} Eureka generates inspectable reward code from simulator code and task descriptions; VLM-RM directly scores observations with CLIP~\citep{ma2023eureka,rocamonde2023visionlanguage}. Both need independent outcome checks. Accuracy on existing records does not ensure reliability as optimization drives the policy into new states or exploits reward shortcuts.

\subsubsection{Offline Reinforcement Learning}\label{sec:5.3.2}

Offline RL uses logged actions and outcomes without new interaction. The methods below distinguish value-based learning, whose estimates need protection against unsupported actions, from return-conditioned action prediction. Conservative penalties, in-support updates, and sequence-based value models are choices within the value-based group; none creates evidence for outcomes absent from the data.

\noindent\textbf{Value-based Learning.} CQL~\citep{arxiv_2006_04779} adds a regularizer to the Bellman error that lowers Q-values for actions outside the dataset, producing conservative value estimates. Its lower-bound and policy-improvement guarantees depend on specific assumptions and update procedures. Excessive conservatism may limit improvement, making regularization strength an important design choice.
IQL instead learns values through upper expectile regression and Bellman backups without evaluating unseen actions during value learning, then extracts a policy through advantage-weighted cloning~\citep{arxiv_2110_06169}. Function approximation can still produce unsupported actions.

The value model can also use a sequence representation. Q-Transformer tokenizes action dimensions for transformer-based Q-learning on heterogeneous records~\citep{chebotar2023qtransformer}. It penalizes unsupported actions and uses Monte Carlo returns as lower bounds; scaling the value model does not remove data-support limits.
Cal-QL~\citep{arxiv_2303_05479} supports offline-to-online fine-tuning by retaining CQL's conservative penalty while calibrating Q-values against a reference policy's returns, preventing excessive pessimism and loss of offline knowledge during online updates. ConRFT adopts this approach for offline initialization.

\noindent\textbf{Return-conditioned Action Prediction.} Decision Transformer~\citep{arxiv_2106_01345} avoids value estimation by training a causal transformer to predict actions conditioned on desired return-to-go and past states and actions. At inference, the target return is updated using received rewards. This avoids bootstrapping but remains limited by behaviors represented in the data.

Offline RL is limited by data support. If logs lack successful cup insertions, more failed trajectories may not reveal the correct contact adjustments. Further improvement depends on whether the data lack state coverage, successful outcomes, or reliable value estimates. Thus, dataset composition by outcome and controller matters, and data collection should target these gaps. Offline methods can generalize from supported transitions, but unsupported outcomes require additional assumptions or new evidence.

\subsubsection{Online Reinforcement Learning}\label{sec:5.3.3}

Online RL adds transitions from current behavior at the cost of physical time, wear, and resets. Parallel simulation, autonomous hardware collection, and human-supported interaction trade fidelity and exploration scale against safety and recovery cost.
Online RL algorithms differ in how they reuse experience. On-policy methods such as PPO update policies using recent data and constrain policy changes, but require frequent new samples. Off-policy actor--critic methods reuse replay-buffer transitions for multiple updates, improving sample efficiency at the risk of value extrapolation errors (\S~\ref{sec:5.3.2}). Thus, on-policy learning suits cheap simulation, while off-policy learning is often preferable for costly robot interactions. RLPD~\citep{arxiv_2302_02948} combines prior and online data equally in each batch, using critic layer normalization and a large ensemble to stabilize frequent updates without offline pretraining.

\noindent\textbf{Massively Parallel Simulation.} Isaac Gym places simulation and learning on the GPU~\citep{arxiv_2108_10470}. Learning to Walk in Minutes uses PPO and a terrain curriculum across thousands of simulated robots, demonstrating rapid learning and hardware transfer~\citep{rudin2021walk}. Cheap interaction favors on-policy updates, while transfer still requires testing the gaps in \S~\ref{sec:5.5.4}.

\noindent\textbf{Autonomous Real-world Interaction.} SERL~\citep{arxiv_2401_16013} combines RLPD-based off-policy learning with automated rewards, reset policies, and robot control infrastructure. It learns real-world assembly, cable routing, and relocation tasks in 25--50 minutes, achieving near-perfect success and recovery under perturbations. DayDreamer~\citep{arxiv_2206_14176} instead learns a world model from real experience and trains policies through imagined rollouts, enabling a quadruped to recover, stand, and walk without resets in one hour. However, imagined rollouts inherit model errors~\citep{bai2026manipulation}. Both highlight that automated rewards, resets, and reliable infrastructure are crucial for efficient real-world learning.

\noindent\textbf{Human-supported Real-world Interaction.} HIL-SERL~\citep{arxiv_2410_21845} combines 20--30 initial demonstrations with human interventions during online RL. Intervention segments enter both demonstration and online replay buffers, which are sampled equally for RLPD updates, while a success classifier supplies sparse rewards. Mean success improves from 49.7\% for the imitation baseline to 100\%, though prolonged interventions can cause value overestimation. ConRFT extends this approach to VLA post-training. Safe exploration still requires human oversight and physical safeguards.


\subsubsection{Policy Learning in Imagined Environments}\label{sec:5.3.4}

Imagined-policy learning separates real collection, predictive-model fitting, and policy optimization in generated trajectories. The policy can then act without deployment-time search, although estimation and inference remain necessary.
DreamerV3~\citep{arxiv_2301_04104} trains a recurrent latent world model from real experience, then generates imagined trajectories to update an actor--critic using predicted states, rewards, and episode continuation. Real trajectories also provide critic targets. It applies a single configuration across over 150 tasks, including Minecraft. DayDreamer~\citep{arxiv_2206_14176} extends this approach to physical robots, continuously updating the world model and actor--critic while collecting new experience with the latest policy. This feedback loop allows learning from policy errors encountered during deployment.

Imagined learning can also use offline data. World Models~\citep{ha2018world} trains behavior in a compact generative environment, Dreamer~4~\citep{hafner2025training} learns long-horizon Minecraft behavior from offline data, and DIAMOND~\citep{alonso2024diffusion} generates pixel-level imagined experience using diffusion models. However, these digital results do not capture real-robot constraints, and imagined trajectories remain limited by model accuracy and data coverage. Online interaction can expand that coverage, while offline learning remains constrained by existing data.

Planning and imagined learning can coexist, but expose different decisions to model error: planning uses predictions for immediate action selection, whereas imagined learning propagates prediction errors through policy and value updates. MBPO~\citep{arxiv_1906_08253} limits error accumulation by generating short rollouts from real replay states. Evaluating imagined learning still requires real or held-out interaction and accounting for data, model-training, and rollout costs.

Across offline, online, and imagined-policy learning, offline improvement depends on supported outcomes; online improvement additionally requires affordable exploration, rewards, and resets (Figure~\ref{fig:overview-learning}). Imagined experience adds model coverage and prediction error to these constraints. Missing transitions may call for demonstrations, corrections, or new trials. Learned rewards need independent validation, and assisted resets must be distinguished from autonomous recovery when judging reduced supervision.

\subsection{Policy Pretraining and Post-Training}\label{sec:5.4}

Pretraining establishes a policy (\S~\ref{sec:5.4.1}); post-training refines it. Both use the objectives of \S\S~\ref{sec:5.1}--\ref{sec:5.3}. The remaining choices concern records (\S~\ref{sec:5.4.2}), source weights (\S~\ref{sec:5.4.3}), update scope (\S~\ref{sec:5.4.4}), supervised or reward-based refinement (\S\S~\ref{sec:5.4.5}--\ref{sec:5.4.6}), and their combination in workflows (\S~\ref{sec:5.4.7}). These choices recur across stages rather than defining a mandatory sequence.

\subsubsection{Policy Pretraining}\label{sec:5.4.1}

Policy pretraining connects general features to robot actions. RT-1 learns from robot trajectories using pretrained image and language encoders; RT-2 co-fine-tunes a web-pretrained VLM for action prediction~\citep{arxiv_2212_06817,arxiv_2307_15818}. Initialization and subsequent robot data must therefore be distinguished in comparisons.
Recent approaches distinguish pretraining and post-training by data diversity and quality. \(\pi_0\)~\citep{arxiv_2410_24164} pretrains on diverse multi-robot data to learn robustness and recovery, then post-trains on curated task data for precision, using a flow-matching action expert. \(\pi_{0.5}\)~\citep{intelligence2025pi05} expands pretraining to mobile and static robots, cross-embodiment data, subtask prediction, web data, and human instructions. It uses discrete action tokens during pretraining and adds a flow-matching expert during post-training. Ablations show that cross-embodiment data improve performance, while web data support generalization to unseen objects.

GR00T~N1 combines web, human-video, synthetic, and robot data, using latent or inverse-dynamics actions for unlabeled videos~\citep{bjorck2025groot}. This widens access while introducing pseudo-label errors.
Open X-Embodiment and Octo pool robot experience before specialization~\citep{arxiv_2310_08864,arxiv_2405_12213}. Coverage, interface compatibility, and sampling exposure shape the pretrained baseline against which target improvement and retention are measured.

\subsubsection{Sample Selection}\label{sec:5.4.2}

Selection allocates limited training resources among valid records, targeting missing conditions or corrections. Scores differ in what they estimate and whether they require target demonstrations, policy fitting, or rollouts.

\noindent\textbf{Relevance by Retrieval.} Retrieval selects prior data based on similarity to target demonstrations. Behavior Retrieval~\citep{arxiv_2304_08742} matches state--action pairs in a learned embedding. STRAP~\citep{arxiv_2412_15182} instead matches sub-trajectories using pretrained visual features and dynamic time warping, enabling reuse of relevant segments from unrelated tasks. RWLA~\citep{arxiv_2410_02995} retrieves scene-relevant demonstrations and upweights segments related to observed failures. All three depend on the prior data containing the required behavior.

\noindent\textbf{Demonstration Quality Without Rollouts.} Belkhale et al. assess expert--policy action divergence and transition diversity, showing that more state diversity need not help~\citep{belkhale2023data}. DemInf instead estimates state--action mutual information from embeddings, favoring predictable yet diverse actions without fitting or executing a policy for scoring~\citep{arxiv_2502_08623}.

\noindent\textbf{Influence on Policy Performance.} DataMIL~\citep{dass2025datamil} selects demonstrations using task-specific surrogate objectives, estimating their value through repeated policy training or metagradients without evaluation rollouts. CUPID~\citep{agia2025cupid} instead uses influence functions and evaluation rollouts to estimate each demonstration's effect on expected return, enabling harmful-data filtering and new-data selection. Their main trade-off is computational cost versus rollout requirements.

Retrieval needs relevant prior behavior; rollout-based influence needs interaction; surrogate scoring trades this for modeling or training cost. Sirius additionally weights intervention records~\citep{arxiv_2211_08416}. Filtering and weighting can combine with replay, but excessive target focus may remove earlier coverage.

\subsubsection{Training Mixtures}\label{sec:5.4.3}

Sampling weights determine exposure after selection. Episode counts alone do not specify it: transition sampling also depends on episode length. Mixtures can be size-proportional, source-uniform, manually weighted, or learned, but require aligned action conventions and control rates.
For \(M\) datasets, the training objective separates source size from sampling weight:

\begin{equation}
\mathcal L(\theta;w)=\sum_{i=1}^{M} w_i\,\mathbb E_{x\sim\mathcal D_i}[\ell_\theta(x)],\qquad w_i\geq0,\quad\sum_{i=1}^{M}w_i=1.\label{eq:3}
\end{equation}

Here, \(\mathcal D_i\) is source \(i\), \(\ell_\theta\) the sample loss, and \(w_i\) its training weight. Weights control source exposure but cannot compensate for missing behaviors. Zero weight excludes a source from the current objective without deleting it, while excessive weighting of small datasets may cause overfitting.
Re-Mix~\citep{arxiv_2408_14037} learns source weights for Open X-Embodiment using group distributionally robust optimization, minimizing worst-case excess behavior-cloning loss relative to a reference policy. This distinguishes reducible error from irreducible source difficulty. Weights are set before training, alongside action normalization, discretization, and early stopping. Re-Mix improves average task success by approximately 38 and 32 percentage points over size-proportional and human-designed RT-X mixtures, respectively. However, limited gains on full BridgeV2 show that its benefits depend on dataset heterogeneity and training budget.

Sim-and-real co-training samples simulated demonstrations with probability \(\alpha\) and real ones otherwise (Eq.~\eqref{eq:3})~\citep{maddukuri2025simandreal}. Its CupPnP study shows non-monotonic performance as \(\alpha\) increases: greater simulated exposure helps only within a useful range. No direct size-proportional baseline is reported.
Larger datasets can hurt performance when dominant domains introduce shortcuts or incompatible action conventions. Xing et al.~\citep{xing2025shortcut} attribute shortcut learning to limited within-dataset diversity and large cross-dataset differences, showing that targeted augmentation can reduce shortcuts in policies such as \(\pi_0\). Reweighting can limit source-specific biases but cannot create missing diversity, which requires augmentation or new data collection.

\subsubsection{Choosing What to Update}\label{sec:5.4.4}

When adaptation updates parameters, it must select both a parameter scope and the objectives to retain. The same choices can arise during initial policy training; here they determine how a pretrained policy is refined.
Freezing a pretrained encoder restricts adaptation to the action head or interface modules, preserving existing features but preventing their correction. Full fine-tuning offers greater task adaptation but risks losing earlier capabilities. LoRA~\citep{arxiv_2106_09685} provides a parameter-efficient alternative by freezing \(W_0\) and learning a low-rank update \(\Delta W=BA\), which can be merged without inference overhead. On Franka tasks, LoRA matches full fine-tuning of a smaller OpenVLA variant while updating only 1.4\% of parameters~\citep{arxiv_2406_09246}. However, low rank does not guarantee preservation of earlier skills.

Sequential adaptation trains on robot tasks after general pretraining, as in OpenVLA, while retained-objective training continues general supervision alongside robot learning, as in RT-2~\citep{arxiv_2307_15818}. Retaining objectives may preserve earlier capabilities but consumes training capacity, so retention must be evaluated. Knowledge insulation~\citep{driess2025knowledge} prevents gradients from a newly initialized action expert from disrupting the pretrained backbone. It trains the backbone with discrete action tokens and vision--language data while the flow-matching expert learns continuous actions from backbone features. Without this separation, \(\pi_0\) required 7.5 times more training steps for comparable table-bussing performance. Thus, objective design determines which parameters each data source updates. Models producing plans or text, such as PaLM-E~\citep{arxiv_2303_03378}, also require different evaluation from those directly controlling robots.

\subsubsection{Supervised Fine-Tuning}\label{sec:5.4.5}

Supervised fine-tuning uses additional action labels to correct target-task gaps. Unlike the general imitation mechanisms of \S~\ref{sec:5.2}, the question here is how much new supervision a pretrained policy needs and what it changes.
Requirements vary by task: \(\pi_0\) post-training uses roughly 5 to over 100 hours, while OpenVLA's Franka adaptations use 10--150 demonstrations. OpenVLA-OFT also changes action representation and decoding during refinement~\citep{arxiv_2502_19645}; gains must therefore be separated from data exposure alone.
RoboCat adapts to new demonstrations, collects trajectories, and incorporates achieved-goal relabeling into later generalist training~\citep{arxiv_2306_11706}. Fitting these action labels remains supervised learning even when the policy collected them. Corrections must cover failure states; successful episodes alone need not teach recovery.

\subsubsection{Reward-Based Fine-Tuning}\label{sec:5.4.6}

Reward-based refinement is useful when outcomes are easier to judge than actions to label. Direct policy updates use likelihood-based or critic-guided objectives~\citep{silver2014dpg}; a frozen generator can instead be steered through learned corrections; and improved specialist behavior can be transferred by distillation. These routes distinguish where improvement enters the pretrained policy, while parameter scope remains a separate choice.

\noindent\textbf{Direct Policy Updates.} Token policies expose action likelihoods. RIPT-VLA uses PPO-style sparse-reward updates, SimpleVLA-RL uses group-relative advantages, and VLA-RL combines PPO with critic and process feedback; RL-VLA\(^3\) adds asynchronous training~\citep{arxiv_2505_17016,li2025simplevlarl,lu2025vlarl,sun2026rlvla3}.
Implicit diffusion and flow policies need another likelihood interface. DPPO treats Gaussian denoising as an inner decision process, while \(\pi_{\texttt{RL}}\) introduces learnable noise or stochastic sampling paths~\citep{arxiv_2409_00588,chen2025pitextttrl}. Both enable reward-based updates of pretrained generative policies.

Critic-guided objectives provide another route to direct policy refinement. ConRFT freezes Octo-small's backbone and fits a consistency head, combining cloning with Cal-QL offline before human-supported online learning; Cal-QL alone fails to initialize a successful policy~\citep{chen2025conrft}. Learning While Deploying likewise uses a critic to refine a flow policy. The learning objective, rather than the head family, distinguishes these updates.

\noindent\textbf{Corrections to a Frozen Policy.} Residual RL learns corrections to a fixed controller~\citep{arxiv_1812_03201}. DSRL instead steers a frozen diffusion or flow generator through its input noise, improving \(\pi_0\) on a held-out LIBERO-90 task with black-box access~\citep{arxiv_2506_15799}. Fixed base weights do not imply unchanged behavior or guaranteed retention.

\noindent\textbf{Distillation from Improved Specialists.} Reinforcement-learned behavior can also improve generalists through supervised distillation. RLDG~\citep{arxiv_2412_09858} trains task-specific policies with HIL-SERL and uses their rollouts to fine-tune generalists such as OpenVLA and Octo. Improvements stem mainly from action quality rather than state coverage, though this requires training a specialist for each task.

A ManiSkill study~\citep{liu2025rl} further shows that RL fine-tuning of OpenVLA, particularly with PPO, generalizes better than supervised fine-tuning under semantic and execution shifts, while performing similarly under visual shifts. PPO also outperforms the tested DPO and GRPO implementations.

However, reward-based updates risk eroding pretrained behavior. One mitigation is to penalize deviation from a reference policy in Eq.~\eqref{eq:2b}:
\begin{align*}
J_{\mathrm{RL}}(\theta)&=J(\theta)-\lambda_{\mathrm{KL}}\,\mathcal K(\theta),\\
\mathcal K(\theta)&=\mathbb E_{(h,g)\sim d_{\mathrm{train}}}\!\left[\mathrm{KL}\big(\pi_\theta(\cdot\mid h,g)\|\pi_{\mathrm{ref}}(\cdot\mid h,g)\big)\right].
\end{align*}
Here \(\lambda_{\mathrm{KL}}\geq0\) controls deviation from the reference over \(d_{\mathrm{train}}\). This requires a tractable KL or suitable estimator; implicit flow and diffusion policies may need alternative surrogates. Such regularization limits drift but cannot guarantee retention, which requires direct evaluation. Overall, reward-based adaptation depends on the generator's available interface—likelihoods, critic gradients, or steerable inputs—and how much pretrained behavior the update is allowed to change.

\subsubsection{Post-Training Workflows}\label{sec:5.4.7}

Two systems illustrate how post-training combines feedback sources rather than selecting a single objective.
\(\pi^{*}_{0.6}\)'s RECAP combines demonstrations, on-policy experience, and expert interventions through advantage conditioning, followed by on-robot specialization~\citep{intelligence2025pi06}. Interventions serve both protection and learning and need records suitable for each use.

Learning While Deploying provides a second workflow. Sixteen dual-arm robots stream autonomous rollouts and executed corrections into a shared buffer, with terminal success/failure labels and retained offline data~\citep{wang2026while}. Distributional Implicit Value Learning preserves rare successes through quantile-based critic targets; Q-learning via Adjoint Matching converts critic gradients to flow-regression targets without explicit action likelihoods. Under a four-hour budget per method (about 60 robot-hours), the workflow scores 0.95 versus 0.76 for supervised fine-tuning across eight real tasks, four scored by binary success and four by step-wise progress. This is a whole-workflow comparison, not an isolated optimizer result.

Budgets must include initialization, demonstrations, interaction, corrections, and optimization. Compare candidate and parent on matched held-out new and retained tasks, with independent benefit and safety checks; reused validation tasks can bias selection.

\begin{figure}[!htbp]
\centering
\includegraphics[width=\linewidth]{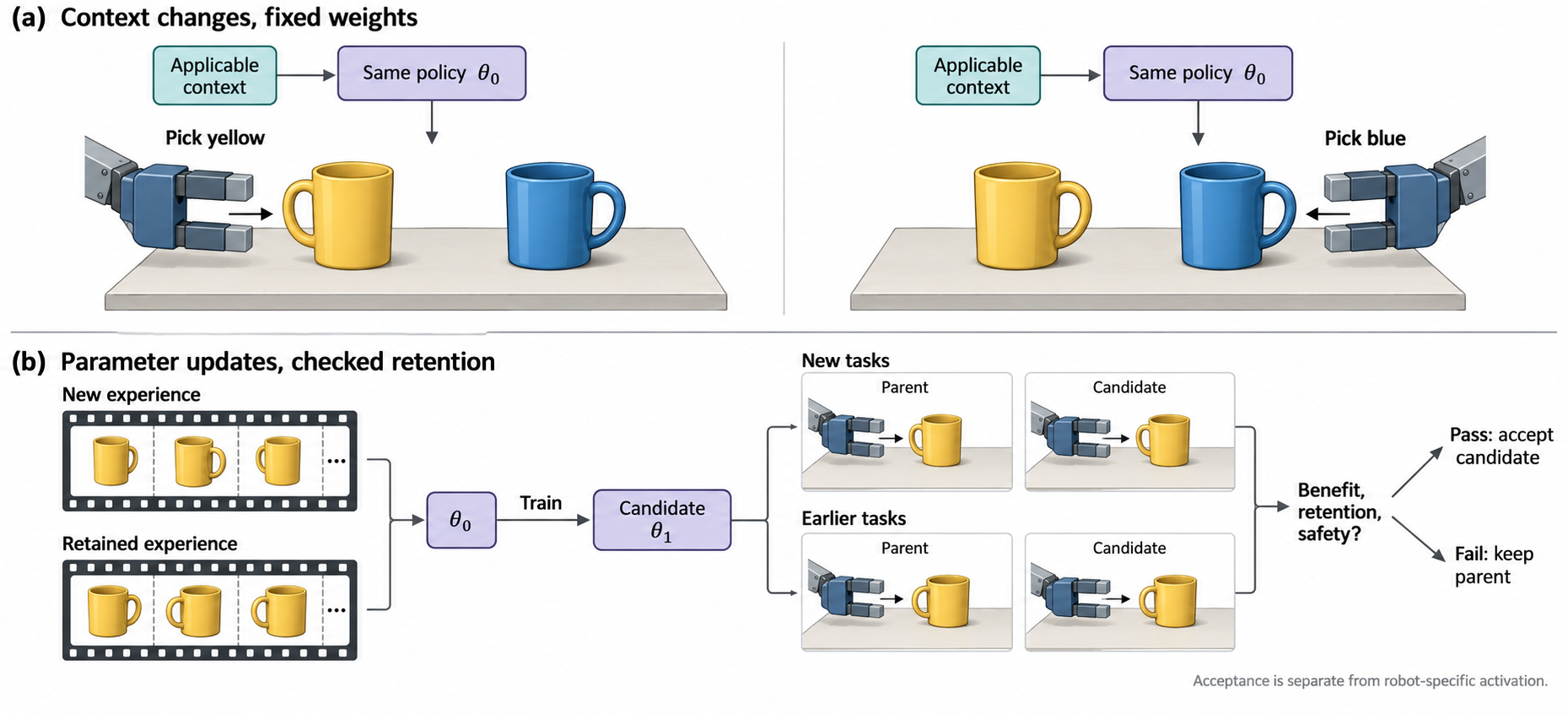}
\caption{Context changes and parameter adaptation. (a) Different instructions select different mugs using the same policy weights; retrieved context also requires applicability checks. (b) New and retained experience produce a candidate policy, which is compared with its parent on matched, held-out new and earlier tasks. Acceptance depends on benefit, retention, and safety criteria; failed checks retain the parent. Replay does not guarantee retention. Candidate acceptance is separate from robot-specific activation, and continual learning requires testing successive changes to the same evolving system.}\label{fig:overview-adaptation}
\end{figure}

Across pretraining and post-training choices, diagnose the failure before choosing an update: incorrect rewards need correction, interface mismatches may need calibration, and missing behaviors need additional supervision. Freezing, adapters, full fine-tuning, or contextual guidance address different limits; parameter efficiency alone does not establish retention. Figure~\ref{fig:overview-adaptation} separates fixed-weight context changes from parameter updates and matched candidate acceptance. Robot-specific activation remains a separate compatibility decision (\S~\ref{sec:7.6.3}).

\subsection{Transfer and Adaptation}\label{sec:5.5}

A robot transferring its cup-handling skills may encounter a new task, such as drying the cup (\S~\ref{sec:5.5.1}); a changed environment, such as another kitchen (\S~\ref{sec:5.5.2}); or a different robot body and action interface (\S~\ref{sec:5.5.3}). Sim-to-real transfer additionally examines mismatches between simulated training and physical deployment in observations, dynamics, and execution (\S~\ref{sec:5.5.4}). The order identifies what changed and what information is available for adaptation; these shifts can coexist. Adaptation may change instructions, retrieved context, calibration, or model parameters, with the post-training methods in \S~\ref{sec:5.4} applying when a parameter update is needed.


\subsubsection{Task Adaptation}\label{sec:5.5.1}

When a robot learns to dry a cup after washing it, adaptation may involve reusing existing skills, selecting them under new instructions, or learning new behaviors, such as wiping with a towel. Evaluation should distinguish skill acquisition from selection or recombination. Adaptation methods differ in whether they modify policy inputs, fine-tuning data, or model parameters through fixed or learned update rules.

\noindent\textbf{Adaptation Through Policy Inputs.} New instructions can enable task adaptation without parameter updates when the required skills already exist. RT-2 follows commands absent from its robot training data, such as placing objects on specified numbers or icons, while \(\pi_{0.5}\) executes long-horizon cleaning tasks from high-level prompts in unseen homes. Code as Policies~\citep{arxiv_2209_07753} composes existing perception and control functions into programs, using variables to organize intermediate results. These methods select or recombine learned behaviors, while retrieved examples can further condition a fixed policy without updating its parameters.
Demonstrations offer another form of input conditioning without parameter updates. ICRT~\citep{arxiv_2408_15980} uses a causal transformer trained on sensorimotor trajectories to predict actions. At inference, it conditions on a randomly selected teleoperated demonstration and the current observation. Franka-arm evaluations show adaptation to new tasks with familiar and unseen objects, but these tasks reuse learned pick-and-place and poking primitives, leaving generalization to unseen motion primitives untested.

Task ambiguity can also be resolved without parameter updates. KnowNo~\citep{arxiv_2307_01928} uses conformal prediction to calibrate candidate plan sets and requests human help when multiple options remain. Its completion guarantee assumes matched calibration and test distributions, successful assistance, and reliable execution; its objective minimizes set size rather than help frequency. YAY Robot~\citep{shi2024yell} instead accepts real-time verbal corrections, such as ``move left'', and immediately adjusts execution. Both modify policy inputs without updating weights, while persistence depends on how corrections are stored.

\noindent\textbf{Adaptation Through Parameter Updates.} When input-only adaptation is insufficient, target-data updates can learn new behaviors. Fine-tuning multitask priors such as Octo and RoboCat~\citep{arxiv_2405_12213,arxiv_2306_11706} exploits adaptability acquired through multitask training, whereas meta-learning explicitly optimizes for rapid adaptation. One-Shot Visual Imitation Learning~\citep{arxiv_1709_04905} learns an initialization that enables adaptation from a single visual demonstration, but its effectiveness depends on the coverage of prior tasks. Fair comparisons should therefore control target demonstrations and update budgets.
Prior records can supplement the target data used for these updates. Behavior Retrieval supplements a small target set with similar state--action records~\citep{arxiv_2304_08742}. As discussed in \S~\ref{sec:5.4.2}, this improves adaptation only when prior data contain relevant behavior.

The appropriate task-adaptation route depends on whether the required behavior already exists. Instructions, programs, or in-context examples can select existing skills; retrieval and target-data updates can support partially covered behaviors; new capabilities require evidence from further learning and task-level execution.

\subsubsection{Environmental Adaptation}\label{sec:5.5.2}

Environmental change can affect appearance, action feasibility, or hidden dynamics. Randomization broadens anticipated coverage; conditional policies infer current conditions; calibration or parameter updates fit remaining mismatches.
RMA first trains a simulated locomotion policy with privileged condition latents, then trains a module to infer those latents from state--action history~\citep{arxiv_2107_04034}. On the real A1 quadruped, only the inferred latent changes, so adaptation to terrain, payload, and wear is conditional inference rather than online parameter learning.

Randomized training broadens coverage, while calibration addresses specific mismatches. RWLA instead adapts to new scenes by fine-tuning a policy copy on retrieved demonstrations. Unlike conditional policies, this changes parameters and requires target data and retention testing.
Appearance and dynamics shifts should be evaluated separately, such as varying lighting and texture while fixing physics, then changing payload or friction. Training exposure and target-time information must be reported, since robustness to visual changes does not imply robustness to dynamics shifts.

\subsubsection{Cross-Embodiment Transfer}\label{sec:5.5.3}

New bodies change observations, action dimensions, reachability, and contact geometry. Multi-robot data broaden coverage, model interfaces accommodate embodiment-specific inputs and outputs, and shared action representations support reuse. These data, model, and interface choices can be combined. Numerical normalization alone cannot establish feasible execution.

\noindent\textbf{Training Coverage.} Open X-Embodiment~\citep{arxiv_2310_08864} standardizes data from 22 robots across 21 institutions, covering 527 skills. RT-X models trained on this dataset demonstrate positive cross-robot transfer, with RT-1-X outperforming the original method on four of five small-data domains but not surpassing embodiment-specific RT-1 on large-data domains. Such transfer depends on compatible observation and action mappings, as standardized data alone cannot resolve embodiment differences. Its effectiveness also depends on how well the target robot is represented in the training pool.

\noindent\textbf{Model Adaptation.} Octo~\citep{arxiv_2405_12213} trains a transformer on multi-robot trajectories and supports language or goal-image instructions, enabling zero-shot transfer and fine-tuning to new observations, action spaces, and embodiments. Related methods differ in how they handle embodiment-specific interfaces. HPT~\citep{wang2024scaling} uses embodiment-specific tokenizers with a shared trunk; CrossFormer~\citep{doshi2024scaling} trains across diverse robots without manual observation or action alignment; X-VLA~\citep{zheng2025xvla} uses embodiment-specific soft prompts; and UniVLA~\citep{bu2025univla} shares task-centric latent actions across robots. These designs reduce cross-embodiment adaptation costs, but differences in target data and control interfaces prevent direct zero-shot performance comparisons.

\noindent\textbf{Executable Interfaces.} UMI~\citep{arxiv_2402_10329} enables cross-robot transfer through handheld-gripper demonstrations represented as relative end-effector trajectories executable by different arms. However, transfer still depends on compatible gripper and contact geometry; reproducing wrist trajectories alone may be insufficient.

Evaluate coverage, model adaptation, and executable interfaces separately: gains attributed to a shared policy may depend on body-specific calibration or control. Comparisons must state target-data access and the physical differences between source and target robots.

\subsubsection{Sim-to-Real Transfer}\label{sec:5.5.4}

Sim-to-real transfer must address observation, dynamics, and execution mismatches. Three complementary choices organize the discussion: which simulated and real conditions training covers, which privileged information must be replaced by deployable observations, and how real feedback calibrates training or conditions execution. These choices can be combined; visual randomization alone cannot resolve dynamics or execution mismatches.

\noindent\textbf{Training Coverage.} Randomization varies simulator parameters \(\xi\), including textures, friction, mass, and latency, under \(p(\xi)\). Tobin et al. demonstrate real-object localization from randomized synthetic images~\citep{tobin2017domain}. A narrow distribution can miss deployment conditions, while an overly broad one wastes capacity; neither covers unmodeled contact or execution effects. Real-world transfer must therefore be tested directly.
Automatic domain randomization~\citep{openai2019solving} adaptively adjusts the simulator parameter distribution \(p(\xi)\). Starting from calibrated real-world values, it expands or contracts parameter ranges based on policy performance at their boundaries. This enabled a robotic hand to solve a Rubik's cube requiring 15 face rotations in 6/10 trials with sensorized face angles and 2/10 with vision-estimated angles. However, adaptation remains limited to the parameters selected for randomization.

Training coverage can also include real demonstrations. Sim-and-real co-training~\citep{maddukuri2025simandreal} combines real demonstrations with simulated data from task-specific digital cousins and task-agnostic multitask environments. Across six real tasks on a robot arm and a humanoid, it improves average success from 45\% to 83\%. While exact simulation replicas are unnecessary, effectiveness depends on sufficient simulated data and an appropriate sampling ratio.

\noindent\textbf{Deployable Observation Interfaces.} Lee et al. distill a simulated teacher with privileged terrain and contact information into a proprioceptive-history student for ANYmal~\citep{lee2020quadrupedal}. RMA similarly trains condition inference from deployable observations. Distillation cannot recover privileged information that those observations do not identify.

\noindent\textbf{Calibration and Condition Inference.} SimOpt uses real rollouts to calibrate the simulator distribution, whereas RMA infers current conditions without changing policy parameters~\citep{arxiv_1810_05687,arxiv_2107_04034}. They adapt training conditions and execution inputs, respectively.
Calibration and condition inference are limited by observability: physical parameters cannot be reliably identified when they produce indistinguishable trajectories. Sim-to-real evaluation should therefore report real-world adaptation budgets, robustness to uncalibrated changes, and separate observation, dynamics, and actuation gaps. Success under calibrated conditions does not guarantee transfer to untested environments.

Different transfer challenges require different adaptation strategies. For new tasks, instructions, in-context examples, or retrieved data may suffice when relevant skills exist, while novel behaviors require demonstrations and parameter updates. For new environments, augmentation and randomization address appearance shifts, whereas conditional policies or online estimation handle hidden dynamics changes. For new embodiments, transfer depends on executable action interfaces or embodiment-specific model adaptation.

\subsection{Continual Learning}\label{sec:5.6}

Continual learning must accumulate and retain capability across successive changes. We first examine where updates persist and how storage choices affect interference (\S~\ref{sec:5.6.1}), then how to test retention in the same evolving system (\S~\ref{sec:5.6.2}; Figure~\ref{fig:overview-adaptation}). Independently adapted copies do not establish cumulative learning.

\subsubsection{Where Changes Are Stored}\label{sec:5.6.1}

Shared updates may improve new tasks while degrading earlier skills. Changes can instead be stored in memory, separate modules, or shared parameters, each balancing interference, storage, retrieval, and validation costs. Continual learning must therefore determine which past experiences to retain, how to protect existing capabilities, and when updates are safe to deploy.

\noindent\textbf{External Records and Skill Libraries.} Memory and skill libraries can preserve knowledge without updating shared policy weights. Corrections should retain their scope, provenance, and validity conditions to support selective reuse. Voyager~\citep{wang2023voyager} demonstrates this approach in Minecraft, storing self-verified programs for retrieval on later tasks without parameter updates. YAY Robot~\citep{shi2024yell} instead uses verbal corrections immediately and later incorporates them through policy retraining. LOTUS~\citep{arxiv_2311_02058} expands a skill library by discovering new skills, updating existing ones, and training a meta-controller to compose them. While modular expansion can reduce interference, retention still requires evaluating both updated skills and the meta-controller's ability to select them appropriately.

\noindent\textbf{Separate Parameter Modules.} Isolation mitigates forgetting by freezing or separating model components. TAIL~\citep{arxiv_2310_05905} freezes pretrained encoders and transformer decoder while training adapters and task-specific fusion layers and policy heads, starting from a policy pretrained on 40 LIBERO kitchen tasks. It stores separate parameters for each task suite or individual task and restores them when needed. On ten long-horizon LIBERO tasks, low-rank adapters achieved higher peak new-task success than full fine-tuning while preserving earlier task parameters. In contrast, full fine-tuning, replay, and EWC showed negative backward transfer. Isolation reduces interference without replay but requires correct parameter selection at inference, while frozen components cannot improve through gradient updates.

\noindent\textbf{Shared Parameter Updates.} Experience replay mitigates forgetting by mixing stored past data into new updates, with effectiveness depending on how representative the retained data are. RWLA~\citep{arxiv_2410_02995} combines replay for skill retention with retrieval-based local adaptation, emphasizing demonstrations associated with observed failures. It operates without explicit task identities or boundaries. However, its evaluation initializes separate adaptation copies from the same checkpoint, demonstrating local adaptation rather than retention across sequential updates. Establishing continual learning requires evaluating the same evolving system on successive tasks and retesting earlier capabilities.
Regularization offers another way to limit interference in shared parameters. Elastic weight consolidation (EWC)~\citep{kirkpatrick2017overcoming} mitigates forgetting without storing past data. It uses the diagonal Fisher information to estimate parameter importance and applies a quadratic penalty that restricts changes to parameters important for earlier tasks. LIBERO includes EWC as a lifelong-learning baseline~\citep{arxiv_2306_03310}. However, EWC cannot eliminate forgetting: TAIL's LIBERO experiments show that it retains earlier skills less effectively and adapts to new tasks less successfully than full fine-tuning.

Storage choices can be combined: expansion can freeze earlier modules and their dependencies while replay or continued training updates other components. Memory and isolated modules reduce parameter interference but add retrieval, routing, and maintenance costs; shared updates enable reuse but retain forgetting risks. Evaluation must cover routing and shared components, storage and optimization costs, and every capability affected by an update. For personalization, test original tasks, the target user, and other users separately, with explicit release criteria such as parent--candidate comparisons (\S~\ref{sec:5.4.7}). Learning While Deploying illustrates the shared-update case: each round of fleet experience updates and redeploys the generalist, requiring retention checks across iterations.

\subsubsection{Capability Retention}\label{sec:5.6.2}
Continual learning requires evaluating retention across task sequences, not just final success rates. Three measures capture different effects: forward transfer assesses benefits to new tasks, forgetting measures degradation relative to earlier peak performance, and backward transfer measures positive or negative changes on previous tasks after subsequent learning.
LIBERO~\citep{arxiv_2306_03310} standardizes task sequences and protocols for these evaluations, while LOTUS~\citep{arxiv_2311_02058} and TAIL~\citep{arxiv_2310_05905} examine skill-based and parameter-efficient retention strategies. Methods should be compared under matched sequences and storage budgets, since LIBERO shows that sequential fine-tuning can outperform lifelong-learning methods in forward transfer despite weaker retention. This trade-off requires tracking new-task gains and earlier-task performance together throughout the sequence, across affected users and environments.

Across continual-learning mechanisms, context and memory retain scoped information; modules isolate learned changes at a routing and storage cost; shared parameters enable reuse but risk interference. Clearing context, removing records, or restoring weights can test where an improvement persists. Delayed tests on earlier tasks reveal forgetting and stale information.

Across the learning objectives and adaptation settings in this chapter, demonstrations establish behavior, corrections cover deployment gaps, and rewards support improvement when outcomes are verifiable. The choice depends on the failure and the cost of feedback, resets, and supervision. Matched experience budgets and repeated new/retained-task tests are needed to distinguish local adaptation from sustained learning.

\section{Embodiment and Control}\label{sec:6}

The body's geometry, dynamics, sensing, and actuation shape how a robot senses and acts. A policy produces actions at a specified interface; action conventions give their numerical values physical meaning, and the execution layer uses feedback and control to produce actuator commands and physical motion. The amount of control assigned to the policy varies with the interface. Differences in robot bodies, sensors, conventions, and controllers can therefore cause the same policy output to perform differently, while also shaping data collection and learning.

The chapter follows these physical dependencies: morphology sets the body's structural possibilities and constraints (\S~\ref{sec:6.1}), sensing and estimation provide information for decisions (\S~\ref{sec:6.2}), action interfaces specify what policy outputs request (\S~\ref{sec:6.3}), and feedback control turns those requests into commands while responding to measured conditions (\S~\ref{sec:6.4}). This order explains how the body, available information, and control responsibilities shape execution together.
In the cup task, reaching the shelf, gripping a wet cup, and detecting slip depend on these physical layers, not just the policy. Thus, some apparent learning failures may be better addressed through changes to the robot's body, sensors, or controller.

\begin{figure}[!htbp]
\centering
\includegraphics[width=\linewidth]{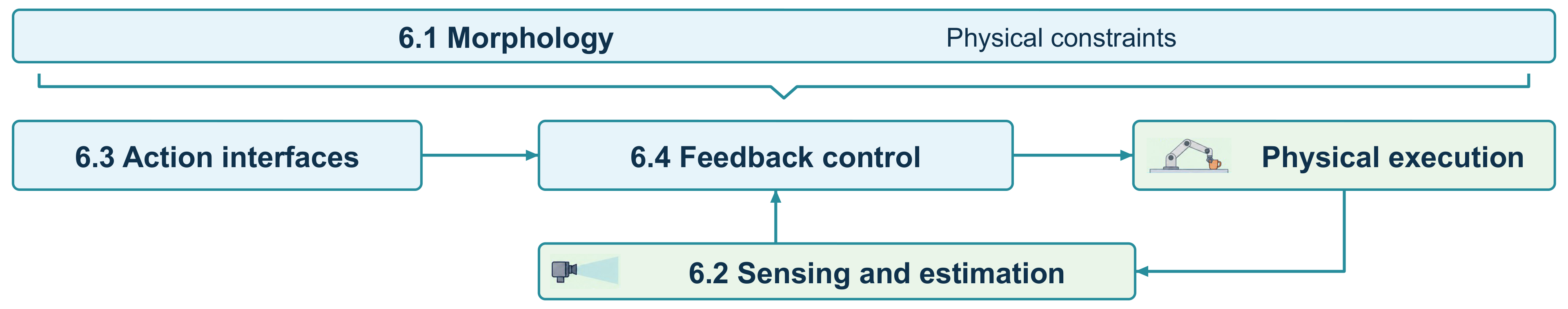}
\caption{Physical execution and feedback. Morphology constrains the system; action interfaces connect requested actions to feedback control. Sensing and estimation close the loop from physical execution back to control. Numbered components correspond to the four subsections.}\label{fig:overview-control}
\end{figure}

{\interlinepenalty=10000
Figure~\ref{fig:overview-control} asks when a policy output can become feasible physical behavior by connecting morphology, sensing, action meaning, and feedback control.\par}

\FloatBarrier

\subsection{Robot Morphology}\label{sec:6.1}

Morphology refers to the robot's physical form and structure, which shape what a policy can learn and execute. We move from kinematic structure, which determines reach and coordination (\S~\ref{sec:6.1.1}), to end effectors, which determine possible contact interactions (\S~\ref{sec:6.1.2}), and then to embodiment interfaces, which describe body-specific capabilities to shared policies (\S~\ref{sec:6.1.3}). This order connects physical constraints to contact and then to their representation for policy transfer. In the cup task, repositioning the base, changing the tool, or moving the shelf may overcome limitations that additional training cannot resolve.

\subsubsection{Kinematic Structure}\label{sec:6.1.1}

Kinematics determines whether action targets are physically feasible. Joint limits, obstacles, and redundancy constrain reachable poses and joint configurations. Adding arms, a mobile base, or legs introduces new capabilities and coordination constraints.
Bimanual manipulation requires coordinated motion under contact and collision constraints, as demonstrated by ACT~\citep{arxiv_2304_13705}. Mobile ALOHA jointly learns base and arm control, extending reach while introducing base-placement decisions~\citep{arxiv_2401_02117}.
Legged manipulation additionally requires maintaining balance and contact. HumanPlus uses learned whole-body shadowing for demonstrations~\citep{arxiv_2406_10454}, while optimization-based control~\citep{kim2019highly} and learned controllers such as HOVER and UMI on Legs~\citep{he2025hover,ha2024umionlegs} coordinate manipulation with support constraints. Thus, identical policy outputs may differ in feasibility and physical meaning across robot bodies.

\subsubsection{End Effectors}\label{sec:6.1.2}

End effectors determine how robots contact objects and adjust grasps. Parallel grippers use simple jaw closure, dexterous hands control multiple contacts, suction cups rely on sealed contact, and compliant components accommodate errors. These designs affect grasp precision, post-contact correction, and sensing requirements. However, more actuators do not necessarily mean greater dexterity; effective control also depends on training data, feedback, and action interfaces.

\noindent\textbf{Parallel Grippers.} PerAct and UMI illustrate manipulation with parallel grippers, where a single opening command makes grasps easy to record and transfer across compatible arms. However, fixed contact points limit post-grasp adjustments, often requiring controlled slip, regrasping, or environmental contact. Grasp success depends on approach pose, friction, grip force, and object properties. Thus, changing grippers affects more than opening range: differences in contact geometry and friction can alter the outcome of identical grasp commands.

\noindent\textbf{Dexterous Hands.} Multi-fingered hands can reposition objects without releasing them, but coordinating their contacts is challenging. A simulation-trained Shadow Hand policy used automatic domain randomization to manipulate a Rubik's cube, following a sequence generated by a conventional solver~\citep{openai2019solving}. Lin et al.~\citep{lin2025simtoreal} transferred vision-based humanoid hand policies from simulation to real-world grasping, lifting, and bimanual handover without further training. Both approaches relied on calibrating or randomizing hand dynamics, while a dynamics model learned from real experience enabled in-hand rotation of two balls. Effective dexterity thus requires training and evaluation across diverse contact configurations.

\noindent\textbf{Suction.} Vacuum grippers lift objects through a single sealed contact, avoiding the need to grasp around them. Success depends on forming a seal and resisting forces and torques during motion. Dex-Net 3.0~\citep{arxiv_1709_06670} models these conditions and trains a grasp-quality network using synthetic point clouds. Success rates dropped from 98\% on basic objects to 58\% on adversarial ones, revealing sensitivity to surface properties and geometry. Its rigid, nonporous object assumption further limits applicability, while poor depth sensing and unsuitable materials can prevent reliable sealing. Thus, suction simplifies grasping but sacrifices finger-based correction and depends heavily on surface conditions.

Across these contact mechanisms, passive compliance provides an additional design choice. Compliance can be built into hardware: soft fingertips conform to objects, while compliant wrists absorb alignment errors during contact. However, deformation introduces additional states that must be sensed or inferred, especially when manipulating deformable objects.
End effectors differ in their ability to correct errors after contact. While compliance reduces geometric errors, it cannot detect slip or measure contact forces on its own, making additional sensing necessary.

\subsubsection{Embodiment Interfaces}\label{sec:6.1.3}

A policy shared across robots must account for differences in reach, joint limits, gripper geometry, sensors, and control, which may not be fully observable from sensor inputs. A shared action layout specifies which quantities can be commanded; learned body-specific components or explicit descriptions can condition the policy on selected body differences; and an execution layer maps selected targets to the particular body. These roles differ: a command layout does not encode physical limits, policy conditioning does not itself make a target feasible, and target conversion does not ensure suitable contact. Cross-embodiment learning also draws on pooled multi-robot data, as in Open X-Embodiment~\citep{arxiv_2310_08864}, and body-specific modules around a shared policy trunk, as in Octo~\citep{arxiv_2405_12213}.

\noindent\textbf{Unified Action Layouts.} The action-vector layout defines what a policy can command. RDT-1B~\citep{liu2024rdt1b} uses a unified 128-dimensional action space with fixed physical meanings, such as joint positions or base velocities. Each robot fills its applicable entries and pads the rest, enabling shared supervision across bodies. However, the layout specifies available quantities, not their limits: an identical joint target may be feasible for one robot but unreachable for another.

\noindent\textbf{Conditioning on the Body.} HPT uses embodiment-specific stems to encode robot observations for a shared trunk~\citep{wang2024scaling}, while X-VLA uses per-source soft prompts~\citep{zheng2025xvla}. Both implicitly capture body differences from data but lack explicit limits for feasibility checks. Adapting to a new robot also requires data to train its body-specific components.
Body conditioning can also encode selected attributes through explicit embodiment descriptions. The Embodied Navigation Foundation Model~\citep{zhang2025embodied} uses identifier tokens to represent camera configurations and temporal context across quadrupeds, drones, wheeled robots, and vehicles. However, these tokens describe sensing configurations rather than joint limits or full-body kinematics.

\noindent\textbf{Execution Layers.} Below the policy, an execution layer translates shared targets into body-specific motions. UMI~\citep{arxiv_2402_10329} uses gripper-relative end-effector trajectories that each robot converts into joint commands. UMI on Legs~\citep{ha2024umionlegs} extends this interface to a quadruped through whole-body control, enabling zero-shot transfer from a fixed-base arm. HOVER~\citep{he2025hover} similarly supports multiple command modes for humanoid whole-body control. These layers reduce kinematic and coordinate differences but do not guarantee target feasibility or resolve differences in gripper properties and workspace.

Across morphology and embodiment interfaces, mechanical capability must be balanced against learning complexity. Passive compliance simplifies force control, while independent contact manipulation may justify additional actuation and coordination. Embodiment interfaces should expose physical limits and reject infeasible targets. Evaluation on unseen robot bodies, materials, and geometries helps determine when simple tools suffice and when greater dexterity is necessary.

\subsection{Sensing and State Estimation}\label{sec:6.2}

Sensor placement determines what a robot can observe. In the cup task, the robot must locate the cup, detect contact, and track it through occlusion. The discussion separates measurement sources from the estimation that uses them: exteroceptive sensing measures the environment (\S~\ref{sec:6.2.1}); proprioceptive and contact sensing provide complementary evidence about body motion and interactions (\S~\ref{sec:6.2.2}); and state estimation integrates these measurements across frames and time (\S~\ref{sec:6.2.3}). Adding sensors, adjusting their placement, and improving estimation address different limitations in observation. The goal is to provide timely state estimates and uncertainty for policy and control.

\subsubsection{Exteroceptive Sensing}\label{sec:6.2.1}

Before contact, the robot must locate the cup, surrounding obstacles, and free workspace. However, images lack direct metric depth, occlusions hide objects, reflective or transparent surfaces challenge depth sensing, and measurements become outdated during motion. Exteroceptive sensing must therefore provide sufficient coverage, geometric accuracy, contact-level resolution, and timely updates.
Sensor modality determines what is measured: RGB cameras capture appearance, depth and RGB-D sensors provide metric geometry, and LiDAR measures longer-range distances. Placement determines the viewpoint: fixed cameras provide stable views, wrist-mounted cameras offer close-up observations, and body-mounted cameras support navigation. Field of view and occlusion further constrain what each sensor can observe.

\noindent\textbf{Appearance and Depth Cameras.} RGB-D sensing enables policies to predict metric 3D actions. PerAct~\citep{arxiv_2209_05451} uses multiview RGB-D observations and language goals to classify actions in a voxel space, while 3D Diffusion Policy~\citep{ze20243d} generates actions conditioned on sparse point clouds. Both depend on reliable depth, which can fail on transparent or reflective surfaces.
ClearGrasp~\citep{arxiv_1910_02550} addresses this limitation by predicting surface normals, transparent regions, and boundaries from RGB images, then reconstructing missing depth through optimization. On transparent objects, depth repair improved suction grasp success from 64\% to 86\% and parallel-jaw grasp success from 12\% to 72\%. These results show that depth errors can create material-dependent failures even when policies perform well on opaque objects.

\noindent\textbf{Range Sensing at Vehicle Scale.} For large-scale navigation, LiDAR provides long-range geometry while cameras capture visual context. TransFuser~\citep{arxiv_2104_09224} combines image and LiDAR features through attention to capture both local geometry and global context, such as traffic lights and approaching vehicles. In CARLA simulations, it reduced collisions compared with geometry-based fusion, demonstrating the value of global context in urban driving.

Beyond sensor modality, viewpoint is a separate configuration choice. Camera placement determines whether observations remain fixed in the workspace or move with the robot. PerAct uses a fixed workspace representation, ViNT~\citep{arxiv_2306_14846} relies on an onboard camera, and UMI~\citep{arxiv_2402_10329} uses a gripper-mounted camera to align demonstration and deployment viewpoints. Hsu et al.~\citep{arxiv_2203_12677} found that hand-centric views improve learning efficiency and out-of-distribution generalization when they provide sufficient task visibility. Otherwise, third-person views may be necessary but can reduce generalization, an effect mitigated by information bottlenecking. Camera placement also affects policy robustness, as examined by LIBERO-Plus~\citep{fei2025liberoplus}.
Fixed cameras provide stable workspace views but may suffer occlusion, while wrist cameras capture local contact details but lack global context. Moving-camera observations also require pose transformations. Sensor placement should therefore match each task stage's visibility needs.

\subsubsection{Proprioception and Contact Sensing}\label{sec:6.2.2}

As the hand approaches the cup, vision may fail to reveal occluded contacts, grasp stability, or small slips. Rapid contact changes and forces require sensing beyond vision. Proprioceptive, wrist force-torque, and tactile sensors provide complementary information about body configuration, overall interaction forces, and local contact states. Sensor choice depends on what remains unobservable and the feedback speed required for control.

\noindent\textbf{Proprioceptive Measurements.} Joint encoders measure positions, with velocities estimated through filtering and differentiation. IMUs measure angular rate and specific force for orientation estimation, while motor currents or torque sensors indicate joint effort. Gripper aperture and current provide indirect evidence of grasping but cannot confirm grasp stability. These signals describe the robot's internal state regardless of contact and typically enter policies as proprioceptive input \(s_t^{\mathrm{prop}}\) (\S~\ref{sec:3.4.4}).
Proprioceptive signals can also detect contact without dedicated sensors. Generalized-momentum observers~\citep{haddadin2017robot} estimate external joint torques from discrepancies between predicted and applied torques, using joint positions, velocities, and motor torques without acceleration measurements. Thresholding the filtered residual detects sufficiently strong contacts, but some forces remain unobservable. Contact localization requires additional geometric assumptions, while model errors can cause false alarms. Thus, detection thresholds balance sensitivity and reliability.

\noindent\textbf{Force and Tactile Sensing.} Wrist force--torque sensors measure overall interaction loads but cannot localize finger contacts, whereas tactile sensors capture local contact information. ReSkin~\citep{arxiv_2111_00071} uses replaceable magnetic elastomer skins with 400\,Hz sensing and self-supervised recalibration to estimate contact location and force. Sparsh~\citep{arxiv_2410_24090} learns transferable visual-tactile representations from over 460k unlabeled images across gel-based sensors. It improves performance on six TacBench tasks, including force, slip, pose, and grasp-stability estimation, compared with task-specific training. However, representation-level improvements do not necessarily imply better manipulation success.

Using these contact measurements in a policy requires task-level validation. Touch Begins Where Vision Ends~\citep{zhao2025touch} combines vision-based object localization with a reusable local policy using egocentric vision and tactile sensing for contact-rich manipulation. Ablations show that tactile feedback substantially improves success, particularly when occlusion limits vision. However, tactile data remain scarce: Sparsh learns from unlabeled tactile images, while DexCap~\citep{wang2024dexcap} records human wrist and finger motions without directly measuring contact forces.

Joint-torque residuals detect external loading, wrist sensors measure net forces, and tactile sensors resolve local contacts. Although noise, saturation, hysteresis, and calibration drift limit reliability, combining these signals can recover contact states that individual sensors cannot observe.

\subsubsection{State Estimation}\label{sec:6.2.3}

Sensors operate in different coordinate frames, at different rates, and with different delays, requiring a consistent state estimate. Calibration aligns coordinate frames, synchronization aligns measurements in time, and state estimation combines sensor data with motion models to handle noise, occlusion, and missing observations. For example, when the wrist camera loses sight of the cup, the robot must propagate its estimate using motion information until new observations become available.
The comparison below distinguishes model-based estimation from learned estimation, while allowing hybrid systems. Contact measurements can constrain the former; history can supply features for the latter. Both can estimate uncertainty, but neither guarantees reliable uncertainty estimates without validation. Estimates must also account for their age, since outdated measurements may no longer reflect the current scene.

Calibration and timing errors systematically distort observations and actions. Hand--eye calibration estimates the wrist camera's pose relative to the gripper using paired motions satisfying \(AX=XB\). Tsai and Lenz~\citep{tsai1989new} solve rotation before translation, requiring motions about at least two non-parallel axes. UMI~\citep{arxiv_2402_10329} addresses temporal alignment by matching inference latency to training conditions (Section~\ref{sec:4.6.1}; Eq.~\eqref{eq:4}). Policies may adapt to consistent alignment errors during training but fail when those errors change across robots or setups.

\noindent\textbf{Model-based Estimation.} Kalman filtering~\citep{kalman1960filtering} predicts state and covariance using a motion model and updates them with measurements. Its uncertainty estimates depend on model accuracy and generally increase during occlusion. Visual--inertial odometry (VIO) combines slower visual observations with high-rate but drifting IMU measurements. VINS-Mono~\citep{arxiv_1708_03852} uses sliding-window optimization, IMU pre-integration, online calibration, and loop closure to improve estimation and reduce drift. Swift~\citep{kaufmann2023swift} uses 100\,Hz VIO for drone racing and corrects motion-blur-induced position and velocity drift with a Kalman filter using detected race gates, while retaining VIO orientation.
Legged robots can use stationary foot contacts to estimate base motion. The contact-aided invariant extended Kalman filter~\citep{arxiv_1904_09251} combines IMU propagation with leg-kinematic measurements, treating foot slip as contact-velocity noise. Its bias-free formulation provides estimate-independent linearized error dynamics and observation Jacobians, though IMU biases weaken this property. Experiments on Cassie demonstrate improved convergence, but significant foot slip can violate the stationary-contact assumption and degrade estimation.

\noindent\textbf{Learned Estimation.} Learned state estimation can use latent representations instead of explicit physical states. Learning Quadrupedal Locomotion over Challenging Terrain~\citep{lee2020quadrupedal} uses proprioceptive history to control ANYmal across unseen terrains, while RMA~\citep{arxiv_2107_04034} explicitly estimates latent body and environment conditions from recent states and actions. Both are trained in simulation. Learning High-Speed Flight in the Wild~\citep{arxiv_2110_05113} directly predicts collision-free trajectories from noisy observations while retaining visual--inertial state estimation.
Learned representations can simplify estimation pipelines but do not guarantee calibrated uncertainty or generalization to unseen dynamics. Runtime monitoring therefore requires explicit confidence estimates or independent consistency checks.

Object tracking illustrates the remaining limits when observations become occluded. The robot must track the cup's pose even under occlusion. FoundationPose~\citep{arxiv_2312_08344} estimates and tracks unseen objects' 6D poses using CAD models or reference images. After initialization, it refines the previous pose by comparing rendered objects with new RGB-D observations. However, severe occlusion, weak textures, and poor edge cues can cause tracking failures. Since FoundationPose lacks explicit object dynamics, maintaining pose estimates through complete occlusion requires additional mechanisms to track uncertainty and observation age.

Across sensing and estimation choices, effective sensing requires deciding what to measure directly, infer, or regulate through feedback. Repositioning cameras can resolve occlusion, while touch and force sensing reveal contact states that vision cannot. Rich tactile policies support complex interactions, whereas simpler force feedback may suffice for narrower tasks. Sensor transfer depends on calibration and the contact states that must be distinguished.
Both explicit and learned state estimates require validated uncertainty, and neither can recover fundamentally unobservable information. Recognizing these limits is essential for timely control and safe intervention.

\subsection{Action Interfaces}\label{sec:6.3}

Once the robot estimates the cup's state, the policy must propose actions that can be interpreted and executed. We first examine physical action spaces (\S~\ref{sec:6.3.1}): the quantities supplied at the interface, such as position, velocity, torque, or impedance targets, assign different responsibilities to the downstream controller, while tokens or latent actions require decoding before they reach a physical interface. We then specify action conventions (\S~\ref{sec:6.3.2}), whose frame, scale, timing, and reference update rule determine how those quantities are interpreted. The runtime determines when targets are used, and controllers and filters determine the commands ultimately applied. Embodiment interfaces in \S~\ref{sec:6.1.3} describe or compensate for body differences across robots; the present section specifies the meaning of each action at a particular execution interface.

\subsubsection{Physical Action Spaces}\label{sec:6.3.1}

A physical action interface specifies both a coordinate space and the quantity requested from the controller. Joint or Cartesian coordinates can carry motion targets, while position, velocity, torque, and impedance outputs assign different feedback responsibilities. These are intersecting choices, not categories determined by vector length. Encoded tokens or latent actions must ultimately be decoded into such an interface.

Joint-space targets specify robot configurations, leaving tracking and constraints to the controller. End-effector targets improve transfer across robots but require inverse kinematics to resolve feasibility, redundancy, and collisions. VICES~\citep{arxiv_1906_08880} compares joint and end-effector control, position and velocity commands, and variable impedance across free-motion and contact-rich tasks. Its end-effector impedance policies demonstrate cross-robot and sim-to-real transfer. Similarly, UMI~\citep{arxiv_2402_10329} transfers end-effector trajectories from handheld demonstrations to UR5e and Franka arms. While end-effector control simplifies policy transfer, downstream controllers must still handle joint configurations and collision avoidance.

Velocity commands specify motion rates but can accumulate position errors without corrective feedback. Diffusion Policy~\citep{arxiv_2303_04137} found that position control consistently outperformed velocity control for its diffusion model, while BCRNN and BET showed the opposite trend. The authors attribute this difference to reduced error accumulation and diffusion models' ability to represent multimodal position actions. Velocity control was also more sensitive to simulated latency. Torque control places greater responsibility for stabilization and compensation on the policy, requiring it to meet control deadlines.

Variable-impedance policies output motion targets alongside stiffness and damping, allowing adaptation to contact forces. Learning Variable Impedance Control~\citep{arxiv_1907_07500} compares learned joint-space impedance with torque and fixed-gain position control in hopping and manipulation tasks. Direct-torque policies failed to learn hopping, while a position-tracking reward, separate from torque smoothness, helped produce interpretable targets and supported hardware transfer. VICES~\citep{arxiv_1906_08880} found variable impedance particularly useful for wiping, but less necessary for path following and door opening. This flexibility improves contact handling at the cost of larger action spaces and additional stability constraints.

Encoded actions, including FAST tokens, OpenVLA-OFT continuous outputs, and LAPA and UniVLA latent actions~\citep{arxiv_2501_09747,arxiv_2502_19645,ye2024latent,bu2025univla}, require decoding into physical commands. Decoding may be deterministic or learned, introducing potential errors; shared latents need not uniquely specify body-specific actions.

No action space is universally optimal. Position control delegates dynamics to the controller, velocity and torque control demand more from the policy, and variable impedance improves contact flexibility but adds complexity and stability constraints. As VICES illustrates, performance depends on the combination of action space and low-level controller, not the action representation alone.

\subsubsection{Action Conventions}\label{sec:6.3.2}

Action conventions define the reference frame, scale, timing, and update rule that give policy outputs physical meaning. The same numerical command can produce different motions depending on whether it uses world, base, or end-effector coordinates, absolute or relative targets, and per-step or per-chunk updates. Calibration errors shift targets, stale references offset action chunks, and timing mismatches alter execution speed and potentially the resulting motion.
Training and execution must use consistent conventions, including resampling and time scaling. Action normalization also requires matching statistics to recover physical units. Importantly, coordinate space (joint or Cartesian) and reference update rules (absolute, measured-relative, or chunk-relative) are separate choices; absolute joint targets need no global Cartesian frame.

\begin{figure}[!htbp]
\centering
\includegraphics[width=\linewidth]{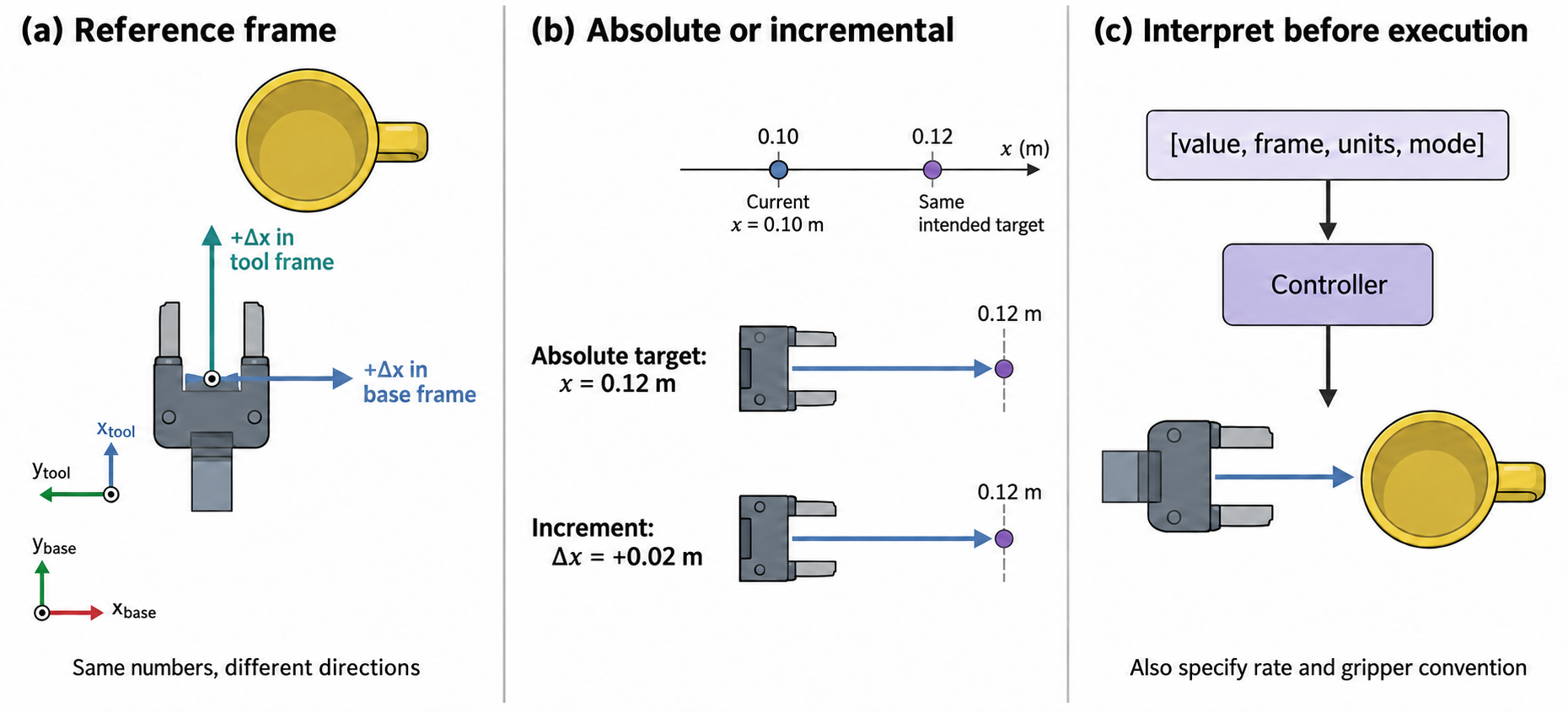}
\caption{Action meaning depends on interface conventions. (a) With a tool frame rotated $90^\circ$ from the base frame, the same positive $x$ increment requests different planar directions; the arrows show alternatives. (b) From $x=0.10$\,m, an absolute target of $0.12$\,m and a measured-relative increment of $+0.02$\,m specify the same target. This equivalence assumes the stated reference and units. (c) The controller must interpret values, frames, units, modes, rates, and gripper conventions consistently. The planar example does not establish dynamic feasibility or prescribe a full pose representation.}\label{fig:action-conventions}
\end{figure}

Figure~\ref{fig:action-conventions} illustrates these dependencies in a planar example. Rotating the tool frame changes the physical direction of an identically numbered increment. Conversely, two different numerical descriptions can request the same target: from a measured position of $0.10$\,m, an absolute target of $0.12$\,m matches a $+0.02$\,m increment in the same frame. This equivalence concerns one specified reference, not arbitrary repeated or chunk-relative updates.

\noindent\textbf{Target References.} Absolute joint targets, as in ACT, are executed by joint controllers, while absolute Cartesian targets also require coordinate transformations. Measured-relative targets use the observed pose as their reference. Repeated relative commands may accumulate errors depending on feedback and reference updates. In contrast, UMI's delta-action baseline defines each action relative to the preceding commanded action rather than the measured pose~\citep{arxiv_2402_10329}.
A trajectory chunk can instead share one reference pose. SRT~\citep{arxiv_2407_12998} predicts waypoints relative to a shared initial end-effector pose, expressing translation in the endoscope frame and rotation relative to the tool frame. This reduces sensitivity to forward-kinematics inconsistencies in surgical tasks, though large-scale clinical training remains a proposed extension. UMI~\citep{arxiv_2402_10329} similarly uses a shared reference pose for each action chunk, with latency alignment. Despite differences in coordinate frames and control interfaces, both approaches remain sensitive to stale or biased reference poses, which affect the entire action chunk.

\noindent\textbf{Rotation Representations.} Orientation representations introduce discontinuities absent from position coordinates. Euler angles suffer from wrapping and gimbal lock, while quaternions have sign ambiguity, complicating regression. Zhou et al.~\citep{arxiv_1812_07035} show that \(\mathrm{SO}(3)\) has no globally continuous representation in four or fewer Euclidean dimensions and propose continuous 5D and 6D alternatives. The 6D representation reconstructs rotations by orthogonalizing two vectors and computing their cross product, with numerical handling for degenerate cases. Thus, rotation conventions must specify both the representation, its decoding, and the reference frame.

Policy performance depends on keeping both reference and rotation conventions consistent during execution. OpenVLA-OFT~\citep{arxiv_2502_19645} reports gains under specific embodiment and action interfaces, but changes in control frequency, gripper geometry, or normalization can alter performance. Such sensitivities may remain hidden in benchmark results, just as camera placement affects observations.
Absolute Cartesian targets depend on calibrated transforms, measured-relative targets on accurate reference updates, and chunk-relative trajectories on their initial reference. None is universally superior. Reliable execution requires consistent frames, scales, rates, references, and rotation conventions between training and deployment.


Across physical action spaces and conventions, the controller translates policy actions into physical motion. Position control handles joint dynamics, supporting target reuse across compatible robots, while torque control places greater responsibility for stabilization on the policy. Impedance and latent actions offer flexibility, but their benefits must be distinguished from those of the low-level controller. Comparing representations with a fixed controller and interfaces with a fixed policy isolates different effects. Action meanings must be preserved in dataset records, while feedback control determines whether the requested actions can be realized.

\subsection{Feedback Control}\label{sec:6.4}

Feedback control determines how actions are executed using updated measurements. We first consider motion and contact control (\S~\ref{sec:6.4.1}), which tracks targets and regulates interaction forces, then learning in the control loop (\S~\ref{sec:6.4.2}), which examines how learning changes or complements those feedback mechanisms. This order establishes the control responsibilities before examining what learning changes. In the cup task, a controller may track the wrist target and regulate contact forces; an interface that supplies joint torques assigns more of that responsibility to the policy and its lower-level loops. Comparing these approaches requires specifying the action interface, base controller, feedback signals, and operating conditions.
Figure~\ref{fig:overview-control} places action interfaces upstream of feedback control; the interface determines which responsibilities belong to the policy and controller.

\subsubsection{Motion and Contact Control}\label{sec:6.4.1}

Feedback control tracks motion targets while regulating contact forces. Stiff position control can produce excessive forces when pose errors occur, especially when contact dynamics are faster than policy updates. Position control minimizes tracking errors, force control regulates contact loads, impedance control converts motion errors into forces, and admittance control converts measured forces into motion. These approaches differ in how they handle contact and distribute control responsibilities. Direct torque policies assume more of this responsibility, although actuator loops and independent limits remain.

For \(n\) revolute joints, let \(q_t\in\mathbb R^n\) denote joint angles and \(\bar a_t=q_t^{\mathrm{des}}\) the policy's desired configuration. A stiffness--damping controller computes joint torques as
\begin{equation}
u_t=\mathsf K\,(\bar a_t-q_t)+\mathsf B\,(\dot{\bar a}_t-\dot q_t)+\mathsf f_t,
\label{eq:compliant-tracking}
\end{equation}
where \(u_t\) and \(\mathsf f_t\) are torques in N\,m, \(\mathsf K\) is stiffness in N\,m/rad, and \(\mathsf B\) is damping in N\,m\,s/rad. The feedback terms resist position and velocity errors, while \(\mathsf f_t\) can compensate gravity, friction, or contact loads. The sans-serif symbols distinguish controller quantities from the executed prefix \(K\) in Eq.~\eqref{eq:1}. This formulation does not explicitly shape inertia, and piecewise-constant targets require a defined desired velocity, such as zero during holds.

Position control reduces tracking errors but can generate excessive contact forces at high stiffness. Force control instead regulates measured forces through force-error feedback. Variable impedance adjusts \(\mathsf K\) and \(\mathsf B\) alongside the target (\S~\ref{sec:6.3.1}), balancing responsiveness and damping. Admittance control converts measured forces into target motion for an inner position controller, whereas direct torque control lets the policy specify \(u_t\). In all cases, stability and constraint satisfaction depend on the complete feedback system.

Three classical formulations underpin these control laws. Hybrid position/force control~\citep{raibert1981hybrid} separates task directions into position- and force-controlled subsets, such as tangential motion and normal force during wiping. Impedance control~\citep{hogan1985impedance} regulates the dynamic relationship between motion and contact force, with Eq.~\eqref{eq:compliant-tracking} providing a joint-space stiffness--damping realization. Operational-space control~\citep{khatib1987unified} formulates end-effector dynamics and maps task-space forces to joint torques through the Jacobian transpose.
Hybrid control requires identifying constrained directions, while operational-space dynamic decoupling relies on an accurate dynamics model. Impedance control's model requirements depend on its implementation. Learned approaches retain these structures while adapting targets or gains, as reviewed in~\citep{arxiv_2010_06246}.

Legged and humanoid robots require whole-body control to maintain balance while tracking commands. Kim et al.~\citep{kim2019highly} combine MPC for planning ground reaction forces with whole-body impulse control for generating joint commands. HOVER~\citep{he2025hover} instead learns unified humanoid control across navigation, loco-manipulation, and tabletop tasks, allowing high-level commands while handling balance below. Model-based constraints and learned balance provide different assurances, but neither guarantees arbitrary target feasibility.

Controllers correct tracking and contact errors but cannot resolve task-level mistakes. In the cup task, the policy selects the target, the controller manages contact, and a separate check verifies the grasp. Control gains, update rates, and enforced limits therefore matter when comparing systems.

\subsubsection{Learning in the Control Loop}\label{sec:6.4.2}

Learning can modify three parts of a feedback system: the controller output or its parameters, the targets supplied to a lower-level controller, or the predictive model used to choose actions. These locations assign different responsibilities to learning and can coexist. In the cup task, engineered compliance can handle small alignment errors while learned components address uncertain friction or contact.

\noindent\textbf{Controller Corrections and Parameters.} Residual Reinforcement Learning~\citep{arxiv_1812_03201} combines conventional feedback control with learned corrections to handle poorly modeled dynamics, demonstrating contact-rich block assembly with unstable objects. ASAP~\citep{he2025asap} instead learns action corrections from real data to improve simulated dynamics before fine-tuning a humanoid tracking policy. However, adding residual corrections does not guarantee stability. Formal assurances require explicit assumptions about system dynamics, feedback, and residual bounds, together with stability or robustness analysis~\citep{arxiv_2108_06266}.
Instead of adding an output correction, learning can change controller parameters. Parameter learning retains the control law in Eq.~\eqref{eq:compliant-tracking} while learning its gains, as in VICES~\citep{arxiv_1906_08880} and Learning Variable Impedance Control~\citep{arxiv_1907_07500}. However, time-varying stiffness and damping can destabilize the system, even without contact. Kronander and Billard~\citep{kronander2016stability} derive stability constraints linking stiffness, its rate of change, and damping. Learned gains therefore require explicit bounds and stability analysis rather than relying on fixed-gain guarantees.

\noindent\textbf{Targets for Lower-level Feedback.} Hierarchical control separates high-level target generation from low-level feedback execution. NaVILA~\citep{cheng2024navila} uses a VLA model to generate spatial language commands, executed by a learned locomotion policy using proprioception and LiDAR-based terrain information. HumanPlus~\citep{arxiv_2406_10454} similarly combines imitation-learned skill policies with a simulation-trained whole-body controller. High-level targets must remain within the low-level controller's validated operating limits, with infeasible commands rejected or replanned. The update rates of both levels determine how quickly errors can be corrected.
The retained feedback layer can also be a conventional servo. RMA~\citep{arxiv_2107_04034} outputs joint-position targets at 100\,Hz to the A1 robot's PD controller, using the estimator described in \S~\ref{sec:6.2.3}. Similarly, VLA policies often rely on position or impedance controllers for tracking and contact response. This resembles hierarchical control but differs from residual learning: the retained controller need not accomplish the task independently.

\noindent\textbf{Predictive Models for Control.} Model-based learning retains the MPC optimizer while learning its predictive model. TD-MPC2~\citep{arxiv_2310_16828} optimizes actions in a learned latent dynamics model using receding-horizon planning (Eq.~\eqref{eq:plan}), with results demonstrated in simulation. PDDM~\citep{arxiv_1909_11652} uses an ensemble of learned dynamics models and sampling-based optimization to control a Shadow Hand at 10\,Hz through a 1\,kHz position controller, enabling in-hand rotation of two Baoding balls using real-world training data. Unlike residual or parameter learning, model learning changes the predictions guiding action selection, but inaccurate contact dynamics can lead to planning errors that low-level tracking cannot resolve.

In the cup task, policies select grasp targets, compliant controllers handle contact deviations, and higher-level decisions manage regrasping. Failures must be traced to perception, calibration, timing, or control rather than inferred from symptoms alone. A competent engineered controller supplies the baseline for local, measurable errors; residual learning can address remaining deviations, while more extensively learned control requires broader validation of complex interactions. Adaptive compliance and better contact sensing may also improve performance without increasing policy complexity. Across these choices, low-level feedback, multirate execution, and runtime filtering remain essential.

Across the physical design choices in this chapter, body and control design can improve capability without increasing policy complexity. Missing contact information may call for better sensing, while reachability or compliance failures may be addressed through tool, workspace, or controller changes. UMI and VICES illustrate these choices without establishing a universal preference for engineering over learning.
Adapting to new robots requires not only training data but also calibration, control adjustment, and maintenance. Shared representations therefore do not eliminate body-specific engineering. Design decisions should balance these recurring costs against the benefits of policy robustness.

\section{Computing Systems}\label{sec:7}

\begin{figure}[!htbp]
\centering
\includegraphics[width=\linewidth]{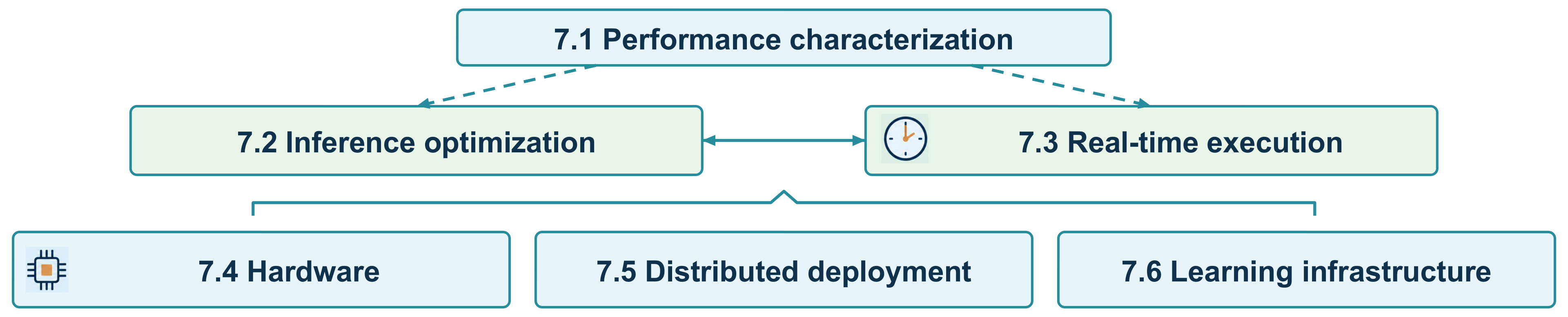}
\caption{Computing for physical systems. Performance characterization guides inference optimization and real-time execution, which constrain one another. Hardware, distributed deployment, and learning infrastructure provide the supporting resources and organization.}\label{fig:guide-07}
\end{figure}

Computation determines when robot decisions and policy updates become usable, within limits on energy, hardware, and network resources. Because the physical state continues changing, faster components improve task performance only when they shorten a relevant response path.
The chapter follows the choices in Figure~\ref{fig:guide-07}: characterize workload and timing (\S~\ref{sec:7.1}), optimize inference (\S~\ref{sec:7.2}), and schedule execution (\S~\ref{sec:7.3}). Hardware (\S~\ref{sec:7.4}) and network placement (\S~\ref{sec:7.5}) constrain these choices, while learning infrastructure supports collection, training, and policy release (\S~\ref{sec:7.6}).

\FloatBarrier

\subsection{Performance Characterization}\label{sec:7.1}

Profiling connects computation to physical response: workload structure identifies what runs and its dependencies (\S~\ref{sec:7.1.1}), timing establishes when results remain useful (\S~\ref{sec:7.1.2}), and resource analysis locates the bottleneck (\S~\ref{sec:7.1.3}). Planning, control, and monitoring may require different rates while competing for the same resources.

\begin{figure}[!htbp]
\centering
\includegraphics[width=\linewidth]{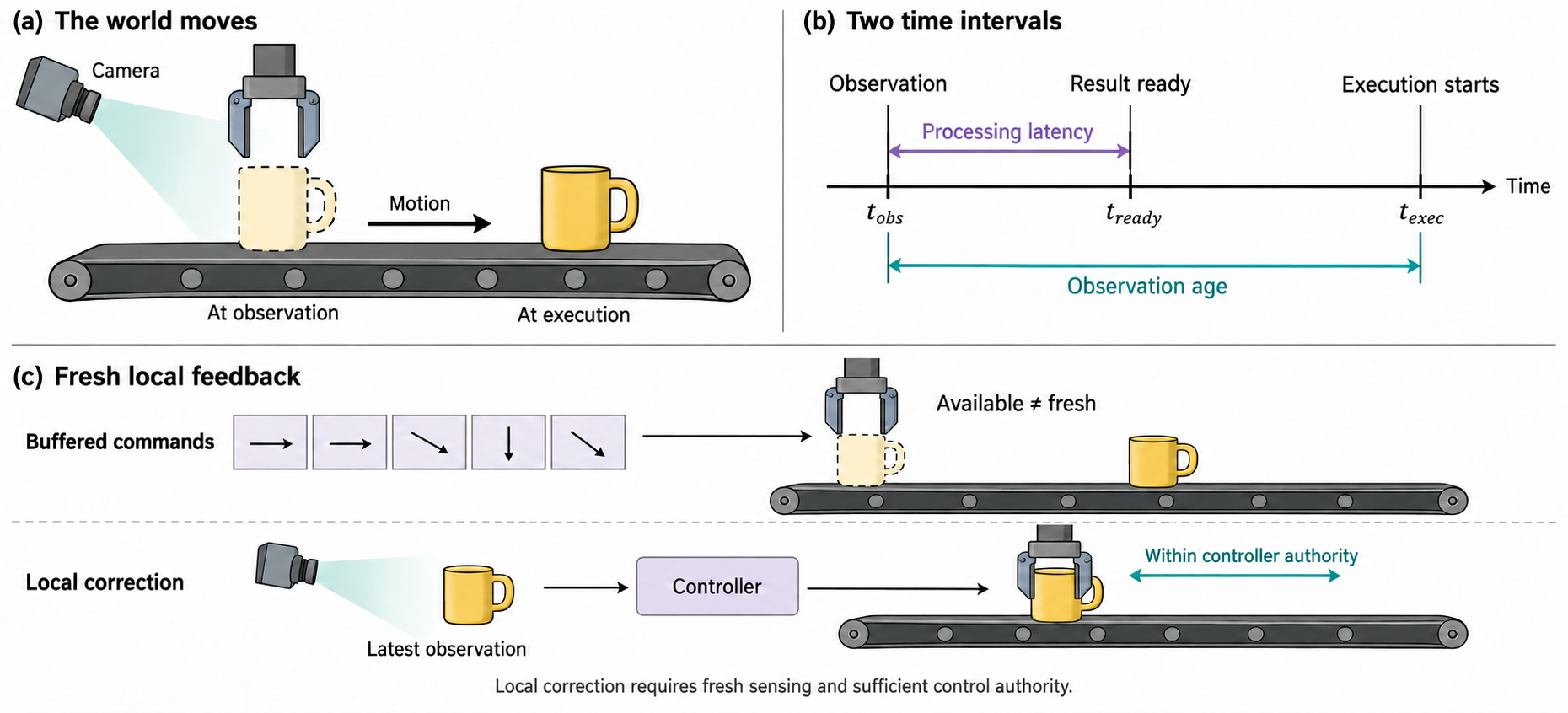}
\caption{Freshness and execution timing. (a) A moving mug leaves the location associated with an earlier observation. (b) Observation, result readiness, and execution start distinguish processing latency from observation age. (c) Separate execution examples show that buffered commands may remain available while their observation ages; fresh local sensing can support corrections within controller authority. These rows are not event-aligned continuations of the timeline. Correction feasibility and physical deadlines depend on the task; continuous action supply alone establishes neither freshness nor safe execution.}\label{fig:overview-timing}
\end{figure}

{\interlinepenalty=10000
Figure~\ref{fig:overview-timing} shows why available actions can still be based on outdated observations and how local feedback supplies a separate correction path.\par}

\subsubsection{Workload Structure}\label{sec:7.1.1}

Sensing, estimation, memory, prediction, planning, monitoring, and action generation may share a model or run across services. Their timing requirements and resource demands differ: contact control needs fresh estimates, whereas planning may tolerate slower updates supported by local feedback. Quantitative evidence here focuses mainly on action generation and learning.
VLA-Perf~\citep{arxiv_2602_18397} separates visual encoding, the vision--language backbone, and action generation. Its models connect architecture, context length, and execution strategy to deployment costs; accelerating one stage can expose another bottleneck.
ROSA~\citep{arxiv_2607_01088} organizes factory tasks as models and services with distinct rates, dependencies, and failure handling. Profiling inputs, invocation triggers, queueing, and shared resources identifies the critical path and contention outside it. Equal model-call latencies can therefore have different effects on physical progress.

\subsubsection{Latency and Freshness}\label{sec:7.1.2}

The response path includes sensing, preprocessing, transmission, queueing, inference, handoff, and actuation. Timestamps distinguish processing delays from waiting and communication. Latency measures response time, observation age measures information freshness, and deadline misses identify late responses; action throughput alone establishes none of these.
For each adopted policy result \(i\), let \(t_i^{\mathrm{obs}}\), \(t_i^{\mathrm{ready}}\), and \(t_i^{\mathrm{exec}}\) denote the reference observation time, result-ready time, and first action execution time, respectively. Define latency \(L_i\) and observation age \(A_i\) as:

\begin{equation}
\begin{aligned}
L_i&=t_i^{\mathrm{ready}}-t_i^{\mathrm{obs}},\\
A_i&=t_i^{\mathrm{exec}}-t_i^{\mathrm{obs}}
=L_i+\big(t_i^{\mathrm{exec}}-t_i^{\mathrm{ready}}\big).
\end{aligned}\label{eq:4}
\end{equation}

Here, \(L_i\) includes preprocessing, transmission, queueing, and inference, while \(A_i\) additionally captures the wait before execution. For action \(j\) from the same result, \(A_{i,j}=t_{i,j}^{\mathrm{exec}}-t_i^{\mathrm{obs}}\), with \(A_{i,0}=A_i\). At uniform action intervals \(\Delta t\), \(A_{i,j}=A_i+j\Delta t\).
These definitions assume synchronized clocks, complete-result handoff, ordered timestamps, and execution measured at controller handoff. Streaming segments and other sensor streams need separate timestamps; actuator delay is measured independently. Overlapping stages cannot be summed. The impact of observation age depends on physical change before execution (Figure~\ref{fig:overview-timing}).

Physical dynamics set response deadlines: a mobile robot must stop within free space despite sensing delay, braking distance, and tracking error, while a manipulator must correct contact before forces exceed tolerance. MAVBench and Krishnan et al. illustrate compute--motion constraints in aerial robots~\citep{boroujerdian2019mavbench,krishnan2020sky}. Rare delays matter even when average latency is low.
For a serial, non-overlapping sensor-to-actuator path, total latency and the deadline requirement are

\begin{equation}
\begin{aligned}
L^{\mathrm{sa}}&=t_{\mathrm{sense}}+t_{\mathrm{prep}}+t_{\mathrm{infer}}+t_{\mathrm{queue}}+t_{\mathrm{net}}+t_{\mathrm{act}},\\
&\Pr\big[L^{\mathrm{sa}}>D\big]\le P_{\mathrm{miss}}.
\end{aligned}\label{eq:deadline}
\end{equation}

Here, the terms represent sensing, preprocessing, inference, queueing, communication, and actuation, with queueing including waits both before inference and before execution. \(D\) is the physical deadline, and \(P_{\mathrm{miss}}\) is the tolerated miss probability under deployed load. Relative to Eq.~\eqref{eq:4}, \(L_i\) covers processing from observation to result readiness, while \(A_i\) additionally includes waiting before execution. Thus, \(L_i^{\mathrm{sa}}\approx t_{\mathrm{sense}}+A_i+t_{\mathrm{act}}\) when sensing begins before the observation timestamp; otherwise, \(t_{\mathrm{sense}}\) is omitted. This deadline applies to the first adopted action, while later actions require per-action ages \(A_{i,j}\) and disturbance-to-response measurements.

Rare delays can dominate task risk~\citep{grover2026gauntlet,dean2013tail}, while ROS~2 middleware introduces additional latency and jitter~\citep{kronauer2021ros2}. For chunked VLAs, FASTER~\citep{lu2026faster} shows that reaction time depends on both time to first action and the executed chunk horizon \(K\) (Eq.~\eqref{eq:1}), producing a distribution of response times. Profiling should therefore report latency percentiles, deadline-miss rates, observation age, disturbance-to-response time, and closed-loop accuracy under realistic load.

\subsubsection{Resource Bottlenecks}\label{sec:7.1.3}

Profiling distinguishes compute limits, memory-bandwidth limits, and launch or synchronization overhead. These require different remedies, and parameter count alone predicts none of them. Observation processing may dominate computation and activation memory, while repeated action generation can be limited by weight traffic~\citep{grover2026gauntlet}.
Batch size, precision, sequence length, caching, and concurrent workloads can change the limiting resource~\citep{yun2013memguard,he2026memflow}. MemGuard provides CPU DRAM bandwidth reservations, but its guarantees do not extend directly to GPUs or NPUs~\citep{yun2013memguard}. Roofline analysis supplies an idealized bound that must be checked against measured stage costs.

The roofline model relates arithmetic intensity \(I=F/B\), measured in FLOPs \(F\) per byte \(B\) transferred, to an idealized throughput bound:

\begin{equation}
P(I)=\min\big(P_{\mathrm{peak}},\ \beta I\big).\label{eq:roof}
\end{equation}

Here, \(P_{\mathrm{peak}}\) is peak compute throughput and \(\beta\) is memory bandwidth. A kernel is memory-bound when \(I<P_{\mathrm{peak}}/\beta\) and compute-bound otherwise; actual throughput may be lower due to utilization and overhead~\citep{williams2009roofline}.
Prefix--expert policies process observation conditioning once and repeatedly evaluate an action expert. VLA-Perf models these architecture-dependent costs~\citep{arxiv_2602_18397}; other architectures require their own execution decomposition. Roofline limits do not capture utilization, launches, synchronization, or overlap.

Analytical models identify potential bottlenecks, while profiling validates them. On a Hopper GPU at batch size one, PhyAI reports that the action expert in \(\pi_{0.5}\) accounts for only 8.8\% of FLOPs but 57.2\% of latency; its latency share drops to 13.5\% at batch size 32 as weight traffic is amortized~\citep{wang2026phyai}. Similarly, profiling MolmoAct~\citep{lee2025molmoact} on Jetson Orin and Thor attributes up to 75\% of end-to-end latency to memory-bound action generation~\citep{vishwanathan2026characterizing}. CrossVLA also identifies iterative denoising as the dominant action-sampling cost, limiting gains from prefix caching alone~\citep{liu2026crossvla}. These results highlight the importance of action-generation bottlenecks, but their quantitative implications depend on the model, hardware, precision, and batch size.
VLA-Perf's reported intensities place prefix components in the compute-bound regime and the action expert in the memory-bound regime on RTX~4090, A100, H100, and B100 configurations~\citep{arxiv_2602_18397}. Its Thor balance point is excluded because peak-throughput precision and dense/sparse conventions are unclear. Cross-device conclusions require consistent arithmetic, traffic, and power assumptions, alongside measured phase timings such as those reported for GPUs and edge NPUs by Zhou et al.~\citep{zhou2026xpus}.


These workload and timing measurements guide optimization: input reduction targets early computation, compression targets memory traffic, and fewer generation steps reduce repeated work. Yet compression can worsen completion time or motion quality without changing success rates~\citep{li2026inference}. Profile again after optimization: the limiting resource may shift, and queue management or local feedback may improve response more than another reduction in model latency.

\subsection{Inference Optimization}\label{sec:7.2}

Inference optimization targets five costs: inputs (\S~\ref{sec:7.2.1}), model size and depth (\S~\ref{sec:7.2.2}), sequential decoding (\S~\ref{sec:7.2.3}), repeated computation (\S~\ref{sec:7.2.4}), and runtime overhead (\S~\ref{sec:7.2.5})~\citep{guan2025efficientvla,yu2025efficientvlasurvey}. These interventions can combine, but speedups cannot simply be multiplied as bottlenecks shift. Unless noted, results below come from each study's own tasks, hardware, and baselines. Scheduling the resulting actions is treated in \S~\ref{sec:7.3}.


\subsubsection{Input Reduction}\label{sec:7.2.1}

Reducing views, resolution, patches, or tokens saves computation but can omit contact cues or small objects. Relevance-based selection differs from reuse across calls: an unchanged feature may still be essential.

\noindent\textbf{Fixed-budget Pruning and Merging.} FastV prunes visual tokens at a chosen depth and retention ratio, while ToMe merges similar attention keys with size-weighted averaging~\citep{chen2024image,arxiv_2210_09461}. Neither explicitly models changing task-critical regions, and FastV's original visual-question-answering evidence requires validation for robotic manipulation.

\noindent\textbf{Task- and Action-aware Selection.} EfficientVLA combines relevance and coverage-based token selection with layer pruning and action-head caching, reporting a 1.93\(\times\) CogACT speedup with a 0.6\% success-rate drop on SIMPLER; token selection's contribution is not isolated~\citep{arxiv_2506_10100}. SpecPrune-VLA adapts history-, attention-, and layer-based pruning to end-effector speed, whereas VLA-InfoEntropy uses image and task-related attention entropy without motion signals or retraining~\citep{wang2025specprunevla,liu2026infoentropy}.

Both selection strategies need validation of their effect on subsequent decisions. Token selection can introduce closed-loop failures invisible in single-step evaluations. The Gate, Not the Cache~\citep{luo2026gate} shows that when token selection relies on previous accelerated outputs, skipped information can cause errors to accumulate. At a 0.9 skip ratio on LIBERO-Object, success drops from 1.00 (dense) to 0.68 with token reuse and 0.31 with deletion, without detection by the tested action-level error checks. An unconditional dense pass executed off the critical path refreshes the selection signal, restoring success to 0.98 while reducing serving latency by 18--22\% relative to dense inference.
Selection is therefore useful only when savings exceed its overhead and closed-loop tests confirm that omitted information does not compromise later decisions.

\subsubsection{Model Compression}\label{sec:7.2.2}

Compression targets memory capacity and weight traffic (Eq.~\eqref{eq:roof}). Quantization lowers precision, compact models reduce capacity, and layer skipping changes depth per call~\citep{he2025adaskip,yue2024deervla}. Their latency gains depend on hardware kernels and execution overhead; their action errors can accumulate through feedback.

\noindent\textbf{Quantization.} Quantization reduces weight traffic but improves latency only when supported by efficient hardware kernels. OpenVLA~\citep{arxiv_2406_09246} reports that 4-bit quantization preserves real-robot success (71.9\% versus 71.3\% with bfloat16 across 8 tasks and 80 rollouts per configuration), while 8-bit quantization reduces success to 58.1\%. The authors attribute this decline to quantization overhead that slows inference and alters closed-loop dynamics.
GPTQ minimizes calibrated layer-output errors using approximate second-order information~\citep{arxiv_2210_17323}. QuantVLA extends post-training quantization to the language backbone and diffusion head while retaining floating-point attention projections; SQAP-VLA jointly optimizes quantization and token pruning to address their incompatibility~\citep{zhang2026quantvla,fang2025sqapvla}.

Low precision can also be incorporated during training. Quantization-aware imitation learning adapts weights to low precision for edge deployment~\citep{park2024quantizationaware}. BitVLA combines a ternary backbone and 1.58-bit vision encoder with full-precision connector and action head, matching full-precision OpenVLA-OFT in its reported evaluation with 11.0\(\times\) less model memory~\citep{wang2025bitvla}. Runtime gains still require efficient low-bit kernels.
Within quantization, bit allocation can prioritize action sensitivity~\citep{xu2026qvla,akbari2026actquant,zou2026acttune}. QVLA assigns channel precision using action sensitivity, retaining 98.9\% of OpenVLA-OFT's LIBERO performance with 29.2\% of its VRAM; ActQuant instead allocates precision per matrix and adjusts scales using action-aware curvature~\citep{xu2026qvla,akbari2026actquant}. Both evaluate quantization in action space rather than relying on language-model metrics.

\noindent\textbf{Reduced Execution Depth.} EfficientVLA statically removes language layers. Dynamic alternatives include budget-constrained exits in DeeR-VLA, spatial--temporal routing and self-distillation in MoLe-VLA, and action-aware skipping in DySL-VLA~\citep{arxiv_2506_10100,yue2024deervla,zhang2025molevla,yang2026dyslvla}. Variable execution depth makes tail latency and deadline misses essential (Eq.~\eqref{eq:deadline}).

\noindent\textbf{Compact Models.} Smaller models reduce deployment cost through transferred knowledge or architecture choices. TinyVLA pairs a small multimodal backbone with a diffusion decoder; SmolVLA targets single-GPU training and consumer deployment; TurboVLA replaces the language-model backbone with separate encoders and lightweight interaction~\citep{wen2024tinyvla,shukor2025smolvla,xie2026turbovla}.

Compression should be compared by task performance and generalization alongside latency, memory, and training cost. Unsupported sparsity or dequantization overhead can erase nominal savings, while runtime optimization cannot restore capacity lost through compaction.

\subsubsection{Efficient Action Decoding}\label{sec:7.2.3}

Action generation repeats backbone passes for autoregressive heads or expert passes for diffusion and flow heads. Prefix--expert architectures reuse encoded context (\S~\ref{sec:7.1.3}), so call count alone does not determine latency. The methods below shorten dependencies, reduce refinement, or distribute it across calls (Figure~\ref{fig:overview-generation}); continuity additionally depends on execution handoff.


\noindent\textbf{Shorter Decoding Dependencies.} Autoregressive action heads generate tokens sequentially, making decoding latency depend on sequence length. OpenVLA-OFT~\citep{arxiv_2502_19645} replaces sequential decoding with parallel chunk prediction, continuous actions, and L1 regression. On an A100, its reported request latency decreases from 0.240 to 0.073 seconds, while action throughput increases from 4.2 to 109.7 actions/s. The larger throughput gain reflects multiple actions per request rather than an equivalent latency reduction.
Other methods retain autoregressive decoding while reducing dependencies. PD-VLA~\citep{song2025pdvla} uses parallel fixed-point iterations, CEED-VLA~\citep{song2025ceedvla} accelerates their convergence through distillation and early exit, and Spec-VLA~\citep{wang2025specvla} uses speculative decoding with distance-based token acceptance. However, token proximity may not reflect physical action similarity, requiring task-level validation.
FAST accelerates \(\pi_0\) training through frequency-space action coding but remains sequential and is slower at inference than the compared flow-matching configuration~\citep{arxiv_2501_09747}. Training and inference efficiency must therefore be separated.

\noindent\textbf{Fewer Refinement Steps.} For diffusion and flow heads, reducing sampling steps decreases the number of expert evaluations; the latency benefit depends on how much total inference time the expert consumes. Consistency Policy~\citep{prasad2024consistency} distills a pretrained Diffusion Policy into a few-step generator, while One-Step Diffusion Policy (OneDP)~\citep{arxiv_2410_21257} uses KL-divergence-based distillation for single-step generation at 2--10\% additional pretraining cost. On a V100, OneDP reduces encoding-plus-generation latency from 669\,ms (100-step DDPM) to 16\,ms, supports 20\,Hz real-robot control, and achieves competitive or improved task success.
Training-free AdaVLA adapts sampling steps and MLP pruning to flow curvature; FASTER prioritizes immediate actions so execution starts before the whole chunk is refined~\citep{han2026adavla,lu2026faster}. DriftingVLA instead learns a one-pass noise-to-action map~\citep{gao2026driftingvla}. These approaches respectively adapt work online or move approximation into training.

\noindent\textbf{Refinement Across Calls.} Streaming methods distribute denoising across successive calls rather than reducing total refinement steps. Streaming Diffusion Policy~\citep{arxiv_2406_04806} maintains a buffer of partially denoised action chunks, executing the cleanest chunk while refining future ones, although its real-world Push-T evaluation pauses execution during sampling. FlashVLA~\citep{li2026flashvla} extends this approach to flow-matching VLAs using multi-noise-level buffers and chunk-wise causal attention, producing one executable chunk per forward pass. It requires fine-tuning and initial warm-up with safe default actions, but achieves at least 30\,Hz in real deployment.
Reflex~\citep{guo2026reflex} improves streaming efficiency through incremental key--value caching, phase-gated normalization for BFloat16 stability, and asynchronous visual encoding. Its evaluations report faster streaming and lower reaction latency. However, eliminating execution pauses also requires appropriate asynchronous scheduling.

\subsubsection{Computation Reuse}\label{sec:7.2.4}

Computation reuse caches inference intermediates rather than task experiences or skills~\citep{gao2024cachedattention,arxiv_2502_02175}. Fixed-prefix key--value reuse within a request can be exact; reuse across denoising steps or observations is generally approximate. The cached tensor, attention dependencies, and refresh conditions determine validity. Prefix caches are distinct from the representation \(c\) in \S~\ref{sec:3.4}, and savings shared with decoding optimization must not be counted twice.

\noindent\textbf{Exact Reuse of Fixed Conditioning.} In \(\pi_0\), the observation prefix cannot attend to noisy action tokens, so its key--value tensors remain fixed across flow-matching steps~\citep{arxiv_2410_24164}. They are computed once per observation and reused without changing model outputs; each new observation requires a new prefix cache.

\noindent\textbf{Approximate Reuse Across Denoising Steps.} DeepCache reuses high-level U-Net features in image diffusion, while EfficientVLA periodically refreshes cached attention and MLP outputs in CogACT's action head~\citep{arxiv_2312_00858,arxiv_2506_10100}. The latter reports a 1.23\(\times\) caching-only speedup on one SIMPLER task with little FLOP reduction, suggesting memory and overhead savings. Zhou et al. combine caching with asynchronous GPU/NPU execution~\citep{zhou2026xpus}. Unlike prefix reuse, these approximations can alter actions and require comparison with reduced-step sampling across deployment-relevant conditions.

\noindent\textbf{Reuse Across Observations.} VLA-Cache recomputes task-relevant tokens while reusing stable visual key--value features; Adaptive Visual Token Caching learns selection and reuse ratios~\citep{arxiv_2502_02175,arxiv_2602_00686}. ActionCache instead retrieves intermediate actions from similar contexts to warm-start generation~\citep{oi2026actioncache}. Savings must include lookup and refresh costs, without counting warm-start step reductions twice.

Across these reuse scopes, cache validity depends on physical and task changes. Motion, instructions, and task transitions can invalidate cached information even when pixels or lookup keys remain similar. Excessive refresh erases savings, while visual caching cannot remove dominant expert costs~\citep{liu2026crossvla}. Hardware exploiting temporal redundancy faces similar limits~\citep{liu2026deltoris,kong2024cambricond,kim2025ditto,heo2025exion}. Matched cached/uncached trials should include sudden changes and measure hit rates, bypassed work, and the combined effects of caches.

\subsubsection{Runtime Implementation}\label{sec:7.2.5}

At batch size one, launches, allocation, synchronization, and Python dispatch can dominate overhead. Kernel fusion reduces traffic, graph replay reduces launches, preallocation avoids repeated allocation, and scheduling overlaps work~\citep{dai2025actionflow}. Compiler and tensor-program autotuning can further improve generated-code performance~\citep{li2026texplorer,li2026carbs}. Comparisons need optimized baselines such as TensorRT or torch.compile, with hardware, shapes, batch size, and preprocessing specified.

\noindent\textbf{Kernels, Fusion, and Graph Replay.} FlashAttention reduces attention memory traffic without changing its mathematical output; its original speedups are not VLA measurements~\citep{arxiv_2205_14135}. Batch-one execution also exposes launch overhead~\citep{kwon2023efficient,ma2025realtimevla}. Ma et al. combine CUDA graphs, graph simplification, and specialized kernels for a \(\pi_0\)-level policy, reporting lower latency than openpi/JAX and naive PyTorch under their tested configuration~\citep{ma2025realtimevla}. Their proposed streaming rate counts buffered action supply, not fresh-observation decisions.
ActionFlow~\citep{dai2025actionflow} instead improves throughput by overlapping decoding from earlier requests with prefill from new ones. This multi-request scheduling increases OpenVLA-7B throughput by 2.55\(\times\) on edge hardware, though it requires concurrent in-flight requests and does not imply an equivalent reduction in single-request latency.

Runtime optimizations are often deployed together with model-level approximations. FlashDrive combines cross-frame caching, speculative reasoning, adaptive denoising, quantization, and compiled kernels for a driving VLA~\citep{li2026flashdrive}. Its end-to-end gains cannot be attributed to runtime optimization alone, and approximation must be checked for effects on action quality.

\noindent\textbf{Portable and Unified Runtimes.} vla.cpp provides PyTorch-free execution and architecture-specific kernels; Embodied.cpp combines modular execution with multirate control; PhyAI uses model adapters over shared kernels, graph execution, and memory management~\citep{jiang2026vlacpp,xu2026embodiedcpp,wang2026phyai}. They report lower latency or memory use than their respective baselines, though specialized runtimes can be faster~\citep{ma2025realtimevla}. Portability still requires fixed-input and closed-loop checks: vla.cpp identifies positional-index precision as behaviorally significant in SmolVLA. Benchmarks must separate model execution from preprocessing and host--device transfers.

Optimizations trade input coverage, precision, capacity, or repeated work against cost. Their value depends on which bottleneck remains and whether approximation errors alter subsequent observations. Matched runtime baselines and closed-loop trials are therefore complementary; neither operation counts nor fixed-input speedups establish physical gains.

\subsection{Real-Time Policy Execution}\label{sec:7.3}

Execution scheduling determines when computed actions affect the body. Receding-horizon execution chooses how much of a chunk to use (\S~\ref{sec:7.3.1}); asynchrony overlaps inference and action supply (\S~\ref{sec:7.3.2}); handoff connects new and committed actions (\S~\ref{sec:7.3.3}); and multirate execution separates slow decisions from fast feedback (\S~\ref{sec:7.3.4}). Together they govern continuity and timely response (Figure~\ref{fig:overview-timing}).

\subsubsection{Receding-Horizon Execution}\label{sec:7.3.1}

Action chunks supply motion during inference: ACT predicts coherent sequences and Diffusion Policy generates them through denoising~\citep{arxiv_2304_13705,arxiv_2303_04137}. Buffered execution avoids waiting but relies on earlier observations.
In receding-horizon execution, the robot executes \(K\) of \(H\) predicted actions before replanning (\(1\le K\le H\)); full-chunk execution corresponds to \(K=H\). Inference may be synchronous or overlapped with execution. For example, \(\pi_0\)~\citep{arxiv_2410_24164} predicts \(H=50\) actions but replans after 16 actions at 20\,Hz or 25 at 50\,Hz, corresponding to 0.8 and 0.5\,s. Thus, chunk horizon, executed prefix, and control rate are distinct. During execution, local feedback and safety monitoring may continue, but new observations do not alter the high-level action sequence unless it is interrupted or regenerated.

For example, with 100\,ms inference latency and a 20\,ms action interval, observation ages increase from 100 to 120, 140\,ms, and beyond, despite a 50\,Hz action rate. Longer executed prefixes amortize inference and reduce synchronous pauses but delay policy updates, while shorter prefixes improve responsiveness at the risk of discontinuities. RTC~\citep{arxiv_2506_07339} demonstrates this trade-off: its dynamic Kinetix experiments show that maintaining action continuity enables shorter prefixes and more frequent corrections, improving task success. Evaluation must therefore consider observation age, action throughput, and task accuracy together.
A disturbance's position within the executed prefix, inference delay, and local correction determine reaction time~\citep{lu2026faster,jiang2026vlacpp}. The choice of \(K\) balances revision frequency against inference and handoff costs; a larger \(K\) delays updates. PolicyTrim uses RL to lengthen prefixes and reduce task steps, with a small success-rate decrease in its LIBERO-Object evaluation~\citep{wang2026policytrim}. The useful prefix therefore depends on both inference cost and available interruption or feedback.

\subsubsection{Asynchronous Execution}\label{sec:7.3.2}

Overlapping inference with execution avoids synchronous pauses when buffered actions cover computation and handoff delays. However, the robot moves while a new chunk is generated, so its actions may no longer match the current state. Action representation and handoff rules determine the consequences (\S~\ref{sec:7.3.3}).
Coverage must be checked at each handoff under variable delays. A prefix-preserving schedule must supply both committed actions and their continuation. Sufficient buffer length alone does not ensure smooth motion: RTC addresses the discontinuities caused by naive overlap through inpainting~\citep{arxiv_2506_07339}.

SmolVLA decouples action prediction and execution through a queue~\citep{shukor2025smolvla,capuano2025lerobot}. It requests a chunk when the remaining fraction falls below \(\eta\), then aggregates the response with queued actions. Setting \(\eta=0\) gives sequential execution; earlier requests reduce depletion risk. Near-duplicate observations are skipped unless the queue is empty.
Runtimes need timestamps, bounded queues, cancellation, and expired-result handling. Evaluation should distinguish timely execution, discarded proposals, and safety-modified actions (\S~\ref{sec:9.3.2}), alongside sensor-to-actuator delay, task progress, and intervention cost. Longer buffers tolerate delays but can prolong outdated behavior.

\subsubsection{Chunk Handoff}\label{sec:7.3.3}

During asynchronous execution, a new action chunk must connect with actions already executed or committed while responding to new observations. Handoff methods intervene at three locations: the input state used for prediction, the generation or selection of a compatible continuation, and the fusion of output trajectories. Naive replacement can cause discontinuities, while averaging distinct behavioral modes can produce invalid motion. All three approaches must preserve actions already executed or committed during inference and supply a compatible continuation; smoothness alone does not establish correct behavior. Buffer and request timing depend on the schedule and delay assumptions, with remaining coverage checked at each variable-delay handoff. Predicting from the expected post-prefix state does not remove the need to match committed motion.


\noindent\textbf{Input-state Prediction.} VLASH predicts future proprioceptive state from the preceding chunk and conditions the next inference on that state~\citep{tang2025vlash}. Temporal-offset fine-tuning reduces reaction latency without architectural changes or additional runtime overhead. Its assumption that preceding actions execute as planned makes it vulnerable to disturbances such as slipping.

\noindent\textbf{Generation and Selection of Continuations.} RTC~\citep{arxiv_2506_07339} handles asynchronous handoff through action inpainting, freezing committed actions and progressively relaxing constraints over the remaining overlap to generate a smooth continuation. It applies to diffusion- and flow-based VLAs without retraining and demonstrates robustness to inference delays across twelve dynamic Kinetix tasks and six real bimanual tasks, including precise manipulation such as lighting a match. However, inpainting adds computation, and its guidance may bias new actions toward outdated trajectories.
Bidirectional Decoding~\citep{liu2024bidirectional} instead maintains continuity through candidate selection. It samples multiple action chunks and selects one consistent with the previous decision while favoring the trained policy over a weaker policy. Both methods balance temporal consistency and responsiveness, but Bidirectional Decoding requires generative policies capable of producing multiple candidates.

Continuity can instead be learned during training. Training-time RTC simulates delays and conditions on action prefixes, exchanging training for lower handoff overhead. It outperforms inference-time RTC at higher simulated delays and matches its real box-building and espresso performance at lower computational cost~\citep{black2025trainingtime}. Legato instead trains flow policies with schedule-dependent mixtures of committed actions and noise; randomized schedules improve continuity under variable delays~\citep{liu2026native}.

\noindent\textbf{Output-trajectory Fusion.} ACT~\citep{arxiv_2304_13705} uses temporal ensembling to combine overlapping action predictions through exponential weights \(w_i=\exp(-mi)\), where \(w_0\) represents the oldest prediction and smaller \(m\) gives greater weight to newer observations. It queries the policy at every control step and reports a 3.3-percentage-point improvement across four tuned ablation settings. However, averaging does not guarantee behavioral consistency, even with deterministic decoding. The \(\pi_0\) authors found temporal ensembling harmful and instead executed chunks open-loop~\citep{arxiv_2410_24164}.
VLA-RAIL~\citep{zhao2025vlarail} instead smooths individual chunks using polynomial fitting and fuses successive trajectories while maintaining position, velocity, and acceleration continuity. This enables smoother execution without pauses in dynamic simulation and real manipulation. However, trajectory fusion assumes compatible behaviors and may fail when the policy switches between distinct action modes.

In a matched-delay comparison, Agouzoul et al.~\citep{agouzoul2026understanding} compare handoff methods under delays of up to 20 control steps, using MLPMixer policies on Kinetix and SmolVLA on LIBERO. A2C2's per-step correction maintains over 90\% success up to a delay of 8 on Kinetix and leads on LIBERO from a delay of 4. Inference-time RTC performs well at low delays but degrades with long chunks (\(H=30\)) and high delays. Training-time RTC is the most robust training-based approach without inference overhead, while VLASH's performance depends on its fine-tuning delay range.
These results favor delay-aware training and per-step correction over inference-time constraints alone at longer delays, within the evaluated policies, tasks, and delay ranges. Handoff assessment must therefore test responsiveness and behavioral correctness alongside trajectory smoothness.

\subsubsection{Multirate Execution}\label{sec:7.3.4}

Multirate execution separates fast disturbance correction from slower decisions. HiRT and Fast-in-Slow use pathways at different rates; RMA asynchronously updates environmental representations for a faster locomotion policy~\citep{zhang2024hirt,chen2025fastinslow,arxiv_2107_04034}. Editing modifies slow outputs, whereas fast inputs or interleaved denoising bring new evidence into policy computation.
An illustrative deployment assigns planning and policy inference to MPUs, GPUs, or NPUs, state and setpoint checks to an RTOS-based MCU, and current control to motor drives~\citep{grover2026gauntlet}. Rates must follow physical dynamics, not a universal hierarchy. Runtime assurance separately rejects expired or inadmissible chunks~\citep{sha2001simplex,ames2019cbf,arxiv_2309_05837}; multirate operation does not itself provide this protection (\S~\ref{sec:9.3.2}). Each layer must meet its relevant deadline (Eq.~\eqref{eq:deadline}).

Real-Time EXPO-FT~\citep{arxiv_2609_18207} pairs slower VLA action chunks with a lightweight, reinforcement-learned editor that corrects actions using recent observations. Across four dynamic real-world tasks (30 trials each), successes increased from 50/120 with supervised fine-tuning to 116/120. Online data collection lasted at most ten minutes, excluding prior demonstrations and setup, and could end earlier upon meeting the success criterion. However, task-specific rewards, human resets, and independent outcome verification were still required, limiting claims of unattended learning. The results demonstrate rapid task-specific improvement, but the gains cannot be attributed to the correction layer alone, as both execution and learning are modified.
A2C2~\citep{sendai2025a2c2} adds a lightweight correction head to an off-the-shelf VLA without retraining the base policy. At each control step, it uses the latest observation, base action, chunk position, and policy features to generate a time-aware correction. Under increasing delays and horizons, it improves success over RTC by 23 percentage points on twelve Kinetix tasks and 7 points on LIBERO Spatial.

\(\pi\mathrm{R}^2\)~\citep{park2026pir2} instead separates fast proprioceptive feedback from slower asynchronous vision--language features. Using diffusion forcing and latency-adaptive flow scheduling, it conditions on in-flight actions and generates outputs in one denoising step. Applied to GR00T~N1.7 on an xArm6 with XHand, it updates actions from fresh observations every 40\,ms on an A5000 GPU, rather than merely executing cached chunks.
TIDAL~\citep{sun2026tidal} interleaves flow denoising with execution, caching semantic features at low frequency while updating each denoising step with fresh state and motion cues. Training with misaligned timing helps compensate for stale intent. In tested settings, it achieves 4\(\times\) higher feedback frequency and a 2.5\(\times\) average gain in dynamic interception over baselines without within-chunk feedback, with slightly lower static success. Together, \(\pi\mathrm{R}^2\) and TIDAL show that fast feedback can be integrated into the policy itself.

Across execution schedules and handoff mechanisms, longer prefixes and overlap sustain action supply, while local feedback improves disturbance response. Neither guarantees that the high-level goal remains valid or that successive trajectories are compatible. Evaluation should vary semantic goals, disturbance timing, delays, and network availability; smooth motion alone cannot establish correct continuation.

\subsection{Hardware Platforms}\label{sec:7.4}

Hardware choices move from distributing workloads across onboard processors (\S~\ref{sec:7.4.1}) to specialized robotics kernels (\S~\ref{sec:7.4.2}) and joint policy--hardware design (\S~\ref{sec:7.4.3}). Each trades flexibility and coordination against efficiency under memory, power, and thermal limits. Network placement is addressed separately in \S~\ref{sec:7.5}.

\subsubsection{Heterogeneous Computing}\label{sec:7.4.1}

CPUs support irregular processing and system services; GPUs accelerate dense computation; NPUs favor supported low-precision operators; MCUs provide predictable control. Splitting a pipeline helps only when gains exceed transfer and coordination costs~\citep{kang2017neurosurgeon,zhang2026saem}. Precision, placement, and shared-resource contention can all move the bottleneck.
GRiD generates parallel rigid-body dynamics and gradient kernels, reporting up to 7.2\(\times\) speedup over a multithreaded CPU but 2.5\(\times\) including transfers~\citep{arxiv_2109_06976}. cuRobo accelerates complete motion generation through collision checking, parallel seeds, and batched optimization~\citep{arxiv_2310_17274}. Policies that output poses can shift substantial costs to these downstream computations.
Temporal reuse, sparsity, mixed precision, and locality-aware designs improve resource efficiency~\citep{guan2025efficientvla,yan2025jsemicond,chen2026aihw2035,bai2026sparrowsnn}. In-memory and near-memory architectures also target language-model inference and retrieval workloads~\citep{gao2024imi,chen2026ragnmp}. Device comparisons need matched precision, power, bandwidth, and operator assumptions (Eq.~\eqref{eq:roof}). Full-task measurements must capture transfer overhead, contention, and thermal throttling missed by short benchmarks.

\subsubsection{Accelerators for Robotics Kernels}\label{sec:7.4.2}

Dedicated hardware exploits structure in perception operators, repeated dynamics, localization, and collision computations~\citep{li2026danmp,neuman2021robomorphic,suleiman2018navion,murray2016microarchitecture}. Its benefit depends on runtime share and transfers, while changing bodies or algorithms can reduce the value of specialization.
Robomorphic Computing~\citep{neuman2021robomorphic} specializes hardware to robot morphology, parallelizing computations across limbs and links while exploiting joint-specific sparsity and constant link inertias to reduce circuit complexity. Its FPGA implementation accelerates rigid-body dynamics gradients, achieving a 1.9--2.9\(\times\) coprocessor speedup including input/output transfers. RoboShape~\citep{neuman2023roboshape} extends this approach across robot morphologies by exploiting shared topology patterns, while GRiD provides a GPU-based alternative.

Hardware specialization can target fixed algorithms, algorithm families, or shared representations. Navion accelerates visual--inertial odometry~\citep{suleiman2018navion}, while other designs target point-cloud search, localization, and collision checking~\citep{xu2019tigris,gan2021eudoxus,liu2021archytas,murray2016microarchitecture,bakhshalipour2022racod}. More flexible approaches include ORIANNA and reconfigurable optimization accelerators~\citep{hao2024orianna,ting2025hiper,xu2026polymorphism}, while GSCore and related designs specialize in Gaussian-splatting representations~\citep{lee2024gscore,li2025rtgs,huang2026splatonic}.
Evidence ranges from fabricated chips (Navion) to FPGA implementations, ASIC synthesis, and simulation, which should not be treated as equivalent. Ultimately, end-to-end gains depend on the accelerated kernel's share of execution time, transfer overhead, and continued relevance as algorithms evolve~\citep{neuman2024magnificent}.

\subsubsection{Co-Design for Learned Policies}\label{sec:7.4.3}

Policy--hardware co-design changes workloads and their implementation together. It must preserve task performance under deployment limits as model changes alter or remove the accelerated computation.
Corki predicts short trajectories and converts them to torque using an FPGA; ReCA targets cooperative embodied agents~\citep{huang2025corki,wan2025reca}. SpecVLA co-designs speculative long-chunk generation and lightweight verification around scene motion, unlike decoding-only Spec-VLA (\S~\ref{sec:7.2.3})~\citep{qi2026specvla}. Focus instead reduces redundant VLM tokens through a streaming datapath~\citep{wei2026focus}.
DiTPA exploits redundancy across action trajectories, denoising steps, and multimodal features, extending beyond image-diffusion accelerators~\citep{zhao2026ditpa,kong2024cambricond,kim2025ditto,heo2025exion}. Deltoris targets consecutive observations through bit-level deltas, speculative loading, and customized arrays~\citep{liu2026deltoris}. VQVLA uses motion-adaptive codebooks and reuses centroid products~\citep{song2026vqvla}. Their gains depend on temporal similarity and reliable decisions about when reuse or reduced precision remains appropriate.

RoS\'E and AutoPilot support pre-silicon hardware--software co-design through full-stack simulation and automated SoC design~\citep{nikiforov2023rose,krishnan2022autopilot}. However, hardware evidence varies and must distinguish fabricated chips, FPGA implementations, RTL simulation, and analytical estimates. Corki uses an FPGA, while Deltoris, SpecVLA, VQVLA, and DiTPA report simulation-based results rather than fabricated chips~\citep{liu2026deltoris,qi2026specvla,song2026vqvla,zhao2026ditpa}. These studies primarily measure inference acceleration, not complete sensor-to-actuator latency (Eq.~\eqref{eq:deadline}).

Across hardware platforms, specialization is valuable while its target remains a significant runtime cost after software optimization. Reusable primitives tolerate model changes better than fixed decoding or denoising hardware~\citep{neuman2024magnificent}. Compare hardware acceleration with workload simplification using task performance, sustained energy use, and deadline compliance; kernel speedups and pre-silicon estimates alone do not establish deployment gains.

\subsection{Distributed Deployment}\label{sec:7.5}

\begin{figure}[!htbp]
\centering
\includegraphics[width=\linewidth]{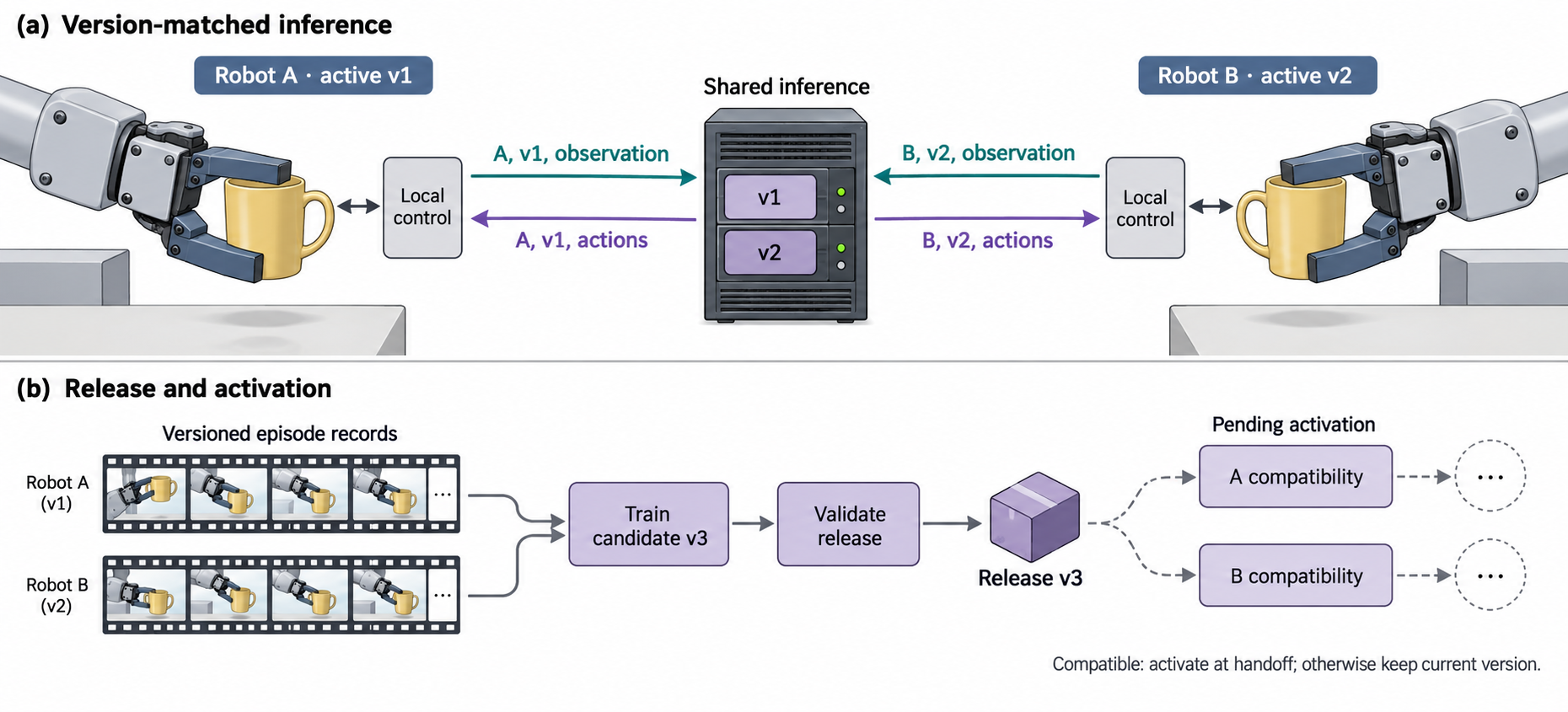}
\caption{Shared inference and versioned deployment. (a) Requests and returned actions identify both the robot and policy version, while control remains local. (b) Episode records retain collection versions through training and release validation. Per-robot compatibility checks precede activation at an appropriate handoff; otherwise the current version remains active. Dashed paths are pending, so robots A and B retain versions v1 and v2. Network availability, buffer coverage, and feasible local fallback require separate validation.}\label{fig:overview-deployment}
\end{figure}

Offboard resources relax onboard power and memory limits but add network delay, queues, and failures. We examine computation placement (\S~\ref{sec:7.5.1}), the information crossing a split-inference link (\S~\ref{sec:7.5.2}), and fleet-wide scheduling (\S~\ref{sec:7.5.3}). These extend the hardware choices of \S~\ref{sec:7.4} from one robot's capacity to communication and shared-resource constraints.
Figure~\ref{fig:overview-deployment} connects shared inference to versioned experience and policy release, while separating server validation from compatibility checks for activation on each robot.

\subsubsection{Computation Placement}\label{sec:7.5.1}

Compute placement must deliver sufficiently fresh actions with defined failure handling, subject to compute, memory, power, traffic, queueing, and network constraints~\citep{gao2022networkplacement,ichnowski2022fogros2}. Server sharing and model partitioning are additional choices: a shared server may host an entire model or selected stages.


\noindent\textbf{Onboard Execution.} Local policies avoid network dependence but face power, thermal, and memory constraints. Cross-platform VLA measurements show nonlinear effects of power caps and configurations where edge devices match older datacenter GPUs~\citep{taherin2025crossplatform}. Appropriately sized edge devices can improve cost and energy efficiency while meeting control rates~\citep{arxiv_2602_18397,zhou2026xpus}.

\noindent\textbf{Edge, Cloud, and Shared Execution.} FogROS2 automates cloud placement for SLAM, grasp planning, and motion planning~\citep{ichnowski2022fogros2}. Continuous policies additionally need delay-aware execution, such as RTC or VLASH~\citep{arxiv_2506_07339,tang2025vlash}; ROSA coordinates fleet models on shared GPUs~\citep{arxiv_2607_01088}. Faster remote inference must compensate for transmission, queues, and availability risk.

Across placement options, compare placement, partitioning, and sharing jointly by cost, energy, deadlines, and physical outcomes under intended load. Split models add representation-dependent transfer costs; local emergency responses must remain feasible during disconnection. Model settings and platform resources introduce further task-dependent choices (\S~\ref{sec:10.2.3}).

\subsubsection{Collaborative Inference}\label{sec:7.5.2}

Collaborative inference can place the whole policy remotely or divide computation across a feature interface. Intermediate tensors and task-level features are different interfaces within the latter choice, while compact task messages may connect separate services rather than layers of one model. The interface changes bandwidth, privacy exposure, latency, and the capability retained locally during delays.

\noindent\textbf{Remote Inference on Raw Observations.} Running the entire policy remotely avoids onboard model storage but makes decisions depend on image transmission and server response~\citep{arxiv_2602_18397}. Local control and protection can remain onboard.

\noindent\textbf{Feature Partitioning.} Neurosurgeon selects DNN split points using computation, feature size, bandwidth, energy, and server load~\citep{kang2017neurosurgeon}. Robot deployments add physical deadlines. RAPID uses body-motion and contact signals to choose between remote inference and cached actions~\citep{arxiv_2603_07949}; EcoVLA dynamically selects energy-efficient partitions under latency constraints~\citep{zhou2026ecovla}. Normal-operation gains do not establish resilience to disconnection.
A task-level split can preserve a local action head with access to fresh observations. Peng et al. pair cloud features from delayed observations with current local vision, training on current/delayed frame pairs with shared action targets~\citep{peng2026latencytolerant}. This improves delay tolerance on four LIBERO suites, provided the slow task context remains valid. FleetAgent uses compact messages for fleet teleoperation assistance~\citep{peng2026fleetagent}.

Intermediate tensors may exceed compressed image sizes, and fewer remote calls can increase staleness. RAPID reports lower latency for cloud-only than collaborative execution, illustrating the energy--response trade-off. Feasibility depends on both deadlines and fresh information; safe local stopping or holding does not establish autonomous task completion after disconnection.

\subsubsection{Multi-Robot Serving}\label{sec:7.5.3}

Shared serving must account for GPU partitioning, cache placement, and request scheduling, as well as robot deadlines and action-buffer consumption~\citep{chen2026migserve,gao2025onlinecontext,arxiv_2608_00337}. Clockwork demonstrates predictable DNN scheduling and deadline-safe batching~\citep{arxiv_2006_02464}; robot serving additionally needs physical response requirements and robot-reported state.

\noindent\textbf{Buffer-aware Scheduling.} Armory schedules lookahead inference using buffer levels and consumption rates~\citep{arxiv_2608_00337}. Heterogeneous execution rates weaken simple arrival-order heuristics; its heterogeneity-aware scheduler improves real-fleet throughput by up to 18\%, with configurable buffer-depletion trade-offs. Coverage must still be distinguished from feedback freshness.


\noindent\textbf{Task-objective Scheduling and Multi-model Pipelines.} ROSA coordinates shared models around factory productivity, task-specific service objectives, and failure handling~\citep{arxiv_2602_18397,arxiv_2607_01088}. It reports higher productivity than dedicated serving in real and synthetic evaluations. Scheduling must balance batching with per-robot rates and deadlines, since queued requests can become stale.

Either scheduling objective also requires reliable admission and failure handling. Admission control protects deadlines, health checks detect stalled streams, and standby services must retain compatible action interfaces. ROSA handles serving failures; STRANDS preserves service progress; RLinf-USER preserves asynchronous learning experience~\citep{arxiv_2607_01088,arxiv_1604_04384,arxiv_2602_07837}. Software restoration alone cannot establish physical resumption. Fault injection must test fresh observations, rejection of stale results, and the continuation conditions in \S~\ref{sec:9.3}.

Across distributed deployment choices, placement trades local capacity against communication, and scheduling trades pooling gains against individual urgency. Aggregate throughput can hide unequal service or repeated buffer depletion~\citep{bai2026faircache,arxiv_2608_00337}. Compare per-robot task progress, deadlines, energy, and behavior under disconnection to determine whether shared computation improves useful service.

\subsection{Learning Infrastructure}\label{sec:7.6}

Learning infrastructure connects rollout collection (\S~\ref{sec:7.6.1}), training (\S~\ref{sec:7.6.2}), and policy release and activation (\S~\ref{sec:7.6.3}). Stages may overlap, but records and deployed policies must remain traceable across versions. Their joint objective is time to validated improvement, not optimizer or environment-step throughput alone.

\subsubsection{Rollout Infrastructure}\label{sec:7.6.1}

Trajectory collection is limited by simulation speed, physical motion and resets, or human attention, depending on the setting. It also requires failure handling and usable records. Useful-episode throughput accounts for rejected and incomplete data, making it distinct from environment-step and optimizer throughput.

\noindent\textbf{Parallel Simulation.} The GPU simulators discussed in \S~\ref{sec:4.4.1} combine physics, rendering, and learning while reducing CPU transfers~\citep{arxiv_2108_10470,nvidia2025isaac,tao2024maniskill3,zakka2025mujoco}. Cheap resets enable scale, but memory and rendering become bottlenecks; RLinf reports memory-limited parallelism with nearly constant per-step environment time~\citep{arxiv_2509_15965}.

\noindent\textbf{Autonomous Physical Interaction.} Hardware collection is limited by motion, resets, and usable records. SERL integrates rewards, control, and resets; DayDreamer sustains model-based learning with reduced reset needs in selected tasks~\citep{arxiv_2401_16013,arxiv_2206_14176}. Their learning mechanisms are covered in \S~\ref{sec:5.3}; the infrastructure requirement is keeping collection operational.

\noindent\textbf{Human-supported Operation.} HIL-SERL and Learning While Deploying combine ongoing collection with human corrections, while AutoRT explicitly schedules supervision across robots~\citep{arxiv_2410_21845,wang2026while,arxiv_2401_12963}. Throughput must include operator capacity, preparation, resets, faults, and maintenance, rather than counting only active execution.

\subsubsection{Training Pipelines}\label{sec:7.6.2}

Training can be limited by data decoding, memory, communication, environment stepping, or checkpoint I/O. Faster input processing addresses idle accelerators, partitioning addresses capacity, and pipeline scheduling balances stages. Live robot data adds collection and persistent-storage constraints.

\noindent\textbf{Supervised VLA Training.} LeRobot's video-backed records require temporal frame decoding during batch construction, which can leave GPUs input-starved~\citep{cadene2026lerobot}. FSDP trades communication for model-state memory by sharding parameters, gradients, and optimizer states; OpenVLA combines it with mixed precision and FlashAttention~\citep{arxiv_2304_11277,arxiv_2406_09246}. KAITIAN addresses heterogeneous accelerator clusters through performance-aware work allocation and communication protocols~\citep{lin2025kaitian}.

\noindent\textbf{Reinforcement-learning Systems.} RLinf separates workflow definition from device placement and uses profiling-guided scheduling to overlap environment stepping, inference, and updates~\citep{arxiv_2509_15965}. Its reported 1.07--2.43\(\times\) training-throughput gains depend on the relative stage costs; configurations optimized for simulation must be reassessed with physical interaction and reset delays.
Physical interaction adds persistent storage requirements to the learning pipeline. RLinf-USER combines a persistent cache-aware replay buffer, heterogeneous robot interfaces, and adaptive communication for cloud learning with edge execution~\citep{arxiv_2602_07837}. Eviction decisions affect learning as well as throughput when they remove rare failures or earlier-task experience. End-to-end iteration time should include collection and transport, with pipeline speedups separated from sample-efficiency gains.

\subsubsection{Policy Update Pipelines}\label{sec:7.6.3}

Overlapping collection, training, evaluation, and release improves utilization but creates policy lag. Each episode must identify its behavior policy, and each robot must identify its active configuration; synchronization frequency alone does not describe end-to-end delay.
HIL-SERL decouples execution and learning. RLinf-USER finds that more frequent weight synchronization does not always improve convergence, partly because updates change the data distribution~\citep{arxiv_2410_21845,arxiv_2602_07837}. IMPALA corrects actor--learner lag with V-trace, requiring behavior-policy probabilities and versioned records~\citep{arxiv_1802_01561}; scalable learning from lagged trajectories is distinct from real-time control.
Learning While Deploying integrates shared collection, policy improvement, and redeployment across sixteen dual-arm robots~\citep{wang2026while}. Its learning workflow is discussed in \S~\ref{sec:5.4.7}; system coordination must preserve episode provenance while multiple policy versions coexist.

To diagnose delays in this pipeline, observation age measures time since an action's source observation (Eq.~\eqref{eq:4}); experience age measures time since collection; deployed-release age measures time since the executing version was released. Aligned clocks are required. Age differs from version lag and release-to-activation delay, and does not itself establish failure: older experience or policies may remain useful.
Activation must check observation and action interfaces, controller conventions, monitors, and filters at controller-consistent boundaries. Rollback must restore a compatible configuration rather than weights alone. Figure~\ref{fig:overview-deployment} separates release validation from per-robot activation; episode and action version logs support retention evaluation (\S~\ref{sec:8.1.3}).

Collection, training, and release improve capability only when experience remains useful and validated policies reach compatible robots. Evaluate sustained collection, successive updates, retained skills, resumed task progress, and support costs. These requirements survive changes in model architecture, whereas gains from specialized decoding or denoising may not. Task-phase-aware computation remains an open comparison against fixed allocation under matched hardware budgets.

\section{Evaluation}\label{sec:8}

Robotic claims should specify capabilities, operating conditions, costs, and human assistance. Since physical testing is slow and variable, success rates alone can obscure limitations in generalization and sustained autonomy. Evaluation therefore has four separate decisions, which organize this chapter: what outcome or cost to measure (\S~\ref{sec:8.1}), where to test it (\S~\ref{sec:8.2}), how observed episodes are scored or unobserved outcomes are predicted (\S~\ref{sec:8.3}), and what protocol makes a comparison credible (\S~\ref{sec:8.4}). The discussion moves from defining the claim and its test conditions to examining assessment methods and the controls needed to interpret results.

\begin{figure}[!htbp]
\centering
\includegraphics[width=\linewidth]{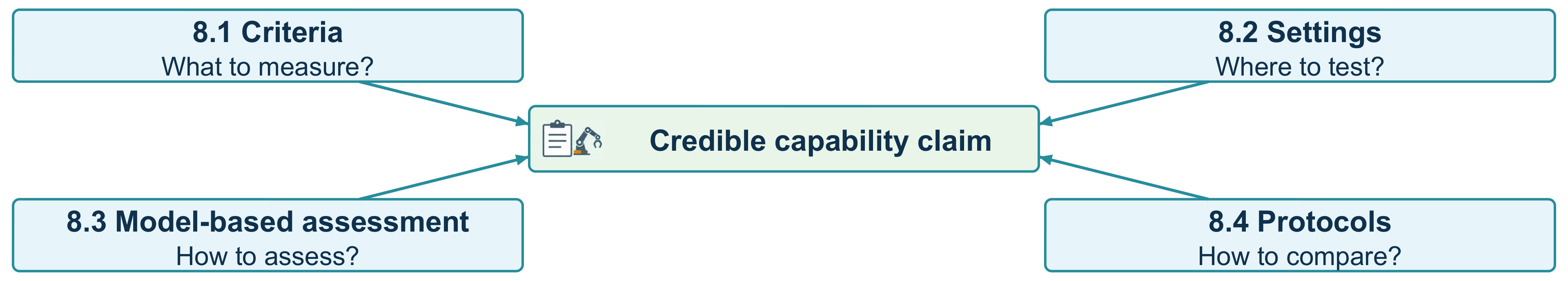}
\caption{Four evaluation decisions. Criteria determine what to measure, settings where to test, model-based assessment how to estimate or assess outcomes, and protocols how to compare results. Together they support a credible capability claim; model-based estimates still require validation against observations.}\label{fig:overview-evaluation}
\end{figure}

Evaluation claims span three scopes: \emph{mechanism validation} tests components, \emph{physical task trials} assess integrated behavior, and \emph{sustained operation} examines repeated execution, recovery, and long-term support. These scopes differ from evaluation settings and ablation levels: an offline, simulated, or physical test can support different claims according to its measured outcome and protocol. Simulation depends on modeling assumptions, physical success does not guarantee sustained operation, and prediction accuracy does not necessarily improve downstream decisions. Paired comparisons isolate component effects, while full-system comparisons must account for data, computation, controllers, and human assistance.
Figure~\ref{fig:overview-evaluation} connects criteria, settings, model-based assessment, and protocols to the capability claim they jointly support.

\FloatBarrier

\subsection{Evaluation Criteria}\label{sec:8.1}

Evaluation criteria should be defined before choosing settings or judges. We move from task performance and its success or failure definitions (\S~\ref{sec:8.1.1}), to generalization across changed conditions (\S~\ref{sec:8.1.2}), to sustained operation and retention over longer timescales (\S~\ref{sec:8.1.3}), and finally to resource and human costs (\S~\ref{sec:8.1.4}). This sequence broadens the assessment from an individual attempt to deployment conditions and the support required to achieve results. A criterion identifies the quantity and denominator; the experimental protocol later specifies which conditions and data access make the resulting claim valid.

\subsubsection{Task Performance}\label{sec:8.1.1}

Success rates depend on the definition of success. In the cup task, binary outcomes may hide wrong placements, retries, assistance, excessive force, or partial progress. Evaluations should therefore specify completion criteria, time limits, and the treatment of retries, aborts, and assistance. A 90\% success rate with retries is not equivalent to 90\% first-attempt success.

\noindent\textbf{Completion and Partial Progress.}
Manipulation benchmarks measure success through goal predicates. LIBERO~\citep{arxiv_2306_03310} reports episode success rates, while CALVIN~\citep{arxiv_2112_03227} evaluates five-instruction sequences using consecutive completion rates, whose sum gives the average chain length. BEHAVIOR~\citep{arxiv_2108_03332} and BEHAVIOR-1K~\citep{li2024behavior1k} measure partial progress as the maximum fraction of satisfied ground literals across valid goal conjunctions. These metrics capture achieved states rather than time or motion quality. For tasks requiring every stage to succeed, overall success depends on the probability of completing each stage conditional on reaching it; high success on individual stages can still yield low end-to-end completion.

\noindent\textbf{Continuous Tracking and Constraint Satisfaction.} Control tasks evaluate trajectories beyond final outcomes. safe-control-gym~\citep{arxiv_2109_06325} measures tracking accuracy, data efficiency, and constraint violations in cart-pole and quadrotor tasks under repeatable disturbances. In the cup task, constraints may include grip force and end-effector speed near the sink. Goal completion, tracking accuracy, and constraint satisfaction measure distinct properties: accurate tracking does not ensure safety.

Beyond scoring the outcome, failure attribution asks which component caused it. Failure categories complement success rates by revealing distinct failure modes. Robot Learning as an Empirical Science~\citep{kressgazit2024robot} recommends quantitative metrics alongside qualitative failure descriptions. In the cup task, wrong-object grasps, drops, and operator interventions suggest different causes but do not confirm them. Component tests can diagnose state estimates, predictions, and memory operations, but only task outcomes establish downstream effectiveness. Automated failure labels also require validation. Failure descriptions should therefore accompany completion criteria, denominators, progress measures, and constraint violations, so that success rates can be interpreted alongside their causes and costs.

\subsubsection{Generalization}\label{sec:8.1.2}

Success rates reflect only tested conditions, while deployment may introduce changes in objects, cameras, environments, instructions, or hardware. Generalization tests should specify what changes, by how much, and what was covered during training to distinguish new settings, skill compositions, and unseen tasks. Data splits, perturbation tests, and shortcut audits assess complementary aspects of generalization.

\noindent\textbf{Declared Held-out Splits and Levels.}
Generalization tests should withhold specific training factors. CALVIN's ABC\(\rightarrow\)D split tests five-step sequences of known tasks in an unseen environment with different textures, object positions, and instructions. VIMA-Bench~\citep{jiang2022vima} defines four levels: new placements (L1), new combinations (L2), unseen objects and textures (L3), and unseen tasks and prompts (L4), with four of 17 tasks held out. These distinguish new conditions, compositional generalization, and zero-shot tasks.
In contrast, LIBERO~\citep{arxiv_2306_03310} evaluates knowledge transfer and forgetting across spatial, object, and goal suites. Its standard protocol tests learned tasks under benchmark initial states, not necessarily held-out factors. Thus, its findings on visual encoders concern these learning protocols rather than unseen-factor generalization.

\noindent\textbf{Perturbation Suites.} Perturbation tests measure generalization through performance degradation under controlled changes. THE COLOSSEUM~\citep{arxiv_2402_08191} evaluates 20 tasks across 14 factors, reporting 30--50\% relative success degradation across five models and at least 75\% under combined perturbations. For PerAct, simulated and real-world responses show moderate agreement across four tasks (\(\bar R^2=0.614\)), indicating sim-to-real consistency rather than model-ranking correlation.
LIBERO-Plus~\citep{fei2025liberoplus} tests seven factors, identifying strong sensitivity to camera viewpoint and robot initial state. Camera perturbations reduce third-view-only OpenVLA-OFT success from 95.3\% to 16.8\% (Table 1). Weak sensitivity to language changes, supported by further experiments, suggests that models often ignore instructions.

\noindent\textbf{Shortcut Audits.} Shortcut audits test whether benchmark scores reflect intended capabilities. LIBERO-PRO~\citep{zhou2025liberopro} varies objects, initial states, instructions, and environments, reducing models' standard LIBERO success rates from above 90\% to 0.0\% under generalized conditions. Unchanged actions despite altered inputs suggest memorized sequences. A broader audit~\citep{jiang2026actually} identifies shortcut solvability, statistical insignificance, overfitting, and data-source dependence, finding that a 90-million-parameter model without a language encoder approaches state-of-the-art LIBERO performance. However, audits cannot rule out all shortcuts.

Generalization claims must match tested shifts: held-out splits test unseen conditions, perturbations measure sensitivity, and audits expose shortcuts. CortexBench~\citep{arxiv_2303_18240} finds no universally dominant visual encoder across 17 embodied tasks. Robustness to individual shifts does not ensure robustness to combinations, and predicting transfer remains an open challenge. Section~\ref{sec:8.4.1} addresses the separate protocol question of whether training and test access actually preserve the claimed boundary.

\subsubsection{Long-Term Performance}\label{sec:8.1.3}

Most benchmarks last seconds or minutes, while deployed robots must handle long tasks, multiple skills, and days of operation. These timescales introduce distinct failures: task-level tracking and recovery errors, forgetting previously learned skills, and deployment disruptions involving batteries, networks, or environments. Reliability at one timescale does not guarantee reliability at another, requiring separate evaluations.

\noindent\textbf{Information Retention Within a Task.}
Long-term memory evaluation should test whether past observations improve future actions. OpenEQA~\citep{majumdar2024openeqa} assesses recognition and retrieval but not downstream action. RoboMemArena~\citep{lei2026robomemarena} evaluates closed-loop manipulation across 26 simulated tasks averaging over 1,000 steps, with 68.9\% of subtasks requiring memory and paired real-world tests. Although reactive \(\pi_{0.5}\) struggles with memory-dependent tasks, it outperforms MemoryVLA overall in simulation (21.5\% vs. 15.0\%). Establishing memory benefits therefore requires matched ablations of task outcomes with and without memory, supported by subtask-level failure analysis.

\noindent\textbf{Capability Retention Across Tasks.}
Continual-learning evaluation measures retention of prior knowledge after learning new tasks, whether stored in model weights, memory, or skill mechanisms. LIBERO~\citep{arxiv_2306_03310} evaluates knowledge transfer and forgetting across task sequences, finding that sequential fine-tuning improves forward transfer over existing lifelong-learning methods, while naive supervised pretraining can hinder learning. Forward transfer, backward transfer, and forgetting depend on task order and reset protocols (Section~\ref{sec:8.4.1}). Final average performance may hide skills forgotten and later relearned.

\noindent\textbf{Sustained Operation.}
Sustained-service evaluation extends beyond task completion. STRANDS~\citep{arxiv_1604_04384} reports 104 combined days of autonomous operation across four deployments, including improvement during service. Herzog et al.~\citep{arxiv_2305_03270} report up to 53\% reduction in waste contamination (39\% under higher occupancy). The \(\pi^{*}_{0.6}\) report~\citep{intelligence2025pi06} describes 13 hours of continuous espresso preparation and over two hours of laundry folding in a new home. However, duration and outcomes alone do not establish uptime or intervention costs. Evaluations must account for failures, maintenance, and human support.

Metrics should match timescales: task-level completion and recovery, continual-learning transfer and forgetting, and deployment uptime, exceptions, and interventions per operating hour. Explicit denominators prevent downtime and assistance from being obscured.

\subsubsection{Resource and Human Costs}\label{sec:8.1.4}

Policies with similar success rates may differ in execution costs and in the learning or adaptation costs incurred before deployment. Single-robot and shared-compute measurements describe different execution settings; human effort contributes to both stages. Comparisons therefore need explicit accounting boundaries, including failed attempts and the reuse of trained policies across tasks.

\noindent\textbf{Execution Costs.} For a single robot, efficiency depends on closed-loop latency and energy, not inference alone. RTC~\citep{arxiv_2506_07339} addresses asynchronous action execution, while FASTER~\citep{lu2026faster} links reaction time to time to first action and action prefix \(K\). Embodied-efficiency studies~\citep{li2026inference} show that reducing model size, computation, or inference latency may increase end-to-end costs or degrade motion quality without changing task success. System-level metrics such as completion time, trajectory smoothness, cumulative joint rotation, and motion energy better capture these trade-offs.
Shared execution additionally incurs fleet-level resource and scheduling costs. For multi-robot serving, efficiency depends on fleet-wide task completion within deadlines. Armory~\citep{arxiv_2608_00337} and ROSA~\citep{arxiv_2607_01088} examine latency--throughput trade-offs. Since request latency, server throughput, and task completion rate are distinct metrics, evaluations must specify the optimization objective.

\noindent\textbf{Learning and Adaptation Costs.} Training and adaptation costs include robot-hours for data collection and practice, human-hours for demonstrations and interventions, and GPU-hours for training. Time to target performance also captures pipeline delays. Since methods trade off these resources, evaluations should report all costs and amortize one-time expenses across robots and tasks reusing the policy. Total and per-task costs may lead to different comparisons.

Human effort cuts across learning and execution. Human assistance adds operational costs through resets, interventions, tool changes, takeovers, and maintenance. Sirius~\citep{arxiv_2211_08416} records interventions as both operational costs and corrective training data. Excluding human effort overstates autonomy, while handover delays further limit intervention effectiveness.

Resource metrics should include failures, retries, and fallback costs within clear measurement boundaries. Energy per attempt differs from total energy per successful completion, which includes failed attempts and support costs. Module-level measurements identify cost sources but cannot replace end-to-end evaluation. Bench tests may also overlook resource contention from concurrent robot processes.

For both cost categories, consistent accounting requires a complete closed-loop record. Policies with similar success rates may differ in latency, deadline misses, motion smoothness, energy, and human assistance. The reported measures should reflect the task and platform. Success rates need confidence intervals; timing reports should give latency percentiles and the fraction of required invocations that miss declared deadlines or time out under load, including failed attempts. Latency and deadline measurements must share endpoints, with cancellations and discarded results reported separately. When motion smoothness matters, the chosen measure needs explicit coordinates, units, sampling, filtering, and duration. Energy per successful completion should include failed attempts and is undefined if no attempt succeeds. Intervention rates need a stated denominator, with resets and assisted successes reported separately. A cost-per-success measure should aggregate these task-level resources, failed attempts, and amortized learning costs within explicit accounting boundaries.

Evaluation criteria use different denominators: episodes, distribution shifts, task sequences, operating hours, or costs per attempt or completion. Gains in one metric may not transfer to others, making a single score insufficient to capture trade-offs among methods.

\subsection{Evaluation Settings}\label{sec:8.2}

Evaluation settings are ordered by the interaction and physical evidence they provide: offline evaluation uses logged data without new robot interaction (\S~\ref{sec:8.2.1}), simulation enables modeled closed-loop tests (\S~\ref{sec:8.2.2}), and real-world evaluation captures physical execution at higher trial and reset costs (\S~\ref{sec:8.2.3}). These are settings for collecting evidence, not interchangeable measures of capability. A task criterion from \S~\ref{sec:8.1} can be applied in more than one setting, but its interpretation changes with data coverage, simulator validity, and physical support costs.

\subsubsection{Offline Evaluation}\label{sec:8.2.1}

Offline evaluation uses demonstrations, logged episodes, or held-out trajectories without new robot interaction. Shared test data supports efficient diagnosis and ablation but cannot capture action-dependent state changes, error accumulation, recovery, or physical contact. Thus, low action error on cup-task demonstrations does not guarantee success when the policy deviates from recorded trajectories.

\noindent\textbf{Action Fitting and Implementation Diagnosis.}
The robomimic study~\citep{arxiv_2108_03298} evaluates six offline algorithms across five simulated and three real-world manipulation tasks, highlighting the effects of algorithm design, demonstration quality, and checkpoint selection. Validation action loss is inexpensive but may poorly predict rollout success when multiple actions are valid. SIMPLER~\citep{arxiv_2405_05941} reports weak correlation between negative action MSE and real-world success across six Google-robot checkpoints and three task categories (average Pearson \(r=0.308\)), compared with \(r=0.924\) for visual-matching simulation. MSE was computed from 25 sampled training demonstrations without a public validation split. Thus, action loss alone is insufficient for reliable checkpoint selection.

\noindent\textbf{Off-policy Evaluation.}
Off-policy evaluation (OPE) estimates target-policy returns from logged data without new rollouts. Importance sampling reweights returns by target-to-logging action-probability ratios, offering unbiased estimates under correct ratios and sufficient support but potentially high variance. Fitted Q evaluation uses Bellman updates but risks approximation and extrapolation errors. Doubly robust estimators~\citep{jiang2016doubly} combine value predictions with importance-weighted corrections, providing unbiasedness under correct ratios and independently fitted models while potentially reducing variance. All methods require adequate data coverage or additional assumptions.
For robotic grasping, off-policy classification~\citep{arxiv_1906_01624} predicts success using learned Q-values. Across 15 models with real success rates of 17--91\%, SoftOPC achieves a Spearman correlation of 0.94 with real success, versus 0.48 for temporal-difference error. However, DOPE~\citep{arxiv_2103_16596} finds no consistently reliable OPE method across tasks. Estimates improve when logging and target policies are similar, while only importance sampling consistently achieves positive rank correlation on AntMaze. Thus, OPE supports policy selection but cannot validate behavior unsupported by logged data.

Offline metrics aid diagnosis and checkpoint selection, but their predictive value depends on data coverage and requires closed-loop validation. Logged episodes can also validate learned judges before replacing human evaluation.

\subsubsection{Simulation Evaluation}\label{sec:8.2.2}

Simulation enables scalable, controlled, and repeatable evaluation through inexpensive resets and systematic perturbations. However, results depend on assumptions about perception, dynamics, contact, and control, which may fail for challenging objects such as wet or transparent surfaces. Benchmarks differ in task diversity and perturbation coverage. Paired simulation–real-world tests are therefore needed to validate transfer to hardware.
Simulation benchmarks standardize tasks, demonstrations, and success criteria. RLBench~\citep{arxiv_1909_12271} provides 100 manipulation tasks with variations and multimodal observations; CALVIN evaluates long-horizon language-conditioned tasks; and ManiSkill3~\citep{tao2024maniskill3} supports GPU-parallel simulation. LIBERO~\citep{arxiv_2306_03310} includes three 10-task transfer suites, 90 pretraining tasks, and 10 long-horizon tasks with demonstrations, subject to limitations in \S~\ref{sec:8.1.2}. Beyond manipulation, Habitat 3.0~\citep{arxiv_2310_13724} supports social navigation and collaborative rearrangement with simulated humanoids or human-controlled avatars, enabling evaluation of policy rankings in human-interaction tasks.

RoboCasa~\citep{arxiv_2406_02523} increases scene diversity through varied kitchens, interactive appliances, generated assets, composite tasks, and automated demonstrations, showing benefits from scaling synthetic data. BEHAVIOR-1K~\citep{li2024behavior1k} covers 1,000 everyday activities in OmniGibson, including rigid objects, deformable materials, and liquids. It highlights long-horizon challenges and demonstrates initial sim-to-real transfer in a matched apartment. Both benchmarks emphasize diverse scenes or tasks rather than controlled perturbations, requiring appropriate evaluation denominators. However, greater diversity does not eliminate simulator assumptions.

Across simulation suites, paired physical tests are needed to assess whether simulated scores transfer. SIMPLER~\citep{arxiv_2405_05941} reduces visual and control mismatches without requiring full-fidelity digital twins. It validates simulation against real-world success using Pearson correlation and mean maximum rank violation (MMRV), which measures worst pairwise ranking errors weighted by real-success differences (lower is better). Its visual-matching simulator achieves \(r=0.924\) and MMRV \(=0.056\) across six Google-robot checkpoints and three task categories, while capturing sensitivity to distribution shifts. However, agreement is limited to tested policies, tasks, and shifts.

\subsubsection{Real-World Evaluation}\label{sec:8.2.3}

Physical evaluation captures perception, actuation, contact, and human assistance under real conditions. However, costly resets, manual scoring, and variations across sites and days limit reproducibility. Centralized facilities, automated evaluation, paired trials, and shared hardware kits address these challenges.

\noindent\textbf{Centralized and Remote Benchmarks.} Centralized benchmarks standardize hardware and tasks. The Real Robot Challenge~\citep{arxiv_2109_10957} used eight TriFinger robots at one institute for automated evaluation of submitted code, with random robot assignment and identical goals. The robots operated for hundreds of hours with minimal failures. This improves comparability but remains limited to one platform design and a narrow task family.

\noindent\textbf{Distributed Paired Evaluation.} RoboArena~\citep{arxiv_2506_18123} uses distributed, double-blind pairwise evaluations instead of fixed tasks and sites. Across seven institutions using the DROID platform, it evaluates seven generalist policies in over 600 real-robot episodes and aggregates preferences into rankings, reportedly outperforming centralized evaluation in ranking accuracy. However, rankings depend on the participating tasks and evaluators, and pairwise preferences do not provide fixed-task success probabilities needed for deployment decisions.

Across evaluation sites, automation can reduce the cost of repeated testing. AutoEval~\citep{arxiv_2503_24278} automates scoring, resets, and safety monitoring for WidowX arms in BridgeData scenes. It deploys learned success classifiers and reset policies after exceeding 95\% accuracy or success. One cell achieves about 850 drawer-task episodes per day, with three human interventions over 24 hours, and closely matches human-run evaluations (average Pearson \(r=0.942\)). However, classifier errors may bias success estimates, while reset policies shape the initial-state distribution.
Hardware standardization provides another means of improving comparability across sites. ManipulationNet~\citep{chen2026manipulationnet} improves cross-site comparability through standardized hardware kits and unified task delivery and result collection. It separates manipulation skills (Physical Skills Track) from multimodal grounding (Embodied Reasoning Track).

Standardized hardware supports absolute capability assessment, paired evaluations reduce local variation, and automated resets and scoring increase trial counts. Reproducibility also requires documenting initial states, retries, and human assistance, with confidence intervals and stopping rules predefined for limited trials (\S~\ref{sec:8.4.3}).
Offline evaluation screens policies using logged data, simulation tests modeled closed-loop behavior, and physical trials assess real-world sensing, contact, and human-support costs. This staged approach reserves costly hardware trials for cases that cheaper evaluations cannot reliably resolve.

\subsection{Model-Based Assessment}\label{sec:8.3}

Manual scoring and hardware trials limit evaluation efficiency. We first compare programmed, human, and learned judges of \emph{observed} episodes (\S~\ref{sec:8.3.1}), then examine predictive assessment, which estimates outcomes through modeled rollouts rather than physical trials of the tested policy (\S~\ref{sec:8.3.2}). This order separates scoring available evidence from predicting outcomes that have not been observed; the distinction concerns the source of an outcome, independent of whether the setting is logged, simulated, or physical. Both approaches risk misclassification or incorrect policy rankings and require validation.

\subsubsection{Learned Outcome Assessment}\label{sec:8.3.1}

Evaluation judges differ in their observations and limitations. Programmed predicates assess explicit conditions but may miss unencoded outcomes or lack real-world state access. Human judges capture broader context but incur costs, fatigue, and disagreement. Learned judges scale scoring from images, videos, or logs but cannot detect unobserved events, such as hidden force violations or off-camera assistance. Reliable evaluation therefore requires predefined success criteria, sufficient observation coverage, and validated label sources.
SuccessVQA~\citep{arxiv_2303_07280} uses a fine-tuned vision--language model (Flamingo) to judge task success from videos across simulated households, real manipulation, and egocentric human settings. It generalizes better than task-specific detectors to unseen tasks and camera changes, but performs worse in distribution. Its balanced accuracy on egocentric videos drops from 99\% on training data to 62\% on test data. RoboReward~\citep{arxiv_2601_00675} emphasizes explicit failure examples, finding unreliable reward assignments by tested vision--language models. OpenEQA~\citep{majumdar2024openeqa} similarly uses language models to evaluate open-vocabulary answers about environments.

Judges and predictors are initially validated against labeled outcomes or observed frames, but recorded episodes cover only observed behavior. High average agreement may hide systematic errors, particularly rare critical failures. Decision thresholds trade off false accepts and false rejects, while human review improves reliability at additional cost. Evaluations should report both error rates by failure type at a specified threshold, automatic scoring coverage, and residual errors after human review. Hidden force violations and operator assistance require additional sensor and intervention records.
A learned judge used as a training reward should not evaluate the final policy, as optimization may exploit its errors. For example, a policy may make a cup appear clean without actually cleaning it. Final evaluation should use independent judges, alternative sensors, or sampled human labels.
Learned judges can replace human scoring when outcomes are observable and false-accept and false-reject rates are validated on relevant failures, independently of training rewards. Otherwise, physical instrumentation or human assessment remains necessary.

\subsubsection{Predictive Assessment}\label{sec:8.3.2}

Predictive assessment estimates policy outcomes without hardware execution, supporting candidate screening and ranking. Learned rollout simulators combine transition models with outcome judges, introducing errors from both. Runtime failure prediction instead guides interventions during execution, as in Model-Based Runtime Monitoring~\citep{arxiv_2310_17552} (\S~\ref{sec:9.3.1}).

\noindent\textbf{Transition Error Under the Policy's Actions.}
Predictive evaluation requires world models that accurately follow policy actions. WorldEval~\citep{li2025worldeval} uses Policy2Vec latent-action conditioning to improve action following for policy ranking and safety assessment. A Veo-based simulator~\citep{geminiroboticsteam2025evaluating} evaluates Gemini Robotics policies through action conditioning, multiview consistency, and scene variations. Ctrl-World~\citep{guo2025ctrlworld} combines action conditioning, multiview prediction, and memory retrieval for long rollouts. Its predictions agree with real-world instruction-following results but underestimate execution success due to contact and motion errors. Out-of-distribution actions may also appear successful despite being physically infeasible.

\noindent\textbf{Outcome-scoring Error.}
A second challenge is scoring predicted outcomes. RoboWorld~\citep{jeon2026roboworld} combines a fast autoregressive video world model, stabilized by Step Forcing for long rollouts, with task-progress-aware vision--language scoring. Predictive evaluation inherits both transition-model and judge errors, including false accepts and false rejects. Thus, even accurately predicted failures may be misclassified as successes.

Both transition and outcome-scoring errors require paired validation. Predictive evaluators require validation against paired real-world outcomes at clearly defined evaluation units. WorldEval~\citep{li2025worldeval} reports per-task Pearson correlations across four policies, measuring policy-level rather than episode-level agreement. The Veo-based study~\citep{geminiroboticsteam2025evaluating} uses over 1,600 real-world evaluations of eight checkpoints across five bimanual tasks to assess rankings under nominal and out-of-distribution conditions and detect safety violations. RoboWorld~\citep{jeon2026roboworld} compares eight policies against RoboArena's preference-based ranking. These results demonstrate agreement only for tested policies and metrics, not reliable episode-level predictions or generalization to unseen policy families. Validation should cover scores, rankings, and failure types, with recalibration after evaluator changes.

Across learned and predictive assessment, learned judges reduce labeling costs when outcomes are observable, uncertain cases receive human review, and additional sensors capture hidden conditions. Predictive models reduce rollout costs when transitions and outcome judgments are validated, with targeted real-world checks for uncertain cases. However, policy updates may produce states outside the surrogate's calibration data, requiring renewed validation even without changes to the surrogate.

\subsection{Experimental Protocols}\label{sec:8.4}

Performance differences between policies may arise from method improvements, information leakage, confounding changes, or chance. The protocol addresses these threats in order: training--test boundaries establish what information was available before evaluation (\S~\ref{sec:8.4.1}), controlled comparisons isolate the conditions behind a difference (\S~\ref{sec:8.4.2}), and statistical reporting quantifies uncertainty across trials and other sources of variation (\S~\ref{sec:8.4.3}). The same safeguards apply whether evidence comes from logs, simulation, or robots and whether a human or model scores it.

\subsubsection{Training and Test Boundaries}\label{sec:8.4.1}

New episodes under the same kitchen, objects, and camera setup do not establish generalization beyond those conditions. Valid generalization claims require separating training and test conditions while accounting for overlap through pretraining, fine-tuning, and checkpoint selection. For foundation models, even an ``unseen cup'' may resemble pretraining examples, making training coverage essential for interpreting results.
Evaluation splits must withhold factors relevant to the generalization claim. CALVIN holds out an entire environment, whereas random episode splits may retain the same objects, scenes, and instructions. Overfitting and data-source dependence can further weaken these boundaries~\citep{jiang2026actually}. Evaluations should report training coverage for claimed factors (objects, scenes, instructions, or robot bodies) and describe pretraining data sources by category.

Repeated checkpoint selection on test data introduces bias, as robomimic~\citep{arxiv_2108_03298} shows that stopping criteria affect performance. Test-time protocols must specify access to demonstrations, exploration, memory updates, and parameter adaptation. LIBERO~\citep{arxiv_2306_03310} defines continual-learning metrics based on task order, data access, and resets, while RWLA~\citep{arxiv_2410_02995} uses retrieved data for local adaptation, which must count toward the test-time budget. Adapted and non-adapted results should be reported separately. The resulting generalization claim is therefore bounded by the declared shifts, data partitions, and target-access budgets.

\subsubsection{Controlled Comparisons}\label{sec:8.4.2}

Cross-study comparisons often differ in cameras, action rates, observation histories, hardware, and test conditions, complicating attribution. Mechanism comparisons isolate component effects under fixed conditions, while system comparisons assess complete configurations, including costs and support, without attributing gains to individual components.
OpenVLA-OFT~\citep{arxiv_2502_19645} illustrates mechanism- and system-level comparisons. On single-image LIBERO, continuous L1 regression (95.3\%) and diffusion (95.4\%) perform similarly, outperforming original OpenVLA (76.5\%) and supporting simpler decoding. Adding wrist images and proprioception raises success to 97.1\%, reflecting system-level gains. Table~I further shows the effects of data filtering and inputs: unfiltered multi-input training achieves 94.5\%, versus 95.3\% for filtered single-image and 97.1\% for filtered multi-input training. Filtering excludes unsuccessful demonstrations and near-zero actions. Different training budgets and checkpoint selection every 50,000 steps limit equal-compute and independent-test claims.

Hardware comparisons should control nuisance factors through matched initial states, randomized or interleaved trials, and operator blinding. RoboArena~\citep{arxiv_2506_18123} uses double-blind policy comparisons across sites, while SIMPLER~\citep{arxiv_2405_05941} pairs simulated and real outcomes for the same policies. Fixed setups improve repeatability, while pairing reduces variation and supports comparisons of absolute outcomes or performance differences. Trial records must include retries and assistance to distinguish autonomous first-attempt success from assisted or repeated completion, connecting the criteria and protocol decisions in Figure~\ref{fig:overview-evaluation}.
Policy rankings depend on action rate, hardware, observation history, reset protocols, and trial count. Changing control rates also alters the evaluated closed-loop system. The performance and cost measures in \S~\ref{sec:8.1} should be chosen to match the comparison. System comparisons and component ablations must report the conditions under which these measures were obtained.

\begin{figure}[!htbp]
\centering
\includegraphics[width=\linewidth]{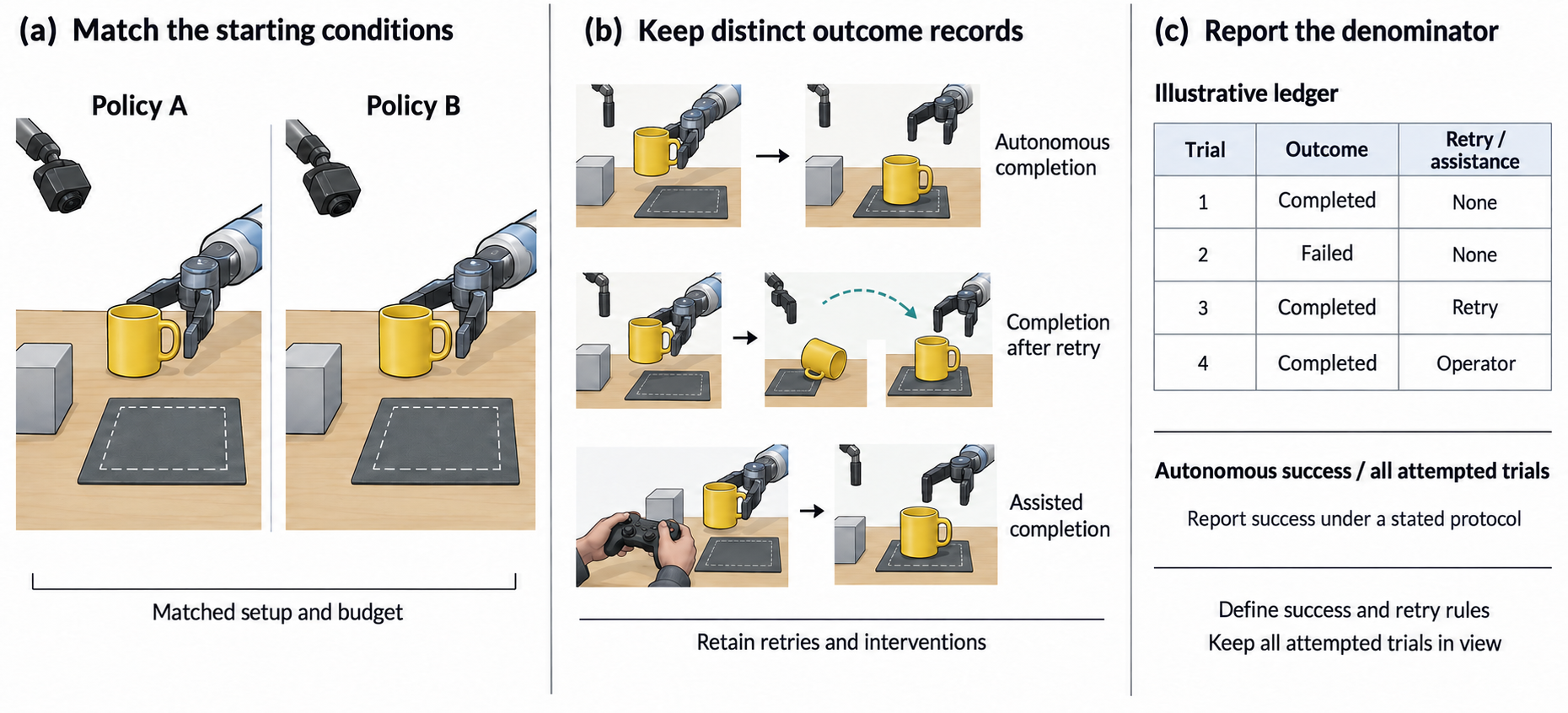}
\caption{Connecting trial conditions to an evaluation claim. (a) Matched initial conditions and budgets support comparison between policies. (b) Records distinguish completion without intervention, completion after retry, and assisted completion. These examples describe provenance rather than a universal set of mutually exclusive outcomes. (c) An illustrative ledger retains failed attempts, retries, and operator assistance. The success definition, trial unit, retry budget, and denominator must be stated; autonomous retries and human interventions are distinct. The entries are schematic, not benchmark results.}\label{fig:evaluation-protocol}
\end{figure}

Figure~\ref{fig:evaluation-protocol} connects matched trials to the reported denominator. A completion record must retain whether retries or human intervention occurred. A protocol may allow autonomous retries within a fixed budget, but that choice must be explicit, and operator-assisted completion must remain identifiable. Reporting only successful terminal scenes discards the information needed to interpret a capability claim; Figure~\ref{fig:intervention-provenance} shows how to preserve it during collection.

\subsubsection{Statistical Reporting}\label{sec:8.4.3}

Success rates from few trials are uncertain: 60\% versus 70\% over 20 trials may reflect sampling noise. Repetition should match the claim: trials capture execution variability, training seeds capture optimization variability, and scenes, robots, sites, or operators introduce additional variation.
For \(k\) successes in \(n\) independent trials,
\begin{equation}
\hat S=\frac{k}{n},\qquad \mathrm{SE}(\hat S)=\sqrt{\frac{\hat S(1-\hat S)}{n}}.\label{eq:se}
\end{equation}

At \(n=20\) and \(\hat S=0.5\), the SE is approximately 0.11, giving an approximate 95\% margin of \(\pm0.22\). This approximation is unreliable near 0 or 1, where Wilson or exact Clopper--Pearson intervals are preferable~\citep{nistbinomial}. With zero failures, the one-sided 95\% upper failure bound is \(1-0.05^{1/n}\approx3/n\), or 13.9\% after 20 trials and 5.8\% after 50. For 20 successes in 20 trials, the two-sided 95\% lower success bound is 83.2\%. These bounds use different confidence conventions and do not establish robustness to distribution shifts.
Trial counts should reflect expected effect sizes and statistical power. Distinguishing success rates of 0.6 and 0.7 requires approximately 356 independent trials per policy at 5\% significance and 80\% power, before clustering or multiple-comparison adjustments. For matched trials, McNemar's test uses discordant pairs, potentially improving efficiency when outcomes are positively correlated.

Trial counts may adapt to intermediate results. A sequential test~\citep{snyder2025your} compares imitation-learning policies using 10--50 trials, enabling early stopping while controlling statistical error and reducing trials by up to 32\% at equal correctness and power. In contrast, Vincent et al.~\citep{arxiv_2405_05439} derive fixed-sample bounds from i.i.d. rollouts. Their randomized binomial bound adds \(U\sim\mathcal{U}[0,1]\) to the success count, tightening Clopper--Pearson bounds, while continuous metrics use distribution-function bounds. For 38 successes in 50 diffusion-policy trials, the one-sided 95\% lower bound is 0.641, depending on the random draw. Unlike sequential tests, these bounds do not support optional stopping.

Robot trials may be correlated through shared setups, operators, or continuous operation. A 1,000-step episode still counts as one trial, so paired or cluster bootstrapping should resample at the task, run, or site level. Latency-tail estimates also require invocation counts and load histories. Deep Reinforcement Learning at the Edge of the Statistical Precipice~\citep{arxiv_2108_13264} highlights the risks of few-run point estimates and recommends confidence intervals, performance profiles, and robust aggregates such as the interquartile mean, implemented in rliable. Multitask comparisons should specify task weights; when site variation dominates, testing more sites or conditions may be more valuable than adding training seeds.

Robot Learning as an Empirical Science~\citep{kressgazit2024robot} identifies inadequate reporting of trial counts, initial conditions, success criteria, uncertainty, and failure modes. A LIBERO audit~\citep{jiang2026actually} similarly finds many improvements lack statistical significance. Such inconclusive differences do not establish equivalence, because precision depends on trial counts and dependencies. Reproducibility requires documenting experimental conditions, metrics, failures, statistical methods, model and code versions, data subsets, action interfaces, and timing. Changes in control rate or observation history alter the evaluated system. Rare-event safety claims also require exposure counts and dedicated validation.

Across experimental protocols, method selection requires verifying training data, controlling comparison conditions, and quantifying statistical uncertainty. Shared implementations and paired trials isolate component effects, while independent end-to-end replication tests system-level reliability. Mechanism descriptions motivate hypotheses, controlled comparisons test component effects, and system trials assess integrated performance. Method choices should therefore rely on evidence appropriate to the intended use.
Comparison records should also include task-specific measures such as surrogate agreement (e.g., SIMPLER), failure detection, recovery, and procedural execution. These measures answer different questions from task success and should retain their own units and denominators.

Across the evaluation criteria, settings, and protocols reviewed in this chapter, the shortcut and perturbation studies in \S~\ref{sec:8.1.2} show that benchmark gains require capability-specific validation, but their findings do not imply that all successful policies lack the claimed capabilities. A score supports a claim only through its criterion, testing setting, outcome assessment, and comparison protocol together.
Interpreting results requires explicit experimental conditions, budgets, and human assistance. Component-level attribution requires controlled comparisons, cross-study comparisons require compatible conditions, and surrogate evaluations require validated agreement for the relevant policies and conditions.

\section{Safety}\label{sec:9}

Robots can cause irreversible harm, such as human injury or equipment damage, or violate task requirements even when motions execute correctly. These risks arise during learning and operation from sensing, control, or timing failures, learning-induced changes, and deliberate attacks on inputs, data, or models. Safety requires limiting these risks and intervening before harm occurs.

Safety constraints can guide learning and planning, while monitoring, action filtering, recovery, and human intervention protect execution. Defenses against deliberate attacks complement these safeguards, but detection alone cannot prevent harm without timely and feasible intervention.

A \emph{hazard} is a condition that may cause harm; \emph{risk} describes its likelihood and severity; a \emph{constraint} specifies a requirement; and a \emph{violation} occurs when that requirement is not met. \emph{Intervention authority} determines who or what can modify, reject, or replace an action. Safety limits unacceptable physical risks, while security protects against deliberate compromise. Safety claims therefore require clearly defined hazards, intervention authority, and evidence covering relevant learning and operating conditions.

Evaluation tests safety claims, whereas safety mechanisms actively limit risk by modifying learning, decisions, or execution. This chapter first defines hazards and constraints (\S~\ref{sec:9.1}), establishing what the subsequent safeguards must address. It then distinguishes protection in learning and planning (\S~\ref{sec:9.2}) from monitoring and intervention during execution (\S~\ref{sec:9.3}), before examining deliberate attacks and defenses (\S~\ref{sec:9.4}). The organization separates where safeguards act from how adversaries can compromise their assumptions; adversarial threats can affect both learning and execution. Because filters may alter actions and supervisors may transfer control, their authority and the physical feasibility of intervention must be explicit.

\begin{figure}[!htbp]
\centering
\includegraphics[width=\linewidth]{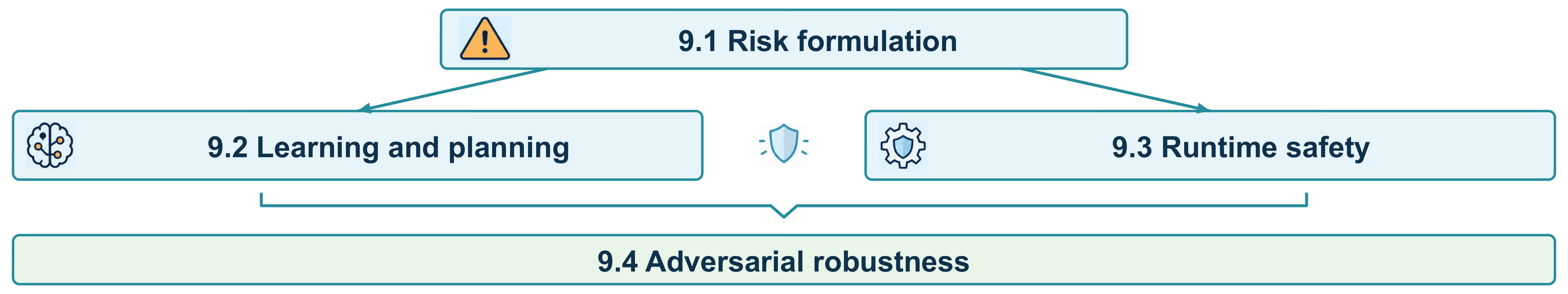}
\caption{Safety across development and operation. Risk formulation informs learning and planning as well as runtime safety. Adversarial robustness spans both, connecting the four subsections without prescribing a single protection architecture.}\label{fig:guide-09}
\end{figure}

{\interlinepenalty=10000
Figure~\ref{fig:guide-09} connects risk formulation to safety during learning, planning, and execution, with adversarial robustness spanning development and operation.\par}

\FloatBarrier

\subsection{Risk Formulation}\label{sec:9.1}

Before a learner, planner, or filter can ensure safety, the harms to prevent must be expressed as checkable requirements. Crushing a cup, grasping at an outdated location, and performing a prohibited action, such as mixing incompatible cleaning agents, involve different causes, signals, and safeguards. The distinction here is between identifying physical and operational hazards, their causes, and observable signals (\S~\ref{sec:9.1.1}), and representing the resulting requirements as constraints that can be checked (\S~\ref{sec:9.1.2}).

\subsubsection{Physical and Operational Risks}\label{sec:9.1.1}

Safety analysis begins by identifying events that can cause harm. Physical hazards include collisions, excessive force or speed, and instability. Operational hazards include stale state estimates, sensor or service failures, and failed control handoffs. Crushing a cup and executing an outdated grasp may cause similar damage but require different safeguards.
A hazard record specifies the affected person, object, or system; triggering conditions; observable signals; and potential severity or cost. Force sensors can detect overload, while timestamps reveal stale estimates, but neither verifies correct task interpretation. Tracing causal paths also distinguishes hazards caused by unsafe plans from those caused by delayed execution, helping identify where intervention is needed.

System-Theoretic Process Analysis (STPA)~\citep{leveson2012engineering} systematically identifies unsafe control actions by examining controllers, commands, and feedback. It considers commands that are omitted, issued unsafely, mistimed or misordered, or applied for an incorrect duration. For example, a grasp based on an outdated pose estimate may become unsafe when a hand enters the workspace. STPA traces such actions to causes such as delayed feedback or incorrect models and derives corresponding constraints, although it cannot guarantee model completeness.

Standards provide measurable limits for specific applications. ISO~10218-1:2025 and ISO~10218-2:2025 address industrial robots and their applications, respectively~\citep{iso10218one2025,iso10218two2025}. ISO/TS~15066:2016 addresses collaborative industrial robot systems~\citep{isots150662016}. Speed and separation monitoring maintains protective distances accounting for motion, reaction and stopping times, intrusion, and uncertainty. Power and force limiting controls contact through mechanical design or active control. ISO/TS~15066 distinguishes transient and quasi-static contact, with Annex~A relating transient-contact speed to force or pressure limits through effective mass, contact area, and body stiffness.
Active contact protection also requires reliable detection. Haddadin et al.~\citep{haddadin2017robot} review proprioceptive collision detection and identification methods. Generalized-momentum observers estimate external joint torques, while contact localization and force reconstruction require additional assumptions.

ISO~13482:2014 instead addresses personal-care robots~\citep{iso134822014}. For a kitchen robot, applicable requirements depend on intended use and identified hazards. Assessing those hazards requires defined events, causal paths, observable signals, and applicable measurable limits; task success alone does not establish safety. Learned policies, semantic guardrails, and control-barrier methods provide component-level evidence, but standards conformity must be assessed for the integrated system under its operating conditions.

\subsubsection{Constraint Representations}\label{sec:9.1.2}

Once a hazard is identified, its requirement must be formulated in a checkable form. Average episode-level bounds differ from conditions that must hold at every step. Requirements may specify numerical limits, such as workspace boundaries or contact forces, or contextual rules involving action order and authorization. Each requires observable evidence or justified judgment to verify compliance.

\noindent\textbf{State Constraints and Safe Sets.} The simplest representation defines admissible states or commands, but satisfying a constraint does not guarantee future safety. For example, a robot within a position limit may be moving too fast to stop before crossing it. A barrier-based filter therefore requires a valid barrier whose nonnegative set satisfies the constraints and remains invariant under available controls (Eq.~\eqref{eq:cbf}). safe-control-gym~\citep{arxiv_2109_06325} supports reproducible comparisons using explicit constraints, symbolic dynamics, and disturbances. Such comparisons require consistent constraints and satisfaction criteria, whether stepwise or in expectation.

\noindent\textbf{Cost Functions and Risk Budgets.} A second representation assigns scalar costs to unsafe events, such as collisions, excessive force, or forbidden-region entry, and bounds their expected cumulative cost over an episode. This budget quantifies the trade-off between task performance and safety and underlies constrained reinforcement learning (RL; Section~\ref{sec:9.2.1}). However, satisfying an expected budget does not guarantee safety on individual trajectories, motivating step-wise runtime protection.

\noindent\textbf{Formal Rules and Semantic Labels.} Some safety requirements concern task rules rather than numerical limits. Temporal logic expresses ordering and prohibitions, while RoboGuard~\citep{arxiv_2503_07885} grounds such rules in the environment to check plans. Semantic labels capture undesirable situations, as evaluated by ASIMOV~\citep{sermanet2025generating}. Safety for generalist robots must also account for context, user intent, and difficult-to-model consequences~\citep{sinha2026rethinking}, such as maintenance-specific permissions or unsafe chemical mixing. Importantly, exact verification of a temporal-logic formula does not guarantee correct grounding of its predicates. Perception and semantic judgments can produce false accepts or rejects, and these errors must be included in safety claims.


Each safety representation defines the scope of its guarantees. Extending them beyond collision and force constraints remains an open challenge. Representing a hazard also leaves an evidence problem: rare violations may not appear in ordinary trials.
Whichever constraint representation is used, rare failures require dedicated validation. Rare failures are difficult to assess through ordinary trials because accurate Monte Carlo estimates require many samples. Black-box safety validation~\citep{arxiv_2005_02979} addresses this by searching simulated disturbances for failures (falsification) or the most likely failure (adaptive stress testing). Importance sampling can estimate failure probabilities more efficiently by oversampling failures and reweighting results. However, probability estimation requires a defined disturbance distribution and statistically valid sampling, while all findings depend on the simulator's dynamics and disturbance model. These tests probe the represented hazard; they do not expand the protection supplied by a learner, plan checker, or runtime filter.

Across risk formulations, timing and shared infrastructure can undermine multiple safeguards, such as when a stale state estimate affects both planning and filtering. Each safety requirement must therefore specify its observation and resource dependencies and feasible responses to their failure. These dependencies inform learning and planning and runtime protection.

\subsection{Safety in Learning and Planning}\label{sec:9.2}

High task rewards do not guarantee safe actions, and language-model-generated plans may violate requirements before execution. The two approaches differ in where they limit risk: constrained learning (\S~\ref{sec:9.2.1}) limits risk across policy training and exploration, while plan validation (\S~\ref{sec:9.2.2}) checks a particular proposed sequence against rules, information, and feasibility. Shields and interventions used during exploration cross into runtime protection when they restrict each executed action. Runtime protection remains necessary because constrained policies may still propose unsafe actions, and valid plans may lose their preconditions during execution.

\subsubsection{Constrained Learning}\label{sec:9.2.1}

Task-only objectives may allow learners to trade safety for reward. In Safety Gym~\citep{ray2019benchmarking}, unconstrained reinforcement learning achieved high returns while violating safety costs. Exploration introduces further risks, as unsafe interactions can damage objects or hardware before a reliable policy emerges. Safety-aware learning addresses these problems by constraining costs during optimization and limiting unsafe exploration, as reviewed in~\citep{arxiv_2108_06266}.

\noindent\textbf{Budgeted Constrained Optimization.} A constrained Markov decision process (CMDP) maximizes task return subject to an expected cumulative risk budget:

\begin{equation}
\max_\theta J_R(\theta)\quad\text{s.t.}\quad
J_C(\theta)=\mathbb E_{\tau\sim\pi_\theta}\!\left[\sum_{t=0}^{T_C-1}\gamma_C^t c_t\right]\leq d.\label{eq:5}
\end{equation}

Here, \(J_R\) is the task return of Eq.~\eqref{eq:2b}, \(c_t\) is the risk cost, \(d\) is the permitted budget, and \(\gamma_C\) and \(T_C\) define cost discounting and horizon, which may differ from those of the reward. Setting \(\gamma_C=1\) gives an undiscounted budget, while discounting penalizes later violations less. Importantly, an expected budget does not guarantee safety on individual trajectories.
Two common approaches solve this CMDP. Lagrangian methods optimize \(\max_\theta\min_{\lambda\ge0}\{J_R(\theta)-\lambda[J_C(\theta)-d]\}\) with an adaptive penalty, increasing \(\lambda\) when estimated cost exceeds the budget. PPO-Lagrangian applies this approach to PPO. Constrained Policy Optimization (CPO)~\citep{arxiv_1705_10528} instead uses trust-region updates to limit policy changes and maintain near-constraint satisfaction theoretically, though practical implementations rely on approximations. In Safety Gym, Lagrangian baselines enforced cost constraints more consistently than CPO under the evaluated settings~\citep{ray2019benchmarking}. Neither method guarantees trajectory-level safety, and approximation or sampling errors may cause expected-cost violations.

SafeVLA~\citep{arxiv_2503_03480} extends CMDP-based safe learning to vision--language--action (VLA) models. It constructs diverse risk scenarios and uses alternating policy and multiplier updates to penalize unsafe behavior. On Safety-CHORES, SafeVLA reports reduced cumulative safety costs while maintaining task success, alongside generalization to out-of-distribution perturbations and real-world Safety-PickUp transfer cases. These findings apply to evaluated hazards and do not establish safety against untested risks.

\noindent\textbf{Restricted Exploration and Intervention.}
Other approaches limit unsafe exploration directly. Shielding~\citep{arxiv_1708_08611} uses a safety game derived from temporal-logic specifications and abstract dynamics to restrict available actions (preemptive shielding) or correct unsafe proposals (post-posed shielding). It can enforce safety during both learning and execution, but its guarantees depend on abstraction fidelity, which is challenging for contact-rich manipulation.
Recovery RL~\citep{arxiv_2010_15920} uses offline data to identify risky states and separates reward optimization from safety control. A task policy pursues rewards, while a recovery policy intervenes when predicted violation risk becomes high. Evaluations across six simulated domains and a physical obstacle-avoidance task reported improved safety–success trade-offs over compared baselines. Human intervention, as in Sirius~\citep{arxiv_2211_08416}, allows operators to take control in difficult situations and reuse interventions as training data, though without formal safety guarantees.

Constrained learning reduces unsafe proposals and reliance on runtime intervention. However, expected-cost constraints cannot certify individual actions, while shielding guarantees depend on valid dynamics abstractions and continuous enforcement.

\subsubsection{Plan Validation}\label{sec:9.2.2}

A plan may fail before execution by violating task rules, exceeding physical capabilities, or relying on missing information. For “wash the cup,” mixing incompatible cleaning agents violates a rule, placing the cup out of reach is infeasible, and an ambiguous cup reference leaves the target unknown. These failures require different responses: rule checking and plan repair, feasibility-based replanning, or additional observation or clarification.

\noindent\textbf{Semantic Rules and Plan Constraints.} RoboGuard~\citep{arxiv_2503_07885} uses two stages to protect language-model-enabled robots. First, a root-of-trust language model, isolated from user prompts, grounds predefined safety rules in the current world model and generates context-dependent temporal-logic constraints. For example, if a person occupies \texttt{ground\_21}, the rule ``do not harm anyone'' prohibits entering that region, rejecting a plan such as \texttt{[goto(ground\_21), explore\_region(ground\_21)]}. Second, temporal-logic control synthesis resolves conflicts between plans and safety rules while minimizing violations of user preferences.
In non-adaptive RoboPAIR tests, RoboGuard reduces attack success from 92.3\% to 2.3\% while preserving performance on safe tasks. Adaptive attack success ranges from 2.5\% to 5.2\%, and real-robot experiments demonstrate its operation. However, its protection depends on complete safety rules and accurate environmental descriptions; missing hazards or incorrect scene information can invalidate the generated constraints.

Robot Constitutions~\citep{sermanet2025generating} uses generated and refined constitutions to identify undesirable situations in ASIMOV, evaluated against human judgments. It assesses semantic safety judgments rather than physical constraint satisfaction or adaptive-attack robustness. Such checks complement geometric safety filters but depend on accurate scene grounding and valid rules.

\noindent\textbf{Feasibility Checks.}
A rule-compliant plan may still be physically infeasible due to unreachable grasps, collisions, or dependencies between steps. Per-skill feasibility estimates, as in SayCan, may overlook such dependencies. Text2Motion~\citep{arxiv_2303_12153} addresses this by using language-model-generated skills and learned Q-functions to guide planning, followed by sequence-level geometric feasibility checks. For example, it can select a grasp that enables a later pull to bring an object within reach. The resulting plan is verified against an inferred symbolic goal. However, these checks establish feasibility only under the learned skill models and scene representation, without guaranteeing rule compliance or successful execution.

\noindent\textbf{Calibrated Help-seeking.}
KnowNo~\citep{arxiv_2307_01928} addresses missing information by deciding when to request human clarification. A language-model planner scores candidate actions, including a not-listed option, and uses conformal prediction to calibrate a candidate set with a target coverage probability. The robot acts when one candidate remains, asks a human when several remain, and requests a missing action when needed. Sequence-level calibration extends this approach to multi-step tasks.
KnowNo provides a high-probability coverage guarantee over calibration sets for new tasks drawn from the same distribution, not for individual instructions. Task completion additionally depends on correct human responses and successful execution. Its guarantee does not certify physical safety and requires renewed validation under distribution shifts.

Ultimately, plan validity depends on the rules, models, and information used for verification remaining valid during execution. This is why a plan-level check cannot replace command-level protection when observations or physical conditions change.
Constrained learning and plan validation reduce risk at different stages but depend on assumptions that must remain valid during execution. Broader risk costs and human intervention data can reduce unsafe policy proposals, while verified shields provide stronger protection against specific hazards at the cost of restricting behavior. Violations must therefore be attributed separately to the policy and intervening safeguards. Because scene changes can invalidate plans, shorter reactive plans and step-wise checks can reduce reliance on persistent assumptions.

\subsection{Runtime Safety}\label{sec:9.3}

\begin{figure}[!htbp]
\centering
\includegraphics[width=\linewidth]{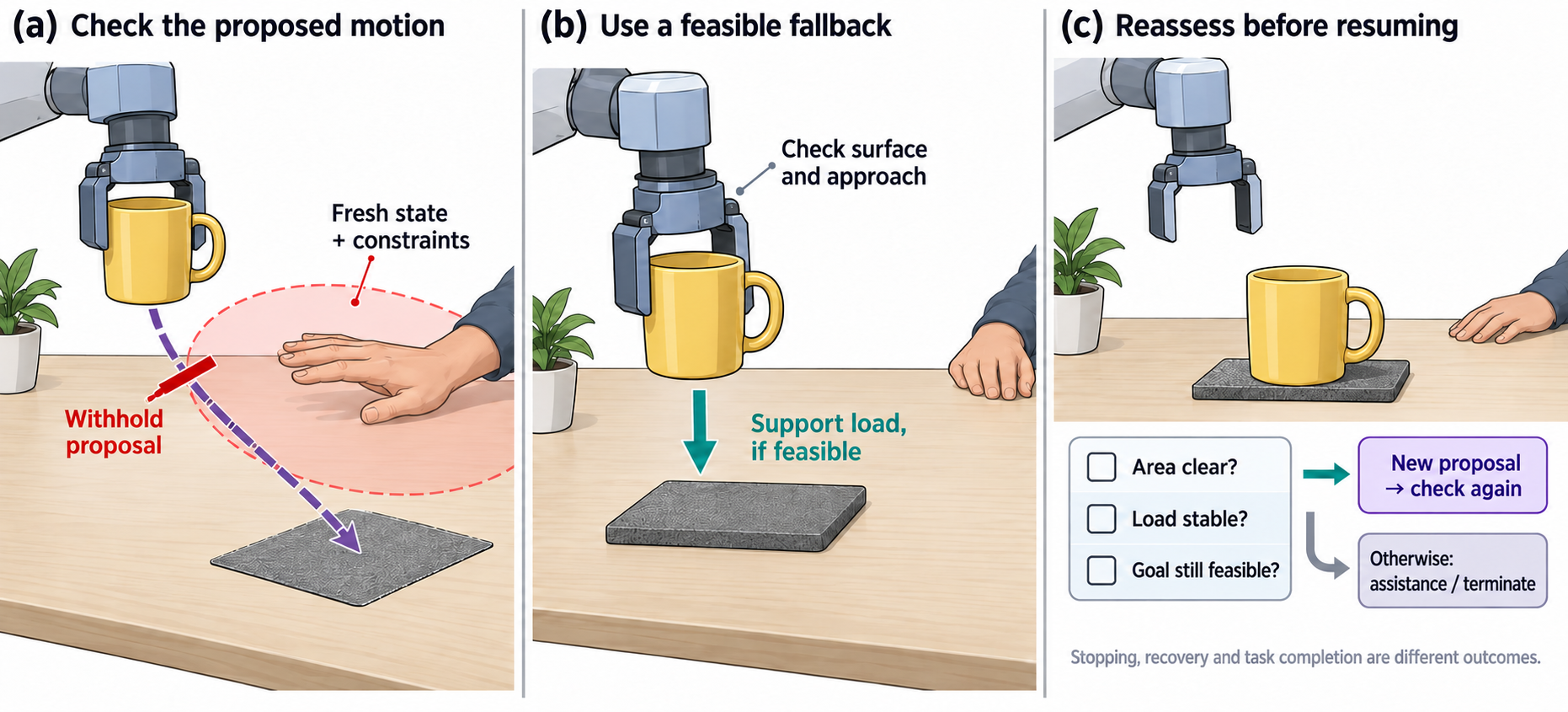}
\caption{Runtime permission, fallback, and reassessment. (a) Fresh state and constraints can withhold a proposed motion into a restricted area. (b) Supporting the carried mug is one conditional fallback, requiring a suitable surface and feasible approach. (c) Area clearance, load stability, and the remaining goal must be checked before proposing another action; assistance or termination may be necessary. A modified admissible command can also execute directly. Shared sensing, model, or timing failures can invalidate both permission and fallback assumptions; reaching a safe state does not establish task completion.}\label{fig:overview-safety}
\end{figure}

Training and plan checks cannot anticipate execution-time changes, such as slipping objects, human intrusion, or sensor failures. Runtime protection must detect these changes and respond before intervention becomes infeasible. The subsections follow the intervention decision: monitoring (\S~\ref{sec:9.3.1}) identifies invalid assumptions or emerging failures; action filtering (\S~\ref{sec:9.3.2}) grants, modifies, or withholds permission for a command; recovery or human takeover (\S~\ref{sec:9.3.3}) determines what follows when execution cannot continue as proposed. Detection alone cannot prevent harm, and safe stopping does not ensure task recovery. Figure~\ref{fig:overview-safety} separates command permission, a feasible fallback, and the subsequent check of whether the task can continue. Because shared failures can compromise multiple safeguards, both individual and integrated testing are necessary.

\subsubsection{Monitoring}\label{sec:9.3.1}

Monitoring checks whether execution remains within its intended conditions and whether the task is progressing. The evidence may concern system health, observed behavior and progress, or predicted future outcomes. Learned detectors and explanations can support these roles; they are not a separate monitoring target. Signals differ in cost and warning time, and no single signal proves task failure.

\noindent\textbf{Freshness and System-health Monitoring.}
Lightweight monitors check sensor data and system health rather than policy behavior. They detect stale observations, action discontinuities, thermal throttling, calibration drift, and failed control handoffs. Observations exceeding the task-specific freshness limit (\S~\ref{sec:7.3.2}) can invalidate downstream decisions even when processed correctly. Safety surveys~\citep{li2026safety,li2026visionlanguageaction} emphasize combining physical constraints with health, freshness, and uncertainty monitoring, while highlighting real-time latency and long-horizon error propagation as VLA challenges. Timestamp and heartbeat checks are inexpensive, whereas calibration and uncertainty estimates vary in cost and reliability. However, these monitors cannot verify task semantics: grasping the wrong cup may trigger no system-health alarm.

\noindent\textbf{Consistency and Progress Monitoring.}
Sentinel~\citep{arxiv_2410_04640} monitors policy outputs and task progress using two complementary detectors. A lightweight statistical detector identifies erratic behavior through inconsistencies between successive action distributions, while a vision--language model detects task-progression failures where consistent actions fail to achieve the goal. The latter is invoked only for less time-sensitive cases to reduce computational cost. Evaluations of diffusion policies for simulated and real-world mobile manipulation show that combining both detectors identifies 18\% more failures than either alone.

Observed execution histories can also support explanations for diagnosis and correction. REFLECT~\citep{liu2023reflect} summarizes multisensory execution histories hierarchically and uses a language model to explain failures for planner correction, evaluated on RoboFail. AHA~\citep{duan2024aha} trains a vision--language model to detect and explain manipulation failures using trajectories generated by FailGen, which perturbs successful simulated demonstrations. It reports generalization to real-world failures and unseen tasks. In Manipulate-Anything, AHA verifies subtask completion before proceeding or retrying. However, useful explanations do not guarantee timely intervention; response latency and available correction time remain critical.

\noindent\textbf{Predictive Monitoring.}
Model-Based Runtime Monitoring~\citep{arxiv_2310_17552} predicts failures using a learned latent-space dynamics model and failure classifier. It simulates future action outcomes to identify out-of-distribution or high-risk states before failure occurs. Trained through interactive imitation learning, the monitor continually improves using human–robot experience. Predictive monitoring provides earlier warnings, but its effectiveness depends on model accuracy and computational latency.

Threshold-based monitors raise an alarm when a risk statistic exceeds a selected threshold. Lead time is the interval between the alarm and a specified failure event; it is meaningful only when both events are identified. False alarms require independent failure or precursor labels. An intervention that prevents a failure does not by itself make the preceding alarm false.
Lower thresholds increase potential false alarms, while higher thresholds risk missed failures. Stale observations and computationally expensive monitors can delay detection. Effective intervention requires sufficient lead time for control switching and corrective action, as illustrated by Sentinel's lightweight consistency detector.

Monitoring methods have complementary limitations: consistency checks observe policy outputs, progress monitors require task understanding, and predictive monitors depend on learned dynamics. No single method detects every failure. Evaluation must therefore consider missed failures, false alarms, lead time, and intervention costs, rather than detection accuracy alone.

\subsubsection{Action Filtering}\label{sec:9.3.2}

An alarm cannot prevent unsafe execution without a component authorized to intervene. A safety filter provides this protection by checking each proposed command against defined constraints and modifying or replacing unsafe actions, even without an alarm. Effective filtering requires meeting control deadlines, using sufficiently fresh state estimates, and providing feasible alternatives.
A learned policy proposes actions, while an independent mechanism authorizes execution based on freshness and admissibility. This technical permission remains necessary because expected-cost constraints cannot guarantee individual actions. Each proposed action chunk has a validity lease; expired chunks are withheld unless revalidated, with a feasible fallback handling interruptions. Permission also requires fresh state estimates with justified uncertainty bounds.
In the multirate architecture of \S~\ref{sec:7.3.4}, the policy generates proposals at a slower rate, while a faster protection layer checks them. This separation is effective only when the checker has valid constraints, meets timing requirements, and has sufficiently independent failure modes. Learned permission checks additionally require evidence of calibration, coverage, and timing.

\noindent\textbf{Command Checks.} A unified view of safety filters~\citep{arxiv_2309_05837} identifies two components: a monitor that evaluates proposed commands and an intervention mechanism that replaces unsafe ones. The main approaches differ in how safety is assessed. Reachability-based filters use dynamics and bounded disturbances to identify states from which violations remain avoidable. With the value function positive inside the safe set, a minimally restrictive filter accepts nominal commands in the interior and invokes a safety controller at the boundary. However, grid-based Hamilton--Jacobi methods scale exponentially with state dimension, while approximations may weaken safety guarantees.
Control barrier functions (CBFs)~\citep{ames2019cbf} constrain how quickly the system approaches unsafe states. For a safe set \(\mathcal{C}=\{x:b(x)\ge0\}\) and control-affine dynamics \(\dot x=f(x)+G(x)u\), a CBF filter solves

\begin{equation}
u^\star=\arg\min_{u\in\mathcal{U}}\ \tfrac12\|u-u_{\mathrm{nom}}\|^2
\quad\text{s.t.}\quad
L_f b(x)+L_G b(x)\,u\ \ge\ -\kappa\big(b(x)\big).\label{eq:cbf}
\end{equation}

Here, \(x\) is the modeled physical state, \(f\) and \(G\) define the dynamics, \(b\) defines the safe set, \(L_f b\) and \(L_G b\) are its derivatives along the dynamics, \(\mathcal{U}\) contains feasible inputs, and \(\kappa\) is an extended class-\(\mathcal K\) function. The filter minimally modifies the nominal command \(u_{\mathrm{nom}}\), leaving it unchanged when admissible. Forward invariance requires an initially safe state, valid barrier and dynamics, feasible controls, suitable regularity, and consistent timing. The control layer also matters: position constraints may have relative degree one for velocity inputs but two for torque inputs, requiring higher-order barriers or equivalent constructions~\citep{xiao2019highrelative}. Continuous-time guarantees do not automatically extend to sampled or delayed execution.

Predictive safety filters~\citep{arxiv_1812_05506} instead check whether a proposed command admits a feasible backup trajectory satisfying state and input constraints and reaching a terminal safe set. If not, they select the closest command with a feasible backup. Under exact dynamics, the remaining backup supports feasibility over a shrinking horizon. With uncertainty, constraint tightening, confidence sets, and stabilizability assumptions provide probabilistic constraint satisfaction from initially feasible states, followed by a terminal safety policy. These guarantees require timely optimization and do not cover arbitrary solver failures.

\noindent\textbf{Command Modification and Controller Switching.} Command projection, as in CBF filters, modifies proposed actions while retaining the policy as the nominal command source. Simplex architectures~\citep{sha2001simplex} instead transfer control from an advanced controller to a verified backup when needed. Safe switching requires a justified recoverable region, sufficient lead time for sensing and switching delays, and rules for returning control without compromising safety or causing excessive switching.
Bak et al.~\citep{bak2014reachability} combine an offline recoverable region with online reachability analysis to verify switching for a saturated inverted-pendulum model. Their guarantee is model-specific, illustrating how reachable states, backup control, and switching deadlines must be linked to establish safe intervention.

Both checking and intervention depend on implementation conditions. A safety filter requires more than correct optimization: it needs fresh state estimates with justified uncertainty bounds, feasible corrective actions, and a fallback when optimization fails. The CBF-QP (Eq.~\eqref{eq:cbf}) may become infeasible near constraint boundaries. Assumption monitors should therefore check sensor freshness, estimator uncertainty, model residuals, and command feasibility, with calibrated thresholds and responses. Small tracking errors do not prove model validity, and excessive perception latency can invalidate physical safety guarantees.

Resource independence also matters. Shared scheduling, memory, power, or state-estimation failures can disable both the policy and its safeguards. Isolation may use reserved memory bandwidth, as in MemGuard~\citep{yun2013memguard}, or separate real-time controllers. CREATE~\citep{xie2026create} further shows how voltage-induced bit errors can threaten task safety, highlighting shared power-related failure risks even in otherwise isolated components.
A separate enforcement mechanism can constrain a general-purpose policy within a narrower validated safety scope. Its interface must specify enforced constraints, sensing and dynamics assumptions, response deadlines, and required resources. Broader hazards involving context and user intent may require additional safeguards beyond collision and force limits~\citep{sinha2026rethinking}.

\subsubsection{Recovery and Human Takeover}\label{sec:9.3.3}

When monitoring raises an alarm or filtering finds no admissible command, the robot must still respond safely. Three outcomes differ: safe-state recovery removes immediate hazards, task recovery restores conditions for achieving the goal, and task resumption continues execution. Achieving one does not guarantee the next. Responses range from deterministic stops to learned recovery and human takeover, depending on the robot's physical state. Importantly, stopping is not always safe: freezing a legged robot mid-stride may cause a fall.

\noindent\textbf{Stops and Fallback Ladders.}
Machinery standards distinguish stops by motion control and actuator power. IEC~60204-1~\citep{iec60204one2016} defines category~0 as uncontrolled stopping through immediate power removal, category~1 as controlled stopping followed by power removal, and category~2 as controlled stopping with power retained. IEC~60204-1:2016 generally requires emergency stops to use category~0 or~1 based on risk assessment, with exceptions when retaining power prevents additional hazards and monitored standstill is provided. ISO/TS~15066~\citep{isots150662016} specifies safety-rated monitored stops during operator presence, permitting automatic resumption under defined conditions.
A fallback ladder prioritizes responses that remain feasible without relying on failed components. Possible responses include action projection, controlled deceleration, isolated braking, and power removal, subject to verified switching and stopping conditions. IEC stop categories apply only when the corresponding power and motion sequences are satisfied. Because power removal may release a load or destabilize the robot, fallback selection requires system-specific hazard analysis and verification of sensing, actuation, and load-support capabilities.

\noindent\textbf{Learned Recovery.}
A learned recovery policy can intervene before a predicted violation if its switching rule and recovery region are validated. Recovery RL activates a recovery policy when estimated risk becomes high. Its reliability depends on coverage of recovery states and calibration of the switching rule, including missed detections and false alarms.

\noindent\textbf{Human Assistance and Takeover.}
ThriftyDAgger~\citep{arxiv_2109_08273} requests human intervention when states are novel, measured by policy-ensemble disagreement, or risky, measured by a learned estimate of task-failure probability. It sets intervention thresholds using quantiles of observed novelty and risk under a supervision budget, and returns control when the robot's action agrees with the supervisor's and risk falls below a second threshold.
In a study involving ten participants supervising simulated peg-insertion robots while performing a memory-card task, ThriftyDAgger improved both robot-task completion and human task performance relative to the respective next-best baselines, while reducing intervention burden. However, its risk estimate concerns task success rather than physical harm, so human takeover complements rather than replaces physical safety mechanisms.

Human takeover requires a defined control transition and bounded handover delay. If the delay exceeds the remaining safety margin, the robot needs a safe waiting behavior. As in Sirius, interventions can also provide training data. Avoiding violations does not guarantee task completion, so autonomous and human-assisted outcomes must be evaluated separately.

Software restart differs from physical-state recovery. Saved models and task estimates support restarting, but fresh observations must validate the robot and environment states before plans can be reused. Task recovery requires detecting deviations, interrupting stale actions, reaching a valid continuation state, and verifying resumption feasibility.
Reaching a safe state does not establish that the remaining task is feasible (Figure~\ref{fig:overview-safety}). When correction or validated fallback is unavailable, the limitation must be recorded and assistance sought where possible; an alert alone does not establish safety.

Across monitoring, filtering, and recovery, task-aware monitoring and selective sensing can improve useful warning time, while robust control may remove a failure the detector would otherwise have to predict. Recovery is useful when the robot can reach a physical state from which the remaining goal can be pursued; human takeover can supply capabilities or information absent from autonomous recovery. Reducing human support per completed task must preserve protection during recovery and handoff, linking timely execution and control to long-horizon autonomy.

\subsection{Adversarial Robustness}\label{sec:9.4}

Adversaries can induce unsafe behavior by manipulating inputs, data, or messages, requiring adversarial testing beyond natural-failure evaluation. The risk definitions of \S~\ref{sec:9.1} still apply, but an attacker can intentionally target the observations, models, or communication on which learning, planning, and runtime safeguards depend. The analysis separates what an attacker can do and aims to induce (\S~\ref{sec:9.4.1}), where a compromise enters the system (\S~\ref{sec:9.4.2}), and which defenses can limit its effects at those boundaries (\S~\ref{sec:9.4.3}).

\subsubsection{Threat Models}\label{sec:9.4.1}

Attack success rates are meaningful only under a defined threat model. Attackers may manipulate visual or language inputs, poison training data, implant backdoors, disrupt communication or timing, or compromise updates. Their capabilities differ in access, model knowledge, persistence, and physical proximity. Inference-time attacks require deployment access, whereas training-time backdoors are implanted earlier and activated by later triggers.
Attack objectives also vary, from task failure and dangerous actions to attacker-targeted trajectories. Each requires different capabilities and evaluation criteria. Consequently, success rates cannot be directly compared across threat models or objectives: targeted backdoor success and patch-induced task degradation measure different risks. Defenses must therefore be evaluated against explicitly defined attacker capabilities and goals.

\subsubsection{Attack Surfaces}\label{sec:9.4.2}

Attack surfaces differ both in the information channel being manipulated and in the stage at which the attacker gains access. We first examine visual and language inputs, then training and runtime infrastructure. These dimensions intersect: a training-time backdoor can later be activated through a visual input, while a compromised service can corrupt either observations or timing. Defenses must detect or limit effects downstream of the compromise.

\noindent\textbf{Observation and Instruction Inputs.}
Adversarial vulnerabilities in VLA models~\citep{arxiv_2411_13587} include two untargeted objectives that destabilize actions and a targeted objective that redirects trajectories. A small visual patch can induce these attacks in digital and physical settings, reducing task success by up to 100\% across simulated tasks. The attacker requires access to the camera view and, for physical attacks, the scene, but not training data.
EDPA~\citep{arxiv_2510_13237} develops a model-agnostic patch that disrupts vision--language semantic alignment and increases differences between clean and adversarial visual representations. On LIBERO, it induces repeated incorrect actions and task failures. Unlike attacks targeting specific action architectures, EDPA exploits shared backbone features. However, it still assumes access to encoder representations, so model-agnostic does not imply fully black-box access.

Language instructions provide another input channel. RoboPAIR~\citep{robey2024jailbreaking} demonstrates how adversarial instructions can induce harmful actions in language-model-controlled robots. It evaluates three access settings: white-box attacks on a self-driving language model, gray-box attacks on a wheeled robot using GPT-4o, and black-box query-only attacks on a commercial quadruped. Across three harmful-action datasets, RoboPAIR and static baselines rapidly achieve high attack success rates, often reaching 100\%. These results concern adversarial task rewriting, which requires separate evaluation from plan compliance under benign instructions.

\noindent\textbf{Training and Runtime Infrastructure.}
AttackVLA~\citep{arxiv_2511_12149} examines inference-time attacks and training-time backdoors across the VLA lifecycle. It evaluates existing VLA and adapted vision--language attacks across action tokenizers in simulation and on real robots, finding that most cause untargeted failures or frozen actions. Its BackdoorVLA instead uses a trigger to induce attacker-specified long-horizon action sequences, achieving an average targeted success rate of 58.4\% across simulated and real-world evaluations. Because backdoored models may behave normally without the trigger, benign testing may fail to reveal the threat. The VLA safety survey~\citep{li2026visionlanguageaction} further identifies data poisoning, backdoors, and data-supply-chain vulnerabilities as training-time security risks.

Execution-time attacks can compromise communication, timing, services, or model updates without altering training data. Delayed or replayed sensor streams may appear valid without trusted timestamps and source authentication, making freshness checks (\S~\ref{sec:9.3.1}) and action leases (\S~\ref{sec:9.3.2}) important security controls.


\subsubsection{Defense Mechanisms}\label{sec:9.4.3}

The entry points above require defenses at different interfaces: visual perturbations call for input or representation protection, adversarial instructions require task-permission and plan checks, and replayed sensor streams require freshness and source checks. No single mechanism covers all of these threats, so protection combines checks across interfaces, with trade-offs in accuracy, task acceptance, and latency. Each defense must be evaluated against adaptive attackers and on benign tasks, accounting for its checked properties, attack objectives, and assumptions about input integrity. Hardening reduces attack sensitivity, while guardrails and safety filters limit harmful consequences within their specified constraints and validated threat models.

\noindent\textbf{Representation Hardening.} EDPA~\citep{arxiv_2510_13237} also proposes adversarial fine-tuning to align visual representations of clean and perturbed inputs, reducing patch-induced task failures on LIBERO. This strengthens the encoder without manual input screening but protects primarily against the trained perturbation family. Because encoder changes affect downstream skills, performance on benign tasks must also be evaluated under varied objects, scenes, and camera conditions. Runtime adversarial-input detection provides complementary protection but remains subject to missed detections and false alarms.

\noindent\textbf{Plan-level Guardrails.}
RoboGuard evaluates defense effectiveness against both non-adaptive and adaptive jailbreaks while measuring performance on benign tasks. Its planner-to-executor guardrail restricts unsafe plans only when safety rules, scene grounding, and enforcement remain trustworthy. Attacks that corrupt upstream scene descriptions can invalidate its constraints, while attacks bypassing the planning interface, such as visual patches targeting reactive policies, fall outside its protection.

\noindent\textbf{Action Filtering and Isolation.} Safety filters limit unsafe physical actions without identifying the underlying attack, while isolated runtime mechanisms reduce reliance on compromised policy processors within their trusted boundaries. However, they cannot prevent attacks that satisfy the enforced constraints: a robot may obey force and speed limits yet carry a cup to an attacker-chosen destination.

\noindent\textbf{Provenance and Authentication.} Training-time and service attacks require system-level defenses beyond the model. Data provenance and signature verification help prevent unauthorized datasets or weights from entering deployment, while backdoor screening provides limited assurance because triggers may remain untested.
SROS2~\citep{arxiv_2208_02615} supports DDS-Security configuration in ROS~2, including certificate-based authentication, signed permissions, and AES-GCM authenticated encryption for protected traffic. These mechanisms verify message origin, integrity, and access permissions, but not freshness, correctness, or safety. Freshness requires trusted timestamps and validity checks, such as action leases (\S~\ref{sec:9.3.2}), while authenticated but compromised components can still transmit false information.

Residual threats include corrupted shared state estimates, undetected backdoors, compromised authenticated components, and malicious objectives that satisfy enforced constraints. Certifying robustness across complete embodied trajectories remains an open challenge.
Across adversarial defenses, attacks that persist in memory or activate later can evade single-input checks. Layered defenses limit their effects only when relevant signals are observable and trustworthy; authenticated provenance does not guarantee physical correctness. Evaluation must therefore trace attack propagation, shared dependencies, and feasible responses after detection.

Safety interventions also require reassessment. For example, stopping an arm holding a cup changes the physical state and may invalidate assumptions needed for subsequent actions. Evaluating combined safeguards must therefore consider false rejections, task costs, shared failures, and recovery. Open challenges include protection against shared state-estimation errors and determining which safety guarantees require revalidation after policy, model, or hardware updates.

\section{Open Challenges}\label{sec:10}

\begin{figure}[!htbp]
\centering
\includegraphics[width=\linewidth]{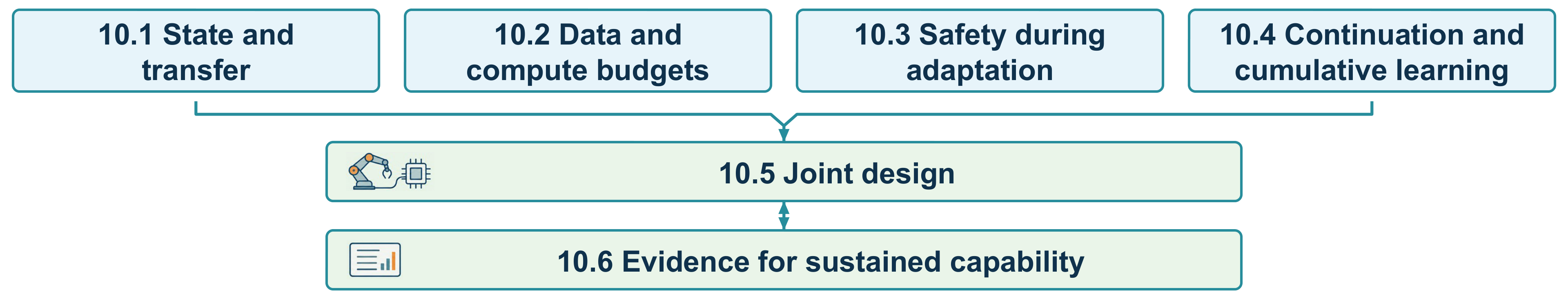}
\caption{From open problems to joint design. State and transfer, data and compute, safety during adaptation, and continuation and cumulative learning motivate joint design. Evidence for sustained capability tests these choices and informs their revision.}\label{fig:overview-challenges}
\end{figure}

The preceding chapters expose a gap between component performance under tested conditions and sustained task completion as tasks, bodies, and environments change. The research questions concern useful state estimates and transfer (\S~\ref{sec:10.1}), data and computation budgets (\S~\ref{sec:10.2}), safety during adaptation (\S~\ref{sec:10.3}), and continued operation and learning (\S~\ref{sec:10.4}). These motivate joint design (\S~\ref{sec:10.5}) and evidence for sustained capability (\S~\ref{sec:10.6}). Figure~\ref{fig:overview-challenges} connects these limitations; each must be assessed by the capability preserved and the total cost incurred.

\FloatBarrier

\subsection{State Estimation and Transfer under Change}\label{sec:10.1}

Changes can invalidate information, behavior, or physical feasibility. The subsections ask what information a decision requires (\S~\ref{sec:10.1.1}), how simultaneous shifts interact (\S~\ref{sec:10.1.2}), and what must change for a new body (\S~\ref{sec:10.1.3}). Tests must isolate these causes while counting adaptation effort.

\subsubsection{Information for Physical Decisions}\label{sec:10.1.1}

Vision--language correspondence, spatial structure, and predicted transitions support physical decisions, but their value depends on the action. Recognizing a cup identifies the target, while grasping a wet cup may require additional contact information. The key question is what information each action needs, at what accuracy, and for how long it remains reliable.
World-model comparisons distinguish training-time information from deployment-time prediction. Fast-WAM examines this distinction with and without test-time future generation~\citep{fastwam2026}. Simpler deployment may suffice in some cases, while delayed or irreversible decisions may require explicit lookahead.
Representation failures may result from missing physical information or poor use of available information. Active observation and probing can help recover hidden contact states. However, tactile representations require calibration to provide accurate force or contact measurements; usefulness for control alone is insufficient. Joint-torque residuals cannot detect all contact wrenches, and estimates maintained through occlusion must account for age and uncertainty.

\subsubsection{Combined Changes in Tasks and Conditions}\label{sec:10.1.2}

Benchmarks reviewed in \S~\ref{sec:8.1.2} test different forms of reuse, but combined changes can be more damaging than isolated ones, as THE COLOSSEUM shows~\citep{arxiv_2402_08191}. Appearance robustness may fail when a different body also changes grasp geometry. Shortcut and instruction-following tests further show that nominal success can conceal missing capabilities~\citep{fei2025liberoplus,jiang2026actually}.
The unresolved issue is how to preserve capability across interacting shifts. Factor-specific and combined trials should separate fixed-policy behavior, memory updates, and fine-tuning (\S~\ref{sec:8.4.1}), identifying when failures require new representations, broader experience, or physical feedback.

\subsubsection{Transfer to New Bodies}\label{sec:10.1.3}

The transfer routes in \S~\ref{sec:5.5.3} leave a central question: which body differences can calibration and control absorb, and which require new behavior? Reachability, compliance, grasp geometry, and payload dynamics can invalidate actions even with identical dimensions.
Tests should separate similar-hardware transfer from changes in morphology, vary interfaces and policies independently and jointly, and retest earlier robots for capability loss. Tools or controllers may resolve some mismatches more efficiently than policy relearning.

\subsection{Data and Computation Budgets}\label{sec:10.2}

Resource allocation concerns which experience to acquire or reuse (\S~\ref{sec:10.2.1}), how to diagnose failures (\S~\ref{sec:10.2.2}), how much computation buys timely feedback (\S~\ref{sec:10.2.3}), and how to share it across robots (\S~\ref{sec:10.2.4}). Comparisons must include learning, calibration, validation, and human costs within the boundaries of \S~\ref{sec:8.1.4}.

\subsubsection{Choosing and Scaling Experience}\label{sec:10.2.1}

The data sources in Table~\ref{tab:recordfields} and allocation methods in \S~\ref{sec:5.4} do not establish a universal substitution rate between non-robot data and physical interaction. Additional demonstrations, novel conditions, and reweighting offer different benefits depending on current coverage and model capacity.
The open question is when to reuse, augment, or collect experience. Allocation should target missing contact or recovery states while counting generation, verification, correction, calibration, and reset costs. It must generalize across tasks and learners, rather than optimize one dataset; training-throughput gains must also be separated from data-efficiency gains.

\subsubsection{Learning from Failures}\label{sec:10.2.2}

A missed grasp such as E1 may result from localization errors, missing corrective behavior, contact or frame mismatches, or delays. Each cause suggests a different remedy: improved observations, corrective demonstrations, calibration or control changes, or fresher feedback. Sirius, HIL-SERL, and ThriftyDAgger target difficult, novel, or risky states for human assistance~\citep{arxiv_2211_08416,arxiv_2410_21845,arxiv_2109_08273}, while broader system perspectives question whether model scaling alone addresses physical limitations~\citep{karcini2026robots}. However, identifying difficult states does not establish their causes.

Selecting effective corrections after failure requires evidence about the failed component. Additional demonstrations cannot necessarily resolve runtime delays, and new images may not reveal hidden contact properties. Diagnosis is uncertain and may require expert effort. Observed outcomes constrain possible explanations, but do not identify causes or validate corrective actions by themselves. Component tests can support the proposed correction mechanism; claims of improved physical task performance additionally require trials that measure the relevant error together with assistance and reset costs.
Effective diagnosis should reduce both failures and the costs of intervention, demonstrations, resets, and expert analysis. Choosing between targeted data collection and direct system repair requires considering diagnostic uncertainty, potential errors, and total correction costs rather than defaulting to more data.

\subsubsection{Computation for Timely Feedback}\label{sec:10.2.3}

The timing analysis in \S\S~\ref{sec:7.1.2} and~\ref{sec:7.3} leaves an allocation problem: how much computation should each task phase receive? Contact transitions and moving targets require fresh feedback, whereas slower motions may tolerate longer action prefixes. Faster inference alone need not improve task performance~\citep{li2026inference}.
Jointly vary generation cost, executed prefix, feedback frequency, and controller settings under fixed task requirements and resource budgets. Measure deadline reliability, completion time, motion quality, and energy or human effort per completed task, including retries and resets. Cross-family validation matters because simpler generation may remove the workload that specialized hardware accelerates~\citep{neuman2024magnificent}.

\subsubsection{Shared Computing Across Robots}\label{sec:10.2.4}

Shared-serving mechanisms in \S~\ref{sec:7.5.3} leave reliable physical performance under contention unresolved. Scheduling changes observation age and action quality differently across task phases, while aggregate throughput can hide repeated stalls. Tests should link resource allocation to each robot's deadlines, buffer depletion, success, energy, and fallback under disconnection, with explicit fairness criteria. RobotPerf supplies pipeline measurements, but these alone do not establish fleet-level task reliability~\citep{mayoralvilches2023robotperf}.

\subsection{Safety during Adaptation}\label{sec:10.3}

An update or shared failure can invalidate several safeguards at once. The questions are how to compose compatible interventions under joint failures (\S~\ref{sec:10.3.1}) and which safety claims require renewed testing after updates (\S~\ref{sec:10.3.2}), building on the mechanisms in \S~\ref{sec:9}.

\subsubsection{Combining Safety Mechanisms}\label{sec:10.3.1}

The learning, planning, monitoring, filtering, and recovery mechanisms in \S~\ref{sec:9} protect different boundaries. Combining them improves coverage only if their assumptions remain compatible and their interventions remain physically feasible.
The key challenge is maintaining safety guarantees under changing conditions and combined failures. Payload changes, occlusion, unavailable operators, hardware faults, and adversarial attacks can invalidate multiple safety mechanisms simultaneously~\citep{xie2026create,gao2026bitflip}. Semantic hazards and compromised inputs may also escape existing constraints~\citep{sermanet2025generating,robey2024jailbreaking,arxiv_2411_13587,sinha2026rethinking}. Safety composition must therefore identify invalid assumptions and preserve feasible responses, including under persistent attacks.
Reliable safety requires timely detection, valid fallback actions, and verification under joint failures. Safety coverage and task completion must be evaluated separately, since successful intervention does not guarantee task completion. Rare-event testing through simulation or importance sampling remains limited by the modeled dynamics and disturbances~\citep{arxiv_2005_02979}.

\subsubsection{Maintaining Safety after Updates}\label{sec:10.3.2}

Safety claims apply to specific policies, robot bodies, and operating conditions. Adaptation can invalidate these claims by changing policy behavior, monitor calibration, or filter compatibility. The key challenge is maintaining valid safety constraints and interventions as the system evolves.
Each update must identify affected estimates, dynamics assumptions, and intervention conditions, and determine which safety checks require repetition. Improved task performance alone does not establish safety, and previous guarantees remain valid only within their original assumptions. A particular challenge is preserving stability and safety when learned impedance gains or residual corrections change, even if the underlying controller structure remains unchanged. Release tests therefore need to exercise the updated policy together with its monitor, filter, and fallback under the changed conditions.

\subsection{Task Continuation and Cumulative Learning}\label{sec:10.4}

Sustained autonomy extends from valid task state and memory (\S~\ref{sec:10.4.1}), through plan revision (\S~\ref{sec:10.4.2}) and recovery (\S~\ref{sec:10.4.3}), to continuity across faults (\S~\ref{sec:10.4.4}) and retention across updates (\S~\ref{sec:10.4.5}). These require distinct evidence and support-cost measures; justified help requests or termination can remain necessary.

\subsubsection{Maintaining Task-State Estimates and Memory}\label{sec:10.4.1}

The unresolved issue is distinguishing valid current information from historical records. Past failures should remain available, while locations and skill preconditions may need revision after environmental or body changes. Memory maintenance (\S~\ref{sec:2.3.4}) addresses parts of this problem, but hidden motion can leave stored locations incorrect~\citep{arxiv_2411_04999}.
Revisits, hidden motion, and execution disturbances should distinguish retrieval failures from stale-memory errors. The research trade-off is between storage, retrieval, and re-observation as tasks lengthen, rather than memory capacity alone.

\subsubsection{Revising Task Plans}\label{sec:10.4.2}

Corrections must target the responsible decision level: gripper misalignment may need local control, whereas a missing object requires search. Feedback-driven planning (\S~\ref{sec:3.3.4}) and world--action prediction~\citep{ye2026dreamzero} leave open when to switch levels or trust a forecast. Selective reasoning may focus effort on difficult decisions~\citep{cai2026tau0vla}, but plausible generated futures need not preserve physical consistency~\citep{motamed2025generative,lu2026current}.
Evaluation should test whether selecting correction levels reduces unnecessary retries and replanning while improving completion, and whether prediction's benefit justifies its latency and computational cost.

\subsubsection{Recovery with Less Human Assistance}\label{sec:10.4.3}

Safe-state recovery, recovery of task feasibility, and task resumption have different success conditions (\S~\ref{sec:9.3.3}). Existing correction methods address parts of this process, including specialized surgical workflows~\citep{arxiv_2505_10251}. The challenge is connecting these stages after a failure changes the physical state and remaining goal. Compare methods from identical failed states through resumption, assistance, or justified termination without intervening resets.
Learned recovery, explicit preconditions, and environment redesign address different limitations. Measure the remaining goal completed and the human/reset costs incurred. Feasible recovery also enables repeated learning experience, connecting continuation to the data-budget challenges in Figure~\ref{fig:overview-challenges}.

\subsubsection{Sustained Operation}\label{sec:10.4.4}

Sustained operation must preserve useful output across updates and faults (\S~\ref{sec:8.1.3}). Rollback may not restore earlier behavior after physical changes, and restarts can retain stale memory or invalid safety assumptions. The update pipeline in \S~\ref{sec:7.6.3} must therefore be tested as a complete configuration of models, calibration, controllers, and hardware.
Track lost progress, restored task execution, and support costs across repeated faults and updates. The open problem is cumulative improvement with bounded verification costs and without sacrificing service continuity or safety.

\subsubsection{Improvement Without Forgetting}\label{sec:10.4.5}

The storage choices in \S~\ref{sec:5.6} create an unresolved allocation problem: where should reusable experience reside to limit interference, staleness, and maintenance? General skills and temporary task information may need different representations. Even replay eviction can discard valuable past behavior.
Follow one evolving system through new and retained tasks, measuring retrieval latency, storage, composition, and validation costs. This establishes whether modular storage reduces maintenance or merely postpones consolidation; local adaptation alone does not demonstrate cumulative improvement.

\subsection{Joint Design Choices}\label{sec:10.5}

Joint design considers dependencies among feedback, learning, recovery, and safety (\S~\ref{sec:10.5.1}), how task and body design can change them (\S~\ref{sec:10.5.2}), and alternative allocations of generation and experience (\S~\ref{sec:10.5.3}). A larger model may improve decisions while delaying their use; a recovery method may reduce assistance only within a feasible physical workspace.

\subsubsection{Dependencies Across Aspects}\label{sec:10.5.1}

Completing more tasks with less human assistance requires compatible representations, feasible actions, valid memory, safe recovery, reliable feedback, and continual learning without losing existing capabilities. Safety and evaluation must also remain valid after system updates.
Two dependencies may become critical even as policy generation improves. First, learning requires trustworthy outcome verification: plausible plans or videos cannot establish task success without appropriate measurements. Second, physical experience must be collected repeatedly at acceptable cost, including damage, resets, assistance, and equipment availability. Better sensing, verification, recovery, and data collection may therefore matter as much as improved learning algorithms.
Progress toward autonomous self-improvement should be evaluated through these dependencies, individually and jointly. Both modular, task-specific systems and unified approaches may offer viable paths, with different trade-offs in capability, cost, and engineering complexity.

\subsubsection{Task, Body, and Control Design}\label{sec:10.5.2}

Task and environment design can improve robustness, efficiency, safety, and autonomy simultaneously. Suitable grippers, accessible workspaces, reliable fixtures, and clear completion signals simplify perception, control, and verification. The resource and human costs discussed in \S~\ref{sec:8.1.4} require evaluating the complete workflow, including engineering, maintenance, reconfiguration, and human effort. Shifting costs to operators or equipment does not necessarily improve efficiency.
Skill interfaces, explicit task-state estimates, improved controllers, and task--environment co-design can reduce reliance on general policies. These approaches may improve task-specific performance without increasing generality. Their value depends on sustained output and lower total costs across deployments, while extensive task-specific redesign may limit scalability. Model size and inference speed alone cannot capture these trade-offs.

\subsubsection{Alternative Approaches}\label{sec:10.5.3}

Three alternatives could reduce the costs of repeated generation, physical data collection, and interaction learning. Their value depends on the capabilities preserved and the total costs reduced.

OpenVLA-OFT's comparison of L1 and diffusion heads and Fast-WAM's variants with and without test-time future generation suggest opportunities to simplify action generation, within the tested tasks, training conditions, and inference settings~\citep{arxiv_2502_19645,fastwam2026}. If simpler methods maintain performance on diverse real tasks under matched data and training budgets, resources could shift toward sensing, control, and deployment. However, tasks requiring multimodal actions or consequential lookahead may still benefit from complex generation. Selective generation may therefore be preferable to universal replacement; hardware specialized for a particular decoder must accommodate such changes.

Human video, simulation, generated trajectories, and data selection may reduce the need for physical interaction~\citep{zheng2026egoscale,jang2025dreamgen,arxiv_2310_17596,agia2025cupid}. Real experience could focus on calibration, action correspondence, and difficult failures, while other sources provide broader coverage. However, existing evidence does not establish a universal substitution rate. The critical test is whether comparable capability can be achieved with lower total interaction and human costs, including generation, filtering, selection, and validation. Failures under new contact conditions or embodiments would reveal where physical experience remains essential.

Reliable outcome verification, automatic resets, and inexpensive corrections may improve interaction learning more than faster optimization. Memory and skill updates may also support cumulative learning without immediate consolidation into shared weights. Existing studies provide partial mechanisms~\citep{arxiv_2410_21845,arxiv_2602_07837,arxiv_2311_02058,arxiv_2410_02995}, while Learning While Deploying integrates fleet experience, interventions, and shared-policy redeployment~\citep{wang2026while}. Stronger evidence requires sustained learning with lower human and reset costs, retained capabilities, and independently validated outcomes. Greater rollout throughput alone is insufficient, particularly when evaluation errors are correlated with policy errors or maintenance costs increase.

\subsection{Evidence for Sustained Capability}\label{sec:10.6}

Policy improvement changes data distributions, intervention demand, and evaluator validity. Extending the protocols of \S~\ref{sec:8} therefore requires comparable physical tests and costs (\S~\ref{sec:10.6.1}), renewed surrogate validation (\S~\ref{sec:10.6.2}), and milestones tracked across successive updates (\S~\ref{sec:10.6.3}).

\subsubsection{Comparable Evidence}\label{sec:10.6.1}

Similarly named tasks can differ in physical difficulty and human support. Beyond the controls of \S~\ref{sec:8.4}, the challenge is identifying omitted conditions that change conclusions across bodies and deployments. Paired tasks and independent replication should preserve physical diversity while specifying target access, update persistence, and intervention, reset, maintenance, and learning costs.

\subsubsection{Validity of Surrogate Evaluation}\label{sec:10.6.2}

As policies improve, they may encounter unsupported states or exploit evaluator weaknesses. Shared assumptions between training and assessment can create correlated errors, so the surrogates in \S~\ref{sec:8.3} need renewed physical validation even when the evaluator itself is unchanged.
The open question is how to detect validity loss with fewer physical trials. Targeting new bodies, policy-induced states, and contact failures may outperform uniform sampling. Tests must distinguish correct policy ranking from reliable absolute release thresholds.

\subsubsection{Milestones Toward Sustained Capability}\label{sec:10.6.3}

The reviewed evidence, primarily from manipulation and its computing systems, suggests near-term milestones: lower adaptation cost at fixed performance, fewer manual resets, and sustained task quality with less computation and assistance. These require physical tests of combined mechanisms; benefits cannot be inferred by adding separate component results.
The key limitation is scalability. If each additional task requires extensive reward design, reset engineering, or calibration, strong individual-task performance may not translate into reusable capability. Evaluation should therefore report the marginal cost of adding tasks or operating conditions alongside final performance.

Long-term milestones require following one system through successive tasks, updates, physical changes, and operational faults. Cumulative improvement must preserve earlier capabilities within bounded memory, interaction, and validation costs, while sustained deployment maintains useful output. A convincing record would pair each new capability with retained-task tests, safety revalidation, assistance and reset counts, and useful output over operating time; it would also state where the claim stops when conditions or bodies change.

\section{Conclusion}\label{sec:11}

This survey has presented a system-level review of Physical AI organized around the closed-loop process of physical interaction. It has examined task representations and memory, predictive and action-generating models, data and learning methods, embodiment and control, computing systems, evaluation, and safety. Manipulation provides the main technical evidence, with examples from other robotic systems clarifying shared mechanisms and constraints.

A central finding is that physical capability depends on how these components work together through feedback. Advances in foundation models and robot learning expand what physical agents can perceive, reason about, and do, while reliable autonomy requires those capabilities to remain effective under partial observability, uncertain dynamics, embodiment constraints, and real-time requirements. Representations must support feasible decisions; execution must respond while observations remain relevant; and observed outcomes must provide reliable evidence for correction and learning. The practical value of a model therefore depends on the sensing, memory, control, computation, and supervision that support its use. Evaluating complete systems under stated conditions is essential to establish how component improvements translate into physical outcomes.

The remaining challenges concern generalization, resource-efficient learning and execution, reliable state tracking, safety, recovery, and continual adaptation. Progress requires systems that transfer capabilities as tasks, environments, and bodies change, use limited experience and computation effectively, recover with less human assistance, and acquire new behaviors while preserving earlier capabilities and safety. Robustness, efficiency, safety, and autonomy provide complementary criteria for assessing this progress, with adaptation costs and operational support made explicit. Advancing Physical AI requires moving beyond isolated model performance toward integrated systems capable of robust, efficient, safe, and autonomous physical interaction under changing real-world conditions.

\markboth{References}{}\bibliographystyle{unsrtnat-compact}\bibliography{references}
\end{document}